\documentclass[final,3p,times,sort&compress]{elsarticle}

\usepackage{graphicx,bm,mdwlist}

\usepackage{amsmath}
\usepackage[titletoc,title]{appendix}
\usepackage{bigstrut,array}

\numberwithin{equation}{section}

\def\be{\begin{equation}}
\def\ee{\end{equation}}
\def\e#1{\label{#1}\end{equation}}
\def\bea{\begin{eqnarray}}
\def\eea{\end{eqnarray}}
\def\ea#1{\label{#1}\end{eqnarray}}
\def\rqn#1{(\ref{#1})}
\def\bes#1{\begin{subequations}\label{#1}}
\def\ese{\end{subequations}}
\def\alt{\lesssim}
\def\agt{\gtrsim}

\journal{Physics Reports}

\begin{document}

\begin{frontmatter}



\title{Nonperturbative theory of weak pre- and post-selected
measurements}


\author{Abraham G. Kofman}
\author{Sahel Ashhab}
\author{Franco Nori}
\address{Advance Science Institute, RIKEN, Wako-shi, Saitama 351-0198, Japan}
\address{Physics Department, The University of Michigan, Ann Arbor, Michigan 48109-1040, USA}

\begin{abstract}
This paper starts with a brief review of the topic of strong and weak pre- and post-selected (PPS) quantum measurements, as well as weak values, and afterwards presents original work. 
In particular, we develop a nonperturbative theory of weak PPS measurements of an arbitrary system with an arbitrary meter, for arbitrary initial states of the system and the meter.
New and simple analytical formulas are obtained for the average and the
distribution of the meter pointer variable.
These formulas hold to all orders in the weak value.
In the case of a mixed preselected state, in addition to the standard weak value, an associated weak value is required to describe weak PPS measurements.
 In the linear regime, the theory provides the generalized
Aharonov-Albert-Vaidman formula.
Moreover, we reveal two {\em new regimes} of weak PPS measurements: the strongly-nonlinear regime and the inverted region (the regime with a very large weak value), where the system-dependent contribution to the pointer deflection {\em decreases} with {\em increasing} the measurement strength.
The optimal conditions for weak PPS measurements are obtained in the strongly-nonlinear regime, where the magnitude of the average pointer deflection is equal or close to the maximum.
 This maximum is {\em independent of the measurement strength}, being typically of the order of the pointer uncertainty.
In the optimal regime, the small parameter of the theory is comparable to the overlap of the pre- and post-selected states.
We show that the amplification coefficient in the weak PPS measurements is generally a product of two qualitatively different factors.
The effects of the free system and meter Hamiltonians are discussed.
 We also estimate the size of the ensemble required for a measurement and identify optimal and efficient meters for weak measurements.
Exact solutions are obtained for a certain class of the measured observables.
These solutions are used for numerical calculations, the results of which agree with the theory.
Moreover, the theory is extended to allow for a completely general post-selection measurement.
We also discuss time-symmetry properties of PPS measurements of any strength and the relation between PPS and standard (not post-selected) measurements.
\end{abstract}


\end{frontmatter}

\tableofcontents



\section{Introduction}
 \label{I'}

The issue of measurement is of fundamental significance in quantum
mechanics (see, e.g., Refs.~\cite{neu55,whe83,bra92,wis10,sch04,gio04,dun06}).
Recent developments in fabricating ever smaller nano-devices as well as in quantum information processing (see, e.g., Refs.~\cite{nie00,you05,bul09,mar09,she10,you11,bul11,maw11,geo,nat}) have made it more important to understand quantum measurement.

Over the past several decades, there has been significant progress in the study of general quantum measurements, which differ from projective (ideal) measurements described in textbooks on quantum mechanics.
In particular, in recent years, there has been increasing interest in quantum measurements with pre- and post-selection as well as in weak or, more generally, non-ideal measurements.

One of the most striking developments in the studies of such measurements was the discovery that measurements that are both weak and pre- and post-selected provide the so called {\em weak value} of the measured observable \cite{aha88}.
Weak values possess unusual properties.
In particular, a weak value can be complex, and its real part can be far outside the range of the eigenvalues of the observable.
The unusual properties of weak values initially gave rise to controversy \cite{leg89,per89}, and the physical meaning and significance of weak values is not understood completely until now.
In spite of this, weak values proved to be very useful in various fields of physics, including fundamentals of quantum mechanics and high-precision metrology.

This paper starts with a brief review in Sec.~\ref{I} and afterwards presents many original contributions. 
Note, however, that this distinction between the review and non-review parts is not absolute, since we have included in Sec.~\ref{I} some original results in order to make the text more self-contained, while in other sections we discuss previous work in order to put our results into appropriate perspective.
Most of this paper is original work, which sometimes is explicitly linked to previous theoretical and experimental work. 
Many special cases are considered in some detail, because the study done here is systematic and quite general, spanning many specific cases---some of which have been studied before, while most are new. 

The present paper is organized as follows.
In Sec.~\ref{I}, we provide a brief review of quantum measurements of arbitrary strength, with and without post-selection, the emphasis being on weak pre- and post-selected (PPS) measurements and weak values.
The subsequent sections are devoted to extensions of the theory introduced in Sec.~\ref{I}, with the emphasis on developing a nonperturbative theory of weak PPS measurements in a very general but simple form.
In particular, in Sec.~\ref{IIA'} we provide a general theory of standard (i.e., not post-selected) measurements of arbitrary strength and discuss in detail weak standard measurements.
In Sec.~\ref{III''} we provide a general theory of PPS measurements of arbitrary strength.
 In Sec.~\ref{IIID'} effects of the free system and meter Hamiltonians are discussed.
In Sec.~\ref{III'} we develop a nonperturbative theory of weak pre- and post-selected measurements for the case of a pure preselected state.
In Sec.~\ref{V} the results of Sec.~\ref{III'} are extended to the
case of a mixed preselected state.
In Sec.~\ref{III} we specialize our general formulas for several types
of meters, including continuous-variable and two-level meters.
 In Sec.~\ref{V'} we discuss the distribution of the pointer
values for various types of meters.
 In Sec.~\ref{VI} we consider in detail weak values and weak PPS measurements for a qubit system.
In Sec.~\ref{VIII} we obtain exact solutions for PPS measurements of
arbitrary strength in the case when the measured quantity has two, possibly degenerate, eigenvalues with equal magnitudes and opposite signs; meters of various types are considered.
In Sec.~\ref{IX} we provide numerical calculations and discussions; in particular, our general simple formulas for weak PPS measurements are shown to approximate the exact solutions very well.
In Sec.~\ref{XI} we show that the recent experiments \cite{dix09,sta10x} are described by two limits of the same formula, obtained in this paper.
In Sec.~\ref{XII} we consider an extension of the theory to the case of a general post-selection measurement described by an arbitrary POVM; we also obtain conditions under which PPS measurements of any strength are equivalent to standard (i.e., not post-selected) measurements and discuss time-symmetry properties of PPS measurements.
 Concluding remarks are given in Sec.~\ref{X}.
The four Appendices supplement the main text and provide some details of the calculations.

Some important symbols used in this paper, with their description and the places where they are defined, are listed in Tables \ref{t6} and \ref{t7}.
The general formulas for different regimes of weak PPS measurements obtained in the present paper are listed in Table \ref{t8}.
Moreover, a number of the main results of the present paper are briefly summarized in Sec.~\ref{IIH}.

\begin{table}[tb]
\begin{center}
\begin{tabular}{lll}
\hline
\multicolumn{1}{c}{Symbol}&\multicolumn{1}{c}{Description} &\multicolumn{1}{c}{Defined in:}\bigstrut[t]\\
\hline
 $A$&Physical quantity for the system $S$&Sec.~\ref{IA}\bigstrut[t]\\
 $\hat{A}$&Operator for $A$&Sec.~\ref{IA}\\
 $A_w$&Weak value of $A$&Eqs.~\rqn{6}, \rqn{56}, \rqn{12.6}\\
 $A_w^{(1,1)}$&Associated weak value of $A$&Eqs.~\rqn{57a}, \rqn{12.10}\\
 $\cal A$&(Proper) amplification coefficient for&Sec.~\ref{IIIH1}\\
&the linear and strongly-nonlinear regimes&\\
 ${\cal A}_T$&Total amplification coefficient for the&Eq.~\rqn{337}\\
&linear and strongly-nonlinear regimes&\\
 ${\cal A}'$&(Proper) amplification coefficient for the inverted region&Eq.~\rqn{3.16}\\
 ${\cal A}_T'$&Total amplification coefficient for the inverted region&Eq.~\rqn{3.17'}\\
 $b$&The parameter characterizing the quadratic phase&After Eq.~\rqn{103}\\
 $E$&POVM operator corresponding to the post-selection&Sec.~\ref{IC2}\\
 ${\cal E}$&Enhancement factor&Eq.~\rqn{335}\\
 $F$&Input meter variable&After Eq.~\rqn{1}\\
 $\hat{F}$&Operator for $F$&After Eq.~\rqn{1}\\
 $F_c$&``Centered'' variable $F$&After Eq.~\rqn{167}\\
 $I_{\rm M}$&Identity operator for the meter $M$&After Eq.~\rqn{3.5}\\
 $I_{\rm S}$&Identity operator for the system $S$&After Eq.~\rqn{1.35}\\
 $N_0$&Minimum ensemble size&Secs.~\ref{IIB}, \ref{IIIB}\\
 $\bar{O}$&Average of the operator $O$ over the initial state of &Footnote 3\\
 & the system or meter&\\
 $p$&Meter momentum&Sec.~\ref{IB2}\\
 $q$&Meter coordinate&Sec.~\ref{IB2}\\
 $R$&Pointer (or output) meter variable&After Eq.~\rqn{355}\\
 $\hat{R}$&Operator for $R$&Footnote 1\\
 $R_c$&``Centered'' variable $R$&Before Eq.~\rqn{44}\\
 $\bar{R}_f$&Average pointer value for a standard measurement&Eq.~\rqn{74}\\
 $\bar{R}_s$&Average pointer value for a PPS measurement&Sec.~\ref{IC4}\\
 $\cal R$&Signal-to-noise ratio for quantum noise&Eq.~\rqn{3.12}\\
 $U$&Unitary transformation due to the system-meter coupling&Eq.~\rqn{2}\\[.5ex]
\hline
 \end{tabular}
\end{center}
\caption{The list of important symbols used in this paper, their description, and the places where they are defined.
Part 1---Latin letters.
}
 \label{t6}\end{table}

\begin{table}[tb]
\begin{center}
\begin{tabular}{lll}
\hline
\multicolumn{1}{c}{Symbol}&\multicolumn{1}{c}{Description} &\multicolumn{1}{c}{Defined in:}\bigstrut[t]\\
\hline
 $\gamma$&Strength of the system-meter coupling&Eq.~\rqn{355}\bigstrut[t]\\
 $\Delta O$&Uncertainty of the observable $O$&Footnote 3\bigstrut[t]\\
 $\Delta R_{\rm max}$&Shift of the maximum of the pointer distribution due to measurement&Sec.~\ref{IC6}\bigstrut[t]\\
 $\zeta(p)$&Phase of the meter state in the momentum space&Eq.~\rqn{96}\bigstrut[t]\\ 
 $\theta$&Argument of the weak value&Eq.~\rqn{3.7}\bigstrut[t]\\
 $\theta_0$&Argument of $\overline{R_cF}$&Eq.~\rqn{3.7}\bigstrut[t]\\
 $\mu$&Small parameter for weak PPS measurements&Eq.~\rqn{12'}\bigstrut[t]\\
 $\mu_0$&Measurement strength&Eq.~\rqn{297}\bigstrut[t]\\
 $\mu_1$&Strength of the unitary transformation due to $\bar{F}$&Eq.~\rqn{3.20}\bigstrut[t]\\
 $\xi(q)$&Phase of the meter state in the coordinate space&Eq.~\rqn{96}\bigstrut[t]\\ 
 $\Pi_\phi$&Projector on the state $|\phi\rangle$&Eq.~\rqn{1.51}\\
 $\rho$&Preselected (possibly mixed) state of the system&Sec.~\ref{IB1}\bigstrut[t]\\
 $\rho_{\rm M}$&Initial (possibly mixed) state of the meter&Sec.~\ref{IB1}\bigstrut[t]\\
 $\sigma_{FR}$&Covariance for the meter variables $F$ and $R$&Eq.~\rqn{262}\bigstrut[t]\\
 $\Phi_s(R)$&Pointer distribution after a PPS measurement&Sec.~\ref{IC4}\bigstrut[t]\\
 $|\phi\rangle$&Post-selected state of the system&Sec.~\ref{IC1}\\
 $|\psi\rangle$&Pure preselected state of the system&Sec.~\ref{IB1}\bigstrut[t]\\
 $|\psi_{\rm M}\rangle$&Pure initial state of the meter&Sec.~\ref{IB1}\bigstrut[t]\\[.5ex]
\hline
 \end{tabular}
\end{center}
\caption{The list of the important symbols used in this paper, their description, and the places where they are defined.
Part 2---Greek and Latin letters.
}
 \label{t7}\end{table}

\begin{table}[tb]
\begin{center}
\begin{tabular}{llll}
\hline
\multicolumn{1}{c}{Case}&\multicolumn{1}{c}{Preselected state}&\multicolumn{1}{c}{Formula}&\multicolumn{1}{c}{Validity condition}\bigstrut[t]\\
\hline
 &Pure&\rqn{13} or \rqn{182}&\rqn{12'}\bigstrut[t]\\[-1ex]
\raisebox{1.5ex}{General nonlinear formula}&Mixed&\rqn{61} or \rqn{183}&\rqn{63}\bigstrut[t]\\[1ex]
Linear-response (AAV) regime&Arbitrary&\rqn{14''}&\rqn{113}\\[1ex]
&Pure&\rqn{13} or \rqn{182}&\rqn{132}\\[-1ex]
\raisebox{1.5ex}{Strongly-nonlinear regime}&Mixed&\rqn{61} or \rqn{183}&\rqn{282}\bigstrut[t]\\[1ex]
&Pure&\rqn{15}&\rqn{233} and \rqn{12'}\\[-1ex]
\raisebox{1.5ex}{Inverse region}&Mixed&\rqn{5.6}&\rqn{4.1} and \rqn{63}\bigstrut[t]\\[1ex]
&Pure&\rqn{169}&\rqn{173} and \rqn{168}\\[-1ex]
\raisebox{1.5ex}{Resonance for $|\bar{F}|\gg\Delta F$}&Mixed&\rqn{221}&\rqn{173} and \rqn{168}\bigstrut[t]\\[.5ex]
\hline
 \end{tabular}
\end{center}
\caption{General formulas for the average pointer deflection in different regimes of weak PPS measurements.
}
 \label{t8}\end{table}

\section{Measurements with and without post-selection, weak values}
 \label{I}

In this section, we provide a brief review of measurements of arbitrary strength, with and without post-selection, some of the results of this section being original.
In particular, we discuss in detail strong PPS measurements; however, the emphasis here is on weak PPS measurements and weak values.

\subsection{Measurement in quantum mechanics}
\label{IA}

The mathematical apparatus of quantum mechanics and its (Copenhagen) interpretation were created about eighty years ago, and since then they were confirmed in a countless number of experiments in various areas of physics.
In spite of this, a complete understanding of quantum mechanics has
not been achieved yet.
From time to time, there occur revelations of phenomena which
illuminate from an unexpected side the nonclassical nature of quantum
mechanics and thus deepen our understanding of this discipline.
Examples include experiments on Bell-inequality violations
\cite{Bell,asp02,ans09,wei05,wei06,kof08}, which show the impossibility of local
hidden-variable theories, and the emerging fields of quantum
computation and quantum communication \cite{nie00}, where tasks which
are believed to be impossible or very difficult to perform in the realm of the classical world were shown to be solvable.
Weak values of physical quantities \cite{aha88} are another example of nonclassical phenomena with unexpected results, as discussed below.

In quantum mechanics, each physical quantity $A$ is described by a
Hermitian operator $\hat{A}$ in the Hilbert space of a quantum system $S$.
Ideal (or projective or strong) measurements of a system $S$ are described by the projection postulate.
Let the operator $\hat{A}$ have discrete, nondegenerate eigenvalues
$a_i$ and the corresponding eigenvectors $|a_i\rangle$, and let the system $S$ be in the state $\rho$.
Then the projection postulate states that a measurement of $A$
yields the value $a_i$ with probability 
 \be
P_i=\langle a_i|\rho|a_i\rangle 
 \e{1.50}
and that due to this measurement the state of the system becomes $|a_i\rangle$ (the so called wave-function collapse).
Such measurements are ``complete'', in the sense that no subsequent measurement can provide any information on the original state of the system, since the states of the system before and after the measurement are not correlated.

When the operator $\hat{A}$ has degenerate eigenvalues, the projection postulate should be generalized as follows \cite{lud51}.
A general Hermitian operator $\hat{A}$ with discrete eigenvalues has the spectral decomposition
 \be
\hat{A}=\sum_ia_i\,\Pi_i,
 \e{1.12}
where $a_i\ne a_j$ for $i\ne j$, and $\Pi_i$ is the projection operator on the subspace of eigenstates with eigenvalue $a_i$. 
The set of all $\Pi_i$ is the projection-valued measure associated with the measurement of $A$, with the projectors $\Pi_i$ possessing the properties
 \be
\Pi_i\Pi_j=\Pi_i\,\delta_{ij},\quad\sum_j\Pi_j=I_{\rm S},
 \e{1.35}
where $\delta_{ij}$ is the Kronecker symbol and $I_{\rm S}$ is the identity operator for the system.
A projective measurement of the quantity $A$ yields an eigenvalue $a_i$ with probability
 \be
P_i={\rm Tr}\,(\Pi_i\,\rho),
 \e{1.28}
leaving the system in the state 
 \be
\rho_i=\frac{\Pi_i\,\rho\,\Pi_i}{{\rm Tr}\,(\Pi_i\,\rho)}.
 \e{1.29}
Equation \rqn{1.29} implies that for a degenerate $a_i$, the state $\rho_i$ may depend on $\rho$ and thus is unknown; hence a subsequent measurement may extract additional information about the initial state of the system.
Thus, measurements of observables with degenerate eigenvalues are generally incomplete.

A peculiar feature of quantum mechanics is that a measurement changes
the state of the measured system [cf.\ Eq.~\rqn{1.29}].
As a result, consecutive measurements of a quantum system result in an evolution of the system, which is basically different from the unitary evolution due to the Hamiltonian.
The measurement-induced evolution is a purely quantum phenomenon.
This evolution is generally random.

Thus, quantum measurements can play, at least, two fundamentally different roles.
One role is proper measurements, i.e., obtaining information on the values of physical observables.
The other role is generating an evolution of the quantum system.
An example of the second role is the possibility to transform an arbitrary state of a quantum system to any other state with a probability arbitrarily close to one by means of a sufficiently large number of projective measurements \cite{neu55}.
Generally, the evolution of a quantum system is generated both by the Hamiltonian and by measurements.
Examples of evolution driven simultaneously by the Hamiltonian and frequent measurements are the quantum Zeno and anti-Zeno effects \cite{mis77,ita90,kof96,gur97,kof00,fis01,fac01,kos05,wan08,zho09,%
cao10,aix10,cao12}.

The situations where measurements play both roles simultaneously are especially interesting.
One example is the conditional evolution due to post-selected measurements.
In this case the information provided by the measurements is used to choose only a subset of realizations of the measurement-induced random evolution.
Post-selection has recently grown in importance as a tool in fields such as quantum information, e.g., for linear optics quantum computation \cite{kni01}, where it is used to implement quantum gates.
Another example where measurements play both roles is one-way quantum computing \cite{rau01,rau03,nie06,tan06,you07,tan09,bri09}, where a series of measurements is employed to achieve the required evolution, each measurement being chosen on the basis of the information provided by the previous measurements.
 As an additional example, we mention the problem of preparing an arbitrary state of a quantum system by a restricted set of measurements \cite{ash10,wis11}.

In recent decades, there have appeared generalizations of the projection postulate to non-ideal and weak measurements \cite{dav76,kra83,bus96,nie00}.
In particular, when the state of the system after the measurement is not important, the measurement in the most general case is described by a positive operator valued measure (POVM) $\{E_k\}$, where $E_k$ are Hermitian operators with nonnegative eigenvalues satisfying the relation
 \be
\sum_kE_k=I_{\rm S}.
 \e{1.48}
The operator $E_k$ determines the probability of the $k$th measurement outcome by
 \be
P_k={\rm Tr}\,(E_k\,\rho).
 \e{1.49}
Note that the above generalizations do not change the postulates of quantum mechanics.
Namely, the most general measurement is {\em equivalent to a projective measurement} of a composite system consisting of the system $S$ and an auxiliary system \cite{nie00,dav76,nai40}.
Experimental realizations of general measurements have been considered, see, e.g., Ref.~\cite{ota12} and references therein.

General measurements are incomplete, in the sense that the state after the measurement depends on the state before the measurement.
The measurement-induced change of the state (measurement backaction) is commensurate with the measurement strength \cite{ban01}, so that weak measurements change the state weakly.
Recently, significant attention has been given to the subject of
multiple and continuous weak measurements, and many interesting topics were touched upon, such as measurement-induced decoherence, interplay of the unitary evolution and measurement backaction, quantum feedback control, entanglement amplification, etc.\
\cite{bel92,bar93,dal92,gar92,car93,kor99,ave05,kat06,%
jac06,ash09,ash09a,ash09b,cle10,pal10,zha10,otaa}.

Pre- and post-selected (PPS) measurements, which are of primary interest here, were introduced by Aharonov, Bergmann, and Lebowitz
(ABL) \cite{aha64} in an attempt to achieve a better understanding of the role of measurements in quantum mechanics.
PPS measurements are performed on ensembles of quantum systems chosen
(pre- and post-selected) in the given initial and final states.
In particular, PPS measurements and the closely related two-wave-functions formalism were applied for an analysis of time-symmetry properties of measurement-induced evolution \cite{aha02,aha05,aha10,aha11a}.
In Ref.~\cite{aha64}, only strong PPS measurements were considered.

As an important extension of the ABL theory \cite{aha64}, Aharonov, Albert, and Vaidman (AAV) \cite{aha88} introduced the concept of weak PPS measurements.
 Such measurements of an observable $A$ produce the so called weak
value $A_w$, which has unusual properties.
In particular, generally a weak value is a complex number, and its
magnitude is unbounded, so that Re$\,A_w$ can be far outside the
range of eigenvalues of the operator $\hat{A}$.
Unusual (or strange) weak values, i.e., weak values that are
complex or outside the spectrum of $\hat{A}$ were observed in a
number of experiments
\cite{rit91,sut93,sut95,par98,sol04,bru04,res04,pry05,wan06,mir07,%
hos08,lun09,yok09,dix09,how10,sta09,sta10a,cho10,gog11,iin11,tur11,%
hog11,koc11,lun11}.

It has been shown that, at least in some cases, unusual weak values cannot be explained classically.
In particular, as shown in Ref.~\cite{joh04}, a negative weak value of the energy of an oscillator contradicts all classical models; Johansen and Luis \cite{joh04} also proposed a method for measuring such a value in a coherent state of the radiation field.
Furthermore, as shown by Williams and Jordan \cite{wil08}, there is a one-to-one correlation between achieving real unusual
weak values $A_w$ for a projection of a spin 1/2 (i.e., $A_w$ such that $|A_w|>1/2$) and violating the Leggett-Garg inequality for a two-level system (a qubit) \cite{leg85,leg02,lam10,lam10a,lam11,chen}, i.e., violating one or both of the assumptions required for classicality: macrorealism and a noninvasive detector.
This relation between weak values and the Leggett-Garg inequality violations was verified experimentally in Ref.~\cite{gog11}.

The unusual properties of weak values initially gave rise to controversy over their meaning and significance \cite{leg89,per89}. 
However, subsequent research has made significant progress in elucidating the interpretation of weak values and indicating a variety of situations where they provide interesting physical insights \cite{res04,aha02,lun09,yok09,mir07,wis07,dre10,koc11}.
Moreover, irrespective of the interpretation of unusual weak values, they have proved useful in such diverse physical areas as quantum paradoxes, high-precision metrology, and superluminal propagation.

In recent years, there has been a fast growing interest in weak values.
 They were discussed extensively \cite{duc89,aha90,kni90,ste95,aha96,wis02,bru03,joh04a,joh04,%
aha05a,lun05,ral06,joz07,wis07,tol07,lor08,bru08,aha09,lob09,%
wum09,ash,hof10,ges10,bru10,ste10,hos10,ked10,tol10,ber10,%
ber11,sim11,wul11,lix11,aha11,nak12,ber12} and reviewed in Refs.~\cite{aha02,aha05,aha10,aha11a}. (After this work was completed, two recent reviews on quantum measurements and weak values have appeared \cite{shi12,sve}.)
Weak values were measured in a number of experiments
\cite{rit91,sut93,sut95,par98,sol04,bru04,res04,pry05,wan06,mir07,%
hos08,lun09,yok09,dix09,how10,sta09,sta10a,cho10,gog11,iin11,tur11,%
hog11,koc11,lun11}, primarily in the field of optics, though one of the early experiments was in NMR \cite{sut93}.
 There are recent proposals for the observation of weak values
using electrons in solids \cite{wil08,rom08,shp08} and photons and atoms \cite{sim11}.
 The experiments performed included applications to metrology \cite{rit91,hos08,dix09,how10,sta09,sta10a,tur11,hog11}, optical communications \cite{sol04,bru04}, and Hardy's paradox \cite{lun09,yok09}.

In the rest of Sec.~\ref{I} we review PPS measurements, whereas in subsequent sections we elaborate on different aspects the theory of PPS measurements, the emphasis being on developing a nonperturbative theory of weak PPS measurements.

\subsection{Standard (preselected only) quantum measurements of variable strength}
\label{IB}

Here the term ``standard quantum measurement'' or ``standard measurement'' refers to measurement of a physical quantity $A$ without post-selection.
Standard quantum measurements may be ideal (i.e., projective or strong) or non-ideal, with an arbitrary measurement strength (e.g., standard measurements can be  weak).
Standard quantum measurements of arbitrary strength are discussed here with the help of the von-Neumann-like measurement scheme.

\subsubsection{Von-Neumann-like measurement scheme}
\label{IB1}

Quantum measurements are usually performed in the laboratory by
bringing the system under study into an interaction with the
measuring apparatus (the meter) and then measuring the meter.
Von Neumann developed a model which describes how the above process
produces projective measurements \cite{neu55}.
Many studies of quantum measurements are based on the von Neumann
measurement model or its generalizations.

Consider the von-Neumann-like measurement scheme, which is a direct
extension of the original von Neumann measurement model \cite{neu55}.
In this scheme, the quantum system $S$ and the measurement apparatus
$M$ (meter) are coupled by the interaction described by the
Hamiltonian
 \be
H=g(t)\,\hat{A}\otimes\hat{F},
 \e{1}
where $g(t)$ is the instantaneous coupling rate, which differs from
zero in the interval $(t_{\rm i},t_{\rm f})$, $\hat{A}$ is the operator representing the measured quantity $A$, and $\hat{F}$ is the operator\footnote{
Generally, we denote a Hermitian operator and the corresponding physical quantity by the same symbol, the exception being only the notation for the operators $\hat{A},\ \hat{F}$ and $\hat{R}$ of the quantities $A,\ F$, and $R$.}
corresponding to the ``input'' meter variable $F$.

\begin{figure}[htb]
\begin{center}
\includegraphics[width=7cm]{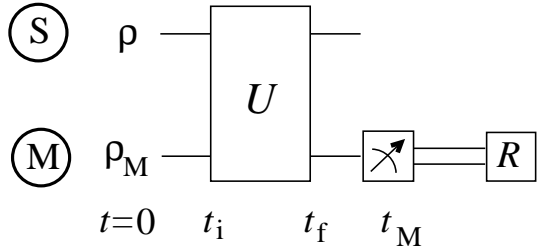}
\end{center}
\caption{Schematic diagram for standard quantum measurements of arbitrary strength.
 The system $S$ becomes correlated with the meter $M$ by the unitary
transformation $U$ in Eq.~\rqn{2}. Then a projective measurement of the meter pointer variable $R$ is performed.
The double line carries classical information.
Initially (at $t=0$), the system state is $\rho$ and the meter state is $\rho_{\rm M}$, and they are uncorrelated.
}
 \label{f9}\end{figure}

We assume that initially (at $t=0\le t_{\rm i}$) the system and the meter are uncorrelated, being in the pure states $|\psi\rangle$ and $|\psi_{\rm M}\rangle$, respectively (the case of arbitrary system and meter states, $\rho$ and $\rho_{\rm M}$, is discussed in Sec.~\ref{IIA'}).
Then for $t\ge t_{\rm f}$ the state of the system and the meter becomes correlated,
 \be
|\psi_f\rangle=U\,|\psi\rangle|\psi_{\rm M}\rangle,
 \e{1.1}
by the unitary transformation
 \be
U=\exp(-i\gamma\,\hat{A}\otimes\hat{F}),
 \e{2}
where $\gamma$ is the coupling strength,
 \be
\gamma=\int_{t_{\rm i}}^{t_{\rm f}} g(t)\,dt
 \e{355}
(we use measurement units in which $\hbar=1$).
Finally, a measurement of the ``output'' meter observable $R$ (the ``pointer variable'') at $t_{\rm M}\ge t_{\rm f}$ provides information about the system.
 This process is depicted schematically in Fig.~\ref{f9}.
 (In Figs.~\ref{f9} and \ref{f8} the standard quantum-circuit notation \cite{nie00} is used: single and double lines carry quantum and classical information, respectively.)
Here we make the common assumption that the free Hamiltonians of the
system and meter can be neglected \cite{neu55,aha88}; the effects of the free Hamiltonians of the system and the meter are discussed in Secs.~\ref{IIID'} and \ref{VIB4}.

\subsubsection{Canonically conjugate meter variables}
\label{IB2}

The problem becomes drastically simplified when the meter is a continuous-variable system, e.g., a free linearly moving particle, whereas $F$ and $R$ are canonically conjugate variables.
We also make the customary assumption that the free Hamiltonian of the meter can be neglected, which implies that the particle mass is very large.
As in the original von Neumann model \cite{neu55}, we assume that $F$
is the momentum $p$ and $R$ is the coordinate $q$,
 \be
F=p,\quad R=q.
 \e{97}
To simplify the problem even more, we assume here that $\hat{A}$ has discrete and nondegenerate eigenvalues.
Then, expanding $|\psi\rangle$ in the basis of the eigenvectors of $\hat{A}$, $|\psi\rangle=\sum_j\alpha_j|a_j\rangle$, Eq.~\rqn{1.1} yields 
 \bea
&|\psi_f(q)\rangle&=\exp(-i\gamma\hat{A}\otimes p)
\sum_j\alpha_j\,|a_j\rangle\,\psi_{\rm M}(q)\nonumber\\
&&=\sum_j\alpha_j\exp(-i\gamma a_jp)\,\psi_{\rm
M}(q)\,|a_j\rangle\nonumber\\
&&=\sum_j\alpha_j\,\psi_{\rm M}(q-\gamma a_j)\,|a_j\rangle,
 \ea{3}
where $|\psi_f(q)\rangle= \langle q|\psi_f\rangle$, and
$\psi_{\rm M}(q)=\langle q|\psi_{\rm M}\rangle$.
A projective measurement of $q$ at $t\ge t_{\rm f}$ results,\footnote{
Since $q$ is a continuous variable, projective measurement of $q$ always has a finite error.
For our purposes, this error should be much less than $|\gamma|\,(\delta a)$.}
to a very good approximation, in a projective measurement of $A$, when different wavepackets $\psi_{\rm M}(p-\gamma a_j)$ practically do not overlap in Eq.~\rqn{3}.
This is realized when the coupling is sufficiently strong,
\be
|\gamma|\,(\delta a)\gg\Delta q,
 \e{1.7}
where $\delta a$ is the minimal distance between different $a_j$'s and $\Delta q$ is the uncertainty\footnote{
The average of an arbitrary operator $O$ in a state $\rho_0$ is given by $\bar{O}= {\rm Tr}\,(O\rho_0)$ (in particular, for a pure state $|\Psi\rangle$, $\bar{O}=\langle\Psi|O|\Psi\rangle$), whereas the uncertainty of an observable $O$ is given by the square root of the variance, $\Delta O=(\overline{O^2}- \bar{O}^2)^{1/2}$.}
of $q$ at $t=0$.

\subsubsection{Non-ideal and weak standard measurements}
\label{IB3}

When the coupling is not sufficiently strong, i.e., when Eq.\ \rqn{1.7} is not satisfied, a measurement is non-ideal (partial).
However, for any $\gamma$ one can still measure the average (expectation value) of $A$ over the initial state $|\psi\rangle$,  
 \be
\bar{A}=\langle\psi|\hat{A}|\psi\rangle, 
 \e{1.43}
since Eq.~\rqn{3} implies that \cite{aha88}
 \be
\bar{q}_f-\bar{q}=\gamma\bar{A},
 \e{80}
where $\bar{q}$ and $\bar{q}_f$ are the averages of $q$ at $t=0$
and $t\ge t_{\rm f}$, respectively.

A standard measurement with a small coupling strength $\gamma$ is called a weak standard measurement or simply a weak measurement.

Note that a weak measurement of one system provides almost no
information, since the average pointer deflection \rqn{80} is much
less than the pointer uncertainty.
Therefore, to obtain $\bar{A}$, one must perform measurements on
each member of a sufficiently large ensemble of systems prepared (preselected) in
the same state and then average the results of the measurements.
The measurement error decreases when increasing the size of the
ensemble and thus can be made arbitrarily small.
The way of extracting the expectation value $\bar{A}$ in weak measurements differs conceptually from that in projective measurements.
Indeed, projective measurements provide probabilities $P_i$ of the eigenvalues $a_i$ of an observable $A$, and $\bar{A}$ is obtained from the standard definition of the expectation value by the formula
 \be
\bar{A}=\sum_ia_iP_i.
 \e{1.47}
In contrast, in weak measurements $\bar{A}$ is extracted by Eq.~\rqn{80} directly, without measuring each $P_i$ individually.

In the general case, when the input and output meter variables $F$ and $R$ are not canonically conjugate to each other, Eq.~\rqn{80} does not hold for arbitrary $\gamma$.
However, for a sufficiently small coupling strength, when the linear response holds, the average deflection of the pointer $R$ is generally proportional to $\bar{A}$, i.e., weak standard measurements are still possible (see Sec.~\ref{IIA'} for further details).

\subsection{Pre- and post-selected measurements}

\subsubsection{General considerations}
\label{IC1}

In classical mechanics, one can completely determine the motion of an isolated system for all future and past times, if its Hamiltonian and state at some moment are known.
In contrast, in quantum mechanics, only a fraction of the observables can be completely determined in a given state of a system, whereas other observables cannot be determined, as demonstrated explicitly by the Heisenberg uncertainty principle.
This makes the evolution of the system non-deterministic, i.e.,
probabilistic.

In the usual approach, the (random) behavior of a quantum system is
studied assuming the knowledge of the state at some initial time
$t_0$.
As an extension of the usual approach, ABL \cite{aha64} asked the
question:
How does the description of a quantum system in the interval $(t_0,t_{\rm S})$ change, when not only the initial state at time $t_0$ but also the final state at time $t_{\rm S}$ are known, so that one has more complete information on the system than in the usual approach?

\begin{figure}[htb]
\centerline{\includegraphics[width=10cm]{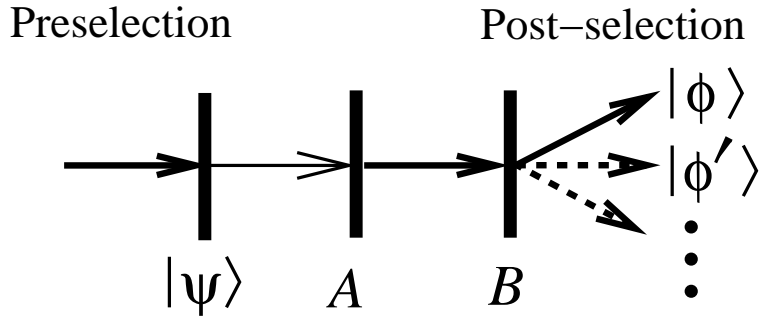}}
\caption{A schematic diagram of pre- and post-selected quantum measurements.
It involves the pre-selection of the identical systems from an ensemble in the initial state $|\psi\rangle$, measurement of a variable $A$ (the thin arrow) for each system, and the post-selection by means of a projective measurement of a variable $B$ and selecting the systems in a final state $|\phi\rangle$.
}
 \label{f11}\end{figure}

To answer this question, ABL \cite{aha64} devised pre- and
post-selected measurements described as follows (see Fig.~\ref{f11}).
Consider an ensemble of quantum systems prepared initially
(preselected) in the same state $|\psi\rangle$.
Each member of the ensemble is subjected to a measurement of the
quantity $A$, which may be strong or weak (the thin arrow before $A$ in Fig.~\ref{f11}).
Then, at a later moment, a final projective measurement of a variable $B$ with a discrete, nondegenerate spectrum is performed, which, in view of the projection postulate (Sec.~\ref{IA}), leaves the system in one of the orthogonal states $|\phi\rangle,|\phi'\rangle,\dots$.
The ensemble of the system can be broken into subensembles with
different final (post-selected) states
$|\phi\rangle,|\phi'\rangle,\dots$; such a subensemble is called a
pre- and post-selected ensemble.
The statistical distribution of the results of the measurement of $A$ for a given subensemble depends on the subensemble and is different from the statistical distribution over the whole ensemble.
Thus, the possible results of the measurement of $A$ depend both on the initial and the final states of the system.
A measurement in a pre- and post-selected ensemble is called a pre- and post-selected measurement.
Above we considered one measurement in a pre- and post-selected ensemble, but there may be two or more such measurements of some observables $A,A',\dots$ \cite{aha64,aha05}.
Note that, though the terms ``pre-selection'' and ``post-selection'' are very similar, they denote conceptually different physical processes: preparation of the initial state and conditioning of the measured statistics on the acquired information, respectively; see also the discussion in Sec.~\ref{XIIID}.

Until now we discussed {\em pure} PPS ensembles, i.e., ensembles with pure initial and final states.
More generally, we will consider also {\em mixed} PPS ensembles, where the pre-selection is incomplete, i.e., the initial state $\rho$ is mixed.
In addition to the aforementioned preselected and PPS ensembles, there is also a third type of ensemble---post-selected only ensembles \cite{aha02}, i.e., ensembles of systems with a pure final state $|\phi\rangle$ and the completely mixed initial state
 \be
\rho_{\rm c.m.}=\frac{I_{\rm S}}{d},
 \e{1.10}
where $d$ is the dimension of the Hilbert space of the system.
Post-selected ensembles are the limiting case $\rho\rightarrow\rho_{\rm c.m.}$ of mixed PPS ensembles.

Another important generalization of PPS ensembles is for the case where the post-selection measurement is performed by a general measurement described by an arbitrary POVM; then the PPS ensemble includes the systems with a certain measurement outcome.
Such a measurement is generally incomplete, i.e., it does not specify a single post-selection state independent of the state of the system before the post-selection.

This generalization allows one to connect PPS and preselected ensembles.
Namely, when $B$ is a multiple of the unity operator, the measurement of $B$ does not provide any new information which could be used for the post-selection, and a PPS ensemble becomes a preselected (only) ensemble.
As a result, in this case PPS measurements of an arbitrary strength coincide with standard measurements.
For the cases of strong and weak measurements with a pure preselected state, this was formally proved in Ref.~\cite{aha02} (see also Sec.~\ref{IC2}), whereas for the general case of measurements of arbitrary strength with an arbitrary preselected state this will be proved in Sec.~\ref{XIIIB}.

PPS measurements have unusual properties, some of which hold only for strong or weak PPS measurements, whereas others hold irrespective of the strength of measurements.
These properties are discussed below.
Here we mention only one of them: in a pure PPS ensemble, any observable with an eigenstate $|\psi\rangle$ or $|\phi\rangle$ has a definite value, equal to the corresponding eigenvalue \cite{aha91}.
This means that PPS measurements of any strength will always yield the above value of such an observable.
As a consequence, when $|\phi\rangle\ne|\psi\rangle$, there are, at least, two non-commuting observables with no common eigenstates (e.g., components of spin 1/2), which have definite values in a PPS ensemble.
This is in sharp contrast with the conventional case of systems preselected only in a state $|\psi\rangle$, for which solely the observables with the common eigenstate $|\psi\rangle$ have definite values.

\subsubsection{Strong (ideal) PPS measurements}
\label{IC2}

Let us discuss PPS measurements in more detail.
Consider first strong (ideal) PPS measurements.

Let an ensemble of quantum systems be prepared (preselected) in a
(pure or mixed) state $\rho$.
According to the projection postulate (Sec.~\ref{IA}), a projective
measurement of an observable $A$ with discrete eigenvalues provides an eigenvalue $a_i$ with probability $P_i$ \rqn{1.28} and leaves the system in the state $\rho_i$ \rqn{1.29}.
In the most general case, the final measurement is characterized by a POVM (see Sec.~\ref{IC1}), and the PPS ensemble includes systems with a certain measurement outcome which is characterized by a POVM operator $E$ and occurs with probability ${\rm Tr}\,(E\rho_i)$ [cf.\ Eq.~\rqn{1.49}].
The joint probability to measure the eigenvalue $a_i$ of $A$ and to observe the outcome corresponding to $E$ is the product of the respective probabilities,
 \be
P_{iE}=P_i\,{\rm Tr}\,(E\rho_i)={\rm Tr}\,(E\,\Pi_i\,\rho\,\Pi_i).
 \e{1.30}
As follows from Eq.~\rqn{1.30} and Bayes' theorem, the probability that
a projective measurement of $A$ yields the value $a_i$, provided the
system is post-selected by means of $E$, is
 \be
P_{i|E}\,=\,\frac{P_{iE}}{\sum_jP_{jE}}
=\frac{{\rm Tr}\,(E\,\Pi_i\,\rho\,\Pi_i)}{\sum_j{\rm Tr}\,(E\,\Pi_j\,\rho\,\Pi_j)}.
 \e{81'}
This equation is an extension of the ABL formula \cite{aha64,aha05} to general $E$ and $\rho$. 
It is obvious from the first equality in Eq.~\rqn{81'} that the probability distribution $P_{i|E}$ is normalized to one,
 \be
\sum_iP_{i|E}=1.
 \e{1.42}
When $E$ is the unity operator, strong PPS measurements become strong standard (not post-selected) measurements, and correspondingly, as can be easily checked, Eq.~\rqn{81'} reduces to Eq.~\rqn{1.28} (compare the discussion in the last but one paragraph of Sec.~\ref{IC1}).

Henceforth (with the exception of Sec.~\ref{XII}), we will assume that the post-selection measurement is a projection on a nondegenerate, discrete eigenvalue of a variable $B$.
Such a post-selection measurement is complete in the sense that it specifies a single post-selection state $|\phi\rangle$.
In this case,
 \be
E=\Pi_\phi\equiv|\phi\rangle\langle\phi|,
 \e{1.51}
and Eq.~\rqn{81'} becomes the probability that a projective measurement of $A$ yields the value $a_i$, provided the system is post-selected in the state $|\phi\rangle$, 
 \be
P_{i|\phi}=\frac{\langle\phi|\,\Pi_i\,\rho\,\Pi_i\,|\phi\rangle}
{\sum_j\,\langle\phi|\,\Pi_j\,\rho\,\Pi_j\,|\phi\rangle}.
 \e{1.31}

Consider now special cases.
For a nondegenerate eigenvalue $a_i$, one has $\Pi_i=|a_i\rangle\langle a_i|$, and Eq.~\rqn{1.31} becomes
 \be
P_{i|\phi}=\frac{|\langle\phi|a_i\rangle|^2\,\langle a_i|\rho|a_i\rangle}
{\sum_j|\langle\phi|a_j\rangle|^2\,\langle a_i|\rho|a_i\rangle}.
 \e{81}
For a pure initial state $\rho=|\psi\rangle\langle\psi|$, Eq.~\rqn{1.31} becomes 
 \be
P_{i|\phi}=\frac{|\langle\phi|\,\Pi_i\,|\psi\rangle|^2} {\sum_j|\langle\phi|\,\Pi_j\,|\psi\rangle|^2}.
 \e{1.11}
Finally, in the case of a pure preselected state $|\psi\rangle$ and a  nondegenerate $a_i$, Eq.~\rqn{1.11} yields the result
 \be
P_{i|\phi}=\frac{|\langle\phi|a_i\rangle\,\langle
a_i|\psi\rangle|^2} {\sum_j|\langle\phi|a_j\rangle\,\langle
a_j|\psi\rangle|^2}.
 \e{1.3}
Equations \rqn{1.11} and \rqn{1.3} were obtained in Refs.~\cite{aha91} and \cite{aha64}, respectively.
In particular, Eq.~\rqn{1.3} is the ABL formula for the case of the vanishing system Hamiltonian, whereas Eqs.~\rqn{81'} and \rqn{1.31}-\rqn{1.11} are extensions of the ABL formula.

When $|\psi\rangle$ or $|\phi\rangle$ coincides with an eigenstate $|a_{i'}\rangle$ of the observable $A$, Eq.~\rqn{1.3} yields $P_{i|\phi}=\delta_{ii'}$, i.e., the eigenvalue $a_{i'}$ is observed with the probability one.
This proves the unusual property, discussed in the last paragraph of Sec.~\ref{IC1}, for the case of strong PPS measurements.

In the case when the pre- and post-selected states are pure, strong PPS measurements are invariant under time reversal \cite{aha64,aha02,aha05}, which is seen from the fact that the ABL formulas \rqn{1.11} and \rqn{1.3} are symmetric with respect to an exchange of the initial and final states.
Moreover,  the idea was suggested that quantum mechanics is time-symmetric in PPS ensembles (at least, when the pre- and post-selected states are pure), and the corresponding two-state vector formalism was developed \cite{aha64,aha02,aha05,aha10,aha11a}.
For the present, this approach is not generally accepted; for recent discussions of time-symmetric quantum mechanics see Refs.~\cite{pht11,pht11a}.

Note, however, that generally PPS measurements are not symmetric with respect to an exchange of the initial and final states.
In Sec.~\ref{XIIID} we will obtain a general time-symmetry relation, which holds for PPS measurements of arbitrary strength, with arbitrary (possibly mixed) preselected states and arbitrary non-ideal post-selection  measurements.
 
In the limit of a completely mixed initial state, $\rho\rightarrow\rho_{\rm c.m.}$ [see Eq.~\rqn{1.10}], Eq.~\rqn{1.31} provides the following probabilities of measurement outcomes for an ensemble post-selected only in the state $|\phi\rangle$,
 \be
P_{i|\phi}=\langle\phi|\Pi_i|\phi\rangle.
 \e{1.32}
This formula coincides with the result \rqn{1.28} with $\rho=|\phi\rangle\langle\phi|$.
Thus, the probability distribution of the outcomes of a measurement performed at time $t,\ t_0<t<t_{\rm S}$, in a {\em post-selected} ensemble is identical to that for the system {\em preselected} in the state $|\phi\rangle$ \cite{aha02}.
A similar property holds also for weak PPS measurements (see paragraph {\em g} in Sec.~\ref{IE2}).

\subsubsection{Contextuality of strong PPS measurements}
\label{IC3}

A peculiar property of strong PPS measurements is their contextuality.
Namely, for systems with the Hilbert-space dimension $d\ge3$, the probabilities given in Eqs.~\rqn{81'} and \rqn{1.31}-\rqn{1.3} are context-dependent, i.e., the probability of an outcome of a strong PPS measurement depends not only on the projector associated with that outcome but on the entire projection-valued measure associated with the measurement \cite{aha91,lei05}.
For a two-level system (qubit), the probabilities of strong PPS measurements are not contextual, since any of the projectors $\Pi_i$ in Eq.~\rqn{1.12} determines the other one by the completeness relation [the second equality in Eq.~\rqn{1.35}] and thus determines the entire projection-valued measure.

The contextuality of strong PPS measurements is illustrated by the three-box problem \cite{aha91}.
Consider a particle which can be located in one of three boxes.
The state of the particle when it is in box $i$ is denoted by $|i\rangle$.
At time $t_0$ the particle is prepared in the state
 \be
|\psi\rangle=\frac{1}{\sqrt{3}}\,(|1\rangle+|2\rangle+|3\rangle),
 \e{1.13}
and at a later time $t_{\rm S}$ the particle is found in the state
 \be
|\phi\rangle=\frac{1}{\sqrt{3}}\,(|1\rangle+|2\rangle-|3\rangle).
 \e{1.14}
We assume that in the time interval $[t_0,t_{\rm S}]$ the Hamiltonian is zero.
Opening box $i$ at time $t,\ t_0<t<t_{\rm S}$, corresponds to measuring the projection operator 
 \be
\Pi_i=|i\rangle\langle i|.
 \e{1.15}
Now Eq.~\rqn{1.11} involves two operators, $\Pi_i$ and
 \be
\tilde{\Pi}_i=\sum_{j\ne i}|j\rangle\langle j|.
 \e{1.17}
Hence, one obtains from Eq.~\rqn{1.11} that the probability to find the particle in box 1, without opening the other boxes, is
 \be
{\rm prob}\,(\Pi_1=1)=\frac{|\langle\phi|1\rangle\langle
1|\psi\rangle|^2} {|\langle\phi|1\rangle\langle
1|\psi\rangle|^2+|\langle\phi|2\rangle\langle
2|\psi\rangle+\langle\phi|3\rangle\langle 3|\psi\rangle|^2}
=\frac{(1/3)^2} {(1/3)^2+(1/3-1/3)^2}=1.
 \e{1.16}
Similarly,
 \be
{\rm prob}\,(\Pi_2=1)=1
 \e{1.18}
and ${\rm prob}\,(\Pi_3=1)=1/5$.
Thus, we obtain a paradoxical result that on opening any of boxes 1 and 2 one is certain to find the particle in the opened box.
The results \rqn{1.16} and \rqn{1.18} were verified experimentally \cite{res04}.

For comparison, consider opening the three boxes simultaneously, which corresponds to measuring a nondegenerate observable with the eigenstates $|1\rangle,\ |2\rangle$, and $|3\rangle$.
Now the probability to find the particle in box $i$ is given by Eq.~\rqn{1.11}, where the projection-valued measure is $(\Pi_1,\Pi_2,\Pi_3)$ [Eq.~\rqn{1.15}], or, equivalently, by Eq.~\rqn{1.3}, yielding 
 \be
P_{1|\phi}=P_{2|\phi}=P_{3|\phi}=\frac{1}{3}.
 \e{1.19}
The probabilities in Eqs.~\rqn{1.16} and \rqn{1.18} differ from, respectively, the probabilities $P_{1|\phi}$ and $P_{2|\phi}$ in Eq.~\rqn{1.19}.
This shows explicitly the contextuality of strong PPS measurements.

\subsubsection{Model for PPS measurements of arbitrary strength}
\label{IC4}

\begin{figure}[htb]
\centerline{\includegraphics[width=7cm]{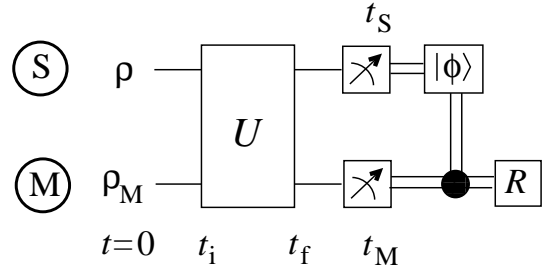}}
\caption{Schematic diagram of a model for pre- and post-selected
quantum measurements.
This approach differs from the von Neumann scheme in Fig.~\ref{f9} in
that the measurement of the pointer variable $R$ is conditioned (``post-selected'') on the measurement of the system $S$ in a state $|\phi\rangle$.}
 \label{f8}\end{figure}

Non-ideal PPS measurements can be discussed by analogy with non-ideal
standard measurements (Sec.~\ref{IB}) \cite{aha88}, with the help of
a suitably generalized von Neumann model, as follows (see
Fig.~\ref{f8}).
Let us consider an ensemble of pairs consisting of a system and a
meter in the pure states $|\psi\rangle$ and $|\psi_{\rm M}\rangle$,
respectively (extensions to the cases of arbitrary states $\rho$ and/or $\rho_{\rm M}$ are given in Secs.~\ref{III''}-\ref{V}).
For each system-meter pair the coupling \rqn{1} is turned on in the
interval $(t_{\rm i},t_{\rm f})$; then a PPS ensemble is formed by performing a projective measurement of a variable $B$ for each system at $t_{\rm S}>t_{\rm f}$ and selecting for further consideration only the systems which are in the eigenstate $|\phi\rangle$ of $B$.
A PPS measurement is completed after measuring the pointer observable $R$ of the meters at $t_{\rm M}>t_{\rm f}$ and performing the statistical analysis of results in the PPS ensemble, with the goal, e,g., to obtain the average pointer value $\bar{R}_s$ or the distribution of the pointer values $\Phi_s(R)$. 

Note that the meters can be measured both after ($t_{\rm M}>t_{\rm
S}$) and before ($t_{\rm M}<t_{\rm S}$) the post-selection.
The only difference is that for $t_{\rm M}>t_{\rm S}$ it is sufficient
to measure only the meters corresponding to the PPS ensemble, whereas
for $t_{\rm M}<t_{\rm S}$ all meters in the initial ensemble should
be measured, but in the statistical analysis after the post-selection (at $t>t_{\rm S}$) only the
meters corresponding to the PPS ensemble should be included.

\subsection{Weak PPS measurements}
\label{ID}

\subsubsection{Simple approach}
\label{IC5}

Let us now consider weak PPS measurements.
Here we describe the simple approach \cite{aha88,duc89,aha90}, which has been used in most studies on weak values; a more general approach is discussed in the following sections.
As in Secs.~\ref{IB2} and \ref{IB3}, we consider the coupling \rqn{1} with $F=p$ (the case $F=q$ is not essentially different, as discussed in Sec.~\ref{IC6}).
Using Eqs.~\rqn{1.1} and \rqn{2}, we obtain that when $t_{\rm M}>t_{\rm S}$, the unnormalized meter state after the post-selection but before the measurement of the meter is ($t_{\rm S}<t<t_{\rm M}$)
 \bea
&\langle\phi|\psi_f(p)\rangle&=\langle\phi|
\exp(-i\gamma\hat{A}\otimes p)\,|\psi\rangle\,\psi_{\rm M}(p) \nonumber\\
&&\approx\langle\phi|1-i\gamma\hat{A}\otimes p\,|\psi\rangle\,\psi_{\rm M}(p)\nonumber\\
&&=\langle\phi|\psi\rangle(1-i\gamma A_wp)\,\psi_{\rm M}(p)\nonumber\\
&&\approx\langle\phi|\psi\rangle\exp(-i\gamma A_wp)\,\psi_{\rm M}(p).
 \ea{1.4}
Here $|\psi_f(p)\rangle=\langle p|\psi_f\rangle$, $\psi_{\rm
M}(p)=\langle p|\psi_{\rm M}\rangle$, and $A_w$ is called the weak
value of $A$ \cite{aha88},
 \be
A_w=\frac{A_{\phi\psi}}{\langle\phi|\psi\rangle},
 \e{6}
where $A_{\phi\psi}=\langle\phi|\hat{A}|\psi\rangle$.
The approximations in Eq.~\rqn{1.4} hold up to first order in
$\gamma$.

Traditionally, in the present approach \cite{aha88,duc89,aha90} the meter wavefunction is taken as a real Gaussian in both $p$ and $q$ spaces, which implies that the averages $\bar{p}$ and $\bar{q}$ vanish.
Here we make a slight generalization, assuming that $\psi_{\rm M}(p)$
is a complex Gaussian with a phase linear in $p$,
 \be
\psi_{\rm M}(p)=Z_p\exp\left[-\frac{(p-\bar{p})^2}{4(\Delta
p)^2}-i\bar{q}p\right],
 \e{1.5}
where $Z_p=(\sqrt{2\pi}/\Delta p)^{1/2}$, and $\bar{p}$ and $\Delta p$
are, respectively, the average and the uncertainty of $p$ at $t=0$.
The Fourier transform of Eq.~\rqn{1.5} yields a Gaussian wavefunction $\psi_{\rm M}(q)$ with a phase linear in $q$,
 \be
\psi_{\rm M}(q)=Z_q\exp\left[-\frac{(q-\bar{q})^2}{4(\Delta q)^2} +i\bar{p}q\right],
 \e{1.20}
where $Z_q=(\sqrt{2\pi}\Delta q)^{-1/2}$ and the uncertainty $\Delta q=(2\Delta p)^{-1}$.
The latter equality implies that in the Gaussian state [Eq.~\rqn{1.5} and/or \rqn{1.20}], the Heisenberg uncertainty relation is saturated,
 \be
\Delta q\,\Delta p\,=\,1/2.
 \e{2.10}
This fact is advantageous for weak measurements, as discussed below (see, e.g., Sec.~\ref{IIIC}).

Inserting Eq.~\rqn{1.5} into Eq.~\rqn{1.4} yields a Gaussian
wavefunction which differs from the initial wavefunction \rqn{1.5}
only by the changes $\bar{q}\rightarrow\bar{q}_s$ and $\bar{p}\rightarrow\bar{p}_s$, where the shifts in the averages $\bar{q}$ and $\bar{p}$ are determined by
 \be
\bar{q}_s-\bar{q}=\gamma\,{\rm Re}\,A_w,
 \e{4}
 \be
\bar{p}_s-\bar{p}=2\gamma\,(\Delta p)^2\,{\rm Im}\,A_w,
 \e{5}
The quantities $\bar{q}_s$ and $\bar{p}_s$ are the post-selected averages of $q$ and $p$.
Thus, weak PPS measurements of the averages of $p$ and $q$ provide the real and imaginary parts of the weak value, respectively.
Note that that the magnitudes of the average pointer deflections $|\bar{q}_s-\bar{q}|$ and $|\bar{p}_s-\bar{p}|$ in Eqs.~\rqn{4} and \rqn{5} are much less than the respective statistical dispersions of the measurement results $\Delta q$ and $\Delta p$, and therefore performing weak PPS measurements requires averaging of the measurement results obtained for many identical systems prepared in the same initial state.
Equation \rqn{4} looks very similar to the result \rqn{80} of weak standard measurements.
However, in contrast to standard measurements (Sec.~\ref{IB}), now
not only $q$ but also $p$ contains information about the system, as shown by Eq.~\rqn{5}.

Equations \rqn{4} and \rqn{5} show that weak PPS measurements in the linear-response (or AAV) regime provide the weak value $A_w$ of the quantity $A$. 
The physical significance of the weak value arises from the fact that in the linear-response regime the backaction of measurements on the system is very small, and therefore the weak value provides, in a sense, information on the unperturbed system.

The weak value in Eq.~\rqn{6} has unusual properties, which drastically distinguish it from the expectation value of a variable \rqn{1.43} resulting from a standard measurement.
The weak value diverges when the overlap $|\langle\phi|\psi\rangle|$ tends to zero.
For instance, the weak value of a component of spin 1/2 can be equal to 100 \cite{aha88}.
Moreover, the weak value can be complex.
Later we will discuss weak values in further detail.

The results \rqn{4} and \rqn{5} were obtained\footnote{Actually Eqs.~\rqn{4} and \rqn{5} differ somewhat from the AAV results \cite{aha88} in that here the roles of $p$ and $q$ are exchanged in comparison to Ref. \cite{aha88}, as in some
optical experiments \cite{rit91,hos08}.
The original AAV results are given below by Eqs.~\rqn{1.45} and \rqn{1.46}.} 
by AAV \cite{aha88} for real Gaussian functions $\psi_{\rm M}(p)$ and $\psi_{\rm M}(q)$, i.e., for $\bar{p}=\bar{q}=0$.
As shown above, Eqs.~\rqn{4} and \rqn{5} also hold for Gaussians with a linear phase.
Jozsa \cite{joz07} considered the case of an arbitrary meter wavefunction and showed that generally there is an additional term, proportional to Im$\,A_w$, on the right-hand side of Eq.~\rqn{4}, whereas Eq.~\rqn{5} remains valid in the general case.

Hosten and Kwiat \cite{hos08} showed experimentally that the term proportional to Im$\,A_w$ may arise in Eq.~\rqn{4} due to the free meter Hamiltonian; they utilized this term to achieve strong amplification in a measurement of a weak optical effect (for details see Sec.~\ref{VIB4}).
Below we show that in the most general case, i.e., for arbitrary meter variables $F$ and $R$, the term proportional to Im$\,A_w$ arises in the linear-response regime whenever there is correlation between $F$ and $R$ (see Sec.~\ref{IIIF1} for more details), the presence of a nonzero meter Hamiltonian being only one of the possible ways to generate this correlation (see Sec.~\ref{VIA}).
Furthermore, below we show that a correlation between $p$ and $q$ arises whenever the phase of the wavefunction $\psi_{\rm M}(p)$ or $\psi_{\rm M}(q)$ is nonlinear in $p$ or $q$, respectively (see Sec.~\ref{VIA}, cf.\ also Sec.~\ref{IC6}).

In addition to the average pointer variable, one can measure also the distribution of the pointer variable.
In the present case the measured distributions of $q$ and $p$ are Gaussian; they are given by the squares of the moduli of the functions \rqn{1.20} and \rqn{1.5}, respectively, with the changes $\bar{q}\rightarrow\bar{q}_s$ and $\bar{p}\rightarrow\bar{p}_s$.
Thus, a weak PPS measurement results in a displacement of the Gaussian distributions in the coordinate and momentum spaces, without changes in their shapes and widths.
In particular, the maxima of the probability distributions of $q$ and $p$ are shifted by, respectively,
 \be
\Delta q_{\rm max}=\gamma\,{\rm Re}\,A_w,
 \e{4'}
 \be
\Delta p_{\rm max}=2\gamma\,(\Delta p)^2\,{\rm Im}\,A_w.
 \e{5'}
Note, however, that generally weak PPS measurements change the shape of the pointer distribution, as discussed below.

The conditions for the validity of Eqs.~\rqn{4} and \rqn{5} were obtained in Ref.~\cite{duc89} for the special case $\bar{q}=\bar{p}=0$.
Namely, the first and second approximations in Eq.~\rqn{1.4} hold,
respectively, for
 \be
|\gamma|\,\left|\frac{(A^n)_{\phi\psi}}{A_{\phi\psi}}\right|^{1/(n-1)} \Delta p
\ll1\quad(n=2,3,\dots)
 \e{1.8}
and
 \be
|\gamma A_w|\,\Delta p \ll1.
 \e{1.9}

The results \rqn{4}-\rqn{5'}, as well as their generalizations mentioned above, hold up to first order in $\gamma$, i.e., in the linear-response regime.
The linear response is important, since in this regime the measurement backaction is minimal, and hence one may expect to reveal in this regime such properties of the system that cannot be probed by strong measurements, which strongly perturb the system state.
Therefore the linear response is the most well studied regime of weak PPS measurements.

In the linear-response regime, the weak value is bounded by the condition \rqn{1.9}.
Correspondingly, for any given $\gamma$, the condition \rqn{1.9} is always violated for a sufficiently large weak value or, equivalently, for sufficiently small overlap $\langle\phi|\psi\rangle$.
In this case, linear-response results are not applicable, even though the condition \rqn{1.8} holds and hence PPS measurements are weak.
It would be of interest to extend the theory of weak PPS measurements 
beyond limits of the linear response.
A simple and general theory of weak PPS measurements, which is correct to all orders in $\gamma A_w$, will be developed and discussed in Sec.~\ref{III'} and henceforth.

In the remainder of this section we will continue to review PPS measurements in the linear-response regime.

\subsubsection{The pointer distribution}
\label{IC6}

In addition to the average value of the pointer variable $R$, it is of interest to consider the probability distribution of the pointer values, since it is measured directly in experiments.
For simplicity, we assume that the initial pointer distribution $\Phi(R)$ has a bell-like shape (e.g., Lorentzian or Gaussian).

Here we consider situations where the pointer distribution $\Phi_s(R)$, resulting from a weak PPS measurement in the linear-response regime, has the following property, which is advantageous for experimental realizations:\\
(i) $\Phi_s(R)$ is displaced with respect to the initial distribution $\Phi(R)=|\psi_{\rm M}(R)|^2$ {\em without a change of the shape} of the distribution (at least, for the central part of the distribution; the tails of the distribution can be deformed by the measurement even in the linear-response regime, see Sec.~\ref{VIB} for details).
This property is equivalent to the following relation,
 \be
\Phi_s(R)=\Phi(R+\bar{R}-\bar{R}_s).
 \e{2.11}

Property (i) implies the following property:\\
(ii) The shift of the maximum of the distribution $\Delta R_{\rm max}$ equals the average pointer deflection,
 \be
\Delta R_{\rm max}=\bar{R}_s-\bar{R}.
 \e{1.25}
Note, however, that property (ii) does not necessarily imply property (i) and Eq.~\rqn{2.11}.
[The general case, where properties (i) and (ii) may not hold, is discussed in Sec.~\ref{VIB}.]

In particular, property (i) and hence Eqs.~\rqn{2.11} and \rqn{1.25} hold in the following cases.

{\em a. Real weak value.} 
Let $A_w$ be real, whereas the meter variables are canonically conjugate, e.g., $F=p$ and $R=q$.
Then from Eq.~\rqn{1.4} rewritten in the coordinate representation we obtain that
 \bea
&\langle\phi|\psi_f(q)\rangle&\approx\langle\phi|\psi\rangle\exp(-i\gamma A_wp)\,\psi_{\rm M}(q)\nonumber\\
&&=\langle\phi|\psi\rangle\,\psi_{\rm M}(q-\gamma A_w).
 \ea{1.21}
Thus, when $A_w$ is real, an arbitrary coordinate distribution is shifted, due to a weak measurement, by the value \cite{joz07,hos08}
 \be
\bar{q}_s-\bar{q}=\gamma A_w.
 \e{1.22}
In particular, if the coordinate distribution is bell-like, we have
 \be
\Delta q_{\rm max}=\bar{q}_s-\bar{q}=\gamma A_w.
 \e{1.26}

{\em b. Conjugate-variable meter in a general complex Gaussian state.}
Consider a meter with canonically conjugate meter variables, e.g., $F=p$ and $R=q$.
We assume that the meter is in a general complex Gaussian state.

The most general form of a complex Gaussian state is given by
 \be
\psi_{\rm M}(p)=Z_p\exp\left[-\frac{(1+ib)(p-\bar{p})^2}{4(\Delta
p)^2}-i\bar{q}p\right].
 \e{103}
Here $b$ is a real parameter characterizing the quadratic phase---the phase of the state \rqn{103} is a quadratic function of $p$, with the quadratic term proportional to the parameter $b$.

 In coordinate space, a general Gaussian state has a similar form,
 \be
\psi_{\rm M}(q)=Z_q\exp\left[\frac{(ib-1)(q-\bar{q})^2}{4(\Delta
q)^2}+i\bar{p}q\right],
 \e{38}
where $\Delta q$ is determined from the equality
 \be
\Delta p\:\Delta q\,=\,\frac{\sqrt{1+b^2}}{2}.
 \e{49}
This equation has the meaning of the generalized uncertainty relation with the equals sign [cf.\ Eq.~\rqn{248} below].
In Eq.~\rqn{38}, similarly to Eq.~\rqn{103}, the phase is a quadratic function of $q$, with the quadratic term proportional to $b$.

Inserting Eq.~\rqn{103} into Eq.~\rqn{1.4} yields a wavefunction of the same form as Eq.~\rqn{103} with the only difference that $\bar{p}$ is shifted by the value \rqn{5'}, whereas $\bar{q}$ is shifted by the value
 \be
\Delta q_{\rm max}=\bar{q}_s-\bar{q}=\gamma\,({\rm Re}\,A_w+b\,{\rm Im}\,A_w).
 \e{1.27}
In other words, a weak PPS measurement shifts the Gaussian distributions of $p$ and $q$, $|\psi_{\rm M}(p)|^2$ and $|\psi_{\rm M}(q)|^2$, without a change of the form, by the values \rqn{5'} and \rqn{1.27}, respectively.

The case of a Gaussian state with a zero or linear phase considered in Sec.~\ref{IC5} [see Eq.~\rqn{4'}] follows from the present case for $b=0$.
The term proportional to ${\rm Im}\,A_w$ in Eq.~\rqn{1.27} arises due to the nonlinear (quadratic) phase.

It is often stated \cite{aha02,hos08,aha10} that the imaginary part of the weak value does not affect the probability distribution of the meter coordinate, and ${\rm Im}\,A_w$ can be observed only in the distribution of the meter momentum.
Equation \rqn{1.27} shows that this statement is not exact, since ${\rm Im}\,A_w$ enters the shift of the coordinate distribution for a general Gaussian wavefunction.
The same holds for a general (non-Gaussian) meter state, as discussed in Sec.~\ref{VIB}.

{\em c. Meter with $R=F$ and a Gaussian distribution of $F$.}
Consider a meter with $R=F$.
Similarly to Eq.~\rqn{1.4}, one obtains that
 \be
\langle\phi|\psi_f(F)\rangle\approx\langle\phi|\psi\rangle\exp(-i\gamma A_wF)\,\psi_{\rm M}(F).
 \e{1.23}
Let $F$ be a continuous variable with a Gaussian initial distribution $\Phi(F)=|\psi_{\rm M}(F)|^2$.
Then the distribution of $F$ after the measurement, given, up to a normalization factor, by the squared modulus of Eq.~\rqn{1.23}, is also a Gaussian, which differs from $\Phi(F)$ only by a shift of the center equal to
 \be
\Delta F_{\rm max}\,=\,\bar{F}_s-\bar{F}\,=\,2\gamma\,(\Delta F)^2\,{\rm Im}\,A_w.
 \e{160}
Equation \rqn{160} was proved in Sec.~\ref{IC5} [see Eq.~\rqn{5'}] for the special case, when $F=p$ and $\psi_{\rm M}(p)$ is a Gaussian state whose phase is constant or linear in $p$.
Here Eq.~\rqn{160} is shown to be valid for states $\psi_{\rm M}(F)$ with a Gaussian modulus and an arbitrary phase.
This result will be extended to the case of a mixed meter state $\rho_{\rm M}$ in Sec.~\ref{VIIB1}.

In the above formulas for weak PPS measurements we assumed that $F=p$ [except for Eq.~\rqn{160}].
It is easy to show that for $F=q$, the above formulas for $\bar{q}_s$, $\bar{p}_s$, $\Delta q_{\rm max}$, and $\Delta p_{\rm max}$ change according to the rule
 \be
p\leftrightarrow q,\quad\quad{\rm Re}\,A_w\,\rightarrow\, -{\rm Re}\,A_w.
 \e{1.44}
For example, Eqs.~\rqn{4} and \rqn{5} become \cite{aha88}
 \bea
&&\bar{p}_s-\bar{p}=-\gamma\,{\rm Re}\,A_w,\label{1.45}\\
&&\bar{q}_s-\bar{q}=2\gamma\,(\Delta q)^2\,{\rm Im}\,A_w.
 \ea{1.46}
Note that Eqs.~\rqn{5} and \rqn{1.46} are special cases of the second equality \rqn{160}.

It is worth mentioning that meters described in paragraphs {\em b} and {\em c} are well suited for weak PPS measurements, as argued below (see, e.g., Secs.~\ref{IIIA6''} and \ref{VIJ2}).

\subsection{Discussion of weak values}
\label{ID3}

\subsubsection{Interpretation of weak values in terms of probabilities}

Until now, we have assumed that the preselected state in a weak PPS measurement is pure.
It is of interest to extend the theory to a mixed preselected state.
In the case of a mixed preselected state $\rho$, the definition of the weak value becomes \cite{shp08,joh04a} (see Sec.~\ref{V} for the derivation)
 \be
A_w=\frac{(A\rho)_{\phi\phi}}{\rho_{\phi\phi}},
 \e{56}
where $\rho_{\phi\phi}=\langle\phi|\rho|\phi\rangle$.
For a qubit with a mixed preselected state, weak values are
always finite (see Sec.~\ref{VI} for more details).
However, for $d$-level systems with $d\ge3$, weak values can be unbounded even with a mixed preselected state, when $\rho$ has one or more zero eigenvalues.
In this case, the weak value diverges, when the pre- and post-selected states approach orthogonal subspaces \cite{wum09}.
Below the definition of the weak value will be further extended to the case of a general post-selection measurement (see Sec.~\ref{XII}).

The results \rqn{6} and \rqn{56} for weak values are surprising in the sense that one might expect, by analogy with the result \rqn{80}-\rqn{1.47} of weak standard measurements, that a weak PPS measurement yields the average of $A$ obtained in a strong PPS measurement,
 \be
A_s=\sum_ia_i\,P_{i|\phi},
 \e{142}
where $P_{i|\phi}$ are given by Eq.~\rqn{81} or \rqn{1.3}.
Equation \rqn{142} is a ``usual value'' of $A$, i.e., a real number
within the range of the eigenvalues of $A$, and hence it generally
significantly differs from the weak value.
Even so, as shown below, there are situations where the weak value
coincides with Eq.~\rqn{142}.

It is possible to obtain an expression for the weak value similar to Eq.~\rqn{142}.
Indeed, inserting Eq.~\rqn{1.12} into Eq.~\rqn{56} yields the weak value in a useful form
 \be
A_w=\sum_ia_i\,(\Pi_i)_w,
 \e{1.36}
where $(\Pi_i)_w$ is the weak value of $\Pi_i$,
 \be
(\Pi_i)_w=\frac{\langle\phi|\Pi_i\,\rho|\phi\rangle}{\langle\phi|\rho|\phi\rangle}.
 \e{1.34}
Equation~\rqn{1.36} implies that, in contrast to the results of strong PPS measurements, weak values do not depend on the measurement context, just as the standard (preselected only) measurements.
Indeed, due to the fact that $A_w$ is linear in $A$, the contributions to $A_w$ from different projectors in Eq.~\rqn{1.36} are independent of each other.

The quantity $(\Pi_i)_w$ in Eq.~\rqn{1.34} can be called the weak probability corresponding to the eigenvalue $a_i$ \cite{res04}.
Summing both sides of Eq.~\rqn{1.34} over $i$ and using the second equality in Eq.~\rqn{1.35}, we obtain that the weak probabilities are normalized \cite{res04},
 \be
\sum_i(\Pi_i)_w=1.
 \e{1.39}
The weak probability distribution $\{(\Pi_i)_w\}$ is generally nonclassical, in the sense that some weak probabilities may be greater than one or negative or even complex; such weak probabilities are unusual weak values of the projectors $\Pi_i$.
Nonclassical discrete \cite{res04,lun09,yok09} and continuous \cite{mir07} probability distributions were measured experimentally.
However, whenever all $(\Pi_i)_w$ are positive or equal to zero, the normalization \rqn{1.39} ensures that the set $\{(\Pi_i)_w\}$ is a classical probability distribution, and hence $A_w$ is a usual value \cite{hos10}.

Equations \rqn{1.36} and \rqn{1.39} imply that the weak value $A_w$ is the average of the observable $A$ over a nonclassical probability distribution which can assume negative and complex values.
It is often stated \cite{bru10,koc11} that the weak value should be understood as the mean value of the observable $A$ when weakly measured between the pre- and post-selected states.
However, we stress that, in view of Eq.~\rqn{1.36}, the ``mean value'' here is not a usual (classical) mean value, since it is taken over a nonclassical probability distribution.

An additional insight into weak values is provided by the fact that the weak probability $(\Pi_i)_w$ in Eq.~\rqn{1.36} can be interpreted as a (nonclassical) conditional probability of the measurement result $a_i$ given that the subsequent measurement result corresponds to the state $|\phi\rangle$ \cite{ste95}.
Indeed, the weak probability \rqn{1.34} can be recast in the form of Bayes' theorem
 \be
(\Pi_i)_w\equiv\tilde{P}_{i|\phi}=\frac{\tilde{P}_{i\phi}}{P_\phi}=\frac{{\rm Tr}\,(\Pi_\phi\Pi_i\,\rho)}{{\rm Tr}\,(\Pi_\phi\,\rho)},
 \e{1.53}
where $\Pi_\phi$ is given in Eq.~\rqn{1.51}.

The quantity
 \be
\tilde{P}_{i\phi}={\rm Tr}\,(\Pi_\phi\Pi_i\,\rho)
 \e{1.54}
in Eq.~\rqn{1.53} is known as the Kirkwood distribution \cite{kir33,joh07}.
It may be interpreted as the joint probability that two (generally non-commuting) observables have the values corresponding to the eigenstates $|a_i\rangle$ and $|\phi\rangle$. 
Indeed, recall that ideal quantum measurements yield an eigenvalue of an observable with the probability given by the average of the corresponding projection operator [see Eq.~\rqn{1.28}].
The quantity $\tilde{P}_{i\phi}$, being the average of a product of projection operators, is a direct generalization of this probability and can be considered as the joint probability.
Generally, $\tilde{P}_{i\phi}$ is a nonclassical probability, i.e., $\tilde{P}_{i\phi}$ may be negative or even complex, since the operator $\Pi_\phi\Pi_i$ is not Hermitian unless $\Pi_\phi$ and $\Pi_i$ commute with each other.

\subsubsection{Sufficient conditions for usual weak values}
\label{IE2}

The most interesting situations occur when the weak value is unusual.
It is not easy to provide necessary and sufficient conditions for unusual weak values.
However, it is easy to list some situations where weak values are usual, i.e., real and within the range of the eigenvalues of $A$.

In particular, when the preselected state is pure, weak values are usual in the following cases {\em a-d}:

{\em a.} In the case $|\phi\rangle=|\psi\rangle$, Eq.~\rqn{6} yields
 \be
A_w=A_{\psi\psi}=\bar{A}.
 \e{143}
Now $A_w$ is equal to the result $\bar{A}$ of a weak standard measurement (see Sec.~\ref{IB3}).
The reason for this is seen from the fact that in the present case, when measurements are sufficiently weak, the post-selection probability equals approximately $|\langle\phi|\psi\rangle|^2=|\langle\psi|\psi\rangle|^2=1$, i.e., the post-selected ensemble almost coincides with the total ensemble.
Hence, now there is practically no difference between weak PPS measurements and weak standard measurements.

{\em b.} 
When $|\psi\rangle$ is an eigenstate of $\hat{A}$ with eigenvalue $a_i$, then
 \be
A_w=a_i.
 \e{1.33}

{\em c.} 
When $|\phi\rangle$ is an eigenstate of $\hat{A}$ with eigenvalue $a_i$, then Eq.~\rqn{1.33} holds.\\
Note that for a pure PPS ensemble, statements {\em b} and {\em c} prove the unusual property, discussed in the last paragraph of Sec.~\ref{IC1}, for the case of weak measurements.

{\em d.} 
When, in a {\em pure} PPS ensemble, a strong measurement yields a particular eigenvalue $a_j$ of a variable $A$ with certainty, then $A_w=a_j$ \cite{aha91,aha02}.
Indeed, then $P_{i|\phi}=\delta_{ij}$ in Eq.~\rqn{1.11}, i.e., for $i\ne j$, $\langle\phi|\Pi_i|\psi\rangle=0$; hence due to Eq.~\rqn{6} $(\Pi_i)_w=0\ (i\ne j)$.
 The latter result implies, in view of Eq.~\rqn{1.39}, that $(\Pi_i)_w=\delta_{ij}$, and hence Eq.~\rqn{1.36} yields $A_w=a_j$.

Moreover, for a {\em mixed} preselected state weak values are usual in case {\em c} and also in the following cases {\em e-g}:

{\em e.} When $|\phi\rangle$ is an eigenstate of $\rho$ with a nonzero eigenvalue $\lambda$, then Eq.~\rqn{56} yields
 \be
A_w=\frac{A_{\phi\phi}\lambda}{\lambda}=A_{\phi\phi}.
 \e{1.55}
The present situation is an extension of case {\em a}.

{\em f.} 
When $\hat{A}$ commutes with $\rho$,
 \be
[\hat{A},\rho]=0,
 \e{157}
then $A_w$ is given by Eq.~\rqn{142}.
Indeed, taking into account that Eq.~\rqn{157} implies $[\Pi_i,\rho]=0$ and using the properties of the projection-valued measure in Eq.~\rqn{1.35}, we obtain that the weak probability in Eq.~\rqn{1.34} equals the probability $P_{i|\phi}$ in Eq.~\rqn{1.31} and hence Eq.~\rqn{1.36} coincides with Eq.~\rqn{142}.
The present situation is an extension of case {\em b}.

{\em g.} 
When the initial state is completely mixed, Eq.~\rqn{1.10}, i.e., the measurement is made on a post-selected only ensemble, then Eq.~\rqn{56} yields
 \be
A_w=A_{\phi\phi}.
 \e{1.38}
The present situation is a special case of paragraph {\em f}.
The result \rqn{1.38} is the same as for a weak standard measurement on an ensemble preselected only in the state $|\phi\rangle$.
A similar property holds also for strong measurements in post-selected only ensembles (see Sec.~\ref{IC2}).
The above two results are special cases of the time-symmetry relation, which states that measurements of any strength in an ensemble {\em post-selected} only in state $|\phi\rangle$ give the same results as in an ensemble {\em preselected} only in the same state (see the proof in Sec.~\ref{XIIIB}).

The usual weak values obtained in the above cases {\em b}-{\em d} and {\em f}-{\em g} (but not in {\em a}, {\em e}) coincide with $A_s$ in Eq.~\rqn{142}.
In cases {\em b}, {\em c}, {\em f}, and {\em g} the reason for this is that in these cases $\Pi_\phi=|\phi\rangle\langle\phi|$ or $\rho$ commutes with $\hat{A}$.
As shown in Sec.~\ref{XIIIB}, this implies that PPS measurements of any strength are formally equivalent to standard measurements with a suitable preselected state.
As a result, in these cases $A_w=A_s$ [see Eq.~\rqn{14.5}].

The requirement that all the above sufficient conditions for usual weak values should be violated provides a necessary (but not sufficient) condition for unusual weak values.
A thorough discussion of the conditions needed to obtain unusual weak values in a qubit is given in Sec.~\ref{VI}.

\subsubsection{Quantum interference in PPS measurements}
\label{IE3}

In order to understand the reason why the weak value generally sharply differs from Eq.~\rqn{142}, it is useful to compare the meter states for standard and PPS measurements just before the measurement of the pointer.
Equation \rqn{3} implies that in standard measurements, there is no
interference between the wave packets $\psi_{\rm M}(q-\gamma a_j)$,
since they are multiplied by mutually orthogonal vectors $|a_j\rangle$.
In contrast, for PPS measurements we obtain from Eq.~\rqn{3} that for $t_{\rm S}<t<t_{\rm M}$ the unnormalized meter wavefunction is
 \be
\langle\phi|\psi_f(q)\rangle=\sum_j\alpha_j\,\langle\phi|a_j\rangle\,
\psi_{\rm M}(q-\gamma a_j),
 \e{1.6}
i.e., the post-selection creates interference between the wave packets $\psi_{\rm M}(q-\gamma a_j)$.

In the case of a large $\gamma$, different wavepackets in Eq.~\rqn{1.6} do not overlap, and interference is practically absent in a measurement of $q$; therefore, a strong PPS measurement results in a projective measurement of $a_i$ with probability 
 \be
P_{i|\phi}=c|\alpha_i\,\langle\phi|a_i\rangle|^2=c|\langle\phi|a_i\rangle\,\langle a_i|\psi\rangle|^2.
 \e{1.52}
Here $c$ is a numerical factor, which can be obtained from the normalization condition $\sum_iP_{i|\phi}=1$ [cf.\ Eq.~\rqn{1.42}].
As a result, Eq.~\rqn{1.52} yields the ABL formula \rqn{1.3}.
In particular, this shows explicitly that the model of PPS measurements introduced in Sec.~\ref{IC4} contains strong PPS measurements as a special case.

In the opposite limit of a small $\gamma$, i.e., for a weak PPS measurement, different wavepackets significantly overlap, and interference strongly affects the measurement results.
Quantum interference is especially strong (and destructive),
when the pre- and post-selected states, $|\psi\rangle$ and
$|\phi\rangle$, are almost orthogonal.
This strong interference effect explains the striking difference between the weak value and Eq.~\rqn{142}.

Thus, it is the interference in the meter state that leads to unusual weak values.
Let us discuss the physical origin of this interference.
As mentioned above, after correlating the system and the meter, there is no interference in the meter state, as implied by \rqn{3}.
The reason for this is that different wave packets $\psi_{\rm M}(q-\gamma a_j)$ are ``tagged'' by the mutually orthogonal system states $|a_i\rangle$ and hence can be completely distinguished by measuring the observable $A$, even though they can significantly overlap each other. 
In contrast, a measurement of system $S$ with post-selection transforms the state of the system and the meter into a product state, thus eliminating any correlation between the system and the meter.
Now, unless the meter is measured, an observer even in principle cannot obtain information in which wave packet the meter is located, which results in interference between different wave packets in Eq.~\rqn{1.6}.

\subsubsection{Sum rule for weak values}

To provide a further insight into how complex and/or very large weak values and amplification result from post-selection, we mention an interesting property of weak values, which can be called the ``sum rule''.

The result of a weak standard measurement [cf.\ Eq.~\rqn{80}] is the expectation value of the linear-response results of weak PPS measurements [cf.\ Eq.~\rqn{4} or \rqn{1.27}] corresponding to all subensembles resulting from the post-selection measurement of the quantity $B$ (cf.\ Fig.~\ref{f11}).
The weights in the above expectation value are provided by the probabilities of different outcomes of the measurement of $B$, given in the linear-response approximation by $P_i^B=|\langle\phi_i|\psi\rangle|^2$, where $|\phi_i\rangle$ are the eigenvectors of $B$.
As a result, we obtain that $\bar{A}$ is an expectation value of the weak values corresponding to different subensembles.
This can be shown also by writing
 \be
\bar{A}=\langle\psi|\hat{A}|\psi\rangle=\sum_i\langle\psi|\phi_i\rangle\langle\phi_i|\hat{A}|\psi\rangle=\sum_i|\langle\phi_i|\psi\rangle|^2\,\frac{\langle\phi_i|\hat{A}|\psi\rangle}{\langle\phi_i|\psi\rangle},
 \e{1.57}
which yields the sum rule \cite{hos10}
 \be
\sum_iP_i^BA_{wi}=\bar{A},
 \e{1.58}
where
 \be
P_i^B=|\langle\phi_i|\psi\rangle|^2,\quad\quad A_{wi}=\frac{\langle\phi_i|\hat{A}|\psi\rangle}{\langle\phi_i|\psi\rangle}.
 \e{1.59}
Equation \rqn{1.58} shows explicitly that, though weak values can be complex and very large, their average over all subensembles is a usual value of $A$.
In the special case $B=A$, the sum rule \rqn{1.58} reduces to Eq. \rqn{1.47}.
The sum rule \rqn{1.58} is equivalent to the following two equalities,
 \be
\sum_iP_i^B\,\text{Re}\,A_{wi}=\bar{A},\quad\quad
\sum_iP_i^B\,\text{Im}\,A_{wi}=0.
 \e{1.61}
The first equality here was obtained in Ref.~\cite{aha05a}.

The magnitudes of the contributions from different subensembles in the sums in Eq.~\rqn{1.57} have the same upper bound, $|\langle\psi|\phi_i\rangle\langle\phi_i|A|\psi\rangle|\le|\langle\phi_i|A|\psi\rangle|\le||A||$, where $||A||$ is the norm of $\hat{A}$ given by the maximum of the magnitudes of the eigenvalues of $\hat{A}$ (here $||A||$ is assumed to be finite). 
In contrast, the weak values given by the fractions in the last sum in Eq.~\rqn{1.57} diverge when the overlap $|\langle\phi_i|\psi\rangle|$ tends to zero.
As a result, generally in subensembles where $|\langle\phi_i|\psi\rangle|$ is small, the pointer deflection, and correspondingly the signal-to-noise ratio (SNR) characterizing quantum noise, are strongly amplified. 
However, since this amplification is achieved in a relatively small number of measurements proportional to $|\langle\phi_i|\psi\rangle|^2$, the SNR with respect to quantum noise in weak PPS measurements is {\em of the same order} as in standard measurements \cite{hos08} (for details see Secs.~\ref{IIIH1} and \ref{IIIK2}).
Still, this amplification is very useful since it increases the SNR with respect to technical noise.

It is easy to see that the sum rules \rqn{1.58} and \rqn{1.61} hold also for a mixed initial state, as one should expect.
In this case, in Eqs.~\rqn{1.58} and \rqn{1.61} we have $\bar{A}={\rm Tr}\,(\hat{A}\rho)$ and [cf.\ Eqs.~\rqn{1.28} and \rqn{56}]
 \be
P_i^B=\langle\phi_i|\rho|\phi_i\rangle,\quad\quad A_{wi}=\frac{\langle\phi_i|\hat{A}\rho|\phi_i\rangle}{\langle\phi_i|\rho|\phi_i\rangle}.
 \e{1.60}

\subsection{Experimental realizations of weak PPS measurements}

The general scheme for performing PPS measurements was described in Sec.~\ref{IC4}.
Note that the choice of systems suitable for use as a meter is much broader for weak (standard and PPS) measurements than for strong measurements.
Indeed, to perform a projective measurement of a variable $A$ with $n_A$ unequal eigenvalues, one requires the meter to be a $d_{\rm M}$-level system with $d_{\rm M}\ge n_A$.
This is necessary for correlating $n_A$ eigenstates of $\hat{A}$ corresponding to the unequal eigenvalues with $n_A$ orthogonal meter states [cf.\ Eq.~\rqn{3}], in order to obtain from the measurement the maximum information allowed by the projection postulate.
In contrast, weak standard and PPS measurements provide such parameters as $\bar{A}$ and $A_w$, respectively, which contain much less information than ideal measurements.
In consequence, for weak standard and PPS measurements of {\em any} system one can use any other system as a meter, including a two-level system (a qubit), see Sec.~\ref{VIC}.
Moreover, the choice of meters is broader for weak PPS measurements than for weak standard measurements.
For example, meters with $R=F$ are not suitable for weak standard measurements (see below Secs.~\ref{IIA} and \ref{IIIA1}) but are suitable for weak PPS measurements [cf.\ Eqs.~\rqn{5} and \rqn{160}].

Below we overview the experimental studies on weak PPS measurements.

In their seminal paper \cite{aha88}, AAV proposed to perform weak PPS measurements using a Stern-Gerlach setup, where the shift of the transverse momentum of the particle, translated into a spatial shift, yields the outcome of the spin-1/2 measurement.
The meter now is a particle performing one-dimensional free motion, the input and output variables being
 \be
F=q,\quad R=p.
 \e{1.37}
Post-selection of the spin state in a certain direction can be performed by another (now strong) Stern-Gerlach coupling which splits the particle beam.
The analysis of the required beam provides the result of the weak PPS measurement.
The meters associated with the two measurements should be implemented by two independent systems (here two transverse translational degrees of freedom).
This is achieved by arranging the spatial shifts due to the two Stern-Gerlach devices to be orthogonal to each other.

An optical analog of the above Stern-Gerlach experiment was proposed in Ref.~\cite{duc89} and realized in Ref.~\cite{rit91}, which is the first experimental study of weak values.
In this experiment, the system of interest is the optical polarization rather than the spin of a spin-1/2 particle.
The polarization of a light beam is weakly coupled to a transverse degree of freedom of the beam by a birefringent plate, whereas the pre- and post-selection are performed by polarization filters.
In this setup, the meter variables are given by Eq.~\rqn{97} rather than\footnote{
We remind that the results of weak PPS measurements with the meter variables \rqn{97} are very similar to the results with the meter variables \rqn{1.37}, as follows from the discussion in the last paragraph of Sec.~\ref{IC6}.}
Eq.~\rqn{1.37}.
Moreover, in Ref.~\cite{rit91} $\hat{A}$ is the projector on a state with a linear polarization, and $\gamma$ is the birefringence-induced separation.
Figure 2(b) in Ref.~\cite{rit91} and Eq.~\rqn{1.22} imply that Ritchie et al. \cite{rit91} obtained a real unusual weak value $A_w\approx20$, which is very far outside the range of the eigenvalues $(0,1)$.

Knight and Vaidman \cite{kni90} proposed a slightly different optical realization of the AAV experiment, which uses a birefringent prism instead of a plate; as a result, the meter variables are given now by Eq.~\rqn{1.37}.
This experiment was performed in Ref.~\cite{par98}, where results similar to those in Ref.~\cite{rit91} were obtained.

\begin{figure}[htb]
\begin{center}
\includegraphics[width=12cm]{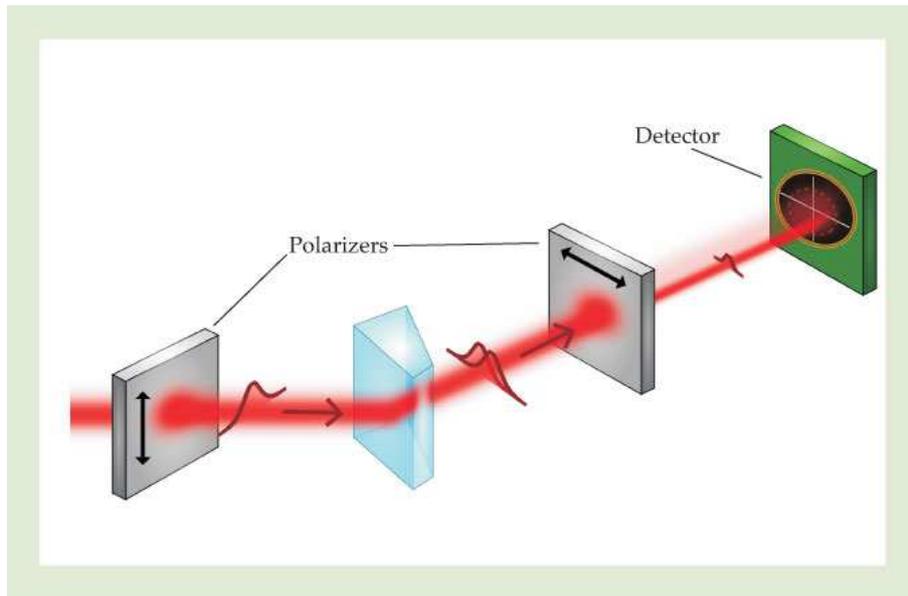}
\end{center}
\caption{(color online). Schematic diagram of the experiment, reported in Ref.~\cite{hos08}, for performing a weak PPS measurement of photon polarization. 
The figure is reprinted from Ref.~\cite{aha10}.
}
 \label{f12}\end{figure}

Figure \ref{f12} depicts a typical scheme of a weak PPS measurement \cite{hos08}.
In this experiment, the system $S$ is the photon polarization, whereas the meter is a transverse degree of freedom of the photon, the meter variables $F$ and $R$ being given by Eq.~\rqn{97}.
In Fig.~\ref{f12}, the pre- and post-selection are performed by the polarizers, which are almost crossed, whereas the prism in between provides a coupling of the system and the meter due to the spin Hall effect of light \cite{ono04,bli06,bli08,nor08}.
The initial wave packet along the meter coordinate is shown on the left in Fig.~\ref{f12}.
After passing through the prism, the wave packet becomes a superposition of two slightly shifted wave packets with mutually orthogonal polarizations [cf.\ Eq.~\rqn{3}].
The post-selection produces a strong destructive interference of the wave packets [cf.\ Eq.~\rqn{1.6}], resulting in a wave packet with a significantly reduced intensity but with a strongly enhanced shift.

In the case studied in Ref.~\cite{hos08}, the above simple picture is complicated somewhat by the presence of a nonzero meter Hamiltonian (see the discussion of the Hamiltonian effects in Sec.~\ref{VIB4}).
Using weak PPS measurements, Hosten and Kwiat \cite{hos08} succeeded to detect the prism-induced wave-packet shift of 1 angstrom and thus to measure the very weak coupling produced by the spin Hall effect of light.
This experiment resolved long-standing controversies concerning the polarization-dependent transverse shifts of light beams \cite{ono04,bli06,fed55,imb72} and confirmed theoretical predictions by Bliokh and Bliokh \cite{bli06,bli07}.
A classical interpretation of this experiment was derived in Ref.~\cite{aie08}.
Further extensions of the weak-measurement approach to the optical spin Hall effect shifts, including employment of both real and imaginary weak values, were discussed in \cite{qin11,dena,gora}.

Until now, a large body of experimental work on weak PPS measurements has been made
\cite{rit91,sut93,sut95,par98,sol04,bru04,res04,pry05,wan06,mir07,%
hos08,lun09,yok09,dix09,how10,sta09,sta10a,cho10,gog11,iin11,tur11,
hog11,koc11,lun11}, and a great variety of physical systems, couplings, and experimental setups have been used.
Most of the experiments were performed in optics, except for one \cite{sut93}, which was done in NMR.
Those optical experiments that use low-intensity light allowing for detection of single photons are evidently non-classical \cite{sut93,pry05,wan06,lun09,yok09,gog11,iin11,koc11,lun11}, whereas other experiments, which use intense optical beams \cite{rit91,sut95,par98,sol04,bru04,res04,mir07,%
hos08,dix09,how10,sta09,sta10a,cho10,tur11,hog11}, admit both classical and quantum interpretations.

A linear classical optical experiment can always be interpreted quantum-mechanically, in terms of single photons.
Indeed, photons in laser beams are prepared in a coherent state and behave independently in linear optical systems; hence the intensity measurements one performs are guaranteed to be the same for coherent states as for single-photon states \cite{gla65}.

Irrespective of the interpretation adopted, the ``weak-value approach'' for designing experiments is not conventional in classical physics and thus can lead to new results for classical systems.
For example, the enhanced shift of the light-beam distribution in the coordinate or momentum space by passage through a (post-selection) filter is essentially a new classical interference effect \cite{kni90,par98}.
Furthermore, recently weak PPS measurements were applied in new classical optical interferometric techniques for beam-deflection \cite{sta09,tur11,hog11}, phase \cite{sta10x}, and frequency \cite{sta10a} measurements.

Note that the theory in Ref.~\cite{sta10x} is purely classical (though the earlier version \cite{sta09ar} of the paper contains a brief discussion in terms of the weak value).
In Sec.~\ref{XI} we provide a quantum interpretation of the experiment in Ref.~\cite{sta10x}; this interpretation is based on the nonlinear theory of weak PPS measurements, which is developed below.

The systems for which weak values were measured involved spin 1/2 \cite{sut93}, photon polarization 
\cite{rit91,sut95,par98,sol04,bru04,pry05,wan06,hos08,cho10,gog11,%
iin11}, photon which-path states in a Sagnac \cite{dix09,how10,sta09,sta10a} and a three-rail Mach-Zehnder \cite{res04} interferometers, a transverse translational degree of freedom of the photon \cite{mir07,koc11,lun11}, and which-path states of two photons in a pair of Mach-Zehnder interferometers \cite{lun09,yok09}.

The meters used in the experiments included, in particular, (a) systems with continuous variables $F$ and $R$: a transverse \cite{rit91,par98,res04,mir07,hos08,dix09,how10,sta09,sta10a,cho10} and the longitudinal \cite{bru04} translational degrees of freedom of the photon, (b) a qubit: spin 1/2 \cite{sut93}, photon polarization \cite{pry05,gog11,koc11,lun11}, and which-path states in a Mach-Zehnder interferometer \cite{sut95,iin11}, and (c) two qubits: the positions of two photons on the two sides of a beam splitter \cite{wan06} in the Hong-Ou-Mandel interferometer \cite{hon87} and the polarizations of the two photons \cite{lun09,yok09}.

The coupling between the system and the meter was created, in particular, by a tilted birefringent plate \cite{rit91,cho10}, a birefringent prism \cite{par98}, retarders (a Soleil-Babinet compensator \cite{sut95}, a birefringent optical fiber \cite{bru04}, and birefringent \cite{wan06,koc11} and half-wave \cite{iin11} plates), a tilted glass plate \cite{res04,mir07}, a tilted mirror \cite{dix09,how10,sta09}, a glass prism \cite{sta10a}, the spin Hall effect of light \cite{hos08}, a nondeterministic photon-entangling circuit \cite{pry05,gog11}, a polarization rotator \cite{lun09,yok09,lun11}, and an Ising-type spin coupling \cite{sut93}.

In the above experiments, the system and the meter were prepared in pure states, except for Ref.~\cite{cho10}, where effects of a mixed meter state were studied.

As discussed above, in the present paper we adopt a conventional approach to weak PPS measurements, based on an extension of the von Neumann model (see Fig.~\ref{f8}).
For completeness, we mention that there exist also somewhat different approaches to weak values, which do not involve explicitly the von Neumann model and employ instead such theoretical tools as POVM and measurement operators \cite{wis02,wil08,ash09}, negative probabilities \cite{joh04,sok07}, and contextual values of observables \cite{dre10}.
In particular, weak values for continuous measurements in quantum optics \cite{wis02} and solid-state systems \cite{wil08} were considered, and
an experiment on cavity quantum electrodynamics \cite{fos00} was interpreted \cite{wis02} in terms of weak values.
A more detailed discussion of these approaches is out of the scope of the present paper.

\subsection{Applications of weak PPS measurements}

Weak PPS measurements possess a number of unique features, which
make possible a host of important applications.
Such measurements can play, at least, two different roles.

First, weak PPS measurements can be employed with the aim of obtaining a weak value of an observable.
The fact that a weak PPS measurement disturbs the system only
slightly in the interval between the pre- and post-selection allows
one to obtain information about the undisturbed behavior of the system
in that interval.
Therefore, weak values have been used to shed new light on a great
variety of quantum phenomena, especially those related to fundamentals
of quantum mechanics.
An example of such phenomena are quantum paradoxes discussed below.

Second, weak PPS measurements can produce strong amplification of the pointer deflection \cite{aha88}, owing to the fact that the weak value can become arbitrarily large when the overlap of the initial an final states $\langle\phi|\psi\rangle$ is sufficiently small, cf. Eq.~\rqn{4}.
Correspondingly, in its second role, a weak PPS measurement acts as a peculiar amplification scheme, rather than a ``proper'' measurement of an observable.
This amplification is one of the most important features of weak PPS
measurements, since it can be exploited for different uses, e.g., to produce superluminal and slow light propagation \cite{sol04,bru04,wan06,aha90,ste95,aha02}.
Moreover, the amplification can yield experimental sensitivity beyond the detector resolution and thus can be used for measuring weak physical effects responsible for the coupling between the system and the meter, as well as for precision measurements of other parameters characterizing the system and the meter.

Furthermore, the weak value can be a complex number, which has important consequences for weak PPS measurements.
It is interesting that a complex weak value is always unusual, irrespective of its magnitude, whereas a real weak value becomes unusual only when it is outside the spectrum of the observable.
The terms proportional to Im$\,A_w$ entering Eqs.~\rqn{5} and \rqn{1.27} have no analogues in standard measurements.
As a result, in particular, the class of meters which can be used for weak PPS measurements is broader than the class of meters appropriate for weak standard measurements.
For instance, meters with commuting $F$ and $R$ (and, in particular, with $F=R$) cannot be used for standard measurements, but they can be used for weak PPS measurements [cf.\ Eqs.~\rqn{5} and \rqn{160}]; this point is discussed in more details in subsequent sections.
Moreover, the terms involving Im$\,A_w$ are proportional to a factor characterizing the classical correlation between $F$ and $R$.
This correlation provides an independent source of enhancement in addition to the amplification due to $A_w$ mentioned above \cite{hos08}, as discussed in detail in subsequent sections.

Experiments on weak PPS measurements have involved various interesting applications, all of them being related to unusual weak values.
In particular, such measurements were used to elucidate quantum retrodiction (i.e., ``prediction'' about the past) paradoxes with pre- and post-selection, such as the three-box problem \cite{aha91} and Hardy's paradox \cite{har92}.
Such problems show vividly that in quantum mechanics it is difficult to answer the question what is the value of a physical quantity in the middle of a time evolution, especially for a PPS ensemble.
Weak PPS measurements are very well suited to answer such questions by two reasons.
First, strong measurements utterly change the time evolution, and hence their results are loosely related to the evolution in question, whereas weak measurements almost do not affect the evolution.
Second, weak values are context-independent, in contrast to the results of strong PPS measurements, as mentioned after Eq.~\rqn{1.34}.

Consider, for instance, the three-box problem \cite{aha91}.
One can ask the question, in which box the particle is located in between the pre- and post-selection.
Strong PPS measurements, being contextual, do not provide an unambiguous answer to this question [cf.\ Eqs.~\rqn{1.16}-\rqn{1.19}].
In contrast, in weak PPS measurements, which are not context dependent, the ``weak probability'' $(\Pi_i)_w$ (with $\Pi_i=|i\rangle\langle i|$) is determined uniquely for each state $|i\rangle$.
In particular, consider the above case, when the pre- and post-selected states are  Eqs.~\rqn{1.13} and \rqn{1.14}, respectively.
In view of paragraph {\em d} in Sec.~\ref{IE2} and Eqs.~\rqn{1.16} and \rqn{1.18}, we obtain that 
 \be
(\Pi_1)_w=(\Pi_2)_w=1.
 \e{1.40}
Moreover, Eq.~\rqn{1.40} and the normalization condition \rqn{1.39}, which becomes now $(\Pi_1)_w+(\Pi_2)_w+(\Pi_3)_w=1$, yield
 \be
(\Pi_3)_w=-1.
 \e{1.41}
Thus, the outcomes \rqn{1.40} of weak PPS measurements are consistent with the paradoxical results \rqn{1.16} and \rqn{1.18}, rather than with Eq.~\rqn{1.19}.
The outcome \rqn{1.41} for box 3 is no less paradoxical, since it is a negative weak probability and hence an unusual weak value.
The results \rqn{1.40} and \rqn{1.41} were verified experimentally in Ref.~\cite{res04}.

\begin{figure}[htb]
\begin{center}
\vspace{-.4cm}
\includegraphics[width=13cm]{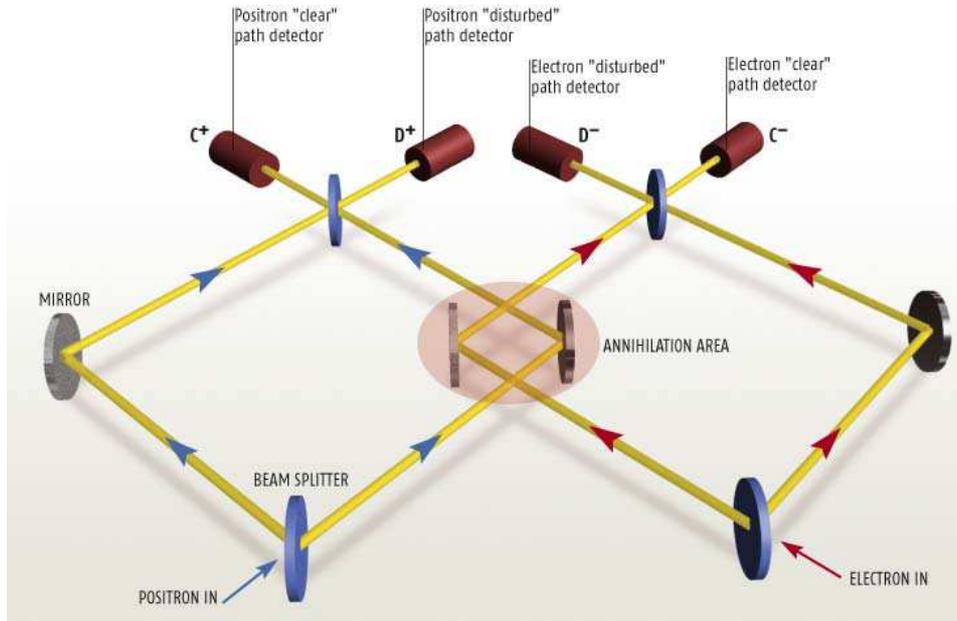}
\end{center}
\caption{(color online). Schematic diagram of Hardy's gedanken experiment.
The setup consists of a pair of overlapping Mach-Zehnder interferometers (MZI).
In each MZI, there is an arm, overlapping with the other MZI, and a non-overlapping arm.
If the arms have the right lengths, in the absence of the other particle, the electron (positron) entering the MZI, as shown in the figure, can only emerge towards the detector $C^-$ ($C^+$).
When both particles enter the setup simultaneously, the presence of one of them in an overlapping arm disturbs the motion of the other particle---the same effect as in interaction-free measurements; as a result, the latter particle may trigger the corresponding detector $D^+$ or $D^-$.
Assuming the existence of ``realistic trajectories'' leads to Hardy's paradox, as follows.
Quantum mechanics predicts a nonzero probability of simultaneous triggering $D^+$ and $D^-$.
One should infer that in this case both electron and positron have gone through the overlapping arms.
However, this is impossible due to the fact that, when traveling along the overlapping arms, electron and positron should meet in the annihilation area and destroy each other.
The figure is reprinted from Ref.~\cite{bro03}.
}
 \label{f13}\end{figure}

Hardy's paradox is a contradiction between classical reasoning and the outcome of measurements on an electron and a positron in a pair of overlapping Mach-Zehnder interferometers (MZI) \cite{har92}.
It is a variation on the concept of interaction-free measurements \cite{eli93}.
The scheme of Hardy's gedanken experiment and a description of Hardy's paradox are given in Fig.~\ref{f13}.
In the case of interest, when the detectors $D^+$ and $D^-$ are triggered simultaneously, the joint probabilities of different paths taken by the two particles in the interferometers can be obtained with the help of weak PPS measurements \cite{aha02a}.
Namely, let $P_{ijw}$ denote the weak probability that the positron and electron go through the arms $i$ and $j$, respectively.
Here $i,j={\rm O,N}$, where O (N) corresponds to the overlapping (non-overlapping) arm of the respective MZI.
Then the theory predicts that \cite{aha02a}
 \be
P_{{\rm OO}w}=0,\quad P_{{\rm ON}w}=P_{{\rm NO}w}=1,\quad P_{{\rm NN}w}=-1.
 \e{1.56}
Here the values of the three latter probabilities are paradoxical.
As in the three-box problem, two of these probabilities equal one, whereas the third probability is negative.
Equation \rqn{1.56} was verified in experiments on photons performed in Refs.~\cite{lun09,yok09}.

An important feature of the experiments \cite{lun09,yok09} on Hardy's paradox is that there the authors performed joint weak measurements, i.e., obtained weak values of two-particle variables, which are products of one-particle variables.
These measurements were performed by two different methods: by calculating the correlations between the pointer variables for the two photons \cite{lun09} and by using an entangled state of the two qubits (the photon polarizations) comprising the meter \cite{yok09}.
There is also a proposal of performing a joint weak PPS measurement of two qubits with a one-qubit meter, using trapped ions \cite{wum09}.
Weak PPS measurements of multiparticle observables can have important applications in the future, e.g., for the probing and characterization of one-way quantum computing, which involves pre- and post-selected multiparticle states, such as cluster states \cite{rau01,rau03,nie06,tan06,you07,tan09}.

Furthermore, the weak-value approach was employed to elucidate the complementarity between wave and particle behavior in Young's double-slit experiment \cite{mir07}.
The measured weak value of the momentum-transfer distribution took both positive and negative values.
As a result, this distribution was shown to be compatible with two conflicting claims concerning the complementarity \cite{scu91,sto94}.
Recently, weak PPS measurements were applied to obtain information on the wavefunction of a quantum particle \cite{koc11,lun11}.
In particular, the proposal in Ref.~\cite{wis07} was realized in Ref.~\cite{koc11}, where weak PPS measurements were used to obtain average trajectories of single photons in a double-slit interferometer.
In Ref.~\cite{lun11}, the transverse spatial wavefunction of a single photon was directly measured with the help of weak PPS measurements.
Lundeen et al.~\cite{lun11} also showed how their technique can be extended for directly measuring the quantum state of an arbitrary quantum system. 

The shift of the pointer distribution due to weak PPS measurements (see Secs.~\ref{IC5}, \ref{IC6}, and \ref{VIB}) in the cases when the weak value is unusual can result in both superluminal propagation and slow light, as was demonstrated experimentally in Refs.~\cite{sol04,bru04,wan06}; see also discussions in \cite{aha90,ste95,aha02}.
Applications of weak values to optical communications were discussed in Refs.~\cite{bru03,sol04,bru04}.
Moreover, weak values are closely related \cite{ste95} to the method of measuring the tunneling time, which involves the so called ``Larmor time'' (a recent review on the tunneling time and superluminality can be found in Ref.~\cite{win06}).

As mentioned above, one of the most important applications of weak PPS measurements is strong amplification of the measurement result in comparison to standard measurements.
In particular, this amplification allows one to measure very small values of the coupling strength $\gamma$ and thus to obtain information on weak effects responsible for the system-meter coupling, as, e.g., small differences in the indices of refraction  \cite{rit91,sut95,par98,bru04,wan06,cho10}, the spin Hall effect of light \cite{hos08}, a mirror tilt \cite{dix09,how10,sta09,tur11,hog11}, and an Ising-type spin coupling \cite{sut93}.
In Ref.~\cite{sta10a} an amplification factor of 80 was achieved in optical frequency measurements; the method developed there can be used in high-resolution relative frequency metrology and for laser locking.

In the early studies, complex weak values attracted much less attention than real weak values.
In particular, until recently, weak PPS measurements were performed only with real weak values.
However, in recent years the situation began to change.
Jozsa \cite{joz07} revealed theoretically a term proportional to Im$\,A_w$ in the coordinate deflection.
Moreover, recently a number of experiments using imaginary weak values were performed \cite{hos08,dix09,how10,sta09,sta10a,tur11,hog11}.
In such experiments, the amplification is enhanced in comparison to experiments with real weak values.
Namely, in experiments with imaginary weak values, the total amplification coefficient is a product of the (proper) amplification coefficient due to a large weak value and the enhancement factor due to correlation between the meter variables $F$ and $R$ (see the discussion in Secs.~\ref{IIIH} and \ref{VIA}).
The total amplification does not increase the SNR due to the quantum noise, but can strongly reduce the effects of technical errors \cite{hos08} (see Sec.~\ref{IIIB}).

Using imaginary weak values, very precise measurements were made.
For example, Hosten and Kwiat \cite{hos08} detected a light-beam
displacement of 1 angstrom by amplifying it by a factor of $10^4$,
whereas Dixon et al.\ \cite{dix09} measured a mirror-actuator travel of $\sim10$ fm and the mirror angular deflection of 400 frad.
Turner et al. \cite{tur11} adjusted the scheme of Ref.~\cite{dix09} for the use in torsion balance experiments in gravity research; they demonstrated picoradian accuracy of deflection measurements.
Hogan et al. \cite{hog11} included a folded optical lever into the scheme of Ref.~\cite{dix09} and achieved a record angle sensitivity of 1.3 prad/$\sqrt{\rm Hz}$; their scheme is potentially applicable for gravitational wave detection.
Brunner and Simon \cite{bru10} showed that in measuring small longitudinal phase shifts, the use of imaginary weak values has the potential to outperform standard interferometry by several orders of magnitude, whereas standard interferometry greatly outperforms weak PPS measurements involving real weak values; see also the discussion in Ref.~\cite{lix11}.

Unfortunately, for a given value of the coupling strength $\gamma$, the amplification cannot be made arbitrarily strong, since the linear-response results, such as Eqs.~\rqn{4}, \rqn{5}, and \rqn{1.27}, fail when $|\gamma A_w|$ becomes sufficiently large [cf.\ the condition \rqn{1.9}].
It would be of interest to develop a theory of weak PPS measurements which holds to all orders in $\gamma A_w$, since such a theory would allow one to perform measurements under optimal conditions, where the magnitude of the pointer deflection is close to the maximum value.
Such a theory is developed in subsequent sections.

\subsection{General theory of weak PPS measurements}
\label{IIH}

In Sec.~\ref{IIA'} and henceforth we consider PPS and standard measurements of arbitrary strength.
The emphasis is made on developing a general theory of weak PPS measurements (see especially Sec.~\ref{III'} and henceforth).
The present theory of weak PPS measurements extends the existing theory in several respects.
 In particular,
\begin{itemize*}
\item[(i)] 
we derive results valid for {\em any} value of $\gamma A_w$,
\item[(ii)] 
we obtain formulas for {\em arbitrary} meter variables $F$ and $R$,
\item[(iii)] 
we consider arbitrary, pure and mixed, initial states of the system and the meter,
\item[(iv)] 
we discuss both the average and the distribution of the pointer variable $R$.
\end{itemize*}

Our main results include the following:
\begin{enumerate*}
\item 
We derive a simple and general formula for the average value of the meter pointer deflection, which holds for all orders in the weak value and for arbitrary system and meter.
\item 
We reveal that there are three qualitatively different regimes of weak PPS measurements.
In addition to the AAV linear-response regime, there exist also the inverted region (the limit of very large weak values) and the intermediate, strongly-nonlinear regime.
\item 
The optical experiment reported in Ref. \cite{sta10x} is interpreted quantum-mechanically as a weak PPS measurement in the inverted region.
\item 
The optimal conditions for weak PPS measurements are obtained in the strongly-nonlinear regime, since then the pointer deflection is maximized, and correspondingly the ensemble size is minimized.
\item 
The maximal pointer deflection
is {\em independent of the coupling strength} $\gamma$, being
typically of the order of the initial uncertainty of the pointer $R$.
\item 
We propose procedures for measuring the coupling strength
$\gamma$ and the weak value $A_w$ in the nonlinear regime.
\item 
The amplification due to weak PPS measurements is generally
a product of the proper amplification, which increases the quantum SNR in post-selected systems, and the enhancement which does not change the quantum SNR.
\item 
The measurement enhancement arises due to the correlation between the meter variables $F$ and $R$.
Moreover, we find that generally the canonically conjugate variables $q$ and $p$ are correlated whenever the phase of the state in the $p$ or $q$ representation is nonlinear in the corresponding variable.
\item 
In the case of a mixed preselected state, in addition to $A_w$, an associated weak value $A_w^{(1,1)}$ is required to describe weak PPS measurements.
\item 
Beyond the linear response, weak PPS measurements significantly depend on the average value $\bar{F}$ of the meter variable $F$.
In particular, the optimal regime is significantly different for $|\bar{F}|\lesssim\Delta F$ and $|\bar{F}|\gg\Delta F$, the amplification being proportional to $\bar{F}$ for $\bar{F}\gg\Delta F$.
\item 
We identified meters which are optimal or just efficient for weak standard and weak PPS measurements.
All meters that are efficient for weak standard measurements are also efficient for weak PPS measurements; however, the converse is not true.
\item 
For continuous-variable meters, we obtain the shift of the
maximum of the pointer distribution for a broad class of initial
states of the meter, and not only for a real Gaussian state, as was
done previously.
\end{enumerate*}

The general formulas for different regimes of weak PPS measurements, obtained in the present paper, are listed in Table \ref{t8}.

Approaches to weak PPS measurements beyond the linear response, resembling some aspects of the present nonperturbative theory, were developed for the special case of a continuous-variable meter in Refs.~\cite{ges10,sim11,wul11}.

\section{Theory of standard measurements of arbitrary strength}
 \label{IIA'}

Here we provide a general theory of standard measurements of arbitrary strength.
Moreover, weak standard measurements are considered in detail.

\subsection{General formulas for standard measurements}
 \label{IIA}

Before we consider PPS measurements, let us discuss
weak measurements without post-selection.
Here we extend the results of Sec.~\ref{IB} to the case of any meter observables $F$ and $R$ and of any, generally mixed, initial states of the system and the meter. 
 As mentioned above, such measurements are performed using the
standard (von-Neumann-like) scheme of quantum measurements (see
Fig.~\ref{f9}).
 We assume that the coupling of the system and the meter is given by
the Hamiltonian \rqn{1}, whereas we neglect the free Hamiltonians of
the system and the meter.
We also assume a general product initial condition: at
$t=0$ the state of the system and the meter is $\rho\otimes\rho_{\rm
M}$, where the system and meter states $\rho$ and $\rho_{\rm M}$,
respectively, can be pure or mixed.
 The average value of a meter observable $R$ (the ``pointer'')
at any time $t\ge t_{\rm f}$ is given by
 \be
\bar{R}_f={\rm Tr}\,[(I_{\rm S}\otimes\hat{R})U(\rho\otimes\rho_{\rm
M})U^\dagger],
 \e{74}
where $\hat{R}$ is the Hermitian operator representing the observable $R$, $U$ is given by Eq.~\rqn{2}, and $I_{\rm S}$ is the identity
operator for the system.

It can be seen from Eq.~\rqn{74} that, irrespective of the
measurement strength $\gamma$, $\bar{R}_f$ in Eq.~\rqn{74} equals
the initial value $\bar{R}={\rm Tr}\,(\hat{R}\rho_{\rm M})$, i.e., a
measurement on the system cannot be performed, whenever
 \be
[\hat{F},\hat{R}]=0\ \ {\rm or}\ \ [\hat{F},\rho_{\rm M}]=0.
 \e{217}
Indeed, the first equality in Eq.~\rqn{217} implies that $[(I_{\rm S}\otimes\hat{R}),U]=0$, so that Eq.~\rqn{74} yields
 \be
\bar{R}_f={\rm Tr}\,[U(I_{\rm S}\otimes\hat{R})(\rho\otimes\rho_{\rm
M})U^\dagger]={\rm Tr}\,[(I_{\rm S}\otimes\hat{R})(\rho\otimes\rho_{\rm M})U^\dagger U]={\rm Tr}\,[(I_{\rm S}\otimes\hat{R})(\rho\otimes\rho_{\rm M})]={\rm Tr}\,\rho\,{\rm Tr}\,(\hat{R}\rho_{\rm M})=\bar{R}.
 \e{2.3}
Moreover, let us show that $\bar{R}_f=\bar{R}$ also in the case when the second equality in Eq.~\rqn{217} holds.
Now the meter initial state is either an eigenstate of $\hat{F}$ or, more generally, $\rho_{\rm M}=\sum_F\lambda_F|F\rangle\langle F|$.
Here $|F\rangle$ is an eigenstate of $\hat{F}$ corresponding to the eigenvalue $F$, so that $\hat{F}|F\rangle=F|F\rangle$.
Then Eq.~\rqn{74} yields
 \be
\bar{R}_f=\sum_F\lambda_F\,{\rm Tr}\,[(I_{\rm S}\otimes\hat{R})U(\rho\otimes|F\rangle\langle F|)U^\dagger]
=\sum_F\lambda_F\,{\rm Tr}\,(\tilde{U}\rho\tilde{U}^\dagger)\,{\rm Tr}\,(\hat{R}|F\rangle\langle F|)
=\sum_F\lambda_F\,{\rm Tr}\,(\hat{R}|F\rangle\langle F|)
={\rm Tr}\,(\hat{R}\rho_{\rm M})=\bar{R}.
 \e{2.4}
Here we took into account that now Eq.~\rqn{2} yields $U=\tilde{U}\otimes I_{\rm M}$ with $\tilde{U}=\exp(-i\gamma F\,\hat{A})$, whereas ${\rm Tr}\,(\tilde{U}\rho\tilde{U}^\dagger)={\rm Tr}\,(\rho\tilde{U}^\dagger\tilde{U})={\rm Tr}\,\rho=1$.
Equation \rqn{217} is equivalent to the statement that a necessary condition for standard measurements to be possible is
 \be
[\hat{F},\hat{R}]\ne0\ \ {\rm and}\ \ [\hat{F},\rho_{\rm M}]\ne0.
 \e{2.5}

\subsection{Weak standard measurements}
\label{IIB}

Equation \rqn{74} can be expanded in powers of $\gamma$, using the
familiar expansion
 \be
U^\dagger CU=C+i\gamma[D,C]+\frac{i^2\gamma^2}{2!} [D,[D,C]]+\dots
 \e{50}
with $D=\hat{A}\otimes\hat{F}$ and $C=I_{\rm S}\otimes\hat{R}$.
As a result, we obtain that the average pointer deflection is given by
 \be
\bar{R}_f-\bar{R}=i\gamma\bar{A}\,\overline{[F,R]} +\frac{i^2\gamma^2\:\overline{A^2}}{2!}\:\overline{[F,[F,R]]}+\dots,
 \e{2.1}
where $\bar{A}={\rm Tr}\,(\hat{A}\rho)$, $\bar{R}={\rm
Tr}\,(\hat{R}\rho_{\rm M})$, ${\overline{[R,F]}=\rm
Tr}\,([\hat{R},\hat{F}]\rho_{\rm M})$, etc.

 For weak coupling, i.e., a small $\gamma$, we can retain in Eq.~\rqn{2.1} only the terms up to the first order in $\gamma$, yielding
 \be
\bar{R}_f-\bar{R}=\gamma\bar{A}\,{\rm Im}\,\overline{[R,F]}.
 \e{75}
Equation \rqn{75} is an extension of the AAV result \rqn{80} for weak
standard measurements to an arbitrary pair of meter variables $R$ and
$F$ and arbitrary initial states of the system and the meter.
For canonically conjugate meter variables, $[\hat{R},\hat{F}]$ is a constant, and the higher-order terms neglected in Eq.~\rqn{75} vanish, as follows from Eq.~\rqn{2.1}, i.e., Eq.~\rqn{75} is exact.
 An example of this case is provided by Eq.~\rqn{80}.
However, generally the higher-order terms do not vanish, and
Eq.~\rqn{75} holds only for sufficiently weak measurements.

Equation \rqn{75} implies that a weak standard measurement can be performed if and only if
 \be
\overline{[R,F]}\ne0.
 \e{2.2}
This necessary and sufficient condition can be shown to be more restrictive than the necessary condition \rqn{2.5}, as one should expect.

\subsection{Optimal and efficient meters for weak standard measurements}
\label{IIIC}

Let us estimate the magnitude of the average pointer deflection in Eq.~\rqn{75}.
According to the Heisenberg-Robertson uncertainty relation
\cite{rob29},
 \be
\Delta F\:\Delta R \ge|\,\overline{[R,F]}\,|/2,
 \e{310}
where $\Delta R$ and $\Delta F$ are the uncertainties of $R$ and $F$
in the state $\rho_{\rm M}$,
so that Eq.~\rqn{75} implies
 \be
|\bar{R}_f-\bar{R}|\le2\,|\gamma\bar{A}|\,\Delta F\,\Delta R.
 \e{294}

As follows from Eq.~\rqn{294}, the upper bound for the magnitude of the average pointer deflection for given $|\gamma\bar{A}|$, $\Delta F$, and $\Delta R$ is 
 \be
|\bar{R}_f-\bar{R}|_{\rm max}=2|\gamma\bar{A}|\:\Delta F\:\Delta R.
 \e{2.7}
This upper bound is achieved when $|\overline{[R,F]}|$ is maximum for given $\Delta F$ and $\Delta R$, that is, when the Heisenberg-Robertson uncertainty relation in Eq.~\rqn{310} becomes an equality,
 \be
|\overline{[R,F]}|=2\Delta F\:\Delta R.
 \e{2.6}
Meters satisfying the condition \rqn{2.6} are {\em optimal for weak standard measurements}.
The class of such meters includes, in particular, meters where $F$ and $R$ are canonically conjugate and the state is a Gaussian, which is real or has a linear phase [cf.\ Eqs.~\rqn{1.5}-\rqn{2.10}].
A qubit meter can be also optimal for weak standard measurements, as shown in Sec.~\ref{VIC}.

More generally, we say that meters are {\em efficient for weak standard measurements}, when the variables $F$ and $R$ and the initial state $\rho_{\rm M}$ are chosen such that both sides of the Heisenberg-Robertson uncertainty relation \rqn{310} are of the same order,
 \be
|\overline{[R,F]}| \sim\Delta F\:\Delta R.
 \e{313}
Such meters provide a pointer deflection whose magnitude is of the order of the upper bound, 
 \be
|\bar{R}_f-\bar{R}|\sim|\gamma\bar{A}|\:\Delta F\:\Delta R.
 \e{292}

Moreover, we determine the minimum size $N_0$ of the ensemble required for weak measurements without post-selection.
For this purpose, we require that the squared shift of the maximum of
the distribution of the sum of $N_0$ pointer values be equal to
the variance of this distribution, $[N_0(\bar{R}_f-\bar{R})]^2=
N_0\Delta R^2$, yielding
 \be
N_0=\left(\frac{\Delta R}{\bar{R}_f-\bar{R}}\right)^2
 \e{2.9}
or, in view of Eqs.~\rqn{75} and \rqn{310},
 \be
N_0=\left(\frac{\Delta R}{\gamma\bar{A}\,|\overline{[R,F]}|}\right)^2.
 \e{295}
The Heisenberg-Robertson uncertainty relation \rqn{310} sets the lower bound on $N_0$ in Eq.~\rqn{295} for given $|\gamma\bar{A}|$ and $\Delta F$,
 \be
N_0\ge(2\gamma\,\bar{A}\,\Delta F)^{-2}.
 \e{3.35}

The lower bound for the ensemble size in Eq.~\rqn{3.35},
 \be
N_0=(2\gamma\,\bar{A}\,\Delta F)^{-2},
 \e{2.8}
is achieved for meters optimal for weak standard measurements [i.e., meters satisfying Eq.~\rqn{2.6}], as one would expect.
More generally, for efficient meters, Eq. \rqn{313}, the ensemble size is of the order of the lower bound,
 \be
N_0\sim(\gamma\,\bar{A}\,\Delta F)^{-2}.
 \e{280}

\section{Theory of pre- and post-selected measurements of arbitrary strength}
\label{III''}

In this section we provide a general theory of PPS measurements,
which holds for an arbitrary measurement strength.
We consider PPS ensembles with a single pure post-selected state.
An extension of this theory to the case of a general post-selection measurement is given in Sec.~\ref{XII}.


\subsection{General formulas for PPS measurements}
\label{IIIA1}

As discussed in Sec.~\ref{IC4}, the pre- and post-selected (conditional) average $\bar{R}_s$ is obtained in experiments by performing a measurement of the pointer variable $R$ for each member of an ensemble of systems prepared (preselected) in the same state $\rho$ and then averaging only the results for the systems obtained (post-selected) in the state $|\phi\rangle$ after a projective measurement at $t\ge t_{\rm f}$.
Figure \ref{f8} shows a schematic diagram illustrating pre- and
post-selected quantum measurements.

The joint probability that after a measurement the system is in the
state $|\phi\rangle$ and the meter is in an eigenstate $|R\rangle$ of the operator $\hat{R}$ is
 \be
{\cal P}_{\phi R}={\rm Tr}\,[(\Pi_\phi\otimes\Pi_R)\rho_f].
 \e{3.1}
Here
 \be
\Pi_\phi=|\phi\rangle\langle\phi|,\quad \Pi_R=|R\rangle\langle R|,\e{3.2'}
 \be
\rho_f=U(\rho\otimes\rho_{\rm M})U^\dagger,
 \e{3.2}
where $\rho_{\rm M}$ is the initial state of the meter.
By Bayes' theorem, the probability to obtain the state $|R\rangle$
provided the system is in the state $|\phi\rangle$ is given by
 \be
{\cal P}_{R|\phi}\:=\:\frac{{\cal P}_{\phi R}}{{\cal P}_\phi}\:
=\:\frac{{\rm Tr}\,[(\Pi_\phi\otimes\Pi_R)\rho_f]} {\langle\Pi_\phi\rangle_f}\,\equiv\,\Phi_s(R),
 \e{3.3}
where ${\cal P}_\phi=\langle\Pi_\phi\rangle_f$ is the probability to find the system in the state $|\phi\rangle$ at $t\ge t_{\rm f}$,
 \be
{\cal P}_\phi=\sum_{R}{\cal P}_{\phi R}={\rm Tr}\,[(\Pi_\phi\otimes I_{\rm M})\rho_f]\equiv\langle\Pi_\phi\rangle_f.
 \e{3.4}
Here in the second equality we used the completeness relation for the
meter,
 \be
\sum_R\Pi_R=I_{\rm M},
 \e{3.5}
where $I_{\rm M}$ is the identity operator for the meter.
The average value of the pointer variable $R$ at $t\ge t_{\rm f}$
conditioned (post-selected) on the measurement of the system in the
state $|\phi\rangle$ is given by
 \be
\bar{R}_s=\sum_{R}R\,{\cal P}_{R|\phi},
 \e{3.6}
where $R$ is the eigenvalue of the Hermitian operator $\hat{R}$ corresponding to the eigenstate $|R\rangle$.
Finally, inserting Eq.~\rqn{3.3} into Eq.~\rqn{3.6} yields that
$\bar{R}_s$ is given by the normalized cross-correlation function
 \be
\bar{R}_s=\frac{\langle\Pi_\phi R\rangle_f}{\langle\Pi_\phi\rangle_f},
 \e{241}
where the cross-correlation function $\langle\Pi_\phi R\rangle_f$ is
an average at $t\ge t_{\rm f}$,
 \be
\langle\Pi_\phi R\rangle_f={\rm Tr}\,[(\Pi_\phi\otimes\hat{R})\rho_f].
 \e{112}
For definiteness, we assumed above that the variable $R$ has a
discrete spectrum; when the spectrum of $R$ is continuous, the sums
in Eqs.~\rqn{3.4}-\rqn{3.6} should be replaced by integrals over $R$.
Note that $\bar{R}_s$ is real, since it is an average of a physical observable; mathematically, this follows from Eq.~\rqn{241}, taking into account that $\hat{R}$ is Hermitian.

The quantity of direct physical interest is the average pointer
deflection $(\bar{R}_s-\bar{R})$ rather than the average pointer
value $\bar{R}_s$ itself.
 On substituting $\hat{R}\rightarrow\hat{R}_c=\hat{R}-\bar{R}$,
where the operator $\hat{R}_c$ corresponds to the ``centered''
quantity $R_c=R-\bar{R}$, i.e., the fluctuating part of $R$,
Eq.~\rqn{241} yields the following expression for the pointer deflection,
 \be
\bar{R}_s-\bar{R}=\frac{\langle\Pi_\phi R_c\rangle_f}{\langle
\Pi_\phi\rangle_f}.
 \e{44}
Equations \rqn{241} and \rqn{44} are the starting point for the present theory of PPS measurements.
These equations are very similar, but one of them can be more convenient than the other for a specific application.

A necessary condition for a PPS measurement to yield a nonvanishing
pointer deflection is
 \be
\Delta F\ne0.
 \e{231}
Indeed, the states $\rho_{\rm M}$ with $\Delta F=0$ are eigenvectors
of $\hat{F}$ or mixtures of eigenvectors of $\hat{F}$ with the same
eigenvalue $\bar{F}$.
 For such cases $\hat{F}\rho_{\rm M}=\rho_{\rm M}\hat{F}=
\bar{F}\rho_{\rm M}$, and Eqs.~\rqn{241}-\rqn{112} yield the zero
pointer deflection,
$\bar{R}_s-\bar{R}=0$, i.e., PPS measurements are impossible.

In contrast to standard measurements, PPS measurements are
generally possible even when the condition \rqn{217} holds.
Hereafter, meters with $\overline{[R,F]}\ne0$ are called ``standard'', since such meters are suitable for weak standard measurements [cf.\ Eq.~\rqn{75}].
Correspondingly, we call meters with $\overline{[R,F]}=0$
``non-standard''.
Examples of non-standard meters are meters with commuting $\hat{F}$
and $\hat{R}$ or with $\Delta R=0$.
As shown below, for weak PPS measurements, non-standard meters
provide almost the same information as standard ones.

It may look paradoxical that meters with commuting $\hat{F}$ and
$\hat{R}$ are suitable for PPS measurements.
Indeed, in this case $\hat{R}$ commutes with the coupling Hamiltonian \rqn{1} and hence is a constant of motion.
As a result, $\bar{R}_f=\bar{R}$, and {\em standard} measurements are
impossible [cf. the first equality in Eq.~\rqn{217}].
However, the post-selection makes the average pointer value proportional to the correlation function $\langle\Pi_\phi R\rangle_f$ [see Eq.~\rqn{241}], which generally {\em does change} under evolution with the Hamiltonian \rqn{1}, for any $F$, unless $|\phi\rangle$ is an eigenstate of $\hat{A}$.

\subsection{Gauge invariance of PPS measurements}
\label{A}

Here we discuss useful invariance properties of PPS measurements (with arbitrary measurement strength) under unitary ``gauge'' transformations of the system and the meter.

\subsubsection{System transformations}
\label{As1}

Equation \rqn{44} shows that $\bar{R}_s-\bar{R}$ is independent of
$\bar{R}$.
 In contrast, $\bar{R}_s-\bar{R}$ generally depends on $\bar{F}$.
 However, it is easy to see that $\bar{R}_s$ in Eqs.~\rqn{241}
and \rqn{44} is invariant under a ``gauge'' transformation
 \be
\hat{F}\ \rightarrow\ \hat{F}'=\hat{F}-F_0,
 \e{24}
where $F_0$ is a real number, if simultaneously the pre- and
post-selected states undergo unitary transformations,
 \bea
&&\rho\ \rightarrow\ \exp(-i\gamma' F_0\hat{A})\,\rho
\exp(i\gamma' F_0\hat{A}),\label{115}\\
&&|\phi\rangle\ \rightarrow\ |\phi'\rangle=\exp(i\gamma''
F_0\hat{A})\,|\phi\rangle,
 \ea{116}
where $\gamma'$ and $\gamma''$ are real numbers satisfying
$\gamma'+\gamma''=\gamma$.
 Note that one can choose $\gamma'=\gamma$ (or $\gamma''=\gamma$)
and $\gamma''=0$ ($\gamma'=0$), leaving thus $|\phi\rangle$ (or
$\rho$) without a change.
In particular, for $\gamma'=\gamma$ and $\gamma''=0$ the
transformation \rqn{115}-\rqn{116} reduces to the change of the initial state,
 \be
\rho\ \rightarrow\ \exp(-i\gamma F_0\hat{A})\,\rho
\exp(i\gamma F_0\hat{A}).
 \e{A6}

The transformation \rqn{24}-\rqn{116} allows one to change $\bar{F}$
according to
 \be
\bar{F}\ \rightarrow\ \bar{F}'=\bar{F}-F_0,
 \e{A7}
where $F_0$ is an arbitrary real number.

\subsubsection{Meter transformations}
\label{As2}

Consider the invariance of the average pointer deflection
under a ``gauge'' transformation of the meter.
 It is easy to see that Eq.~\rqn{44} yields that in PPS measurements of arbitrary strength the pointer deflection is invariant,
 \be
\bar{R}_s-\bar{R}\:=\:\overline{\tilde{R}}_s-\overline{\tilde{R}},
 \e{A1}
under the following class of transformations of the meter initial state and the pointer variable,
 \be
\rho_{\rm M}\ \rightarrow\ \tilde{U}_{\rm M}\,\rho_{\rm M}\,\tilde{U}_{\rm M}^\dagger,\quad\quad
\hat{R}\ \rightarrow\ \hat{\tilde{R}}=\tilde{U}_{\rm
M}^\dagger\,\hat{R}\,\tilde{U}_{\rm M}+C,
 \e{A2}
where $\tilde{U}_{\rm M}$ is a unitary operator commuting with
$\hat{F}$ and $C$ is a real constant.
 When $C=0$, then not only $\bar{R}_s-\bar{R}$ but also
$\bar{R}_s$ and $\bar{R}$ individually are invariant under the
change \rqn{A2}.

For example, if $F=p$ and $R=q$, one can use in Eq.~\rqn{A2}
operators of the form
 \be
\tilde{U}_{\rm M}=\exp[-i\zeta_0(p)],
 \e{A4}
where $\zeta_0(p)$ is an arbitrary real function of $p$.
The operator \rqn{A4} changes the phase of the meter initial state in the momentum representation.
 In this case, Eq.~\rqn{A2} implies the following change of the pointer,
 \be
q\ \rightarrow\ \tilde{R}=q+\zeta_0'(p)+C,
 \e{A5}
where the prime denotes the differentiation with respect to $p$.

\section{Effects of the system and meter Hamiltonians on quantum measurements}
\label{IIID'}

In the previous sections, we have neglected the system and the meter Hamiltonians, $H_{\rm S}$ and $H_{\rm M}$.
Let us now discuss the effects of these Hamiltonians on PPS and
standard measurements.
The results in this section hold for PPS and standard measurements of {\em arbitrary strength}.

We take into account the possibility that the Hamiltonians during the time of the system-meter interaction may differ from the Hamiltonians after the interaction, so that $H_{\rm S}(t)=H_{\rm S1}$ and $H_{\rm M}(t)=H_{\rm M1}$ for $0<t<t_{\rm f}$, whereas $H_{\rm S}(t)=H_{\rm S2}$ and $H_{\rm M}(t)=H_{\rm M2}$ for $t>t_{\rm f}$.
 These assumptions include, in particular, the case of time-independent Hamiltonians,
 \bea
&&H_{\rm S1}\:=\:H_{\rm S2}\:\equiv\:H_{\rm S},\label{5.12}\\
&&H_{\rm M1}\:=\:H_{\rm M2}\:\equiv\:H_{\rm M},
 \ea{351}
A measurement scheme with $H_{i1}\ne H_{i2}\ (i={\rm S,M})$ was discussed in Ref. \cite{rom08}.

When $H_{\rm S1}$ and/or $H_{\rm M1}$ is nonzero, the correlation of the system and the meter is performed by means of the Hamiltonian \rqn{1} transformed into the interaction picture.
In this paper, we consider only the cases when the coupling
Hamiltonian \rqn{1} is the same in the Schr\"{o}dinger and
interaction pictures.
This holds when $H_{\rm S}$ and $H_{\rm M}$ commute with the coupling
Hamiltonian \rqn{1}, so that in the interval $(0,t_{\rm f})$, $H_{\rm
S}(t)=H_{\rm S1}$ [$H_{\rm M}(t)=H_{\rm M1}$] commutes with $\hat{A}$
($\hat{F}$) or vanishes, 
 \be
[H_{\rm S1},\hat{A}]=0,\quad\quad[H_{\rm M1},\hat{F}]=0.
 \e{5.10}
However, for $t>t_{\rm f}$, the Hamiltonians $H_{\rm S}(t)=H_{\rm S2}$ and $H_{\rm M}(t)=H_{\rm M2}$ can be arbitrary.

The first relation in Eq.~\rqn{5.10} ensures that the measured observable is $A$ and not some other quantity; measurements satisfying this relation are often called quantum nondemolition measurements \cite{bra92}.
The second relation in Eq.~\rqn{5.10} is assumed for convenience; when it is violated, the input meter variable can deviate from $F$ and depend on time, which complicates calculations.

Instead of assuming the conditions \rqn{5.10}, one can require that the coupling is instantaneous (impulsive), 
 \be
t_{\rm f}-t_{\rm i}\rightarrow0.
 \e{5.11}
Then the effects of the Hamiltonians $H_{\rm S1}$ and $H_{\rm M1}$ are negligibly small.

Without loss of generality, we will assume below that
 \be
t_{\rm i}=0.
 \e{5.13}
This means that the ``initial'' states $\rho$ and $\rho_{\rm M}$ are the states of the system and meter immediately before the coupling Hamiltonian in Eq.~\rqn{1} is switched on.


\subsection{Effects of the Hamiltonians on pre- and post-selected measurements}

First, consider PPS measurements.
Assume that the post-selection is made at $t_{\rm S}>t_{\rm f}$ and the measurement of the meter is performed at $t_{\rm M}>t_{\rm f}$.
 Then the effects of the system and meter Hamiltonians are taken
into account by the change
 \be
U\rightarrow(U_{\rm S}\otimes U_{\rm M})U
 \e{119}
in the general formulas for PPS measurements in Sec.~\ref{IIIA1}, especially in Eqs.~\rqn{3.2}, \rqn{3.4}, and \rqn{241}-\rqn{44}.
In Eq.~\rqn{119}, we use the notation
 \bes{87'}\bea
&&U_{\rm S}\,=\,\exp[-iH_{\rm S2}(t_{\rm S}-t_{\rm f})]\,
\exp[-iH_{\rm S1}t_{\rm f}],\label{87a}\\
&&U_{\rm M}\,=\,\exp[-iH_{\rm M2}(t_{\rm M}-t_{\rm f})]\,
\exp[-iH_{\rm M1}t_{\rm f}].
 \ea{87}\ese

Instead of changing $U$, one can equivalently make the following
replacements.
A nonzero system Hamiltonian can be accounted for by changing the initial and final states of the system,
 \bes{88'}
 \be
\rho\rightarrow U_{\rm S1}\,\rho\, U_{\rm S1}^\dagger,\quad
|\phi\rangle\rightarrow U_{\rm S2}^\dagger|\phi\rangle,
\e{88a}
whereas a nonzero meter Hamiltonian can be accounted for by changing the meter state and the pointer variable,
 \be
\rho_{\rm M}\rightarrow U_{\rm M1}\,\rho_{\rm M}\, U_{\rm M1}^\dagger, \quad\hat{R}\rightarrow U_{\rm M2}^\dagger\hat{R}\,U_{\rm M2}.
 \e{88}
 \ese
The unitary operators $U_{\rm S1}$, $U_{\rm S2}$, $U_{\rm M1}$, and $U_{\rm M2}$ in Eqs.~\rqn{88'} are not unique.
They only should be such that $U_{\rm S1}$ ($U_{\rm M1}$) commutes with $\hat{A}$ ($\hat{F}$) and
 \be
U_{\rm S2}U_{\rm S1}=U_{\rm S},\quad U_{\rm M2}U_{\rm M1}=U_{\rm M}.
 \e{120}
Thus, there is a freedom in selecting $U_{\rm S1}$, $U_{\rm S2}$,
$U_{\rm M1}$, and $U_{\rm M2}$, which is a convenient feature facilitating considerations of the Hamiltonian effects.

\subsection{Effects of the Hamiltonians on standard measurements}

For standard measurements, it is easy to see that the system Hamiltonian does not affect the measurement result in the general formula \rqn{74}, at least, in the case of quantum nondemolition measurements [the first relation in Eq.~\rqn{5.10}].
The effects of the meter Hamiltonian are taken into account by the change
 \be
U\rightarrow(I_{\rm S}\otimes U_{\rm M})U
 \e{350}
in Eq.~\rqn{74}.
Instead of changing $U$, a nonzero meter Hamiltonian can be accounted for by changing the meter state and the pointer variable, as shown in Eq.~\rqn{88}.

\subsection{Special cases for the meter Hamiltonian}
\label{VB'}

Let us consider in more detail the effects of the meter Hamiltonian.
The results shown below hold both for PPS and standard measurements, irrespective of the measurement strength.
The effects of the system Hamiltonian on PPS measurements can be similarly considered.

In the general case, when $H_{\rm M2}$ does not necessarily commute
with $\hat{F}$, the change of the pointer variable in Eq.~\rqn{88} cannot be eliminated.
Then it may be convenient to include all the effects of the meter Hamiltonian into the effective pointer variable by the relation
 \be
\hat{R}\ \rightarrow\ \hat{R}(t_{\rm M})=U_{\rm M}^\dagger\hat{R}\,U_{\rm M},
 \e{349}
where $R(t)$ is the quantity $R$ in the Heisenberg representation, while $\rho_{\rm M}$ is left unchanged [see Eq.~\rqn{88} with $U_{\rm
M1}=I_{\rm M}$].
Consider now several simple cases.

When the meter Hamiltonian is time-independent, Eq.~\rqn{351}, then
Eq.~\rqn{87} simplifies to
 \be
U_{\rm M}\,=\,\exp(-iH_{\rm M}t_{\rm M}).
 \e{5.1}
In this case the operator $\hat{R}(t)$ obeys the equation
 \be
\frac{\partial\hat{R}(t)}{\partial t}=i[H_{\rm M},\hat{R}(t)]
 \e{5.3}
with the initial condition 
 \be
\hat{R}(0)=\hat{R}.
 \e{5.4}

An alternative simplification exists in the special case, when not only $H_{\rm M1}$ but also $H_{\rm M2}$ commutes with $\hat{F}$,
 \be
[H_{\rm M2},\hat{F}]=0.
 \e{5.5}
In this case, the effect of the meter Hamiltonian can be
taken into account without a change of $R$, by changing only
$\rho_{\rm M}$ in the formulas of the present theory.
Indeed, now, $U_{\rm M}$ in Eq.~\rqn{87} and hence $U_{\rm M2}$ in Eq.~\rqn{120} commute with $\hat{F}$.
Hence, we can choose in Eq.~\rqn{120} $U_{\rm M1}=U_{\rm M}$ and
$U_{\rm M2}=I_{\rm M}$.
Then Eq.~\rqn{88} yields that $R$ is unchanged, whereas $\rho_{\rm M}$ is changed by the relation
 \be
\rho_{\rm M}\ \ \rightarrow\ \ \rho_{\rm M}(t_{\rm M})\:=\:U_{\rm M}\,\rho_{\rm M}\,U_{\rm M}^\dagger.
 \e{242}
Here $\rho_{\rm M}(t)$ is the meter state in the Schr\"{o}dinger representation; the state $\rho_{\rm M}(t_{\rm M})$ would be obtained at the moment $t_{\rm M}$ of the measurement of the meter if the system-meter coupling were absent.

When the condition \rqn{351} holds, Eq.~\rqn{242} is independent of $t_{\rm f}$ [cf.\ Eq.~\rqn{5.1}].
This means that now, as far as the measurement is concerned, it is
not important in which part of the interval $(0,t_{\rm M})$ the
coupling  \rqn{1} is nonzero.

Finally, when the meter measurement is instantaneous (impulsive),  $t_{\rm M},t_{\rm f}\rightarrow0$, the measurement results are independent of the meter Hamiltonians $H_{\rm M1}$ and $H_{\rm M2}$, since then $U_{\rm M}\rightarrow1$ [see Eq.~\rqn{87}].

As shown in this section, if necessary, Hamiltonian effects can be easily taken into account in the theory of measurements where the Hamiltonians are neglected, at least, in the case \rqn{5.10}.
Therefore below, as a rule, we neglect the Hamiltonians.
An exception is Sec.~\ref{VIB4}, where we discuss some applications of the general theory developed in this section.

\section{Nonlinear theory of weak pre- and post-selected measurements}
\label{III'}

Using Eqs.~\rqn{241} and \rqn{3.3} one can try to obtain expressions both for the average pointer value and for the distribution of the pointer values that hold for an {\em arbitrary} coupling strength $\gamma$.
However, such expressions can be obtained in a closed analytical form only for some simple special cases (see examples in Sec.~\ref{VIII} and Refs.~\cite{duc89,kni90,rit91,sut93,sut95,%
bru03,bru04,sol04,pry05,mir07,wil08,koi11,nak12}).
The resulting expressions significantly differ for different cases.
Moreover, they are rather complicated and usually can be analyzed only numerically.
In contrast, the linear-response results discussed above in Sec.~\ref{ID} (see also Sec.~\ref{IIIF1}) are simple and general.
However, they hold only for sufficiently small values of $\gamma A_w$.
Fortunately, as shown below, it is possible to obtain for weak PPS measurements simple and general expressions, which involve $A_w$ explicitly and hold for {\em arbitrarily large} values of $|\gamma A_w|$.

In this section we develop a nonperturbative theory of weak PPS measurements for the case of a pure preselected state.
This theory serves as a basis for discussion of many aspects of weak PPS measurements in the following sections.
Extensions of this theory to the cases of a mixed preselected state and a general post-selection measurement are given in Secs.~\ref{V} and \ref{XII}.

\subsection{Expansions in the coupling parameter}
\label{IIIB'}

To obtain the description of weak PPS measurements for arbitrary values of $\gamma A_w$, we expand the numerator and denominator of Eq.~\rqn{44} in the parameter $\gamma$, as follows.

From Eq.~\rqn{112} with $R$ replaced by $R_c$, we obtain that
 \be
\langle\Pi_\phi R_c\rangle_f=
{\rm Tr}\,[U^\dagger CU(\rho\otimes\rho_{\rm M})]
 \e{320}
with $C=\Pi_\phi\otimes\hat{R}_c$.
Then we use the expansion \rqn{50}, where consecutively embedded
commutators are expanded by the formula (see Appendix \ref{B})
 \be
\underbrace{[D,\dots[D,}_n C\underbrace{]\dots]}_n\:=\:\sum_{k=0}^n\,(-1)^k{n\choose k}D^{n-k}CD^k,
 \e{B4}
to obtain the formula
 \be
\langle\Pi_\phi R_c\rangle_f\:=\:\sum_{n=1}^\infty\frac{i^n\gamma^n}{n!} \sum_{k=0}^n\,(-1)^k{n\choose k}\,(A^k\rho A^{n-k})_{\phi\phi}\; \overline{F^{n-k}\,R_c\,F^k},
 \e{3.23}
where the overbar stands for the average over $\rho_{\rm M}$, so that $\bar{O}={\rm Tr}\,(\hat{O}\rho_{\rm M})$ for a meter operator $\hat{O}$.
The quantity $\langle\Pi_\phi\rangle_f$ is calculated similarly to Eq.~\rqn{3.23} with $C=\Pi_\phi\otimes\hat{R}_c$ replaced by $C=\Pi_\phi\otimes I_{\rm M}$, yielding
 \be
\langle\Pi_\phi\rangle_f\:=\:\sum_{n=0}^\infty\frac{i^n\gamma^n\, \overline{F^n}}{n!}\,
\sum_{k=0}^n\,(-1)^k{n\choose k}\,(A^k\rho A^{n-k})_{\phi\phi}.
 \e{3.24}
For completeness, we show also the expansion of $\langle\Pi_\phi R\rangle_f$ obtained similarly to Eq.~\rqn{3.23},
 \be
\langle\Pi_\phi R\rangle_f\:=\:\sum_{n=0}^\infty\frac{i^n\gamma^n}{n!} \sum_{k=0}^n\,(-1)^k{n\choose k}\,(A^k\rho A^{n-k})_{\phi\phi}\; \overline{F^{n-k}\,R\,F^k}.
 \e{3.34}
An advantage of using Eq.~\rqn{44} instead of Eq.~\rqn{241} is that in the expansion \rqn{3.23}, in contrast to Eq.~\rqn{3.34}, the term with $n=0$ vanishes.

In the next section we will consider the important case when the system is preselected in a pure state $|\psi\rangle$, so that
$\rho=|\psi\rangle\langle\psi|$, whereas the meter state $\rho_{\rm M}$ is generally mixed (the case of a mixed preselected state is discussed in Sec.~\ref{V}).
In this case Eqs.~\rqn{3.23} and \rqn{3.24} yield that
 \bes{18'}
 \be
\langle\Pi_\phi R_c\rangle_f=|\langle\phi|\psi\rangle|^2\{
2\gamma\,{\rm Im}(\overline{R_cF}A_w)+\gamma^2\{\overline{FR_cF}|A_w|^2-{\rm Re}[\overline{R_cF^2}(A^2)_w]\}-\gamma^3{\rm Im}[\overline{F^2R_cF}A_w(A^2)_w^*+\overline{R_cF^3}(A^3)_w/3]\}+\dots, \e{18}
 \be
\langle\Pi_\phi\rangle_f=|\langle\phi|\psi\rangle|^2\{
1+2\gamma\bar{F}\,{\rm Im}\,A_w+\gamma^2\overline{F^2}[|A_w|^2-{\rm Re}\,(A^2)_w]-\gamma^3\overline{F^3}\,{\rm Im}\,[A_w(A^2)_w^*+
(A^3)_w/3]\}+\dots.
 \e{9}
 \ese
Here $(A^n)_w=(A^n)_{\phi\psi}/\langle\phi|\psi\rangle$ [cf.\
Eq.~\rqn{6}] and the dots denote the terms of fourth and higher
orders in $\gamma$.

\subsection{Validity conditions for weak PPS measurements}
\label{IIIA3'}

As mentioned above, weak PPS measurements in the AAV (linear-response) regime are limited by the two conditions in Eqs.~\rqn{1.8} and \rqn{1.9}.
However, only the condition \rqn{1.8} is necessary for the measurements to be weak, whereas the stronger condition \rqn{1.9} is required to ensure the linearity of the theory in $\gamma$.
The condition \rqn{1.9} limits the magnitude of $A_w$ or, equivalently, bounds from below the overlap $|\langle\phi|\psi\rangle|$.
Below we will show that under a condition ensuring the weakness of PPS measurements, simple general formulas can be obtained which hold for arbitrary values of $A_w$. 

The crucial point that allows us to treat weak PPS measurements nonperturbatively in $A_w$ is the fact that, in the limit
$\langle\phi|\psi\rangle\rightarrow0$, the terms of zero and first orders in $\gamma$ vanish in Eqs.~\rqn{18'}, whereas higher-order terms survive.
 Therefore for a sufficiently weak system-meter coupling, one can neglect the third- and higher-order terms in Eqs.~\rqn{18'}, whereas the second-order terms should be retained, since they may dominate the zero- and first-order terms, at least, in the most interesting case $|\langle\phi|\psi\rangle|\ll1$.
 As a result, for this situation of weak PPS measurements, we are able to obtain a simple analytical formula valid for arbitrarily large weak values, as described in Sec.~\ref{IIIA3}.

Here we derive validity conditions for weak PPS measurements.
 For this purpose, we estimate the terms in the expansions \rqn{18'}.
To obtain the validity conditions in a simple form, we will make several simplifying assumptions, which hold, at least, for some typical cases.

We begin with the expansion \rqn{9}.
The magnitudes of $n$th-order terms in Eq.~\rqn{9} are of the order of\be|\gamma^n\,(A^k)_{\phi\psi}\,(A^{n-k})_{\phi\psi}\,\overline{F^n}|\ \
\ \ (0\le k\le n),
 \e{58'}
$k$ being an integer.
In weak PPS measurements $|A_{\phi\psi}|$ is typically sufficiently large.
For simplicity, we assume that $|A_{\phi\psi}|$ is so large that
 \be
|(A^n)_{\phi\psi}|\lesssim|A_{\phi\psi}|^n.
 \e{11}
 This holds, e.g., when $|A_{\phi\psi}|\sim||A||$, where the norm
$||A||$ of $\hat{A}$ is the maximum of the magnitudes of the
eigenvalues of $\hat{A}$.
(Of course, the latter remark applies only to quantities $A$ with a finite $||A||$. An extension of this remark to unbounded quantities is out of the scope of the present paper.) 
 We also assume that
 \be
|\,\overline{F_c^n}\,|\,\lesssim\,(\Delta F)^n,
 \e{167}
where $F_c=F-\bar{F}$.
 Equation \rqn{167} implies that
 \be
|\,\overline{F^n}\,|\,\lesssim\, (|\bar{F}|+\Delta F)^n.
 \e{232}
Using Eqs.~\rqn{58'}, \rqn{11}, and \rqn{232}, we find that the
terms in Eq.~\rqn{9} of orders higher than two are negligibly small
in comparison with the
second-order terms under the weak-coupling condition,
 \be
\mu\,\equiv\,|\gamma A_{\phi\psi}|\,(|\bar{F}|+\Delta F)\,\ll\,1,
 \e{12'}
where $\mu$ is the small parameter of the present theory.

Consider now the expansion \rqn{18}.
The magnitudes of $n$th-order terms in Eq.~\rqn{18} differ from
Eq.~\rqn{58'} only by the replacement
 \be
\overline{F^n}\rightarrow\overline{F^kR_cF^{n-k}}\quad(0\le k\le n).
 \e{319}
The estimation of the moments $\overline{F^kR_cF^{n-k}}$ is simple for canonically conjugate $F$ and $R$, but is rather intricate in the general case.
As shown in Appendix \ref{D}, the terms in Eq.~\rqn{18} of orders
higher than two are negligibly small under the above condition
\rqn{12'}.
The assumptions used to derive this result are given in Appendix \ref{D}.

The small parameter of the theory $\mu$ in Eq.~\rqn{12'} has a simple physical meaning: $\mu$ is an estimation of the exponent $\gamma\hat{A}\otimes\hat{F}$ of the unitary transformation \rqn{2}.
Thus the validity condition of the present theory \rqn{12'} is a requirement for the weakness of the unitary transformation \rqn{2}.
The condition \rqn{12'} simplifies in two cases,
 \bes{12}
 \bea
&&|\gamma A_{\phi\psi}|\,\Delta F\:\ll\:1\quad\quad(\,|\bar{F}|\lesssim\Delta F\,),\label{12a}\\
&&|\gamma A_{\phi\psi}\bar{F}|\:\ll\:1\quad\quad\ \ (\,|\bar{F}|\agt\Delta
F\,).
 \ea{12b}
 \ese

\subsection{Quantifying the strength of a measurement}

In the experiments performed so far, the conditions were chosen in such a way that $\bar{F}$ was zero, either exactly or effectively.
Here we allow for $\bar{F}\ne0$.
To understand the effects of a nonzero $\bar{F}$, we recast the Hamiltonian \rqn{1} in the form
 \be
H=g(t)\,\hat{A}\otimes\hat{F}_c+g(t)\,\bar{F}\hat{A}.
 \e{296}
Here obviously only the first term on the right-hand side can correlate the system and the meter, whereas the second term is responsible for a unitary transformation of the system alone.
Correspondingly, the unitary transformation \rqn{2} splits into two factors relating to the two types of the evolution.

The evolution due to $\bar{F}$ occurs simultaneously with the evolution due to the coupling and hence affects the results of PPS measurements, the effect of $\bar{F}$ increasing with $|\bar{F}|$.
In particular, weak PPS measurements are in a qualitatively different regime for $|\bar{F}|\gg\Delta F$ than for $|\bar{F}|\alt\Delta F$ (see Sec.~\ref{IIIA6}).
However, as shown below, it may be beneficial for experimentalists that $\bar{F}$ have a nonzero value, such as, e.g., $|\bar{F}|\sim\Delta F$ or even $|\bar{F}|\gg\Delta F$.

The two conditions \rqn{12} ensure the weakness of the two types of evolution shown above.
Thus, the small parameter of the theory $\mu$ in Eq.~\rqn{12'} is the sum of two small parameters, which have different physical meanings.
Namely, 
 \be
\mu_0=|\gamma A_{\phi\psi}|\,\Delta F
 \e{297}
quantifies the degree of correlation between the system and the meter or, in other words, {\em the measurement strength}, while 
 \be
\mu_1=|\gamma A_{\phi\psi}\bar{F}| 
 \e{3.20}
characterizes the strength of the unitary transformation of the system due to $\bar{F}$.

\subsection{General nonlinear formula for the average pointer deflection}
\label{IIIA3}

Under the condition \rqn{12'} the terms of orders higher than two can
be neglected in Eqs.~\rqn{18'}, as was discussed in Sec.~\ref{IIIA3'}.
Moreover, the terms involving $(A^2)_w$ can also be neglected in Eqs.~\rqn{18'}, since, in view of Eq.~\rqn{11}, $|(A^2)_w|\ll |A_w|^2$ in the most interesting case $|\langle\phi|\psi\rangle|\ll1$, whereas for $|\langle\phi|\psi\rangle|\sim1$ all second-order terms are
negligibly small due to the conditions \rqn{11} and \rqn{12'}.
 Thus, for weak PPS measurements Eqs.~\rqn{18'} become 
 \bes{6.41}
 \bea
&&\langle\Pi_\phi R_c\rangle_f\:=\:|\langle\phi|\psi\rangle|^2[2\gamma\,{\rm Im}\,(\,\overline{R_cF}A_w)
+\gamma^2\,\overline{FR_cF}\,|A_w|^2],\label{6.41a}\\
&&\langle\Pi_\phi\rangle_f\:=\:|\langle\phi|\psi\rangle|^2(1+2\gamma\bar{F}\,{\rm Im}\,A_w
+\gamma^2\,\overline{F^2}\,|A_w|^2).
 \ea{6.41b}
 \ese
As mentioned above, the second-order terms here generally cannot be neglected, since they dominate for sufficiently small $\langle\phi|\psi\rangle$.

Inserting Eqs.~\rqn{6.41} into Eq.~\rqn{44}, we obtain
 \be
\bar{R}_s-\bar{R}\:=\:\frac{2\gamma\,{\rm Im}\,(\,\overline{R_cF}A_w)
+\gamma^2\,\overline{FR_cF}\,|A_w|^2} {1+2\gamma\bar{F}\,{\rm Im}\,A_w
+\gamma^2\,\overline{F^2}\,|A_w|^2}.
 \e{13}
The approximation \rqn{13} may fail when both terms in the numerator are vanishing or anomalously small or if they cancel, exactly or
approximately; then the $(A^2)_w$ terms and perhaps higher-order
terms [see Eq.~\rqn{18'}] should be taken into account.
 However such cases are of little interest, since then
$\bar{R}_s-\bar{R}$ is very small.

It is easy to see that Eq. \rqn{13} can be recast in another form,
 \be
\bar{R}_s\:=\:\frac{\bar{R}+2\gamma\,{\rm Im}\,(\,\overline{RF}A_w)
+\gamma^2\,\overline{FRF}\,|A_w|^2} {1+2\gamma\bar{F}\,{\rm Im}\,A_w
+\gamma^2\,\overline{F^2}\,|A_w|^2},
 \e{182}
which can sometimes be useful.
The general nonlinear formula \rqn{13} [or \rqn{182}] {\em is one of the main results of the present paper}.
A large portion of the remainder of the paper is devoted to a discussion and to extensions of this formula.

Equation \rqn{13} holds to {\em all orders in the weak value}.
It is remarkable that, according to Eq.~\rqn{13}, weak PPS
measurements depend on $A$ only through one parameter $A_w$ (at least, for a pure preselected state).
Moreover, weak PPS measurements depend on $\gamma$ and $A_w$ through
the product $\gamma A_w$.
Equation \rqn{13} shows that the average pointer deflection as a
function of $\gamma|A_w|$ can have Lorentzian and dispersive
lineshapes as well as linear combinations thereof (see also numerical calculations in Sec.~\ref{IX}).

\subsection{Regimes of weak PPS measurements}
\label{IIIF}

Consider the limiting cases of Eq.~\rqn{13}.

\subsubsection{Linear response}
\label{IIIF1}

 In first-order (linear in $\gamma$) approximation, Eq.~\rqn{13} yields the result, which we write in three equivalent forms,
 \bes{14''}
 \bea
&\bar{R}_s-\bar{R}&=\:2\gamma\,{\rm Im}\,(\,\overline{R_cF}A_w)
\label{14}\\
&&=\:2\gamma\,|\overline{R_cF}A_w|\sin(\theta+\theta_0)\label{14'}\\
&&=\:\gamma\,{\rm Im}\,\overline{[R,F]}\,{\rm Re}\,A_w+
2\gamma\,\sigma_{FR}\,{\rm Im}\,A_w,\quad
 \ea{79}
 \ese
where 
 \be
\theta=\arg A_w,\quad\theta_0=\arg\overline{R_cF},
 \e{3.7}
 \be
\sigma_{FR}\:=\:\frac{\overline{\{R_c,F\}}}{2}\:=\:
\frac{\overline{\{R,F\}}}{2}-\bar{R}\bar{F},
 \e{262}
and \{,\} denotes the anticommutator.
Equation \rqn{79} results from Eq.~\rqn{14}, on writing
$\overline{R_cF}$ as a sum of a real and an imaginary terms,
 \be
\overline{R_cF}\:=\:\sigma_{FR}+\frac{\overline{[R,F]}}{2}.
 \e{76}
Equation \rqn{76} also implies that the quantity $|\,\overline{R_c\,F}\,|$ in Eq.~\rqn{14'} is given by
 \be
|\,\overline{R_cF}\,|\:=\:
\sqrt{\left|\,\frac{\overline{[R,F]}}{2}\,\right|^2+\sigma_{FR}^2}.
 \e{261'}
The quantity $\sigma_{FR}$ in Eq.~\rqn{262} is a measure of the correlation between $F$ and $R$ \cite{gar00}; indeed, $\sigma_{FR}$ is the quantum analogue of the covariance, which is a measure of the correlation between classical random variables.
In particular, when $F=R$, the covariance $\sigma_{FR}$ equals the variance $(\Delta F)^2$.

 Under the present assumptions, each of Eqs.~\rqn{14''} describes in
the most general form the linear response for weak pre- and
post-selected measurements.
 In particular, Eq.~\rqn{79} contains the previous results on the weak value \cite{aha88,joz07} as special cases.
Equation \rqn{14'} shows that the magnitude of the linear response is
maximized,
 \be
\bar{R}_s-\bar{R}\:=\:2\,(-1)^k\gamma\,|\overline{R_cF}\,A_w|,
 \e{4.6}
when the weak-value argument $\theta$ assumes the values
 \be
\theta=-\theta_0+\left(k+\frac{1}{2}\right)\pi\quad\quad(k=0,\pm1,\dots).
 \e{356}
In contrast, the linear response vanishes for
 \be
\theta=-\theta_0+k\pi\quad\quad(k=0,\pm1,\dots).
 \e{358}

The first term in Eq.~\rqn{79}, involving ${\rm Re}\,A_w$, is an
analog of Eq.~\rqn{75}, differing only by the replacement
$\bar{A}\rightarrow {\rm Re}\,A_w$.
 This term is due to quantum properties of the meter, since it vanishes for commuting $F$ and $R$.
 In contrast, the second term, involving ${\rm Im}\,A_w$, has no
analog in weak measurements without post-selection; it arises for
correlated $F$ and $R$, and hence it generally does not vanish for
commuting variables.

Equation \rqn{79} implies that in the linear-response regime, meters with the zero covariance provide only the real part of $A_w$.
The class of such meters includes, in particular, meters optimal for weak standard measurements, Eq.~\rqn{2.6}.
In contrast, meters unsuitable for weak standard measurements, i.e., non-standard meters ($\overline{[R,F]}=0$), provide in the linear-response regime only the imaginary part of $A_w$.

Recall that for standard measurements, the average pointer deflection was estimated above with the help of the Heisenberg-Robertson uncertainty relation \rqn{310}.
Similarly, weak PPS measurements are closely related to the {\em generalized} uncertainty relation given by (see Appendix \ref{AC1})
 \be
\Delta R\,\Delta F\ \ge\ |\,\overline{R_c\,F}\,|
 \e{C14}
or, in view of Eq.~\rqn{261'},
 \be
\Delta R\:\Delta F\:\ge\:
\sqrt{\left|\,\frac{\overline{[R,F]}}{2}\,\right|^2+\sigma_{FR}^2}.
 \e{261}
The generalized uncertainty relation will be used repeatedly below.

Since $|\,\overline{R_cF}\,|$ can be much greater than $|\,\overline{[R,F]}\,|$ [cf.\ Eq.~\rqn{261'}], a comparison of Eqs.~\rqn{75} and \rqn{14'} shows that the average pointer deflection in weak PPS measurements can be {\em strongly enhanced} relative to that in standard measurements.
Notice that this enhancement is independent of the amplification due to a large weak value discussed by AAV \cite{aha88} (for further details see Sec.~\ref{IIIH}).

The necessary condition for the validity of linear response is that the denominator of Eq.~\rqn{13} is close to one, i.e., the post-selection probability in Eq.~\rqn{6.41b} approximately equals the unperturbed value $|\langle\phi|\psi\rangle|^2$ (the value in the absence of measurements in between the pre- and post-selections).
This holds for
 \be
|\gamma A_w|\,(\overline{F^2})^{1/2}\ll1.
 \e{22}
The same condition allows one to neglect the quadratic term in the
numerator of Eq.~\rqn{13} if the linear term is not too small.
 Since
 \be
\overline{F^2}=\bar{F}^2+(\Delta F)^2,
 \e{236}
Eq.~\rqn{22} is equivalent to the condition
 \be
|\gamma A_w|\,(|\bar{F}|+\Delta F)\:\ll\:1.
 \e{113}

\subsubsection{Inverted region (the limit of very large weak values)}

 In the opposite limit of very large weak values,
 \be
|\gamma A_w|\,(|\bar{F}|+\Delta F)\:\gg\:1,
 \e{233}
the average pointer deflection can be approximated by the expansion of Eq.~\rqn{13} up to first order in $(\gamma A_w)^{-1}$,
 \be
\bar{R}_s-\bar{R}\:=\:\frac{\overline{FR_cF}}{\overline{F^2}}
+\frac{2}{\gamma\,\overline{F^2}}\,{\rm Im}\,\frac{\overline{R_cF}}
{A_w^*}+\frac{2\bar{F}\,\overline{FR_cF}}{\gamma\,(\overline{F^2})^2}
\,{\rm Im}\,\frac{1} {A_w}.
 \e{15}

When the overlap $\langle\phi|\psi\rangle$ tends to zero, i.e., $A_w\rightarrow\infty$, the average pointer deflection tends to the value [cf.\ Eq.~\rqn{15}]
 \be
\bar{R}_{s,\infty}-\bar{R}\:=\:\frac{\overline{F\,R_c\,F}}{\overline{F^2}},
 \e{3.10}
where the limiting value of the average meter variable is
 \be
\bar{R}_{s,\infty}\:=\:\lim_{A_w\rightarrow\infty}\bar{R}_s\:=\:\frac{\overline{FRF}}{\overline{F^2}}.
 \e{4.11}

It is remarkable that this quantity depends only on the meter but not on the system or the coupling.
Thus, the case of orthogonal pre- and post-selected states provides {\em no information} on the system.
Note that this holds only for weak measurements, whereas higher-order corrections to Eq.~\rqn{3.10} still can depend on the system.

Now the quantity of interest, which directly provides information on the system, is not the average pointer deflection $\bar{R}_s-\bar{R}$ but the (average) adjusted pointer deflection $\bar{R}_s-\bar{R}-(\bar{R}_{s,\infty}-\bar{R})=\bar{R}_s-\bar{R}_{s,\infty}$.
Only when in the meter $\overline{FR_cF}=0$, the adjusted pointer deflection $\bar{R}_s-\bar{R}_{s,\infty}$ coincides with the average pointer deflection $\bar{R}_s-\bar{R}$.
As follows from Eq.~\rqn{15},
 \be
\bar{R}_s-\bar{R}_{s,\infty}\:=\:\frac{2}{\gamma\,\overline{F^2}}\,{\rm Im}\,\frac{\overline{R_cF}}
{A_w^*}+\frac{2\bar{F}\,\overline{FR_cF}}{\gamma\,(\overline{F^2})^2}
\,{\rm Im}\,\frac{1} {A_w}.
 \e{4.7}
The region \rqn{233} can be called {\em the inverted region}, since here the adjusted meter deflection \rqn{4.7} is inversely proportional to the weak value, {\em decreasing with the increase of the measurement strength}.

In the traditional case $\bar{F}=0$ or for $\overline{FR_cF}=0$, the magnitude of the adjusted pointer deflection is maximized under the condition \rqn{356}, when
 \be
\bar{R}_s-\bar{R}_{s,\infty}\:=\:(-1)^k\,\frac{2\,|\overline{R_cF}|}{\gamma\,\overline{F^2}\,|A_w|}.
 \e{4.8}

Equation \rqn{15} can be also considered as a first-order expansion of
the pointer deflection in the overlap $\langle\phi|\psi\rangle$.
Correspondingly, the condition \rqn{233} can be recast as
 \be
|\langle\phi|\psi\rangle|\:\ll\:|\gamma A_{\phi\psi}|\,(|\bar{F}|+\Delta F).
 \e{361}
The regime \rqn{233} is well suited for measuring very small values of
$\langle\phi|\psi\rangle$ (see also Secs.~\ref{IVF4} and \ref{IIIK3}).
The inverted region \rqn{233} has not been discussed explicitly until now.
However, the interferometric method of phase measurements
demonstrated experimentally in Ref.~\cite{sta10x} can be shown to admit a quantum-mechanical interpretation as a weak PPS measurement in the inverted region (see Sec.~\ref{XI}).

In the inverted region, the dependence on the coupling strength and the system parameters is quite different from that in the linear and the intermediate nonlinear regimes, which hold in the region
 \be
|\gamma A_w|\,(|\bar{F}|+\Delta F)\lesssim1.
 \e{288}
Therefore below in many cases the regions \rqn{233} and \rqn{288} are discussed separately.

\subsubsection{Intermediate (strongly-nonlinear) regime}
\label{IIIF3}

Consider now the region intermediate between linear response and the inverted region,
 \be
|\gamma A_w|\,(|\bar{F}|+\Delta F)\sim1.
 \e{132}
We refer to this region as the strongly-nonlinear (or intermediate) regime.
In this region, Eq.~\rqn{13} cannot be simplified since the dependence of the average pointer deflection on $\gamma$ is significantly nonlinear.
 The condition \rqn{132} can be recast in the form
 \be
\mu\sim|\langle\phi|\psi\rangle|,
 \e{286}
i.e., in the intermediate regime the small parameter of the theory is
of the order of the overlap of the pre- and post-selected states.

In the important case $|\bar{F}|\lesssim\Delta F$, which is of primary interest in most of the present paper, the condition of the strongly-nonlinear regime in Eq.~\rqn{132} becomes
 \be
|\gamma A_w|\,\Delta F\sim1
 \e{303}
or, equivalently,
 \be
\mu_0\sim|\langle\phi|\psi\rangle|.
 \e{3.9}

Equations \rqn{286} and \rqn{12'} imply that the strongly-nonlinear
regime \rqn{132} can be obtained for weak PPS measurements only when
the initial and final states are almost orthogonal,
 \be
|\langle\phi|\psi\rangle|\ll1.
 \e{258}
In the case \rqn{258}, the weak value \rqn{6} is anomalously large,
at least, when $A_{\phi\psi}$ is not too small, as in Eq.~\rqn{11}.


Below (Sec.~\ref{IIIG2}) it is shown that optimal conditions for weak
PPS measurements are obtained in the nonlinear intermediate regime.
Thus, Eq.~\rqn{132} or \rqn{286} [or, equivalently, \rqn{3.9}] provides the optimality condition, at least, for $|\bar{F}|\lesssim\Delta F$, whereas for $|\bar{F}|\gg\Delta
F$, the condition \rqn{132} [or \rqn{286}] is necessary but not sufficient for the optimal regime (see Sec.~\ref{IIIA6}).

\subsection{Estimation of the average pointer deflection}
\label{IIIG1}

Let us now estimate the typical magnitude of the average pointer deflection $\bar{R}_s-\bar{R}$ for the linear and strongly-nonlinear regimes.

\subsubsection{Linear response}
\label{IIIA6''}

In the linear-response regime, Eq. \rqn{14'} implies that the
magnitude of the pointer deflection satisfies the inequality
\be
|\bar{R}_s-\bar{R}|\:\le\:2|\gamma A_w\,\overline{R_cF}\,|.
 \e{4.19}
Combining the generalized uncertainty relation \rqn{C14} and Eq.~\rqn{4.19} yields 
 \be
|\bar{R}_s-\bar{R}|\;\le\;2|\gamma A_w|\,\Delta F\,\Delta R.
 \e{4.16}
Equation \rqn{4.16} implies that the upper bound for the magnitude of the average pointer deflection for given $|\gamma A_w|$, $\Delta F$, and $\Delta R$ is 
 \be
|\bar{R}_s-\bar{R}|\;=\;2|\gamma A_w|\,\Delta F\,\Delta R.
 \e{4.20}
This upper bound is achieved when $|\,\overline{R_cF}\,|$ equals the maximum allowed by the generalized uncertainty relation \rqn{C14},
\be
|\overline{R_cF}|\:=\:\Delta R\,\Delta F.
 \e{4.15}
As discussed below in Sec.~\ref{IIIB}, meters satisfying Eq.~\rqn{4.15} are {\em optimal for weak PPS measurements}, at least, in the linear-response regime.
The class of such meters includes, in particular, two important types of meters: (a) meters where $F$ is a linear function of $R$ (e.g., $F=R$) and (b) meters with canonically conjugate $F$ and $R$ and a general complex Gaussian state [cf.\ Eq.~\rqn{49}].

As follows from Eq.~\rqn{14'}, the estimate of a typical value of $|\bar{R}_s-\bar{R}|$ for a given $|\gamma A_w|$ is
\be
|\bar{R}_s-\bar{R}|\:\sim\:|\gamma A_w\,\overline{R_cF}\,|.
 \e{333}
This estimate holds unless $\theta+\theta_0$ is close to $k\pi$, i.e., unless $|\theta+\theta_0-k\pi|\ll1$ for some integer $k$.
Furthermore, usually both sides of the generalized uncertainty relation Eq.~\rqn{C14} are of the same order,
\be
|\overline{R_cF}|\:\sim\:\Delta R\,\Delta F.
 \e{334}
Inserting Eq.~\rqn{334} into Eq.~\rqn{333} yields 
 \be
|\bar{R}_s-\bar{R}|\;\sim\;|\gamma A_w|\,\Delta F\,\Delta R.
 \e{243}
In the present paper we call meters satisfying the condition \rqn{4.15} or, at least, \rqn{334} {\em regular meters}.
Such meters provide the magnitude of the average pointer deflection which is equal to or of the order of the upper bound not only in the linear regime but, as shown below, in most cases beyond the linear regime.
As discussed in Sec.~\ref{IIIB}, regular meters are efficient for weak PPS measurements.

The average pointer deflection in the linear-response region 
vanishes or becomes very small when the weak-value phase $\theta$ is
close to $-\theta_0$ [cf.\ Eq.~\rqn{14'}] or when $\Delta R$ is vanishing or very small [cf.\ Eq.~\rqn{4.16}].
For such cases, weak PPS measurements cannot be performed in the AAV
(linear) regime.
However, as shown below, weak PPS measurements are possible in the
nonlinear regime, even when the linear response is vanishing or very
weak.

\subsubsection{Strongly-nonlinear regime}
\label{IIIG2}

Consider now the strongly-nonlinear regime \rqn{132}.
 In this regime, measurements of $A_w$ and/or $\gamma$ are optimal,
since now the dependence of $(\bar{R}_s-\bar{R})$ on $A_w$ and
$\gamma$ is the strongest and, moreover, as we will show now, in the
regime \rqn{132} $|\bar{R}_s-\bar{R}|$ achieves the maximum value
$|\bar{R}_s-\bar{R}|_{\rm max}$ or, at least, values of the order of
$|\bar{R}_s-\bar{R}|_{\rm max}$.
 Let us estimate $|\bar{R}_s-\bar{R}|_{\rm max}$.

Here we consider the most important case $|\bar{F}|\lesssim\Delta F$
(the case $|\bar{F}|\gg\Delta F$ is discussed in Sec.~\ref{IIIA6}).
 First, we estimate the maximum value of $|\bar{R}_s-\bar{R}|$ for
the strongly-nonlinear regime \rqn{303} and then show that the
resulting estimate holds for all parameters.
 Under the condition \rqn{303}, the denominator of Eq.~\rqn{13} is of the order of one, so that the maximum value of $|\bar{R}_s-\bar{R}|$ is given by
 \be
|\bar{R}_s-\bar{R}|_{\rm max}\;\sim\;\frac{|\,\overline{R_cF}\,|}
{\Delta F}+\frac{|\,\overline{FR_cF}\,|}{(\Delta F)^2}.
 \e{234}
Taking into account that
 \be
\overline{FR_cF}\;=\;\overline{F_cR_cF_c}+2\bar{F}\,{\rm Re}
\,\overline{R_cF},
 \e{170}
Eq.~\rqn{234} becomes finally
 \be
|\bar{R}_s-\bar{R}|_{\rm max}\;\sim\;\frac{|\,\overline{R_cF}\,|}
{\Delta F}+\frac{|\,\overline{F_cR_cF_c}\,|}{(\Delta F)^2}.
 \e{4.12}
This quantity is of the order of or greater than the magnitude of the
result \rqn{15}, which means that Eq.~\rqn{4.12} provides an estimate
of the maximum of the magnitude of the pointer deflection \rqn{13}
over all values of $\gamma A_w$.
Equation \rqn{4.12} holds for any $\bar{F}$, since the same result
is obtained also for $|\bar{F}|\gg\Delta F$ (see Sec.~\ref{IIIA6}).
For regular meters [Eq.~\rqn{334}], Eq.~\rqn{4.12} simplifies,
 \be
|\bar{R}_s-\bar{R}|_{\rm max}\;\sim\;\Delta
R+\frac{|\,\overline{F_cR_cF_c}\,|}{(\Delta F)^2}.
 \e{235}

In the peculiar case when $\Delta R$ vanishes or is very small, whereas $\Delta F$ is bounded (which is possible, e.g., for finite-dimensional Hilbert spaces), $\overline{R_cF}$ also vanishes or is very small [cf.\ the uncertainty relation in Eq.~\rqn{C14}].
Then the first term on the right-hand side of Eq.~\rqn{4.12} can be neglected, but the pointer deflection generally does not vanish,
 \be
|\bar{R}_s-\bar{R}|_{\rm max}\:\sim\:
\frac{|\,\overline{F_cR_cF_c}\,|}{(\Delta F)^2} \quad\quad 
\left(|\,\overline{R_cF}\,|\ll\frac{|\,\overline{F_cR_cF_c}\,|}{\Delta F}\right).
 \e{311}
This shows that PPS measurements may be performed even when $\Delta
R=0$.

However, usually $\overline{R_cF}$ is sufficiently large to obey
 \be
|\,\overline{R_cF}\,|\;\agt\;\frac{|\,\overline{F_cR_cF_c}\,|}{\Delta F}.
 \e{4.14}
Then the second term on the right-hand side of Eq.~\rqn{4.12} can be dropped without changing the order of the magnitude of the result, yielding
 \be
|\bar{R}_s-\bar{R}|_{\rm max}\;\sim\;\frac{|\,\overline{R_cF}\,|}
{\Delta F}.
 \e{4.13}
In this case, the pointer deflection is of the order of the maximum for a given $\Delta R$ if the meter is regular [Eq.~\rqn{334}],
 \be
|\bar{R}_s-\bar{R}|_{\rm max}\;\sim\;\Delta R.
 \e{238}
This result holds under the condition [which is a special case of Eq.~\rqn{4.14} for a regular meter]
 \be
\Delta R\;\agt\;\frac{|\,\overline{F_cR_cF_c}\,|}{(\Delta F)^2}.
 \e{239}

Note that Eq.~\rqn{238} holds for {\em any} $\Delta R$, when
$[\hat{F},\hat{R}]$ is a $c$-number.
This occurs, e.g., in the cases when the variables $F$ and $R$ are
commuting or canonically conjugate (e.g., $F=p,\ R=q$) or are linear
combinations of such variables.
 Indeed, then
 \be
|\,\overline{F_cR_cF_c}\,|\:=\:|\,\overline{F_c[R,F]}
+\overline{F_c^2R_c}\,|\:=\:|\,\overline{F_c^2R_c}\,|\:\lesssim\:(\Delta F)^2\Delta R
 \e{321}
[see Eq.~\rqn{D3}], yielding Eq.~\rqn{239} and hence Eq.~\rqn{238}.

Equations \rqn{4.12} and \rqn{235} provide simple estimates of the maximum magnitude of the pointer deflection over all possible $A_w$ and all allowed $\gamma$.
Moreover, since the pointer deflection \rqn{13} depends on $A_w$ and
$\gamma$ through the product $\gamma A_w$, the maximum \rqn{4.12} or \rqn{235} results also when only one of the parameters $A_w$ and $\gamma$ is varied, while the other is fixed.
Thus, we obtain that the maximum of the pointer-deflection magnitude
over all $A_w$ for a given $\gamma$ is independent of
$\gamma$ and hence {\em remains finite for} $\gamma\rightarrow0$.

This result may look paradoxical.
Note, however, that Eqs.~\rqn{4.12} and \rqn{235} hold for a {\em subensemble} of the measured systems with the relative size $\langle\Pi_\phi\rangle_f$.
In the present case of a strongly-nonlinear regime, Eq.~\rqn{303}, with $|\bar{F}|\lesssim\Delta F$, the quantity $\langle\Pi_\phi\rangle_f$ in Eq.~\rqn{6.41b} becomes
 \be
\langle\Pi_\phi\rangle_f\:\sim\:|\langle\phi|\psi\rangle|^2.
 \e{287'}
This quantity decreases with the decrease of $\gamma$ as $\gamma^2$ [cf.\ Eq.~\rqn{3.9}], $\langle\Pi_\phi\rangle_f\sim|\langle\phi|\psi\rangle|^2\sim\mu_0^2\propto\gamma^2$.
Correspondingly, the average over the whole ensemble would yield the result \rqn{75}, which vanishes with $\gamma\rightarrow0$.

\subsubsection{Inverted region}
\label{IVE3}

Let us estimate the magnitude of the adjusted pointer deflection $|\bar{R}_s-\bar{R}_{s,\infty}|$ in the inverted region for the important case $|\bar{F}|\lesssim \Delta F$.
Taking into account Eq.~\rqn{170}, we obtain from Eq.~\rqn{4.7} that
 \bea
&|\bar{R}_s-\bar{R}_{s,\infty}|&\sim\ \frac{|\,\overline{R_cF}\,|}{|\gamma A_w|\,(\Delta F)^2} +\frac{|\,\bar{F}\,\overline{F_cR_cF_c}\,|}{|\gamma A_w|\,(\Delta F)^4}
\ \sim\ \frac{1}{|\gamma A_w|\,(\Delta F)^2}\left[|\,\overline{R_cF}\,|+\frac{|\,\bar{F}\,\overline{F_cR_cF_c}\,|}{(\Delta F)^2}\right]\nonumber\\
&&=\ \frac{|\langle\phi|\psi\rangle|}{|\gamma A_{\phi\psi}|\,(\Delta F)^2}\left[|\,\overline{R_cF}\,|+\frac{|\,\bar{F}\,\overline{F_cR_cF_c}\,|}{(\Delta F)^2}\right].
 \ea{4.9}
The latter expression shows that weak PPS measurements in the regime of very large weak values (the inverted region) can be used to measure small overlaps. In the usual case $\bar{F}=0$ or, more generally, for
 \be
|\,\overline{R_cF}\,|\:\agt\:\frac{|\,\bar{F}\,\overline{F_cR_cF_c}\,|}{(\Delta F)^2}
 \e{4.23}
Eq.~\rqn{4.9} becomes,
 \be
|\bar{R}_s-\bar{R}_{s,\infty}|\ \sim\ \frac{|\,\overline{R_cF}\,|\,|\langle\phi|\psi\rangle|}{|\gamma A_{\phi\psi}|\,(\Delta F)^2}.
 \e{4.24}

Similarly to the linear and strongly-nonlinear regimes above, in the inverted region the pointer deflection is of the order of the maximum for regular meters [Eq.~\rqn{334}], at least, when $\Delta R$ is not too small.
Namely, for regular meters, when
 \be
\Delta R\:\agt\:\frac{|\,\bar{F}\,\overline{F_cR_cF_c}\,|}{(\Delta F)^3},
 \e{4.10}
Eq.~\rqn{4.9} yields,
 \be
|\bar{R}_s-\bar{R}_{s,\infty}|\ \sim\ \frac{\Delta R\,|\langle\phi|\psi\rangle|}{|\gamma A_{\phi\psi}|\,\Delta F}.
 \e{3.14}
Note that in the present limit the overlap $\langle\phi|\psi\rangle$ is very small.
However, Eq.~\rqn{3.14} implies a strong amplification of the adjusted pointer deflection, as discussed below in Sec.~\ref{IVF4}.
Therefore, weak PPS measurements in the inverted region are suitable for measuring the overlap.

It is of interest also to estimate the value of the post-selection probability $\langle\Pi_\phi\rangle_f$ in the inverted region.
In the present limit, Eq.~\rqn{233}, we obtain that Eq.~\rqn{6.41b} becomes
 \be
\langle\Pi_\phi\rangle_f\ \approx\ |\langle\phi|\psi\rangle|^2\gamma^2\,\overline{F^2}\,|A_w|^2
=\ \gamma^2\,|A_{\phi\psi}|^2\,\overline{F^2}.
 \e{4.2}
In the common case where $\bar{F}$ is zero or small, $|\bar{F}|\lesssim \Delta F$, Eq.~\rqn{4.2} yields
 \be
\langle\Pi_\phi\rangle_f\:\sim\:\gamma^2\,|A_{\phi\psi}|^2\,(\Delta F)^2.
 \e{4.3}

\subsection{Amplification in weak PPS measurements}
\label{IIIH}

Weak PPS measurements can result in very significant amplification
of the average pointer deflection in comparison with weak standard
measurements.
There are two different types of amplification peculiar to weak PPS measurements, the (proper) amplification due to a large weak value and the enhancement due to correlation between the meter variables $F$ and $R$.
Note that there can be also enhancement of the pointer deflection due to the free meter Hamiltonian $H_{\rm M}$; this effect generally holds for both standard and PPS measurements, as discussed in Sec.~\ref{VIB4}.

Until now, amplification was discussed in the literature only for the linear response, however it takes place also in the two other regimes---the strongly-nonlinear regime and the inverted region.
Here we discuss amplification for all the three regimes in the typical case $|\bar{F}|\alt\Delta F$.
Amplification for the case $|\bar{F}|\gg\Delta F$ will be considered in Sec.~\ref{IIIA6}.
In the discussion of amplification we will assume that the meter is regular [Eq.~\rqn{334}], since such meters are efficient in PPS measurements, as noted in Sec.~\ref{IIIA6''}.

\subsubsection{Proper amplification due to a large weak value}
\label{IIIH1}

Here we consider amplification for the linear-response and strongly-nonlinear regimes, i.e., for the region \rqn{288}, given now by
 \be
|\gamma A_w|\,\Delta F\lesssim1.
 \e{3.26}
We estimate amplification in weak PPS measurements by comparing the magnitudes of the average pointer deflection in weak PPS and standard measurements.
In this comparison, we assume that $\gamma$ and $F$ are the same in both types of measurements.
Moreover, we require that the magnitudes of $A_{\phi\psi}$ and $\bar{A}$ be equal or, at least, of the same order of magnitude,
 \be
|A_{\phi\psi}|\sim|\bar{A}|.
 \e{3.8}

In the region of interest \rqn{3.26}, we define the coefficient $\cal A$ of the proper amplification by the order of magnitude with the help of the relation
 \be
{\cal A}\;\sim\;\frac{|\bar{R}_s-\bar{R}|}{\mu_0\,\Delta R}.
 \e{300}
In Eq. \rqn{300} the quantity
 \be
\mu_0\,\Delta R\;=\;|\gamma A_{\phi\psi}|\,\Delta F\,\Delta R
 \e{302}
is of the order of a typical value of the pointer deflection for
standard measurements [cf.\ Eqs.~\rqn{294} and \rqn{3.8}].

In the linear regime, inserting Eqs.~\rqn{243} and \rqn{302} into Eq.~\rqn{300} with the account of Eq.~\rqn{6} yields \cite{aha88,rit91,hos08}
 \be
{\cal A}\;\sim\;|\langle\phi|\psi\rangle|^{-1}.
 \e{301}
In the nonlinear regime, we assume the validity of Eq.~\rqn{238} (amplification in the peculiar case of a vanishing or very small $\Delta R$ is out of the scope of the present paper).
Inserting Eq.~\rqn{238} into Eq.~\rqn{300} and taking into account Eq.~\rqn{3.9} yields the relation 
 \be
{\cal A}\,\sim\,\mu_0^{-1}\,\sim\,|\langle\phi|\psi\rangle|^{-1},
 \e{3.28}
which results again in Eq.~\rqn{301}.
Thus, the result \rqn{301} holds in the whole region \rqn{3.26}.

The inequality \rqn{3.26} with the account of Eq.~\rqn{301} can be rewritten in the form
 \be
{\cal A}\,\mu_0\alt1,
 \e{3.27}
where the similarity sign is achieved under the optimal conditions.
Thus, though $\cal A$ can be very large, for a given $\mu_0$ the amplification coefficient $\cal A$ has the upper bound equal to $\mu_0^{-1}$.

An important quantity characterizing quantum noise in PPS measurements is the {\em SNR in the post-selected ensemble per one measurement}, i.e., the ratio of the magnitude of the average pointer deflection to the pointer uncertainty after the measurement $\Delta R_s$,
 \be
{\cal R}_0=\frac{|\bar{R}_s-\bar{R}|}{\Delta R_s}.
 \e{4.21}
In the typical case considered here, where $\Delta R$ is not vanishing or too small, we have $\Delta R_s\approx\Delta R$ in the linear-response regime, whereas beyond the linear response in weak PPS measurements $\Delta R_s\sim\Delta R$.
Thus, generally in the typical case ${\cal R}_0\simeq|\bar{R}_s-\bar{R}|/\Delta R$.

Equation \rqn{300} implies that, for a fixed $\mu_0$, the quantity ${\cal R}_0$ increases in direct proportion with the amplification coefficient $\cal A$,
 \be
{\cal R}_0\:\simeq\:\frac{|\bar{R}_s-\bar{R}|}{\Delta R}\:\sim\:{\cal A}\,\mu_0\:=\:{\cal A}\,|\gamma A_{\phi\psi}|\,\Delta F.
 \e{340}
In view of Eq.~\rqn{238}, the maximum value of this ratio is {\em of the order of one}, 
 \be
({\cal R}_0)_{\rm max}\:\sim\:1,
 \e{4.22}
which is achieved under the optimal conditions, i.e., in the strongly-nonlinear regime.
Note, however, that if one takes into account the {\em total} ensemble, the amplification cannot increase the quantum SNR; correspondingly, the latter is the same in weak PPS and standard measurements \cite{hos08,sta09} (see Sec.~\ref{IIIK2} for further details.)
Thus, Eqs.~\rqn{340} and \rqn{3.27} imply the remarkable fact that irrespective of how small the coupling strength $\gamma$ is, the amplification $\cal A$ can be made so large that ${\cal R}_0$ achieves the maximum \rqn{4.22}; this makes measuring very small $\gamma$ values possible, as discussed in Sec.~\ref{IIIA6'}.
The price for this is that the size of the total ensemble increases as ${\cal A}^2$ [see below Eq.~\rqn{307}].

The amplification coefficient satisfies an important relation, as follows.
It is easy to see that in the region \rqn{3.26} $\langle\Pi_\phi\rangle_f\sim|\langle\phi|\psi\rangle|^2$.
As a result, expression \rqn{301} can be recast in the form
 \be
{\cal A}\:\sim\:\langle\Pi_\phi\rangle_f^{-1/2}.
 \e{298}
This relation holds in all cases for which the proper amplification is considered here (see also Secs.~\ref{IVF4} and \ref{IIIA6}).
It shows that the proper amplification is closely related to the post-selection, increasing with the decrease of the post-selection probability $\langle\Pi_\phi\rangle_f$ and disappearing in the limit $\langle\Pi_\phi\rangle_f\rightarrow1$.

\subsubsection{Proper amplification in the inverted region}
\label{IVF4}

Weak PPS measurements in the inverted region involve a strong amplification.
Here we estimate this amplification for the typical case \rqn{4.10}.
In this case we can use Eq.~\rqn{3.14}, which can be recast as 
 \be
\frac{|\bar{R}_s-\bar{R}_{s,\infty}|}{\Delta R}\ \sim\ \frac{|\langle\phi|\psi\rangle|}{|\gamma A_{\phi\psi}|\,\Delta F}
={\cal A}'\:|\langle\phi|\psi\rangle|.
 \e{4.4}
Here the factor
 \be
{\cal A}'\:=\:(|\gamma\,A_{\phi\psi}\,|\,\Delta F)^{-1}\:\gg\:1
 \e{3.16}
provides the proper amplification for the measurement of the overlap; this factor is large in weak PPS measurements, as implied by Eq.~\rqn{12a}.
As follows from Eqs.~\rqn{4.3} and \rqn{3.16}, the amplification coefficient ${\cal A}'$ satisfies the same relation as ${\cal A}$ [cf.\ Eq.~\rqn{298}],
 \be
{\cal A}'\:\sim\:\langle\Pi_\phi\rangle_f^{-1/2}.
 \e{3.32}

The factor ${\cal A}'$ describes the increase of the quantum SNR in the PPS ensemble (per one measurement)
 \be
{\cal R}_0=\frac{|\bar{R}_s-\bar{R}_{s,\infty}|}{\Delta R_s}\sim\frac{|\bar{R}_s-\bar{R}_{s,\infty}|}{\Delta R}
 \e{3.19}
relative to the case when the small parameter approaches the limit $|\gamma\,A_{\phi\psi}|\:\Delta F\sim1$, where the measurement becomes not weak and hence the present theory breaks.
Indeed, Eq.~\rqn{4.4} yields
 \be
\frac{|\bar{R}_s-\bar{R}_{s,\infty}|}{\Delta R}\:\sim\:|\langle\phi|\psi\rangle| 
 \e{3.18}
when $|\gamma A_{\phi\psi}|\,\Delta F\sim1$.
Equation \rqn{3.16} shows that now, paradoxically, {\em the decrease of the measurement strength} $|\gamma A_{\phi\psi}|\,\Delta F$ {\em increases the magnitude of the average adjusted pointer deflection} $|\bar{R}_s-\bar{R}_{s,\infty}|$ (for given $\Delta R$ and $|\langle\phi|\psi\rangle|$) and hence increases the measurement accuracy with respect to technical errors.

However, ${\cal A}'$ cannot be increased indefinitely, since Eq.~\rqn{3.14} holds only until ${\cal A}'|\langle\phi|\psi\rangle|\ll1$ [cf.\ Eq.~\rqn{233}].
When ${\cal A}'$ becomes so large that ${\cal A}'|\langle\phi|\psi\rangle|\sim1$, the inverted-region case (the limit of very large weak values) is not applicable any more.
Instead, the measurement is performed in the strongly-nonlinear regime [cf.\ Eq.~\rqn{132}], which provides the highest accuracy and hence is optimal, as mentioned above.

\subsubsection{Enhancement due to correlation between the meter variables}
\label{IIIH2}

The pointer deflection in a measurement depends both on the system and the meter.
Correspondingly, in weak PPS measurements there exists not only the effect of amplification due to a large weak value discussed above but also the effect of enhancement due to correlation between the meter variables.

Indeed, as follows from Eqs.~\rqn{14'}, \rqn{4.13}, and \rqn{4.24}, in all regimes of weak PPS measurements there is the effect of enhancement, i.e., an increase of the average pointer deflection due to an increase of $|\sigma_{FR}|$ for given $\Delta F$ and $\overline{[R,F]}$.
This effect can be characterized by the enhancement coefficient
 \be
{\cal E}\:=\:\left|\frac{2\,\overline{R_cF}\,}{\,\overline{[R,F]}\,}
\right|\:=\:\left[1+\left(\frac{2\sigma_{FR}}
{|\,\overline{[R,F]}\,|} \right)^2\right]^{1/2},
 \e{335}
which is equal to or of the order of the factor by which the magnitude of the average pointer deflection for a nonzero $\sigma_{FR}$ is increased with respect to the case $\sigma_{FR}=0$ [cf.\ Eqs.~\rqn{14'}, \rqn{4.13}, and \rqn{4.24}].
In the second equality in Eq.~\rqn{335} we used Eq.~\rqn{261'}.
The enhancement coefficient \rqn{335} can be also recast in the form
 \be
{\cal E}\:=\:|\csc\theta_0|,
 \e{4.5}
where $\theta_0$ is defined in Eq.~\rqn{3.7}.
Equation \rqn{335} implies that {\em the enhancement is large if and only if there is a strong correlation between} $R$ {\em and} $F$, i.e., the covariance is large,
 \be
{\cal E}\:=\:\left|\frac{2\sigma_{FR}} {\,\overline{[R,F]}\,} \right|\:\gg\:1
\quad\quad\quad{\rm when}\quad|\sigma_{FR}|\gg|\,\overline{[R,F]}\,|/2.
 \e{336}

This enhancement occurs always in the linear regime, whereas beyond the linear regime it occurs, at least, in typical situations, when Eq.~\rqn{4.13} and \rqn{4.24} hold.
The coefficient ${\cal E}$ in Eq.~\rqn{335} characterizes also an increase of the magnitude of the average pointer deflection in the linear regime for PPS measurements with respect to that for standard measurements due to a nonzero covariance, as follows from a comparison of Eqs.~\rqn{75} and \rqn{14'}.

An increase of $|\sigma_{FR}|$ for given $\overline{[R,F]}$ and $\Delta F$ yields an increase of not only the pointer deflection but also the pointer uncertainty $\Delta R$, as implied by the generalized uncertainty relation \rqn{261}.
However, for regular meters [Eq.~\rqn{334}], which are of interest here, the pointer deflection is typically proportional to $\Delta R$ in all regimes [see Eqs.~\rqn{243}, \rqn{238}, and \rqn{3.14}].
Therefore, in this case, for given $\Delta F$ and $\overline{[R,F]}$, the SNR in the PPS ensemble per one measurement ${\cal R}_0$, Eq.~\rqn{340}, is independent of $\sigma_{FR}$.
Though an increase of $|\sigma_{FR}|$ does not improve the quantum limit of the measurement accuracy, even for the PPS ensemble, an enhancement of the pointer deflection increases the accuracy of the readout of the measurement result with respect to technical noise and thus is beneficial.

Equation \rqn{79} implies that in the linear regime, the enhancement
is possible only when Im$\,A_w\ne0$ \cite{hos08}.
However, in the nonlinear regime, the enhancement is possible not only
for Im$\,A_w\ne0$ but also for a real $A_w$, when $\bar{F}\ne0$.
 For a discussion of specific cases, see below Sec.~\ref{VIB3}.

\subsubsection{Discussion}
\label{IVF3}

Since the proper amplification ${\cal A}$ (or ${\cal A}'$) and the enhancement ${\cal E}$ are independent of each other, we can write the total amplification coefficient in the linear-response and strongly-nonlinear regimes as
 \be
{\cal A}_T={\cal A}{\cal E}
 \e{337}
and in the inverted region as
 \be
{\cal A}_T'\,=\,{\cal A}'{\cal E}.
 \e{3.17'}
The amplification by a weak PPS measurement, described by Eqs.~\rqn{337} and \rqn{3.17'}, does not amplify technical noise \cite{hos08,sta09} and hence is an important effect with promising applications in ultra-sensitive measurements and precision metrology.

It is worth noting here an important advantage of the enhancement $\cal E$ over the proper amplification $\cal A$.
As seen in Eq.~\rqn{298}, there is a trade-off between $\cal A$ and the post-selection probability, that is, $\cal A$ cannot be made large without making the post-selection probability small.
In contrast, the enhancement $\cal E$ {\em can attain arbitrary large values in cases where the post-selection probability is of order unity}.

For non-standard meters ($\overline{[R,F]}=0$) the enhancement cannot be described by the parameter ${\cal E}$, because in this case Eq.~\rqn{335} yields ${\cal E}=\infty$, which is an expression of the fact that standard measurements are impossible for non-standard meters.
Moreover, the parameters ${\cal A}_T$ and ${\cal A}_T'$ are also infinite and hence do not make much sense, but the proper amplification is still a meaningful notion.


\subsection{Measuring weak values and coupling strengths}
\label{IIIA6'}

Equation \rqn{13} allows one to measure any parameter entering this
expression provided the other parameters are known.
 Here we discuss measuring the weak value $A_w$ and the coupling
strength $\gamma$.
Both linear and nonlinear regimes of the measurement are considered.

Since $\gamma$ is real and $A_w$ is complex, in principle one or two
weak PPS measurements are sufficient for obtaining $\gamma$ or $A_w$,
respectively.
Below we discuss extracting $\gamma$ and $A_w$ for the above minimal
number of measurements.
Alternatively, to increase the accuracy of the value of $A_w$, one
may perform more than the minimal number of measurements and then fit
the measurement results to Eq.~\rqn{13}.
 Note, however, that an increase in the number of measurements
requires an increase in time and resources.

\subsubsection{Measuring the coupling strength $\gamma$}
\label{IIII1}

The parameter $\gamma$ can be measured in either the linear or nonlinear regime.
 As discussed above, optimal measurements of $\gamma$ are obtained
in the nonlinear regime \rqn{132}, where $|\bar{R}_s-\bar{R}|$ is of
the order of its maximum value.

Note a difference between the measurement procedures in the linear
and nonlinear cases: 
Given a measured value of the pointer deflection
$(\bar{R}_s-\bar{R})$, in the linear (nonlinear) regime $\gamma$
results from a solution of a
linear (quadratic) equation [cf.\ Eqs.~\rqn{14''} and \rqn{13}].
 The roots of the quadratic equation can be obtained in analytic form; only one of them yields the correct solution.
 Namely, the correct root $\gamma_0$ is determined uniquely by the
condition $\gamma_0\rightarrow0$ for $\bar{R}_s\rightarrow\bar{R}$.

Thus, the nonlinear Eq.~\rqn{13} allows one to optimize the
measurement of the coupling strength.
 Another advantage of Eq.~\rqn{13} is that it allows for measurement
of $\gamma$, even when the first-order result \rqn{14} vanishes or
is very small; in this case, it is required that
$\overline{FR_cF}\ne0$.

\subsubsection{Measuring weak values: One unknown parameter}
\label{IIIA8}

Consider now the measurement of weak values.
 $A_w$ is a complex quantity, and generally both the magnitude
$|A_w|$ and the argument $\theta$ of $A_w$ are unknown.

We first discuss the simple situation, where $\theta$ is known, at
least, with an accuracy of up to $\pi$.
Then $A_w$ and $|A_w|^2$ can be presented in the form,
 \be
A_w=A_{w0}\exp(i\theta'),\quad|A_w|^2=A_{w0}^2,
 \e{244}
where $A_{w0}$ is real but its sign may be unknown, so that $\theta'$ equals $\theta$ for $A_{w0}>0$ or $\theta+\pi$ for $A_{w0}<0$.
 We assume that $\theta'$ in Eq.~\rqn{244} is known.
 (For example, when $A_w$ is known to be real
or imaginary, one can set $\theta'=0$ or $\pi/2$, respectively,
i.e., $A_w=A_{w0}$ or $A_w=iA_{w0}$, respectively.)
 Then, on inserting Eq.~\rqn{244} into Eq.~\rqn{14} or \rqn{13},
$A_{w0}$ [and hence $A_w$ in Eq.~\rqn{244}] can be obtained similarly to $\gamma$ (see Sec.~\ref{IIII1}), from a linear or quadratic equation, respectively.

\subsubsection{Tomography of weak values}
\label{IIIJ3}

Consider now tomography of weak values, i.e, measuring a complex
$A_w$ in the absence of any preliminary information on $A_w$.
A complex $A_w$ depends on two real parameters and hence to obtain
$A_w$ from Eq.~\rqn{13}, it is sufficient to perform two measurements
with different values of the coupling strength $\gamma$ and/or of one or more of the meter parameters.
The meter parameters which can be varied include the observables $R$
and $F$ \cite{shp08} and the meter state $\rho_{\rm M}$ \cite{cho10}.
Hereafter, we regard meters with different parameters as different
meters, even if they are realized with identical physical systems.

 As mentioned above, the nonlinear regime is optimal for
measurements.
However, the nonlinear regime can be achieved only for sufficiently large weak values or, in other words, for sufficiently small values of the overlap $\langle\phi|\psi\rangle$ (cf.\ Sec.~\ref{IIIF3}).
 Therefore the linear regime is also important, since this is the only regime of weak PPS measurements achievable for not too large weak values.
Another reason why the linear regime is of interest is that the linear regime is somewhat easier to analyze than the nonlinear regime.
 Below we show how to extract $A_w$ with two measurements for both the
linear and nonlinear regimes.

\subsubsection{Tomography of weak values: Linear regime}

Since the linear response \rqn{14''} is proportional to $\gamma$, in
the linear regime a variation of $\gamma$ cannot be used for tomography of weak values.
Thus, in the two required measurements, the meters should necessarily
differ by one or more parameters (e.g., the meters may have different pointer variables), so that the parameter $\theta_0$ in Eq.~\rqn{3.7} has different values in the two measurements.
 We require that these values, denoted by $\theta_0$ and $\theta_0'$, obey the condition
 \be
|\theta_0-\theta_0'|\:\ne\:0,\pi,2\pi,\dots.
 \e{245}
 The two measurements yield by Eq.~\rqn{14'} the quantities
 \be
\xi=|A_w| \sin(\theta+ \theta_0),\quad\quad\xi'=|A_w| \sin(\theta+
\theta_0').
 \e{246}
The equalities \rqn{246} can be recast as a set of two linear
equations for Re$\,A_w=|A_w|\cos\theta$ and Im$\,A_w=|A_w|\sin\theta$,
which has a solution under the condition \rqn{245}.
As a result, we obtain
 \be
A_w\,=\,\frac{\xi\exp(-i\theta_0')-\xi'\exp(-i\theta_0)}
{\sin(\theta_0-\theta_0')}.
 \e{247}
For instance, in the experiment \cite{lun11}, the tomography of weak values was realized with $\theta_0=\pi/2$ and $\theta_0'=0$.
In this case, Eq.~\rqn{246} implies that $\xi={\rm Re}A_w$ and $\xi'={\rm Im}A_w$; correspondingly, Eq.~\rqn{247} yields $A_w=\xi+i\xi'$.

The condition \rqn{245} requires that, at least, in one of the two
measurements $\overline{[R,F]}\ne0$, while in the other measurement
$\sigma_{FR}\ne0$.
If $\sigma_{FR}=0$ ($\overline{[R,F]}=0$) in the measurements, only
Re$\,A_w$ (Im$\,A_w$) can be measured in the linear regime.
Finally, when $\overline{[R,F]}=\sigma_{FR}=0$, the linear response
vanishes and hence cannot be used for measurements.
The situation is drastically different in the nonlinear regime, as
follows.

\subsubsection{Tomography of weak values: Nonlinear regime}
 \label{IIIA8d}

Consider now tomography of weak values in the nonlinear regime
\rqn{303} for the important case $|\bar{F}|\lesssim\Delta F$.
As in the linear case, two measurements with different meters can be
performed.
However, now one has also an alternative possibility: to perform the
two measurements with different values of the coupling strength
$\gamma$, using only {\em one meter for all measurements}.
The condition \rqn{245} is generally not required now.
The coupling strength $\gamma$ can be varied by changing the duration
of the interaction and/or the amplitude of the coupling rate $g(t)$,
cf.\ Eq.~\rqn{355}.

 Inserting the measurement results on the left-hand side of
Eq.~\rqn{13} yields two second-order algebraic equations for Re$\,A_w$
and Im$\,A_w$.
 Indeed, multiplying both sides of Eq.~\rqn{13} by the
denominator of the fraction on the right-hand side, transferring all the terms
to the left-hand side and simplifying the expression yields 
 \be
D_{0i}+D_{1i}\,{\rm Re}\,A_w+D_{2i}\,{\rm
Im}\,A_w+D_{3i}\,|A_w|^2\:=\:0\quad\quad(i=1,2),
 \e{254}
where the coefficients $D_{ki}$ can be easily expressed through the
parameters of the problem and $i$ denotes the two measurements.

 Equations \rqn{254} can be solved analytically, as follows.
Multiplying Eqs.~\rqn{254} for $i=1$ and 2 by $D_{32}$ and $-D_{31}$, respectively, and summing the resulting equations cancel
the nonlinear terms and yield a linear relation between ${\rm Re}\,A_w$ and ${\rm Im}\,A_w$.
 This relation allows one to express ${\rm Re}\,A_w$ through
${\rm Im}\,A_w$, thus reducing the problem to obtaining ${\rm Im}\,A_w$.
 Finally, inserting the above expression for ${\rm Re}\,A_w$ into one
of Eqs.~\rqn{254} yields a quadratic equation for ${\rm Im}\,A_w$.
The root of this equation, which tends to zero for $\bar{R}_s\rightarrow\bar{R}$, gives ${\rm Im}\,A_w$, which then is used to find ${\rm Re}\,A_w$.

\begin{table}[tb]
\begin{center}
\begin{tabular}{c|ccc|ccc}
\hline
&\multicolumn{3}{c|}{Measurement type}&\multicolumn{3}{c}{Measurability of:}\bigstrut[t]\\
 Case&$\ \overline{[R,F]}$&$\ \sigma_{FR}$&$\bar{F}$&Re$\,A_w$&Im$\,A_w$&$|A_w|$\bigstrut\\
\hline
 1&$\ne0$&$\ne0$&\ any\ &yes&yes&yes\bigstrut[t]\\
 2&$\ne0$&0&0&yes&$\ |{\rm Im}\,A_w|$&yes\bigstrut[t]\\
 3&$\ne0$&0&$\ne0\ $&yes&yes&yes\bigstrut[t]\\
 4&0&$\ne0$&any&$\ |{\rm Re}\,A_w|$&yes&yes\bigstrut[t]\\
 5&0&0&$\ne0\ $&$\ |{\rm Re}\,A_w|$&yes&yes\bigstrut[t]\\
 6&0&0&0&no&no&yes\bigstrut[t]\\[.5ex]
\hline
 \end{tabular}
\end{center}
\caption{Measurability of ${\rm Re}\,A_w$, ${\rm Im}\,A_w$, and $|A_w|$ for different types of measurements.
Here we use the following conventions: (a) the word ``any'' means any $\bar{F}$ satisfying $|\bar{F}|\lesssim\Delta F$, (b) the parameters denoted as nonzero should be nonzero, at least, in one of the two measurements; (c) the parameters denoted as zero vanish in all the measurements; (d) when both $\overline{[R,F]}$ and $\sigma_{FR}$ vanish, we assume that $\overline{F\,R_c\,F}\ne0$.
Note that $|A_w|$ is generally measurable in all the cases listed here.
}
 \label{t4}\end{table}

Table \ref{t4} shows the feasibility of measuring the weak value
parameters ${\rm Re}\,A_w$, ${\rm Im}\,A_w$, and $|A_w|$ for different
types of measurements.
Generally, Eq.~\rqn{13} involves $A_w$ through terms proportional to Re$\,A_w$, Im$\,A_w$, and $|A_w|^2$.
To determine $A_w$ completely, i.e., to obtain Re$\,A_w$ and Im$\,A_w$
with correct signs, each of Re$\,A_w$ and Im$\,A_w$ should enter linearly in Eq.~\rqn{13} for, at least, one of the two measurements.
In particular, this happens when both $\overline{[R,F]}\ne0$ and
$\sigma_{FR}\ne0$ do not vanish for, at least, one of the two
measurements (Table \ref{t4}, case 1).

In case 2, a term linear in ${\rm Im}\,A_w$ is absent in Eq.~\rqn{13},
and hence ${\rm Im}\,A_w$ can be determined only up to a sign.
In case 3, $A_w$ can be determined completely, since Im$\,A_w$ enters
linearly in the denominator of Eq.~\rqn{13} when $\bar{F}\ne0$.
 Actually, the optimal values of $\bar{F}$ are those satisfying
$|\bar{F}|\sim\Delta F$, since then in the present nonlinear case
\rqn{303} all the three terms in the denominator of Eq.~\rqn{13} are
generally of the same order.

In cases 4 and 5, ${\rm Re}\,A_w$ can be obtained only up to a sign,
whereas in case 6, only $|A_w|$ can be measured.
Note that in cases 5 and 6 the linear response is absent, but still
information on $A_w$ can be extracted (such a case takes place, e.g.,
for $\Delta R=0$, see Sec.~\ref{VIAb}).

In summary, generally $A_w$ can be obtained completely when, at
least, one of the two measurements is performed with a standard meter
($\overline{[R,F]}\ne0$).
Measurements using only non-standard meters ($\overline{[R,F]}=0$) typically yield Im$\,A_w$ and $|{\rm Re}\,A_w|$, but not the sign of Re$\,A_w$.

\subsection{Peculiar case: Large average input variable, $|\bar{F}|\gg\Delta F$}
\label{IIIA6}


Until now we have focused mainly on the typical situation when
$|\bar{F}|\lesssim\Delta F$.
Usually in experiments on weak PPS measurements, the conditions are chosen to make $\bar{F}$ to vanish, exactly or effectively.
However, as shown above, a moderately large value of $\bar{F}$, $|\bar{F}|\sim\Delta F$, may be useful in measuring weak values and coupling strengths.

 Here we discuss the case of a large $\bar{F}$, $|\bar{F}|\gg\Delta F$. This case differs significantly from the above situation $|\bar{F}|\lesssim\Delta F$.
Now the pointer deflection is generally small, except for the optimal regime which has the form of a narrow resonance whose width decreases with increasing $|\bar{F}|$.
Let us consider the linear response and the optimal region.

The linear-response regime is independent of $\bar{F}$, and it is described in the above Sec.~\ref{IIIF1}.
The region of its validity is given by Eq.~\rqn{113}, which now becomes
 \be
|\gamma\,A_w\,\bar{F}|\,\ll\,1.
 \e{3.33}
Since now $\bar{F}$ is large, the maximum value of $|\gamma A_w|$ allowed by Eq.~\rqn{3.33} is small, and hence the average pointer deflection in the linear-response regime is very small in the present case.
The amplification coefficient in the linear response is given by the above Eq.~\rqn{301}.

Consider now the optimal regime.
When $|\bar{F}|\gg\Delta F$, and ${\rm Re}\,A_w$ is small or
vanishes,
 \be
|{\rm Re}\,A_w|\,\ll\,|{\rm Im}\,A_w|,
 \e{173}
then $\bar{R}_s$ versus $\gamma$ has a narrow resonance at
 \be
\gamma\,\bar{F}\:{\rm Im}\,A_w\,\approx\,-1.
 \e{168}
In the vicinity of this resonance, Eq.~\rqn{13} simplifies to
 \be
\bar{R}_s-\bar{R}\:=\:\frac{\overline{F_c\,R_c\,F_c}/\bar{F}- \epsilon\,{\rm Im}\,\overline{[R,F]} -2\,\sigma_{FR}\,x}
{\bar{F}\,\left[x^2+\epsilon^2 +(\Delta F/\bar{F})^2\right]},
 \e{169}
where
 \bea
&&x\,=\,1+\gamma\bar{F}\,{\rm Im}\,A_w,\label{174}\\
&&\epsilon\,=\,\frac{{\rm Re}\,A_w}{{\rm Im}\,A_w}\quad{\rm or}\quad \epsilon\,=\,{\rm sgn}\,({\rm Im}\,A_w)\,\frac{\pi}{2}-\theta.
 \ea{218}
The two expressions for $\epsilon$ in Eq.~\rqn{218} are equivalent in the approximation where Eq.~\rqn{169} holds.
 In the derivation of Eq.~\rqn{169} we used Eqs.~\rqn{170} and
\rqn{262}; moreover, in the numerator and denominator of
Eq.~\rqn{169} we neglected terms of higher orders in $x,\ \epsilon$,
and $\Delta F/\bar{F}$.


Equation \rqn{169} describes a resonance in $\bar{R}_s$ as a function
of the two variables $x$ and $\epsilon$ or, in other words, versus any of the parameters ${\rm Im}\,A_w$, ${\rm Re}\,A_w$, $\gamma$, and $\bar{F}$.
Depending on the parameter values, the resonance as a function of $x$
or $\epsilon$ can have either Lorentzian or dispersive shape or a
linear combination thereof.
The resonance arises due to the fact that destructive quantum
interference results in a strongly reduced post-selection probability,
 \be
\langle\Pi_\phi\rangle_f\:\approx\:|\langle\phi|\psi\rangle|^2
[x^2+\epsilon^2+(\Delta F/\bar{F})^2].
 \e{291}
The resonance as a function of $x$ and $\epsilon$ has one or two extrema; the extremum with the largest magnitude lies in the optimal region,
 \be
x^2+\epsilon^2\,\lesssim\,\left(\frac{\Delta F}{\bar{F}}\right)^2.
 \e{290}

Let us obtain the maximum of the magnitude of the average pointer deflection in Eq.~\rqn{169} as a function of $x$ and $\epsilon$.
The second and third terms in the numerator of Eq.~\rqn{169} can be written in the form $-2|\,\overline{R_cF}\,|\sqrt{x^2+\epsilon^2}\cos(\theta_0-\theta_2)$, where $|\,\overline{R_cF}\,|$ is given by Eq.~\rqn{261'} and $\theta_2$ is determined by the equalities $\cos\theta_2=x/\sqrt{x^2+\epsilon^2}$ and $\sin\theta_2=\epsilon/\sqrt{x^2+\epsilon^2}$.
Thus, we obtain that for a given value of $x^2+\epsilon^2$ the quantity $|\bar{R}_s-\bar{R}|$ assumes the maximum value
 \be
|\bar{R}_s-\bar{R}|\,=\,\frac{|\,\overline{F_cR_cF_c}\,|+2|\bar{F}|\,|\,\overline{R_cF}\,| \sqrt{x^2+\epsilon^2}}{\bar{F}^2\,\left[x^2+\epsilon^2+(\Delta F/\bar{F})^2\right]},
 \e{3.29}
which is attained under the condition $|\cos(\theta_0-\theta_2)|=1$; when $\overline{F_c\,R_c\,F_c}\ne0$, there is an additional condition that the signs of $\cos(\theta_0-\theta_2)$ and $-\overline{F_c\,R_c\,F_c}/\bar{F}$ coincide.
The maximum of Eq.~\rqn{3.29} as a function of the variable $\sqrt{x^2+\epsilon^2}$ can be easily found, and we obtain that the maximum of $|\bar{R}_s-\bar{R}|$ is
 \be
|\bar{R}_s-\bar{R}|_{\rm max}\,=\,\frac{\sqrt{4(\Delta F)^2|\,\overline{R_cF}\,|^2+(\,\overline{F_cR_cF_c}\,)^2}+|\,\overline{F_cR_cF_c}\,|}{2(\Delta F)^2}.
 \e{6.18}
The maximum magnitude in Eq.~\rqn{6.18} satisfies the simple estimate \rqn{4.12} obtained above for the case $|\bar{F}|\lesssim\Delta F$.
Thus, the estimate \rqn{4.12} holds irrespective of the value of $\bar{F}$.

The validity conditions of Eq.~\rqn{6.18} depend on whether $\overline{F_cR_cF_c}$ vanishes or not.
When $\overline{F_cR_cF_c}=0$, the maximum magnitude \rqn{6.18} is achieved for
 \be
x\,=\,\pm\frac{\Delta F\,\sigma_{FR}}{\bar{F}\,|\,\overline{R_cF}\,|} \,=\,\pm\frac{\Delta F}{\bar{F}}\cos\theta_0,\quad\quad
\epsilon\,=\,\pm\frac{\Delta F\,{\rm Im}\,\overline{[R,F]}}{2\bar{F}\,|\,\overline{R_cF}\,|}\,=\,\pm\frac{\Delta F}{\bar{F}}\sin\theta_0.
 \e{6.23}
Inserting Eq.~\rqn{6.23} into Eq.~\rqn{169} yields
 \be
\bar{R}_s-\bar{R}\,=\,\mp\frac{|\,\overline{R_cF}\,|}{\Delta F}.
 \e{6.24}
In Eqs.~\rqn{6.23} and \rqn{6.24} the upper (or lower) signs should be taken simultaneously.

When $\overline{F_cR_cF_c}\ne0$, the maximum \rqn{6.18} is achieved for 
 \bea
&&x\,=\,\frac{\sigma_{FR}\,\overline{F_cR_cF_c}}{2\bar{F}\,|\,\overline{R_cF}\,|^2}
\left\{1-\left[1+\left(\frac{2\,\Delta F\,|\,\overline{R_cF}\,|}{\overline{F_cR_cF_c}}\right)^2\right]^{1/2}\right\},\nonumber\\
&&\epsilon\,=\,\frac{{\rm Im}\,\overline{[R,F]}\:\overline{F_cR_cF_c}}{4\bar{F}\,|\,\overline{R_cF}\,|^2}
\left\{1-\left[1+\left(\frac{2\,\Delta F\,|\,\overline{R_cF}\,|}{\overline{F_cR_cF_c}}\right)^2\right]^{1/2}\right\}.
 \ea{6.21}
Indeed, one can check that inserting Eq.~\rqn{6.21} into Eq.~\rqn{169} yields
 \be
\bar{R}_s-\bar{R}\,=\,{\rm sgn}(\,\overline{F_cR_cF_c}\,)\,|\bar{R}_s-\bar{R}|_{\rm max},
 \e{6.22}
where $|\bar{R}_s-\bar{R}|_{\rm max}$ is given in Eq.~\rqn{6.18}.

As a special case of Eqs.~\rqn{6.21}-\rqn{6.22}, we obtain that for non-standard meters with $\overline{R_cF}=0$ (cf.\ Sec.~\ref{VIAb}) the maximum in Eq.~\rqn{6.18} is achieved for 
 \be
x\,=\,\epsilon\,=\,0,
 \e{6.25}
when
 \be
\bar{R}_s-\bar{R}\,=\,\frac{\overline{F_cR_cF_c}}{(\Delta F)^2}.
 \e{6.26}

To obtain the amplification coefficient in the optimal region \rqn{290}, we use Eqs.~\rqn{300} and \rqn{302}, which yield
 \be
{\cal A}\:\sim\:\frac{|\bar{R}_s-\bar{R}|}{|\gamma A_{\phi\psi}|\,\Delta F\,\Delta R}
\:\sim\:\frac{\Delta R}{|\langle\phi|\psi\rangle/\bar{F}|\,\Delta F\,\Delta R}
\:=\:\frac{|\bar{F}|}{\Delta F\,|\langle\phi|\psi\rangle|}.
 \e{3.30'}
Here we took into account that, in view of Eqs.~\rqn{168} and \rqn{173}, $|\gamma A_{\phi\psi}\bar{F}|\approx|\langle\phi|\psi\rangle|$; we also assumed the typical situation \rqn{238}.
Thus, we obtain
 \be
{\cal A}\:\sim\:\frac{|\bar{F}|}{\Delta F\,|\langle\phi|\psi\rangle|}.
 \e{3.30}
In the optimal region \rqn{290}, we obtain that Eq.~\rqn{291} yields
 \be
\langle\Pi_\phi\rangle_f^{-1/2}
\:\sim\:\frac{|\bar{F}|}{\Delta F\,|\langle\phi|\psi\rangle|}
\:\sim\:{\cal A}.
 \e{305}
Hence, the amplification coefficient in Eq.~\rqn{3.30} satisfies the same relation \rqn{298}, as the amplification coefficient in Eq.~\rqn{301}, though the two coefficients significantly differ from each other.
An estimation of the magnitude of Eq.~\rqn{169} for $x^2+\epsilon^2 \gg\Delta F^2/\bar{F}^2$ also can be shown to yield Eq.~\rqn{298}.

Note that in the present case when $|\bar{F}|\gg\Delta F$, the optimality condition \rqn{290} is much stricter than the condition for the strongly-nonlinear regime \rqn{132} [or, equivalently, \rqn{286}].

An advantage of the present case $|\bar{F}|\gg\Delta F$ is that the
optimal regime occurs at a much smaller value of $|\gamma|$ than for
$|\bar{F}|\lesssim\Delta F$ [cf.\ Eqs.~\rqn{168} and \rqn{303},
respectively].
Correspondingly, the amplification coefficient \rqn{305} is much
higher than the value \rqn{301} obtained in the case
$|\bar{F}|\lesssim\Delta F$, for a given overlap magnitude $|\langle\phi|\psi\rangle|$.
This allows for an increase of the measurement precision in the present case as compared to the case $|\bar{F}|\lesssim\Delta F$, when, by some reason, the overlap $\langle\phi|\psi\rangle$ cannot be made too small.

Moreover, the fact that Eq.~\rqn{169} is a narrow resonance as a
function of a number of parameters makes this resonance very sensitive
to small perturbations of the parameters of the problem.
This sensitivity of the resonance can be used for precise measurements of these parameters.

\subsection{The minimum size of the ensemble, the signal-to-noise ratio, and efficient meters}
\label{IIIB}

Here we estimate the minimum size $N_0$ of the ensemble required for a weak PPS measurement, as well as the signal-to-noise ratio (SNR) characterizing quantum noise.
We also discuss optimal and efficient meters for weak PPS and standard measurements.

\subsubsection{General formulas}
\label{IIIK1}

Consider an ensemble of $N$ pairs consisting of a system and a meter.
Only $\langle\Pi_\phi\rangle_f N$ of the $N$ pairs are taken into account in a PPS measurement.
A measurement produces a shift of the maximum of the distribution of the sum of $\langle\Pi_\phi\rangle_f N$ pointer values, equal to $\langle\Pi_\phi\rangle_f N\,(\bar{R}_s-\bar{R})$.
The SNR ${\cal R}$ characterizing quantum noise equals the ratio of the magnitude of the above shift (the signal) to the standard deviation of the above distribution $\Delta R_s \sqrt{\langle\Pi_\phi\rangle_f\, N}$ (the noise level), yielding
 \be
{\cal R}\:=\:\frac{|\bar{R}_s-\bar{R}|}{\Delta R_s}\sqrt{\langle\Pi_\phi\rangle_fN}
 \e{6.28}
or, equivalently,
 \be
{\cal R}\:=\:{\cal R}_0\sqrt{\langle\Pi_\phi\rangle_fN},
 \e{6.28'}
where ${\cal R}_0$ is defined in Eq.~\rqn{4.21}.

We determine the minimum size of the ensemble $N_0$ by requiring that at $N=N_0$ the signal and the noise levels be equal (${\cal R}=1$), which yields, in view of Eq.~\rqn{6.28},
 \be
N_0\:=\:\frac{(\Delta R_s)^2}
{\langle\Pi_\phi\rangle_f\:(\bar{R}_s-\bar{R})^2}.
 \e{82}
Now Eq.~\rqn{6.28} can be recast as
 \be
{\cal R}=\sqrt{\frac{N}{N_0}}.
 \e{3.12}
Thus, the quantity $N_0$ determines the SNR for a given ensemble of size $N$ through Eq.~\rqn{3.12}.

\subsubsection{The linear response. Efficient and optimal meters}
\label{VIJ2}

In the linear-response regime, we have $\Delta R_s\approx\Delta R$ and $\langle\Pi_\phi\rangle_f\approx\langle\Pi_\phi\rangle={\rm Tr}\,(\Pi_\phi\rho)$.
Then Eq.~\rqn{82} becomes
 \be
N_0\:=\:\frac{(\Delta R)^2}
{\langle\Pi_\phi\rangle\:(\bar{R}_s-\bar{R})^2},
 \e{82'}
We insert Eq.~\rqn{4.16} into Eq.~\rqn{82'}, taking into account Eq.~\rqn{6} and the fact that in the present case of a pure initial state $\langle\Pi_\phi\rangle=|\langle\phi|\psi\rangle|^2$.
Then we obtain the lower bound on $N_0$ for a given value of the measurement strength $\mu_0=|\gamma A_{\phi\psi}|\Delta F$ [Eq.~\rqn{297}],
 \be
N_0\:\ge\:(2\gamma\,|A_{\phi\psi}|\,\Delta F)^{-2}\:=\:(2\mu_0)^{-2}.
 \e{6.42}

Meters for which the value of $N_0$ achieves the lower bound are called here {\em optimal}.
Equivalently, for optimal meters, the SNR achieves the upper bound.
From Eq.~\rqn{82'}, we obtain that the lower bound on $N_0$,
 \be
N_0\:=\:(2\gamma\,|A_{\phi\psi}|\,\Delta F)^{-2},
 \e{306}
is achieved when Eq.~\rqn{4.20} is valid, i.e., when the condition \rqn{4.15} holds.
Thus, meters satisfying Eq.~\rqn{4.15} are {\em optimal for weak PPS measurements}, at least, in the linear-response regime.

More generally, we call meters {\em efficient} if the value of $N_0$ (the SNR) for measurements with such meters is equal to or of the order of the lower (upper) bound. 
For regular meters, i.e., meters satisfying Eq.~\rqn{334}, $N_0$ is of the order of the lower bound,
 \be
N_0\:\sim\:(\gamma\,|A_{\phi\psi}|\,\Delta F)^{-2}.
 \e{6.44}
Thus, regular meters [Eq.~\rqn{334}] are efficient for weak PPS measurements, at least, in the linear-response regime.


Beyond the linear-response regime, the minimal ensemble size $N_0$ is determined by Eq.~\rqn{82}, which is more complicated than Eq.~\rqn{82'}.
In this case, it is impossible to obtain an exact general result for the lower bound on $N_0$, and hence conditions for optimal meters are different for different types of meters.
However, we can obtain a simple general condition for {\em efficient} meters for weak PPS measurements, as follows.

Our analysis shows that beyond the linear-response regime, 
 \be
\Delta R_s\:\sim\:\Delta R.
 \e{6.45}
This holds, at least, in the common case when $\Delta R$ is not too small, Eq.~\rqn{239}; here we restrict our consideration to this case.
Now we can insert Eq.~\rqn{6.45} into Eq.~\rqn{82} to obtain
 \be
N_0\:\sim\:\frac{(\Delta R)^2}
{\langle\Pi_\phi\rangle_f\:(\bar{R}_s-\bar{R})^2}.
 \e{82''}
This result implies that $N_0$ is of the order of its lower bound when $|\bar{R}_s-\bar{R}|$ is of the order of its maximum.
As follows from the results of Secs.~\ref{IIIG2}, \ref{IVE3}, and \ref{IIIA6}, $|\bar{R}_s-\bar{R}|$ is of the order of its maximum for not too small $\Delta R$ when meters are regular, i.e., when the condition \rqn{334} holds.
Thus, regular meters are efficient beyond the linear-response regime.
But, as mentioned above, regular meters are efficient also for the linear-response regime.
Hence, regular meters are efficient for all regimes of weak PPS measurements.

\subsubsection{The strongly-nonlinear regime}

As follows from the results of Secs.~\ref{IIIG2} and \ref{IIIA6}, in the strongly-nonlinear regime the upper bound of the magnitude of the average pointer deflection for not too small $\Delta R$ is given by Eq.~\rqn{238}. 
Inserting Eq.~\rqn{238} into Eq.~\rqn{82''} yields
 \be
N_0\:\sim\:\langle\Pi_\phi\rangle_f^{-1}\:\sim\:{\cal A}^2.
 \e{307}
Here the second relation follows from Eq.~\rqn{298}.
Thus, in the optimal regime the ensemble size equals roughly the
inverse post-selection probability or, equivalently, the squared proper amplification coefficient.

This statement is valid irrespective of the value of $\bar{F}$.
We note that in the optimal regime the following estimate of the post-selection probability can be used for any $\bar{F}$,
 \be
\langle\Pi_\phi\rangle_f\:\sim\:\frac{|\langle\phi|\psi\rangle|^2}{1+(\bar{F}/\Delta F)^2}.
 \e{3.31}
Indeed, Eq.~\rqn{3.31} reduces to Eqs.~\rqn{287'} for $|\bar{F}|\alt\Delta F$ and to the first relation in Eq.~\rqn{305} for $|\bar{F}|\gg\Delta F$.
Inserting Eq.~\rqn{3.31} into Eq.~\rqn{307} yields an explicit expression for $N_0$,
 \be
N_0\:\sim\:\frac{1+(\bar{F}/\Delta F)^2} {|\langle\phi|\psi\rangle|^2}.
 \e{85}
This result implies that the ensemble size is minimal, 
 \be
N_0\:\sim\:|\langle\phi|\psi\rangle|^{-2},
 \e{151}
in the usual case $|\bar{F}|\lesssim \Delta F$, whereas in the case $|\bar{F}|\gg\Delta F$, the ensemble size increases as $\bar{F}^2$,
 \be
N_0\:\sim\:\frac{\bar{F}^2} {(\Delta F)^2|\langle\phi|\psi\rangle|^2}.
 \e{6.27}

\subsubsection{Inverted region}
\label{IIIK3}

We estimate $N_0$ in the inverted region in the important case where $|\bar{F}|\lesssim \Delta F$.
Now the relevant part of the average pointer deflection, which directly depends on the system, is the adjusted pointer deflection $\bar{R}_s-\bar{R}_{s,\infty}$ (cf.\ Sec.~\ref{IVE3}).
Correspondingly, in the above Eqs.~\rqn{6.28}, \rqn{82}, and \rqn{82''} one should perform the substitution
 \be
\bar{R}_s-\bar{R}\ \ \rightarrow\ \ \bar{R}_s-\bar{R}_{s,\infty}.
 \e{3.11}

We consider the typical case \rqn{4.10}.
Inserting Eqs.~\rqn{3.14}, \rqn{4.3}, and \rqn{3.11} into Eq.~\rqn{82''} yields the minimal ensemble size
 \be
N_0\,\sim\,|\langle\phi|\psi\rangle|^{-2},
 \e{3.13}
which has the same form as in the optimal regime, Eq.~\rqn{151}.
Note, however, that in the inverted region $\langle\phi|\psi\rangle$ is very small, and therefore the value of $N_0$ is significantly greater than in the optimal regime.
In view of Eqs.~\rqn{3.12} and \rqn{3.13}, for measurements in the inverted region the SNR is
 \be
{\cal R}\:\sim\:|\langle\phi|\psi\rangle|\,\sqrt{N}.
 \e{3.21}

It is quite remarkable that the quantum SNR \rqn{3.21}, which was obtained for weak PPS measurements, is of the same order as for {\em strong (projective) measurements} of a small overlap $|\langle\phi|\psi\rangle|$.
Indeed, for a system in the state $|\psi\rangle$, the overlap can be determined by measuring the projection operator\footnote{ 
The method considered here is not the only conceivable projective measurement of $|\langle\phi|\psi\rangle|$.
Another projective-measurement scheme, a balanced homodyne detection, is discussed in Sec.~\ref{XIC}; however for both schemes the SNR values are of the same order of magnitude.}
$\Pi_\phi=|\phi\rangle\langle\phi|$.
Such a measurement results in the eigenvalue 1 with the probability $P_1=|\langle\phi|\psi\rangle|^2$ and the eigenvalue 0 with the probability $P_0=1-|\langle\phi|\psi\rangle|^2$.
After the measurement of an ensemble of $N$ systems, the sum of the obtained eigenvalues is described by the binomial distribution and, correspondingly, has the average $NP_1$ and the standard deviation $(NP_1P_0)^{1/2}$.
The equality between the two latter quantities is attained for the minimal ensemble size
 \be
N_0\:=\:\frac{P_0}{P_1}\:\approx\:|\langle\phi|\psi\rangle|^{-2},
 \e{3.17}
where the approximation holds for $|\langle\phi|\psi\rangle|\ll1$.
As follows from Eqs.~\rqn{3.12} and \rqn{3.17}, now the SNR is given by
 \be
{\cal R}\:=\:|\langle\phi|\psi\rangle|\,\sqrt{N}.
 \e{3.22}

Thus, weak PPS measurements in the regime of very large weak values can be used to measure small overlaps with the same quantum SNR as the ideal measurements.
Moreover, when the measurement accuracy is limited by technical noise, weak PPS measurements can provide a higher measurement accuracy than ideal measurements, since weak PPS measurements involve a strong amplification (cf.\ Sec.~\ref{IVF4}).

An experiment on weak PPS measurements in the inverted region is discussed in Sec.~\ref{XI}.

\subsubsection{Comparison of weak measurements with and without post-selection}
\label{IIIK2}

As mentioned above, the conditions for meters to be efficient for weak standard and PPS measurements are given by Eqs.~\rqn{313} and \rqn{334}, respectively.
The condition \rqn{334} is less restrictive than \rqn{313}; in other words, meters efficient for weak standard measurements are also efficient for weak PPS measurements, but the converse generally is not true.

More specifically, when the covariance $\sigma_{FR}=0$, Eqs.~\rqn{2.6} and \rqn{4.15} are equivalent, i.e., meters optimal for weak standard measurements are also optimal for weak PPS measurements and vice versa.
Moreover, when $\sigma_{FR}$ vanishes or is sufficiently small,
 \be
|\sigma_{FR}|\,\alt\,|\,\overline{[R,F]}\,|,
 \e{4.17}
Eqs.~\rqn{313} and \rqn{334} are equivalent, i.e., meters efficient for weak standard measurements are also efficient for weak standard measurements and vice versa.
In other words, {\em weak measurements with and without
post-selection require ensembles of comparable sizes} (for comparable
magnitudes of $A_{\phi\psi}$ and $\bar{A}$).

In contrast, when the magnitude of the covariance is relatively large,
 \be
|\sigma_{FR}|\,\gg\,|\,\overline{[R,F]}\,|,
 \e{4.18}
then the generalized uncertainty relation \rqn{261} implies $|\,\overline{[R,F]}\,|\,\ll\,\Delta R\,\Delta F$, i.e., such meters are not efficient for standard measurements.
Indeed, in the case \rqn{4.18} the ensemble size $N_0$ in Eq.~\rqn{295} is much greater than $N_0$ for efficient meters in Eq.~\rqn{280}.
On the other hand, regular meters [i.e., meters satisfying Eq.~\rqn{334}] are efficient for weak PPS measurements irrespective of the value of $\sigma_{FR}$.

This can be understood in the following way.
Assume that we increase $|\sigma_{FR}|$, whereas $\overline{[R,F]}$ and $\Delta F$ [and hence the measurement strength \rqn{297}] are fixed.
For a regular meter [Eq.~\rqn{334}], this increase of $|\sigma_{FR}|$ results in an increase of $\Delta R$.
Moreover, in this case for weak PPS measurements the pointer deflection increases proportionally to $\Delta R$, at least, in typical cases [cf.\ Eqs.~\rqn{243}, \rqn{238}, and \rqn{3.14}], but for weak standard measurements it stays the same [cf.\ Eq.~\rqn{75}].
As a result, with an increase of $|\sigma_{FR}|$ the minimum ensemble size $N_0$ for weak PPS measurements in Eq.~\rqn{82''} is not changed, whereas $N_0$ for weak standard measurements in Eq.~\rqn{2.9} increases. 

Recall that in the case \rqn{4.18}, the enhancement coefficient ${\cal E}$ is large [see Eq.~\rqn{336}].
It is the effect of enhancement present in weak PPS measurements and absent in standard measurements which explains why meters not efficient for weak standard measurements can be efficient for weak PPS measurements.

Finally, we recall that in the limiting case $\overline{[R,F]}=0$ (non-standard meters), weak standard measurements cannot be performed at all, while weak PPS measurements are still generally efficient.

\section{Mixed preselected state}
\label{V}

\subsection{The general nonlinear formula}

Here we extend the above results to take into account the cases where the initial (``preselected'') state of the system $\rho$ is mixed.
Now Eqs.~\rqn{3.23} and \rqn{3.24} imply that the expansions for $\langle\Pi_\phi R_c\rangle_f$ and $\langle\Pi_\phi\rangle_f$ have the same form as in Eq.~\rqn{18'} with the changes
\bes{325t}
 \bea
&&|\langle\phi|\psi\rangle|^2\ \rightarrow\ \rho_{\phi\phi},\label{325}\\
&&(A^{k})_w (A^{l})_w^*\ \rightarrow\ A_w^{(k,l)}\:\equiv\:\frac{(A^k\rho A^l)_{\phi\phi}}{\rho_{\phi\phi}}\quad\quad(k,l\ge0),
 \ea{324}
\ese
where $A^0=(A^0)_w=1$.
As a result, Eqs.~\rqn{6.41} become now
 \bes{7.9}
 \bea
&&\langle\Pi_\phi R_c\rangle_f\:=\:\rho_{\phi\phi}[2\gamma\,{\rm Im}\,(\,\overline{R_cF}A_w)+\gamma^2\,\overline{FR_cF}\,A_w^{(1,1)}],\label{7.9a}\\
&&\langle\Pi_\phi\rangle_f\:=\:\rho_{\phi\phi}(1+2\gamma\bar{F}\,{\rm Im}\,A_w+\gamma^2\,\overline{F^2}\,A_w^{(1,1)}),
 \ea{7.9b}
\ese
whereas Eqs.~\rqn{13} and \rqn{182} yield, respectively, 
 \be
\bar{R}_s-\bar{R}\,=\,\frac{2\gamma\,{\rm Im}\,(\,\overline{R_cF}A_w)
+\gamma^2\,\overline{FR_cF}\,A_w^{(1,1)}} {1+2\gamma\,\bar{F}\:{\rm
Im}\,A_w +\gamma^2\,\overline{F^2}\,A_w^{(1,1)}}
 \e{61}
and
 \be
\bar{R}_s\,=\,\frac{\bar{R}+2\gamma\,{\rm Im}\,(\,\overline{RF}A_w)
+\gamma^2\,\overline{FRF}\,A_w^{(1,1)}} {1+2\gamma\,\bar{F}\:{\rm
Im}\,A_w+\gamma^2\,\overline{F^2}\,A_w^{(1,1)}}.
 \e{183}
Formally, Eqs.~\rqn{7.9}-\rqn{183} follow from Eqs.~\rqn{6.41}-\rqn{182} on replacing the definition \rqn{6} of the weak value by Eq.~\rqn{56} and replacing
 \be
|A_w|^2\ \rightarrow\ A_w^{(1,1)}\:=\:\frac{(A\rho
A)_{\phi\phi}}{\rho_{\phi\phi}}.
\e{57a}
As shown by Eq.~\rqn{61} or \rqn{183}, in the case of a mixed initial
state, the results of weak PPS measurements depend on {\em two
weak-value parameters}, $A_w$ [given now by Eq.~\rqn{56}] and the
associated weak value $A_w^{(1,1)}$, Eq.~\rqn{57a}.

\subsection{Validity conditions for weak PPS measurements}
\label{IVB}

The validity conditions for weak PPS measurements with a mixed
preselected state can be derived as in Sec.~\ref{IIIA3'}, the only
difference being that the $A$-dependent factors in Eq.~\rqn{58'} are
changed now, in view of Eqs.~\rqn{325t}, as follows,
 \be
(A^k)_{\phi\psi}(A^{n-k})_{\psi\phi}\ \ \rightarrow\ \ (A^k\rho
A^{n-k})_{\phi\phi}\quad\quad(0\le k\le n).
 \e{308}
As a prerequisite to an estimation of these factors, we need to
derive several inequalities, as follows.

The spectral expansion of $\rho$ has the form
 \be
\rho=\sum_i\lambda_i\,|\psi_i\rangle\langle\psi_i|,
 \e{59}
where $\langle\psi_i|\psi_j\rangle=\delta_{ij}$, $\lambda_i\ge0$,
and $\sum_i\lambda_i=1$.
 In view of Eq.~\rqn{59}, we can write 
 \bea
&|(A^{k}\rho\,A^{n-k})_{\phi\phi}|^2&=\ \left|\sum_i\lambda_i\,
(A^k)_{\phi\psi_i}\,(A^{n-k})_{\psi_i\phi}\right|^2\:\le\:\sum_i\lambda_i\,|(A^k)_{\phi\psi_i}|^2 \sum_j\lambda_j\,
|(A^{n-k})_{\psi_j\phi}|^2\nonumber\\
&&=\ (A^{k}\rho A^{k})_{\phi\phi}\,(A^{n-k}\rho A^{n-k})_{\phi\phi},
 \ea{65}
where the Cauchy-Schwarz inequality is used.
 Thus, we obtain the inequality
 \be
|(A^{k}\rho\,A^{n-k})_{\phi\phi}|^2\:\le\:(A^{k}\rho\,
A^{k})_{\phi\phi}\,(A^{n-k}\rho\,A^{n-k})_{\phi\phi}.
 \e{60}
In particular, for $n=k=1$, Eq.~\rqn{60} implies that
 \be
|(A\,\rho)_{\phi\phi}|^2\:\le\:(A\,\rho\,A)_{\phi\phi}\,\rho_{\phi\phi}
 \e{326}
or, in view of Eqs.~\rqn{56} and \rqn{57a},
 \be
|A_w|^2\:\le\:A_w^{(1,1)}.
 \e{64}
For a pure state $\rho$, Eqs.~\rqn{60}-\rqn{64} become equalities.
 Moreover, for $k\ge0$, we have the inequality
 \bea
&(A^{k}\rho A^{k})_{\phi\phi}&=\:\sum_i \lambda_i\:
|(A^k)_{\phi\psi_i}|^2\nonumber\\
&&\le\:\lambda_{\rm max}\sum_i |(A^k)_{\phi\psi_i}|^2\:=\:\lambda_{\rm
max}\,(A^{2k})_{\phi\phi},\quad\quad
 \ea{67}
where $\lambda_{\rm max}=\max\{\lambda_i\}$.

To estimate the quantity on the right-hand side of Eq.~\rqn{308}, we assume that
 \be
(A^{2k})_{\phi\phi}\:\lesssim\:[(A^2)_{\phi\phi}]^k
 \e{71}
and that the left and right sides in Eq.~\rqn{67} (with $k=1$) are
comparable, i.e.,
 \be
(A\,\rho\,A)_{\phi\phi}\:\sim\:\lambda_{\rm max}\,(A^2)_{\phi\phi}.
 \e{69a}
Combining Eqs.~\rqn{60}, \rqn{67}, and \rqn{71} yields 
 \be
|(A^{k}\,\rho\,A^{n-k})_{\phi\phi}|\:\lesssim\:\lambda_{\rm max}\,
[(A^2)_{\phi\phi}]^{n/2}.
 \e{284}
Using the relations \rqn{69a}, \rqn{284}, and \rqn{D5}, we obtain
that the omission of higher-order terms in the numerator and
denominator of Eq.~\rqn{61} is justified under the condition
 \be
\mu'\:\equiv|\:\gamma|\,[(A^2)_{\phi\phi}]^{1/2}\,(|\bar{F}|+\Delta
F)\:\ll\:1,
 \e{63}
where $\mu'$ is the small parameter in the case of a mixed
initial state.

The small parameter $\mu'$ in Eq.~\rqn{63} differs from $\mu$ in
Eq.~\rqn{12'} by the $A$-dependent factor.
 The latter is obtained under the assumptions \rqn{71} and \rqn{69a}.
 Note that Eq.~\rqn{71} holds, e.g., when
$(A^2)_{\phi\phi}\sim||A||^2$.
 It is of interest to compare the present validity conditions
\rqn{71}, \rqn{69a}, and \rqn{63} with the respective conditions
\rqn{11} and \rqn{12'} obtained for the case of a pure initial state
$\rho=|\psi\rangle\langle\psi|$.
 In this case $\lambda_{\rm max}=1$, so that Eq.~\rqn{69a} becomes the relation
 \be
|A_{\phi\psi}|\:\sim\:[(A^2)_{\phi\phi}]^{1/2},
 \e{281}
which implies the equivalence of the conditions \rqn{63} and \rqn{12'}.
 Note that the conditions \rqn{71} and \rqn{281} are generally
stricter than Eq.~\rqn{11}.
However, this difference can be negligible in some cases, as, e.g., in
the important case $|A_{\phi\psi}|\sim||A||$.

\subsection{Discussion}

As mentioned above, the equality in Eq.~\rqn{64} is obtained for all pure preselected states.
Let us consider whether there exist also mixed states $\rho$ for which the equality in Eq.~\rqn{64} is obtained.
Equation \rqn{64} is equivalent to Eq.~\rqn{326}, which, in view of Eq.~\rqn{59}, can be recast as
 \be
\left|\sum_i\lambda_i\,A_{\phi\psi_i}\,\langle\psi_i|\phi\rangle\,\right|^2
\:\le\:\sum_j\lambda_j\,|A_{\phi\psi_j}|^2\sum_i\lambda_i\,|\langle\phi|\psi_i\rangle|^2 .
 \e{5.7}
This relation directly follows from the Cauchy-Schwarz inequality \cite{nie00}, if the sum on the left-hand side is interpreted as a scalar product of the vectors $\{\sqrt{\lambda_i}\,A_{\phi\psi_i}\}$ and $\{\sqrt{\lambda_i}\,\langle\phi|\psi_i\rangle\}$.
The equality in Eq.~\rqn{5.7} and hence in Eq.~\rqn{64} holds if and only if the above vectors differ from each other by a scalar factor, i.e., if and only if for all $i$ for which $\lambda_i\ne0$
 \be
A_{\phi\psi_j}=\alpha_0\langle\phi|\psi_i\rangle,
 \e{5.8}
where $\alpha_0$ is some complex number independent of $i$.
Note that in this case Eq.~\rqn{56} implies that $A_w=\alpha_0=A_{\phi\psi_j}/\langle\phi|\psi_i\rangle$.
In other words, $|A_w|^2=A_w^{(1,1)}$ if and only if the initial state $\rho$ is a mixture of pure orthogonal states such that any of them taken as the initial state would produce the same weak value.

The condition \rqn{5.8} is trivial, i.e., always holds (unless $\langle\phi|\psi_i\rangle=0$), when there is only one allowed value of $i$, i.e., when $\rho$ is pure.
But the condition \rqn{5.8} is nontrivial when there are two or more allowed values of $i$, i.e., when $\rho$ is mixed.
In this case, Eq.~\rqn{5.8} always holds when $\hat{A}$ is proportional to the unity operator, at least, in the subspace spanned by the states $|\phi\rangle$ and $|\psi_i\rangle$ with the allowed values of $i$; however, Eq.~\rqn{5.8} is very unlikely to hold for an arbitrary physical quantity $A$.
Thus, we obtain that, as a rule, the equality $A_w^{(1,1)}=|A_w|^2$ implies that the initial state is pure, the exceptions being the cases for which $A$ and the mixed preselected state $\rho$ satisfy the condition \rqn{5.8}.

When $A_w^{(1,1)}=|A_w|^2$, Eqs.~\rqn{61} and \rqn{183} reduce to Eqs.~\rqn{13} and \rqn{182}, respectively, i.e., the present theory of weak PPS measurements developed for a pure preselected state is applicable also for the cases where $\rho$ is mixed.
However, when $|A_w|^2<A_w^{(1,1)}$, weak PPS measurements are generally affected by the fact that $\rho$ is mixed.
Note that the effect of the mixedness of $\rho$ increases with decreasing the ratio $|A_w|^2/A_w^{(1,1)}$.

\subsection{Measurement regimes}

{\em a. The linear response.}
The linear approximation to Eq.~\rqn{61} has the same form of Eq.~\rqn{14''} as for a pure preselected state, but $A_w$ is now given by Eq.~\rqn{56}.
 Consider the validity conditions of the linear response.
 The denominator in Eq.~\rqn{61} is close to one and hence can be
omitted, when
 \be
\gamma^2\,A_w^{(1,1)}\,[(\Delta F)^2+\bar{F}^2]\:\ll\:1.
 \e{100}
If, moreover, $A_w$ is sufficiently large, 
 \be
|\gamma\,\overline{R_cF}\,A_w|\gg
\gamma^2|\,\overline{FR_cF}\,|A_w^{(1,1)},
 \e{5.9}
then generally also the quadratic term in the numerator of Eq.~\rqn{61} can be neglected.
Thus, now the linear response holds under the conditions \rqn{100} and \rqn{5.9}.
When $\overline{FR_cF}=0$, the condition \rqn{5.9} always holds. When $\overline{FR_cF}\ne0$, one can use Eq.~\rqn{170} to show that for regular meters [Eq.~\rqn{334}] in the typical case of Eq.~\rqn{4.14} the condition \rqn{5.9} is equivalent to the requirement that $|A_w|^2/A_w^{(1,1)}$ is much greater than a number which is much less than 1.
Hence, in particular, when $|A_w|^2\sim A_w^{(1,1)}$ the condition \rqn{5.9} holds, i.e., the condition \rqn{100} is sufficient for the linear regime; moreover, now the condition \rqn{100} is equivalent to Eq.~\rqn{22}.

{\em b. Beyond the linear response.}
In the present case of a mixed initial state, the weak-value
parameters $A_w$ and $A_w^{(1,1)}$ generally cannot be made
infinitely large (see Sec.~\ref{VC}).
 Still, if $A_w^{(1,1)}$ is sufficiently large, the
nonlinear Eq.~\rqn{61} should be used.

 In particular, when the condition \rqn{100} is inverted,
 \be
\gamma^2A_w^{(1,1)}\,[(\Delta F)^2+\bar{F}^2]\:\gg\:1, 
 \e{4.1}
measurements are performed in the regime of the inverted region.
Then Eq.~\rqn{61} yields Eq.~\rqn{15} with the change $A_w\ \rightarrow\ A_w^{(1,1)}/A_w^*$, so that we obtain
 \be
\bar{R}_s-\bar{R}\approx\frac{\overline{FR_cF}}{\overline{F^2}}
+\frac{2\,{\rm Im}\,(\overline{R_cF}\,A_w)}{\gamma\,\overline{F^2}\,A_w^{(1,1)}}-\frac{2\bar{F}\,\overline{FR_cF}\:{\rm Im}\,A_w}{\gamma\,(\overline{F^2})^2\,A_w^{(1,1)}}.
 \e{5.6}

{\em c. Measuring the coupling strength and weak values.}
The dependence of $\bar{R}_s$ on $\gamma$ in Eq.~\rqn{61} is similar
to that in the case of a pure initial state, though the maximum of
$|\bar{R}_s-\bar{R}|$ is now generally reduced.
 The measurements of $\gamma$ and the weak values $A_w$ and
$A_w^{(1,1)}$ are now performed similarly to the case of a pure
initial state (see Sec.~\ref{IIIA6'}), the measurements being optimal
in the strongly-nonlinear regime [cf.\ Eq.~\rqn{132}],
 \be
\gamma^2\,A_w^{(1,1)}\,[(\Delta F)^2+\bar{F}^2]\:\sim\:1.
 \e{282}
Since now there are three real weak-value parameters, Re$\,A_w$,
Im$\,A_w$, and $A_w^{(1,1)}$, measuring them requires, at least, three
weak PPS measurements with different values of the meter or coupling
parameters, rather than two as for a pure initial state.

\subsection{Peculiar case: Large average input variable, $|\bar{F}|\gg\Delta F$}
\label{IVC3}

In the case $|\bar{F}|\gg\Delta F$, it is convenient to characterize
the effects of the mixedness of the preselected state by the parameter
 \be
v\:=\:\frac{(A_w^{(1,1)}-|A_w|^2)^{1/2}}{|{\rm Im}\,A_w|}.
 \e{228}
 Note that for pure preselected states $v=0$.
 When $\nu$ is small, $v\ll1$, the pointer value is resonantly enhanced under the conditions \rqn{173}-\rqn{168}, the resonance being approximately described by the expression
 \be
\bar{R}_s-\bar{R}\:=\:\frac{\overline{F_cR_cF_c}- \epsilon\bar{F}\,{\rm Im}\,\overline{[R,F]} -2x\bar{F}\sigma_{FR}}
{\bar{F}^2\left[x^2+\epsilon^2+(\Delta F/\bar{F})^2+v^2\right]},
 \e{221}
which differs from Eq.~\rqn{169} by the term $v^2$ in the denominator.
The effect of $v\ne0$ is to broaden the resonance \rqn{221} and to
decrease its amplitude, so that the maximum possible pointer
deflection \rqn{4.12} can be achieved only for
 \be
v\:\lesssim\:\frac{\Delta F}{|\bar{F}|}\:\ll\:1,
 \e{315}
whereas for $v\gg\Delta F/|\bar{F}|$ the maximum magnitude of
Eq.~\rqn{221} is much less than Eq.~\rqn{4.12}, decreasing with $v$.
Equation \rqn{315} provides a limitation on the ratio
$|\bar{F}|/\Delta F$ under which the optimal regime \rqn{4.12} is
possible, namely, $|\bar{F}|/\Delta F\lesssim v^{-1}$.
The sensitivity of the resonance \rqn{221} to the quantity $v$ for
$v\agt\Delta F/|\bar{F}|$ can be used to measure $v$ when $v$ is very
small.

\subsection{The minimum size of the measurement ensemble}

Let us estimate the minimum size of the ensemble needed for weak PPS
measurements in the linear and strongly-nonlinear regimes.
We also assume for simplicity that $|A_w|^2\sim A_w^{(1,1)}$.

{\em a. Linear regime.}
Now the condition for the linear regime is given by Eq.~\rqn{100}.
Inserting Eq.~\rqn{4.16} into Eq.~\rqn{82'} and taking into account that $\langle\Pi_\phi\rangle=\rho_{\phi\phi}$, we obtain
 \be
N_0\,\ge\,[4\gamma^2A_w^2(\Delta F)^2\rho_{\phi\phi}]^{-1}.
 \e{7.6}
The lower bound for $N_0$,
 \be
N_0\,=\,[4\gamma^2A_w^2(\Delta F)^2\rho_{\phi\phi}]^{-1},
 \e{7.7}
is obtained for optimal meters [Eq.~\rqn{4.15}], whereas measurements with effective meters [Eq.~\rqn{334}] require $N_0$ of the order of the value in Eq.~\rqn{7.7}.

{\em b. Strongly-nonlinear regime.}
The strongly-nonlinear regime occurs under the condition \rqn{282}.
We assume that the impurity of the preselected state is sufficiently small, so that the upper bound of the magnitude of the average pointer deflection is of the same order as for a pure preselected state, i.e., $|\bar{R}_s-\bar{R}|_{\rm max}\sim\Delta R$ [see Eq.~\rqn{238}; here $\Delta R$ is assumed to be not too small].
Moreover, we use the estimate for $\langle\Pi_\phi\rangle_f$ in Eq.~\rqn{3.31} with the substitution \rqn{325}, i.e., 
 \be
\langle\Pi_\phi\rangle_f\:\sim\:\frac{\rho_{\phi\phi}}{1+(\bar{F}/\Delta F)^2}.
 \e{7.8}
Then Eq.~\rqn{82''} yields [cf.\ Eq.~\rqn{85}]
 \be
N_0\,\sim\,\langle\Pi_\phi\rangle_f^{-1}\,\sim\,\frac{1+(\bar{F}/\Delta F)^2}{\rho_{\phi\phi}}.
 \e{109}
In the usual case $|\bar{F}|\lesssim\Delta F$, Eq.~\rqn{109} reduces to a simple form [cf.\ Eq.~\rqn{151}],
 \be
N_0\,\sim\,(\rho_{\phi\phi})^{-1}.
 \e{283}

\section{Examples of meters}
 \label{III}

The above theory is very general and holds for meters with finite-
and infinite-dimensional Hilbert spaces.
Below we consider the average pointer deflection for various types of
meters.
In particular, we will obtain formulas for the meter parameters which are important for weak PPS measurements.
Such parameters, which are discussed below, include the following mixed moments of the meter variables: $\overline{R_cF}$, $\overline{FR_cF}$, and $\overline{F_cR_cF_c}$ [compare, e.g., Eqs.~\rqn{13}, \rqn{169}, and \rqn{61}].

\subsection{Non-standard meters}
 \label{VIA'}

Here we consider two types of non-standard meters, namely, meters with $R=F$ and those with $\Delta R=0$.

\subsubsection{Meters with coinciding input and output variables, $R=F$}
\label{VIAa}

The above theory significantly simplifies when $R$ and $F$ commute.
 For the simplest such case, $R=F$, the moments of meter variables used in the present theory are shown in Table \ref{t3} for different types of the initial state of the meter.
In particular, configuration 2 in Table \ref{t3} corresponds to the general case of an arbitrary initial state $\rho_{\rm M}$, whereas meter configuration 1 describes the special case $\overline{F_c^3}=0$, which occurs, e.g., for a symmetric  distribution $\Phi(F)= \langle F|\rho_{\rm M}|F\rangle$ (e.g., a Gaussian or a Lorentzian) centered at $\bar{F}$, so that $\Phi(F)=\Phi(2\bar{F}-F)$.
In Table \ref{t3} we used Eq.~\rqn{170}.

\begin{table}[tb]
\begin{center}\setlength{\extrarowheight}{2pt}
\begin{tabular}{c|l|ccc}
\hline
&&\multicolumn{3}{|c}{Moments of the meter variables:}\bigstrut[t]\\
 No.&Meter configuration&\ \ $\overline{R_cF}$&\ \ $\overline{FR_cF}$&\ $\overline{F_cR_cF_c}$\\
\hline
1&$\overline{F_c^3}=0$&\ $(\Delta F)^2$&\ \ $2\bar{F}(\Delta F)^2$&0\bigstrut[t]\\
2&Arbitrary $\rho_{\rm M}$\ &$\ (\Delta F)^2$&$\ \ \overline{F_c^3}+2\bar{F}(\Delta
F)^2$&$\overline{F_c^3}$\bigstrut[t]\\[.5ex]
\hline
\end{tabular}
\end{center}
\caption{Moments of the meter variables used in the present theory,
for an arbitrary meter with $R=F$ and different types of the initial state $\rho_{\rm M}$.}
 \label{t3}\end{table}

Thus, for meters with $R=F$, the general nonlinear formula \rqn{13} becomes (cf.\ configuration 2 in Table \ref{t3}) 
 \be
\bar{F}_s-\bar{F}\:=\:\frac{2\gamma\,(\Delta F)^2\,{\rm Im}\,A_w
+\gamma^2[\overline{F_c^3}+2\bar{F}(\Delta F)^2]\,|A_w|^2}
{1+2\gamma\,\bar{F}\,{\rm Im}\,A_w +\gamma^2\,\overline{F^2}\,|A_w|^2}.
 \e{21}
Consider two important special cases.
 In the linear regime [Eq.~\rqn{22}], Eq.~\rqn{21} yields
 \be
\bar{F}_s-\bar{F}\:=\:2\gamma\,(\Delta F)^2\,{\rm Im}\,A_w,
 \e{27}
which is an extension of the second equality in Eq.~\rqn{160} to the general meter state.

Now let us consider the inverted region, assuming for simplicity that $\bar{F}=\overline{F_c^3}=0$.
We insert the meter moments listed in Table \ref{t3} for configuration 1 into Eq.~\rqn{15}, taking into account that now $\overline{F^2}=(\Delta F)^2$ [cf.\ Eq.~\rqn{236}].
This yields the simple expression
 \be
\bar{F}_s\:=\:\frac{2}{\gamma}\,{\rm Im}\,\frac{1}{A_w^*}\quad\quad{\rm for}\ \ |\gamma A_w|\,\Delta F\gg1.
 \e{363}
It is worth noting that Eq.~\rqn{363} provides an interpretation for the result of a classical theory in Eq.~(9) of Ref.~\cite{sta10x} in terms of weak values (see Sec.~\ref{XI}).

\subsubsection{Meters with zero pointer uncertainty, $\Delta R=0$.}
\label{VIAb}

Furthermore, consider the case when $\Delta R=0$ and $\hat{R}$ has a discrete spectrum.\footnote{Here we do not consider the case $\Delta R=0$, when $\hat{R}$ has a continuous spectrum, since then states with $\Delta R=0$ are generally unphysical.}
 In this case
 \be
\hat{R}\,\rho_{\rm M}\:=\:\rho_{\rm M}\,\hat{R}\:=\:\bar{R}\,\rho_{\rm M}
 \e{327}
[cf. the remark after Eq.~\rqn{231}], and hence
 \be
\hat{R}_c\,\rho_{\rm M}\:=\:\rho_{\rm M}\,\hat{R}_c\:=\:0.
 \e{328}
Equation \rqn{328} implies that
 \be
\overline{R_cF}={\rm Tr}\,(\rho_{\rm M}\hat{R}_c\hat{F})=0.
 \e{7.5}
As a result, now the linear approximation \rqn{14''} vanishes.
However the weak value is still measurable in the nonlinear regime.
 Indeed, now Eq.~\rqn{13} becomes, in view of Eqs.~\rqn{7.5} and \rqn{170},
 \be
\bar{R}_s-\bar{R}\ =\ \frac{\gamma^2\,\overline{F_cR_cF_c}\,|A_w|^2}
{1+2\gamma\,\bar{F}\,{\rm Im}\,A_w +\gamma^2\,\overline{F^2}\,|A_w|^2}.
 \e{237}
Note that Eq.~\rqn{237} differs from zero only when $\hat{F}$ and
$\hat{R}$ do not commute.
Indeed, for commuting $\hat{F}$ and $\hat{R}$, we have
$\overline{F_c\,R_c\,F_c}=\overline{F_c^2\,R_c}={\rm
Tr}\,(\hat{F}_c^2\,\hat{R}_c\,\rho_{\rm M})=0$, in view of Eq.~\rqn{328}.

 In both cases \rqn{21} and \rqn{237} one can measure Im$\,A_w$
and $|{\rm Re}\,A_w|$, as discussed in the end of Sec.~\ref{IIIA8d}.

\subsection{Continuous-variable meters}
 \label{VIA}

 The standard measurement theory \cite{neu55,aha88,aha05} involves
a continuous-variable meter and canonically conjugate variables.
 Correspondingly, the bulk of the literature on weak values involves
such meters.
 Here we apply the above theory to the important case of continuous-variable meters.

First, we remind that meters with $R=F$ (including
continuous-variable meters) were discussed in Sec.~\ref{VIA'} (see,
especially, Table \ref{t3}).
 The case of commuting $F$ and $R$ is essentially similar to the
case $R=F$.
 Consider now a continuous-variable meter with non-commuting $F$ and
$R$.

\subsubsection{Canonically conjugate variables}
\label{VIB1}

The present theory is applicable to arbitrary meter variables,
however here we focus on canonically conjugate variables given by Eq.~\rqn{97}.
Since $[q,p]=i$, now Eq.~\rqn{76} becomes
 \be
\overline{R_cF}\ \equiv\ \overline{q_cp}\ =\ \sigma_{pq}+i/2,
 \e{6.1}
and the general linear-response formula \rqn{79} reduces to the result of Ref.~\cite{joz07}, which is a direct extension of Eq.~\rqn{4},
 \be
\bar{q}_s-\bar{q}=\gamma\,({\rm Re}\,A_w+2\,\sigma_{pq}\,{\rm Im}\,A_w).
 \e{20}

The covariance $\sigma_{pq}$ is an important parameter, since it affects the result \rqn{20} of weak PPS measurements.
Moreover, it enters the generalized uncertainty relation for the canonically conjugate meter variables $p$ and $q$, which, as follows from Eq.~\rqn{261}, has the form
 \be
\Delta p\:\Delta q\,\ge\, \sqrt{1/4+\sigma_{pq}^2}.
 \e{268}
Therefore, it is of interest to obtain the conditions under which $\sigma_{pq}\ne0$.

Assume now for simplicity that at $t=0$ the meter is in a pure state
$|\psi_{\rm M}\rangle$.
 Presenting $\psi_{\rm M}(q)=\langle q|\psi_{\rm M}\rangle$ and
$\psi_{\rm M}(p)=\langle p|\psi_{\rm M}\rangle$ in the forms
 \bea
&&\psi_{\rm M}(q)= f_q(q)\,\exp[i\xi(q)],\label{6.12}\\
&&\psi_{\rm M}(p)=f_p(p)\,\exp[-i\zeta(p)],
 \ea{96}
where $f_q(q)$, $\xi(q)$, $f_p(p)$, and $\zeta(p)$ are real and
continuous functions, we obtain two equivalent expressions for the anticommutator of $q$ and $p$ (see Appendix \ref{C}),
 \be
\overline{\{q,p\}}\:=\:2\,\overline{q\,\xi'(q)}\:=\:
2\,\overline{p\,\zeta'(p)},
 \e{36}
where the prime denotes differentiation.
The second equality in Eq.~\rqn{36} is an interesting and nontrivial relation between the phases $\xi(q)$ and $\zeta(p)$.
Combining Eqs.~\rqn{262} and \rqn{36} yields two equivalent expressions for $\sigma_{pq}$,
 \be
\sigma_{pq}\:=\:\overline{p\zeta'(p)}-\bar{q}\,\bar{p}\:
=\:\overline{q\xi'(q)}-\bar{q}\,\bar{p}.
 \e{6.36}

A consequence of Eq.~\rqn{6.36} is that $\sigma_{pq}=0$ if, at least, one of the phases $\zeta(p)$ and $\xi(q)$ is constant or linear, because linear $\zeta(p)$ and $\xi(q)$ imply, 
 \be
\zeta'(p)=\bar{q},\quad\xi'(q)=\bar{p},
 \e{359'}
as follows from the general expressions derived in Appendix \ref{C},
 \bea
&&\overline{\zeta'(p)}=\bar{q},\label{C3}\\
&&\overline{\xi'(q)}=\bar{p}.
 \ea{C12}

Consequently, we conclude that the covariance $\sigma_{pq}$ for a coordinate $q$ and the canonically conjugate moment $p$ is nonzero if and only if one of the two equivalent conditions holds: (a) the phase $\zeta(p)$ is nonlinear in $p$ and
 \be
\overline{p\,\zeta'(p)}\:\ne\:\bar{q}\,\bar{p}
 \e{6.34}
or (b) the phase $\xi(q)$ is nonlinear in $q$ and
 \be
\overline{q\,\xi'(q)}\:\ne\:\bar{q}\,\bar{p}.
 \e{6.35}

Inserting Eq.~\rqn{6.36} into Eq.~\rqn{20}, we obtain two
equivalent expressions for the linear response,
 \bes{253t}\bea
&\bar{q}_s-\bar{q}&=\ \gamma\,\{{\rm Re}\,A_w+2[\,\overline{p\zeta'(p)}
-\bar{q}\,\bar{p}\,]\,{\rm Im}\,A_w\}\label{253}\\
&&=\ \gamma\,\{{\rm Re}\,A_w +2[\,\overline{q\xi'(q)}
-\bar{q}\,\bar{p}\,]\,{\rm Im}\,A_w\}.
 \ea{253b}\ese
It is usually noted \cite{aha88,joz07,wum09} that Eq.~\rqn{20} reduces
to Eq.~\rqn{4} for a real $\psi_{\rm M}(q)$.
The above discussion of $\sigma_{pq}$ implies a more general result:
Equation \rqn{4} holds whenever the phase of either $\psi_{\rm M}(q)$ or $\psi_{\rm M}(p)$ is a linear function (or, as a special case, a constant or zero).
 In contrast, a nonlinear phase $\zeta(p)$ or $\xi(q)$ generally
results in a non-vanishing correlation between $p$ and $q$,
$\sigma_{pq}\ne0$, so that both terms in Eqs.~\rqn{20} do not vanish
(see also Sec.~\ref{VIB2}).

When weak values are large, one should use the general nonlinear
Eq.~\rqn{13} [or \rqn{61}, for a mixed preselected state] or Eqs.~\rqn{169} and \rqn{221} for the case $|\bar{F}|\gg\Delta F$.
 In particular, Eq.~\rqn{13} with the account of Eq.~\rqn{6.1} now becomes
 \be
\bar{q}_s-\bar{q}\:=\:\frac{\gamma\,({\rm Re}\,A_w +2\sigma_{pq}\,{\rm
Im}\,A_w) +\gamma^2\,\overline{pq_cp}\:|A_w|^2}
{1+2\gamma\bar{p}\:{\rm Im}\,A_w +\gamma^2\,\overline{p^2}\:|A_w|^2},
 \e{98}
whereas for a mixed preselected state one should replace $|A_w|^2
\rightarrow A_w^{(1,1)}$ in Eq.~\rqn{98}.

\begin{table}[tb]
\begin{center}\setlength{\extrarowheight}{1.5pt}
\begin{tabular}{c|l|ccc}
\hline
&&\multicolumn{3}{c}{Moments of the meter variables:}\bigstrut[t]\\
 No.&Meter configuration&$\sigma_{pq}$&$\overline{pq_cp}$ &$\overline{p_cq_cp_c}$\bigstrut\\
\hline
 1&Constant or linear $\zeta(p)$&0&0&0\bigstrut[t]\\
2&Quadratic $\zeta(p)$, $\overline{p_c^3}=0$&$b/2$&$b\bar{p}$&0\bigstrut[t]\\
3&Quadratic $\zeta(p)$&$b/2$&\ $b\,\overline{p^2p_c}/2(\Delta p)^2$
&\ \ $b\,\overline{p_c^3}/2(\Delta p)^2$\bigstrut[t]\\
 4&Arbitrary $|\psi_{\rm M}\rangle$&\ $\overline{p\zeta_c'(p)}$ or $\overline{q\xi_c'(q)}$&
$\overline{p^2\zeta_c'(p)}$ &$\overline{p_c^2\zeta_c'(p)}$\bigstrut[t]\\[.5ex]
\hline
 \end{tabular}
\end{center}
\caption{Moments of meter variables used in the present theory, for
a meter with $F=p$ and $R=q$; different cases correspond to different types of the pure initial state $|\psi_{\rm M}\rangle$.
 Here, $\zeta_c'(p)=\zeta'(p)-\bar{q}$, and $\xi_c'(q)=\xi'(q)-\bar{p}$.}
 \label{t1}\end{table}

The expressions for the meter parameters entering the present
theory for the meter \rqn{97} with an arbitrary pure initial state are shown in Table \ref{t1}, configuration 4.
 To derive these expressions, we used Eqs.~\rqn{262}, \rqn{170}, \rqn{36}, \rqn{C3}, the equality
 \be
\overline{p\,q_c\,p}\:=\:\overline{p\,q\,p}-\bar{q}\,\overline{p^2},
 \e{329}
and the following relation obtained in Appendix \ref{C},
 \be
\overline{p\,q\,p}\:=\:\overline{p^2\,\zeta'(p)}.
 \e{99}
The case when $\psi_{\rm M}(p)$ is real or has a constant or linear phase $\zeta(p)$, is especially simple, since then the real meter parameters vanish, as shown in Table \ref{t1}, configuration 1 (see also Sec.~\ref{VIB3}).

Consider now the simplest case of a nonlinear phase in the momentum
representation: a quadratic $\zeta(p)$.
Using Eq.~\rqn{C3}, it is easy to show that in the general case a
quadratic $\zeta(p)$ satisfies the equation
 \be
\zeta'(p)\:=\:\bar{q}+\frac{b\,(p-\bar{p})}{2(\Delta p)^2},
 \e{216}
where $b$ is a real dimensionless parameter characterizing the
quadratic phase modulation.
 Inserting Eq.~\rqn{216} into the formulas for configuration 4 in
Table \ref{t1} yields configuration 3 in Table \ref{t1}.
 In particular, the linear-response result \rqn{14} becomes 
 \be
\bar{q}_s-\bar{q}\:=\:\gamma\,({\rm Re}\,A_w+b\,{\rm Im}\,A_w).
 \e{159}
Now we obtain $\sigma_{pq}=b/2$ (see Table \ref{t1}, configuration 3), and the generalized uncertainty relation \rqn{268} now becomes
 \be
\Delta p\:\Delta q\:\ge\:\frac{\sqrt{1+b^2}}{2}.
 \e{248}

When $\overline{p_c^3}=0$, which holds, e.g., for the function
$\Phi(p)=|\psi_{\rm M}(p)|^2$, which is symmetric with respect to
$\bar{p}$, $\Phi(p)=\Phi(2\bar{p}-p)$, the formulas for the case of
the quadratic $\zeta(p)$ simplify: see configuration 2 in Table \ref{t1}.

An example of a state with a quadratic phase and a symmetric
$\Phi(p)$ is a general complex Gaussian state given by Eqs.~\rqn{103} and \rqn{38}.
The parameters $\Delta p$ and $\Delta q$ in Eqs.~\rqn{103} and \rqn{38} are related by Eq.~\rqn{49}, which is essentially the generalized uncertainty relation \rqn{248} with the equals sign.
A general Gaussian state implies the formulas in Table \ref{t1},
configuration 2.

\subsubsection{Invariance with respect to a meter gauge
transformation}
\label{VIB2}

As mentioned above (see also Table \ref{t1}), $p$ and $q$ are generally correlated (i.e., $\sigma_{pq}\ne0$) whenever the phase $\zeta(p)$ is nonlinear.
To understand better this result, we make the following remark.

The formulas in Table \ref{t1}, configurations 1 and 4, imply that the average pointer deflection in the presence of a nonlinear $\zeta(p)$ will not change if the meter is modified, as follows: (i) $\zeta(p)$ is replaced by a phase $\tilde{\zeta}(p)$ which is vanishing or at most linear in $p$ and (ii) the pointer is changed according to
 \be
q\ \rightarrow\ \tilde{R}=q+\zeta'(p)+C,
 \e{250}
where $C$ is an arbitrary real constant.
 Equation \rqn{250} is a special case of the invariance property
of the average pointer deflection under a gauge transformation of
the meter, discussed in Sec.~\ref{As2} [see, in particular,
Eq.~\rqn{A5}].
 In the case of a quadratic $\zeta(p)$ as given in Eq.~\rqn{216}, Eq.~\rqn{250} becomes
 \be
q\ \rightarrow\ \tilde{R}=q+bp+C.
 \e{251}
The modified pointer variable $\tilde{R}$ in Eq.~\rqn{250} is obviously correlated with $p$ when $\zeta(p)$ is nonlinear in $p$.

Note that $\tilde{R}$ is canonically conjugate to $p$, since $[\tilde{R},p]=[q,p]=i$, for any $\zeta(p)$.
This is not surprising, since the canonically conjugate variable is known to be determined not uniquely \cite{sch30}.

\subsubsection{Measuring physical parameters}
\label{VIB3}

Here we will consider weak PPS measurements involving several examples of meters with canonically conjugate variables \rqn{97} in the usual case where $|\bar{p}|\lesssim\Delta p$.
As follows from the general discussion in Sec.~\ref{III'}, the optimal conditions for measurements of physical parameters, such as $\gamma$ and $A_w$, are obtained in the strongly-nonlinear regime, the condition for which in Eq.~\rqn{303} becomes now
 \be
|\gamma A_w|\,\Delta p\,\sim\,1.
 \e{8.2}
Here we will discuss the optimal conditions for two cases: (i) the phase $\zeta(p)$ in Eq.~\rqn{96} is constant or linear and (ii) $\zeta(p)$ is nonlinear.

{\em Case (i). A constant or linear $\zeta(p)$.}
In this case Eq.~\rqn{98} is especially simple (cf.\ Table \ref{t1}, configuration 1),
 \be
\bar{q}_s-\bar{q}\:=\:\frac{\gamma\,{\rm Re}\,A_w}{1+2\gamma\bar{p}\,{\rm Im}\,A_w+\gamma^2\,\overline{p^2}\:|A_w|^2}.
 \e{271}
This quantity differs from zero only when ${\rm Re}\,A_w$ is
nonvanishing.
 The magnitude of the pointer deflection \rqn{271} is maximal for
 \be
\gamma A_w\:=\:\frac{\pm\Delta p-i\bar{p}}{\overline{p^2}},
 \e{273}
when Eq.~\rqn{271} becomes, respectively,
 \be
\bar{q}_s-\bar{q}\:=\:\pm\frac{1}{2\Delta p}.
 \e{272}
 Thus, for $\bar{p}=0$ ($\bar{p}\ne0$), the optimal $A_w$
should be real (complex).
 The Heisenberg uncertainty relation [Eq.~\rqn{268} with
$\sigma_{pq}=0$] and Eq.~\rqn{272} imply that now the maximum magnitude of the pointer deflection satisfies the relation
 \be
|\bar{q}_s-\bar{q}|_{\rm max}\:\le\:\Delta q,
 \e{274}
which is in agreement with the general Eq.~\rqn{238}.
The equality in Eq.~\rqn{274} is achieved, e.g., for meters with a Gaussian initial state.

{\em Case (ii). A nonlinear $\zeta(p)$.}
This case can be analyzed similarly to case (i), though generally
Eq.~\rqn{98} is more complicated than Eq.~\rqn{271}.
In contrast to case (i), this case shows the enhancement discussed in
Sec.~\ref{IIIH2}.
In particular, in the optimal regime the maximum magnitude of the
pointer deflection is of the order of $\Delta q$ in both cases (i) and (ii), in agreement with the general result \rqn{238}.
However, for a given $\Delta p$, the quantity $\Delta q$ in case (ii)
is greater than that in case (i) due to a non-zero covariance
$\sigma_{pq}$ [cf.\ Eq.~\rqn{268}].
An increase of the nonlinear $\zeta(p)$ leads to an increase of
$|\sigma_{pq}|$, which in turn yields an enhancement of the maximum
pointer deflection and thus an increase of the measurement accuracy.

The enhancement of the pointer deflection occurs also in the
linear-response regime.
In contrast to the nonlinear regime, this enhancement occurs only when ${\rm Im}\,A_w\ne0$ [see Eq.~\rqn{20}], being maximized when $A_w$ is purely imaginary \cite{hos08,dix09,how10,sta09,sta10a,tur11,hog11}.
In both regimes, the increase of the pointer deflection is
characterized by the enhancement coefficient ${\cal E}$ given in Eq.~\rqn{335}.
 However, the optimal conditions are obtained only in the nonlinear
regime.

Let us consider two examples.

{\em Example 1. Complex} $A_w$.
First, we consider the case of a quadratic $\zeta(p)$ with
$\bar{p}=\overline{p^3}=0$.
 Then, as follows from Eq.~\rqn{98} and Table \ref{t1}, configuration 2, we obtain
 \be
\bar{q}_s-\bar{q}\:=\:\frac{\gamma\,({\rm Re}\,A_w +b\,{\rm
Im}\,A_w)}{1+\gamma^2\,(\Delta p)^2\,|A_w|^2}.
 \e{275}
The magnitude of Eq.~\rqn{275} is maximum for the following values of $\gamma A_w$,
 \be
\gamma A_w\:=\:\pm\frac{1+ib}{\Delta p\sqrt{1+b^2}}.
 \e{277}
Substituting these values back into Eq.~\rqn{275} yields, respectively, 
 \be
\bar{q}_s-\bar{q}\:=\:\pm\frac{\sqrt{1+b^2}}{2\Delta p}.
 \e{276}
In view of Eq.~\rqn{276} and the generalized uncertainty relation \rqn{248}, we again obtain the upper bound for the magnitude of the pointer deflection as in Eq.~\rqn{274}.
The equality in Eq.~\rqn{274} is achievable now, e.g., for a general complex Gaussian meter wavefunction.

Note that the pointer deflection in Eq.~\rqn{276} is enhanced relative to Eq.~\rqn{272} by the factor given {\em exactly} by Eq.~\rqn{335}, which now becomes
 \be
{\cal E}=\sqrt{1+b^2}.
 \e{352}
In particular, for $|b|\gg1$ we obtain that
 \be
{\cal E}\:=\:|b|\:\gg1.
 \e{338}

In the above example, just as in the linear case \rqn{20}, the
enhancement cannot be obtained with a real weak value, since then the
$b$-dependent term disappears in Eq.~\rqn{275}.
The following example shows that in the nonlinear case a strong
enhancement is possible even for a real $A_w$, when $\bar{p}\ne0$.

{\em Example 2. Real} $A_w$.
Let $A_w$ be real, $\zeta(p)$ quadratic, and $\overline{p_c^3}=0$.
 Then Eq.~\rqn{98} and Table \ref{t1}, configuration 2, yield
 \be
\bar{q}_s-\bar{q}\:=\:\frac{\gamma A_w
+\gamma^2\,b\,\bar{p}\,A_w^2}{1+\gamma^2\,\overline{p^2}\,A_w^2}.
 \e{341}
Now the pointer-deflection magnitude is maximal when
 \be
\gamma A_w=2(\bar{q}_s-\bar{q}).
 \e{7.4}
In this case the pointer deflection is for $b\ne0$
 \be
\bar{q}_s-\bar{q}\:=\:\frac{b\,\bar{p}}{2\overline{p^2}}\left\{1+ \left[1+\frac{\overline{p^2}}{(b\bar{p})^2}\right]^{1/2}\right\},
 \e{342}
whereas for $b=0$
 \be
\bar{q}_s-\bar{q}\:=\:\pm\frac{1}{2(\overline{p^2})^{1/2}} \quad\quad(b=0).
 \e{7.3}
For $|b|\gg1$ the magnitude of Eq.~\rqn{342} as a function of $\bar{p}$ is maximum for $|\bar{p}|=\Delta p$,
 \be
\bar{q}_s-\bar{q}\:=\:{\rm sgn}(\bar{p})\,\frac{b}{2\Delta p} \quad\quad(|b|\gg1,\ |\bar{p}|=\Delta p).
 \e{344}
This result is easily obtained, if one takes into account that for $|b|\gg1$ and for not too small $|\bar{p}|$, i.e., $|\bar{p}|\gg\Delta p/|b|$, the fraction in the square brackets in Eq.~\rqn{342} can be neglected, yielding $\bar{q}_s-\bar{q}=b\bar{p}/[\bar{p}^2+(\Delta p)^2]$.

The magnitude of the pointer deflection in Eq.~\rqn{344} is the same as in Eq.~\rqn{276}  with $|b|\gg1$.
Comparing Eq.~\rqn{344} and Eq.~\rqn{7.3} with $|\bar{p}|=\Delta p$ shows that now enhancement is given by the factor $\sqrt{2}|b|$, which is of the order of the enhancement coefficient $\cal E$ in Eq.~\rqn{338}, in agreement with the general discussion in Sec.~\ref{IIIH2}.
It is of interest also to compare Eq.~\rqn{344} with the special case $\bar{p}=0$ of Eq.~\rqn{7.3}, $\bar{q}_s-\bar{q}=\pm(2\Delta p)^{-1}$, where the magnitude of Eq.~\rqn{7.3} as a function of $\bar{p}$ is maximum.
As a result, we obtain that the maximum of $|\bar{q}_s-\bar{q}|$ as a function of both $\gamma A_w$ and $\bar{p}$ is increased by the factor $|b|$ for $|b|\gg1$ in comparison to that for $b=0$.

\subsubsection{Effects of the meter Hamiltonian}
\label{VIB4}

The meter Hamiltonian is often nonzero in experiments.
Therefore, let us consider the effects of the meter Hamiltonian.
For simplicity, we assume that the meter is described by the same
Hamiltonian as a free particle \cite{hos08,lor08},
 \be
H_{\rm M}\:\equiv\:H_{\rm M1}\:=\:H_{\rm M2}\:=\:\frac{p^2}{2m_p},
 \e{130}
where $m_p$ is the ``particle'' mass.
In this subsection, in paragraph {\em a} we consider the case of the canonically conjugate meter variables \rqn{97}, whereas in paragraph {\em b} we consider a more general case.

{\em a. Effective initial state.}
Here we consider the case \rqn{97}.
Now the Hamiltonian \rqn{130} commutes with $F=p$, therefore, as discussed in Sec.~\ref{VB'}, the effects of the meter Hamiltonian can be taken into account by two equivalent ways: either through the effective initial state or through the effective pointer.

Here we describe the effects of the meter Hamiltonian by the effective initial state [see Eqs.~\rqn{242} and \rqn{5.1}],
 \be
\psi_{\rm M}(p,t_{\rm M})\:=\:\exp(-iH_{\rm M}t_{\rm M})\,\psi_{\rm M}(p).
 \e{6.4}
The Hamiltonian \rqn{130} generates a quadratic contribution to
the phase of $\psi_{\rm M}(p,t_{\rm M})$, with $b$ given by the
quantity
 \be
b(t_{\rm M})\:=\:\frac{2(\Delta p)^2\,t_{\rm M}}{m_p},
 \e{278}
which increases with $t_{\rm M}$.
Thus, the effective initial state can have a nonlinear phase modulation due to the free meter Hamiltonian, even when the phase of
the initial meter state $\psi_{\rm M}(p)$ is constant or linear in $p$.

Generally, the initial state $\psi_{\rm M}(p)$ has a nonlinear phase $\zeta(p)$.
As a result, the meter parameters for weak PPS measurements are the sums of the contribution due to $\zeta(p)$ (see Table \ref{t1}, configuration 4) and the contribution due to the meter Hamiltonian, i.e., due to the quadratic phase modulation determined by the parameter $b$ in Eq.~\rqn{278} (see Table \ref{t1}, configuration 3).

In particular, in the simple case, when the initial meter phase in the momentum space is constant or linear, the effect of the meter Hamiltonian is to change case 1 in Table \ref{t1} to configuration 2 or 3.
For a sufficiently long $t_{\rm M}$, this results in a large
pointer-deflection enhancement [cf. Eq.~\rqn{338}],
 \be
{\cal E}\:=\:\frac{2(\Delta p)^2t_{\rm M}}{m_p}\:\gg\:1,
 \e{339}

When $\psi_{\rm M}(p)$ is a Gaussian, Eq.~\rqn{339} can be recast also
as
 \be
{\cal E}\:=\:\frac{2(\Delta q_{\rm M})^2\,m_p}{t_{\rm M}},
 \e{353}
where $\Delta q_{\rm M}$ is the uncertainty of $q$ at the moment
$t_{\rm M}$.
To derive Eq.~\rqn{353}, we took into account that $\Delta p\Delta
q_{\rm M}=b(t_{\rm M})/2\gg1$ [cf.\ Eq.~\rqn{49}], which yields, in
view of Eq.~\rqn{278},
 \be
\Delta q_{\rm M}\:=\:\frac{\Delta p\:t_{\rm M}}{m_p}.
 \e{354}
Equations \rqn{339} and \rqn{353} were obtained and checked
experimentally in Ref.~\cite{hos08} (where the enhancement factor ${\cal E}$ is denoted by $F$) for the special case of linear response (see also Ref.~\cite{lor08}).

Note, however, that the same enhancement is obtainable also in the optimal regime, as discussed in Sec.~\ref{IIIH2} and shown by a direct calculation in Sec.~\ref{VIB3}.
It is advantageous to perform experiments in the optimal
regime, since the proper amplification and hence the total
amplification [Eq.~\rqn{337}] are greater in the optimal regime by,
at least, an order of magnitude than those in the linear regime.

{\em b. Effective pointer variable.}
Consider the case when $F$ is arbitrary, whereas $R=q$.
Now the Hamiltonian $H_{\rm M}$ does not necessarily commute with $\hat{F}$.
In this case, as discussed in Sec.~\ref{VB'}, the effects of $H_{\rm M}$ can be taken into account through the effective pointer variable.
From Eqs.~\rqn{349}, \rqn{5.3}, \rqn{5.4} and \rqn{130} we obtain that the effective pointer variable is
 \be
q(t_{\rm M})\:=\:q+\frac{t_{\rm M}}{m_p}p,
 \e{6.5}
both for PPS and standard measurements.
Correspondingly, now the pointer deflection equals (irrespective of the measurement strength)
 \be
\bar{q}_{s,f}-\bar{q}\:=\:(\bar{q}_{s,f}-\bar{q})_0+\frac{t_{\rm M}}{m_p} (\bar{p}_{s,f}-\bar{p})_0.
 \e{6.6}
Here the subscript $s$ ($f$) corresponds to PPS (standard) measurements, whereas the two terms in the parentheses denoted by the subscript ``0'' are the unperturbed results of the measurements of the coordinate and the momentum, respectively, i.e., the results obtained in the absence of the meter Hamiltonian.

Equation \rqn{6.6} implies that, when $t_{\rm M}$ is very small, the effect of the meter Hamiltonian is negligible,
 \be
\bar{q}_{s,f}-\bar{q}=(\bar{q}_{s,f}-\bar{q})_0.
 \e{6.9}
In the opposite limit, when $t_{\rm M}$ is sufficiently large, the unperturbed contribution from the coordinate can be neglected in Eq.~\rqn{6.6}, and the measurement of the coordinate provides the unperturbed momentum deflection,
 \be
\bar{q}_{s,f}-\bar{q}=\frac{t_{\rm M}}{m_p} (\bar{p}_{s,f}-\bar{p})_0.
 \e{6.7}
In this case, the measurement of the momentum is ``translated'' into the measurement of the coordinate \cite{aha02}.
This is a very useful feature, since it is usually much easier to measure the position of a particle than its momentum.

Note that the factor $t_{\rm M}/m_p$ in Eq.~\rqn{6.7} increases with $t_{\rm M}$ and hence can provide a strong enhancement.
In the case \rqn{97}, this enhancement is equivalent to that mentioned above, which is due to the correlation between $F$ and $R$.
However, generally (e.g., for $F=q$) the enhancement due to the meter Hamiltonian in Eq.~\rqn{6.7} differs from the enhancement  discussed in Sec.~\ref{IIIH2}.

Let us discuss special cases.

(i) Consider measurements with 
 \be
F=R=q.
 \e{6.8}
This case is realized, e.g., in the Stern-Gerlach experiment \cite{aha88}, as well as in some optical experiments \cite{par98,dix09,how10,sta09,sta10a,tur11,hog11}.
We note a difference between standard and weak PPS measurements for the case \rqn{6.8}.
For standard measurements, the first term on the right-hand side of Eq.~\rqn{6.6} vanishes [cf.\ Eq.~\rqn{75} with $F=R=q$], i.e., the ``translation'' \rqn{6.7} is exact for any $t_{\rm M}$.
Hence, effectively the meter variables are given by [cf.\ the case \rqn{1.37}] 
 \be
F=q,\quad R=\frac{t_{\rm M}}{m_p}p.
 \e{6.10}
In comparison, for PPS measurements both terms in Eq.~\rqn{6.6} are generally nonzero.
Now the ``translation'' \rqn{6.7} is approximate; it occurs only when $t_{\rm M}$ is sufficiently long, whereas in the opposite limit of a short $t_{\rm M}$ Eq.~\rqn{6.9} holds \cite{sta10a}. 

(ii) Consider the meter with the canonically conjugate variables given by Eq.~\rqn{97}, $F=p$ and $R=q$.
Now for standard measurements, Eq.~\rqn{6.9} is exact, i.e., effects of the meter Hamiltonian vanish.
In contrast, for weak PPS measurements with the meter \rqn{97}, effects of the meter Hamiltonian do not vanish.
Now Eq.~\rqn{6.6} yields the results discussed above in paragraph {\em a}.
This can be checked by inserting Eq.~\rqn{98} and Eq.~\rqn{21} with $F=p$ on the right-hand side of Eq.~\rqn{6.6} and using Table \ref{t1}.
Thus, the two seemingly different approaches developed for this case in paragraphs {\em a} and {\em b} are equivalent, in agreement with the discussion in Sec.~\ref{VB'}.
This equivalence implies that the quadratic phase characterized by the parameter \rqn{278} can be equivalently replaced by the effective pointer \rqn{6.5}.
In turn, the latter equivalence is a consequence of the invariance of PPS measurements with respect to gauge transformations of the meter (Sec.~\ref{As2}, see also Sec.~\ref{VIB2}).

{\em c. The covariance and the spatial spread.}

Consider a meter modeled as a particle moving in a potential, and let $F=p$ and $R=q$.
Due to nonzero meter Hamiltonian $H_{\rm M}$, the meter state is changing in time.
We now assume, for simplicity, that the system Hamiltonian is zero.
In the special case of instantaneous (impulsive) measurements, $t_{\rm M},t_{\rm f},t_{\rm i}\rightarrow0$, Jozsa \cite{joz07} obtained Eq.~\rqn{20}, where the covariance $\sigma_{pq}$ is related to the rate at which the meter distribution is spreading in space by the equality
 \be
\sigma_{pq}\:=\:\left.\frac{m_p}{2}\frac{d\{[\Delta q(t)]^2\}}{dt} 
\right|_{t=0},
 \e{309}
where $\Delta q(t)$ is calculated for the free-evolving meter state $\rho_{\rm M}(t)=e^{-iH_{\rm M}t}\rho_{\rm M}e^{iH_{\rm M}t}$.
As mentioned in Sec.~\ref{VB'}, in the present case $t_{\rm M},t_{\rm f},t_{\rm i}\rightarrow0$, the measurement results are not modified by the meter Hamiltonian $H_{\rm M}$; hence also Eq.~\rqn{98} with $\sigma_{pq}$ obeying Eq.~\rqn{309} holds now.

Consider now whether it is possible to extend Eq.~\rqn{309} to measurements with a finite duration, $t_{\rm M}-t_{\rm i}>0$.
For weak PPS measurements with a finite duration, even when the coupling is impulsive, $t_{\rm f}-t_{\rm i}\rightarrow0$, the effects of the meter Hamiltonian generally cannot be taken into account by a relation, which, as Eq.~\rqn{309}, depends only on the meter state.
Indeed, when the meter Hamiltonian $H_{\rm M}$ does not commute with the coupling Hamiltonian \rqn{1}, the effects of $H_{\rm M}$ include necessarily a change of the pointer variable (see Sec.~\ref{VB'}).

An exception is the case of the free-particle meter Hamiltonian \rqn{130}.
Indeed, since the Hamiltonian \rqn{130} commutes with the coupling
Hamiltonian \rqn{1}, then, as shown in Sec.~\ref{VB'}, the meter
Hamiltonian can be taken into account in measurements with a finite duration simply by replacing the initial meter state $\rho_{\rm M}$ with the state $\rho_{\rm M}(t_{\rm M})$ [see Eq.~\rqn{242}].
This means that now Eqs.~\rqn{20} and \rqn{98} are not changed, but Eq.~\rqn{309} should be modified by the substitution $t=0\rightarrow t=t_{\rm M}$, i.e.,
 \be
\sigma_{pq}=\left.\frac{m_p}{2}\frac{d\{[\Delta q(t)]^2\}}{dt}\right|_{t=t_{\rm M}}.
 \e{265}

\subsection{Two-level meter}
\label{VIC}

Until recently, pre- and post-selected measurements were studied mainly employing a continuous-variable meter, the exception being two early experimental works \cite{sut93,sut95}.
 In a number of recent theoretical papers, measuring weak values of a
qubit with a qubit (two-level) meter was discussed \cite{rom08,ral06,bru08}, whereas measurement of an arbitrary system with a spin and a qubit meters was considered in Refs.~\cite{lun05} and \cite{wum09}, respectively.
A qubit meter was used in experiments for weak PPS measurements of a qubit \cite{sut93,sut95,pry05,gog11,iin11} and a continuous-variable system \cite{koc11,lun11}.
 Here we discuss weak PPS measurements of an arbitrary system
with a qubit meter, {\em beyond the linear-response regime}.

For a two-level (qubit) meter, the operators $\hat{F}$ and
$\hat{R}$ can be written in the form,
 \bea
&&\hat{F}=\hat{F}_1+f_0,\quad
\hat{F}_1=\vec{\sigma}\cdot\vec{n}_F,\label{6.16}\\
&&\hat{R}=\vec{\sigma}\cdot\vec{n}_R,
 \ea{190}
where $\vec{n}_F$ and $\vec{n}_R$ are unit vectors and $f_0$ is a real number.
The operators \rqn{6.16}-\rqn{190} are not the most general ones.
However, the most general situation easily reduces to the case \rqn{6.16}-\rqn{190} with the help of simple substitutions.\footnote{
Indeed, in the most general case $\hat{R}$ has the form
 \be
\hat{R}=r_1\vec{\sigma}\cdot\vec{n}_R+r_0, 
\e{6.17}
where $r_0$ and $r_1$ are real.
As implied by Eq.~\rqn{44}, replacing Eq.~\rqn{190} by Eq.~\rqn{6.17} results in multiplying the expression for $\bar{R}_s-\bar{R}$ by $r_1$.

Similarly, the most general $\hat{F}$ has the form
 \be
\hat{F}=f_1(\hat{F}_1+f_0), 
\e{6.19}
where $f_1$ is real. 
However, when the factor $f_1\ne1$, it can be absorbed in the parameter $g(t)$ in the Hamiltonian \rqn{1} and, hence, also in the coupling strength $\gamma$.
Hence, using Eq.~\rqn{6.19} instead of Eq.~\rqn{6.16} results in the following substitution in the formulas of the present theory,
 \be
\gamma\rightarrow f_1\gamma.
\e{6.20}}

The meter parameters in the formulas of the present theory [see,
e.g., Eqs.~\rqn{13}, \rqn{169}, \rqn{61}, and \rqn{221}]
are now given by
 \be
\bar{F}=\bar{F}_1+f_0,\quad\overline{F^2}=1+2f_0\bar{F}_1+f_0^2, \quad\Delta F=\sqrt{1-\bar{F}_1^2},
\e{222}
 \be
\Delta R=\sqrt{1-\bar{R}^2},
\e{240}
 \bea
&&\overline{R_cF}=M_R-\bar{R}\bar{F}_1+iM_I,\nonumber\\
&&\overline{FR_cF}=M-\bar{R}+2f_0(M_R-\bar{R}\bar{F}_1),\nonumber\\
&&\overline{F_cR_cF_c}=M-2\bar{F}_1M_R+\bar{R}(2\bar{F}_1^2-1),
 \ea{223}
where
 \be
M_R={\rm Re}\,\overline{RF_1},\quad M_I={\rm
Im}\,\overline{RF_1},\quad M=\overline{F_1RF_1}.
 \e{224}
From Eqs.~\rqn{6.16}, \rqn{190}, and \rqn{224} we obtain that
 \be
M_R=\cos\eta,\quad M_I=\bar{F}_2\sin\eta,
\quad M=\bar{F}_1\cos\eta-\bar{F}_3\sin\eta.
 \e{195}
Here $\eta$ ($0\le\eta\le\pi$) is the angle between $\vec{n}_F$ and
$\vec{n}_R$,
 \be
\bar{F}_1={\rm Tr}\,[(\vec{\sigma}\cdot\vec{n}_F)\rho_{\rm M}],\quad
\bar{F}_{2,3}={\rm Tr}\,[(\vec{\sigma}\cdot\vec{n}_{2,3})\rho_{\rm M}],
 \e{6.13}
whereas $\vec{n}_{2,3}$ are unit vectors defined by
 \be
\vec{n}_2=\frac{\vec{n}_R\times\vec{n}_F}{\sin\eta},\quad
\vec{n}_3=\vec{n}_F\times\vec{n}_2.
 \e{196}
Note that for noncommuting $\hat{R}$ and $\hat{F}$ (i.e.,
for $\eta\ne0$), $\{\vec{n}_F,\vec{n}_2,\vec{n}_3\}$ is an orthonormal
basis in the Bloch sphere of the meter.

 The general initial condition for a two-level meter is
 \be
\rho_{\rm M}=(I+\vec{\sigma}\cdot\vec{s}_{\rm M})/2,
 \e{205}
where $\vec{s}_{\rm M}$ is the pseudospin.
 Using Eqs.~\rqn{205} and \rqn{6.13}, we obtain that in Eqs.~\rqn{240}, \rqn{223}, and \rqn{195}
 \be
\bar{R}=\vec{s}_{\rm M}\cdot\vec{n}_R,\quad\bar{F}_1=\vec{s}_{\rm
M}\cdot\vec{n}_F,\quad\bar{F}_{2,3}=\vec{s}_{\rm M}\cdot\vec{n}_{2,3},
 \e{206}
the quantities $\bar{F}_i$ being the components of the pseudospin in
the orthonormal basis $\{\vec{n}_F,\vec{n}_2,\vec{n}_3\}$.

When $\hat{R}$ and $\hat{F}$ commute, then $\eta=0$ or $\pi$, i.e., $\vec{n}_F=\pm\vec{n}_R$.
In this case, the quantities $\bar{F}_{2,3}$ are not defined, but they drop from the expressions, and Eqs.~\rqn{195} and \rqn{206} yield that
 \be
M_R=\pm1,\quad M_I=0,\quad M=\bar{R}=\pm\bar{F}_1,
 \e{197}
where the choice of the sign on the right-hand sides of the equations coincides with that in the equality $\vec{n}_F=\pm\vec{n}_R$.

In the present case of a two-level meter there are a number of free
parameters, variation of which allows one to obtain desirable values
of the meter moments.
Several possible configurations of the qubit meter are listed in Table \ref{t2}.
To obtain the values of the moments of the meter variables shown in Table \ref{t2}, we used Eqs.~\rqn{223}, \rqn{195}, \rqn{196}, \rqn{206}, and \rqn{197}.

\begin{table}[tb]
\begin{center}
\begin{tabular}{c|l|ccccc}
\hline
&&\multicolumn{5}{|c}{Moments of the meter variables:}\bigstrut[t]\\
No.&Meter configuration&$\bar{R}$&$\bar{F}_1$&$\overline{R_cF}$ &$\overline{FR_cF}$&$\overline{F_cR_cF_c}$\bigstrut[t]\\
\hline
1&$\vec{n}_F\perp\vec{n}_R,\ \vec{s}_{\rm M}=\vec{n}_2$&0&0&$i$&0&0\bigstrut[t]\\
2&$\vec{n}_F=\vec{n}_R,\ \vec{s}_{\rm M}\cdot\vec{n}_F=0$&0&0&1&$2f_0$&0\bigstrut[t]\\
3&$\vec{s}_{\rm M}=\vec{n}_2$&0&0&$e^{i\eta}$&$2f_0\cos\eta$&0\\
4&$\vec{n}_F=\vec{n}_R$&$\vec{s}_{\rm M}\cdot\vec{n}_F$ &$\vec{s}_{\rm M}\cdot\vec{n}_F$&$f_2$&$2f_0f_2$&$-2\bar{F}_1f_2$\bigstrut[t]\\
5&$\vec{s}_{\rm M}=\vec{n}_R$&1&$\cos\eta$&0&
$-\sin^2\eta$&$-\sin^2\eta$\bigstrut[t]\\
6&$\vec{s}_{\rm M}=\vec{n}_R,\ \vec{n}_F\perp\vec{n}_R$&1&0&0&$-1$&$-1$\bigstrut[t]\\
7&$\vec{s}_{\rm M}=0$&0&0&$\cos\eta$&$2f_0\cos\eta$&0\bigstrut[t]\\[.5ex]
\hline
 \end{tabular}
\end{center}
\caption{Moments of meter variables used in the present theory, for various configurations of a two-level meter.
The parameter $\bar{F}_1$ determines $\bar{F}$, $\overline{F^2}$, and
$\Delta F$ by Eq.~\rqn{222}, whereas $f_2=1-(\vec{s}_{\rm M}\cdot\vec{n}_F)^2=(\Delta F)^2$.
 }
 \label{t2}\end{table}

A simple, but important, case is obtained when the initial meter state is pure (i.e., $\vec{s}_{\rm M}$ is a unit vector) and $\{\vec{n}_R,\vec{n}_F,\vec{s}_{\rm M}\}$ is a right-handed basis in the Bloch sphere of the meter, see meter configuration 1 in Table \ref{t2}.
This situation is similar to configuration 1 in Table \ref{t1}.
To obtain the explicit expression, we combine the data of configuration 1 in Table \ref{t2} with Eqs.~\rqn{222} and \rqn{13}, yielding
 \be
\bar{R}_s^{(1)}=\frac{2\gamma\,{\rm Re}\,A_w} {1+2\gamma f_0\,{\rm Im}\,A_w+\gamma^2(1+f_0^2)|A_w|^2}.
 \e{6.14}
Another simple situation, which is especially suitable for the case of an imaginary weak value, is given by configuration 2 in Table \ref{t2}, for which we obtain
 \be
\bar{R}_s^{(2)}=\frac{2\gamma\,{\rm Im}\,A_w+2\gamma^2f_0|A_w|^2} {1+2\gamma f_0\,{\rm Im}\,A_w+\gamma^2(1+f_0^2)|A_w|^2}.
 \e{6.31}
The superscripts ``(1)'' and ``(2)'' remind that Eqs.~\rqn{6.14} and \rqn{6.31} relate to cases 1 and 2 in Table \ref{t1}.
Equations \rqn{6.14} and \rqn{6.31} simplify in the linear regime, where $\gamma A_w$ is small, yielding respectively [cf.\ Eqs.~\rqn{4} and \rqn{5}]
 \be
\bar{R}_s^{(1)}=2\gamma\,{\rm Re}\,A_w,
 \e{6.15}
 \be
\bar{R}_s^{(2)}=2\gamma\,{\rm Im}\,A_w.
 \e{6.32}
Note that in configuration 2 in Table \ref{t2}, the meter may be in a pure or mixed state with $\vec{s}_{\rm M}\perp\vec{n}_F$ or {\em even in the completely mixed state}, $\vec{s}_{\rm M}=0$.
The fact that the purity of the meter state is not important in this case may be used to simplify experiments on weak PPS measurements, which employ configuration 2 in Table \ref{t2}.

When the weak value is complex, joint measurements with meter configurations 1 and 2 allow one to perform weak-value tomography (i.e., to obtain the real and imaginary parts of the weak value).
One can work in the linear-response regime, using Eqs.~\rqn{6.15} and \rqn{6.32}, or in the strongly-nonlinear regime, using Eqs.~\rqn{6.14} and \rqn{6.31} (see Sec.~\ref{IIIA6'}).
The linear-response version of this method was demonstrated\footnote{More specifically, in Ref.~\cite{lun11} in both configurations 1 and 2, $\hat{F}$ is the operator of a spin component, so that the parameters in Eq.~\rqn{6.19} are given by $f_0=0$ and $f_1=1/2$.
As a result, Eq.~\rqn{6.20} implies that Eqs.~\rqn{6.15} and \rqn{6.32} become now, respectively,
 \be
\bar{R}_s^{(1)}\:=\:\gamma\,{\rm Re}\,A_w,\quad\quad \bar{R}_s^{(2)}\:=\:\gamma\,{\rm Im}\,A_w.
 \e{6.40}
}
 experimentally in Ref.~\cite{lun11}.

 Table \ref{t2} lists also several other possible meter configurations.
In particular, configuration 3 resembles configurations 2 and 4 in Table \ref{t1}, configuration 1 in Table \ref{t2} being a special case of configuration 3 for $\eta=\pi/2$.
 Configurations 2 and 4-7 in Table \ref{t2} are examples of non-standard meters.
In configurations 2 and 4, $\hat{R}$ and $\hat{F}$ commute, configuration 2 being a special case of configuration 4.
In configurations 5 and 6, $\Delta R=0$ since then $\rho_{\rm M}$ is a pure state which is an eigenstate of $\hat{R}$ (see also Sec.~\ref{VIA'}); case 6 is a special case of case 5.
Finally, configuration 7 is the case of a completely mixed state of the meter; this demonstrates the possibility of weak PPS measurements
with meters in a completely mixed state.
The special case of configuration 7 with $\vec{s}_{\rm M}=\vec{n}_R$ is included in configuration 2.

It is easy to show that meters with configurations 1-3 in Table \ref{t2} satisfy the condition \rqn{4.15}; thus, they belong to the class of meters optimal for weak PPS measurements in the linear-response regime, as discussed in Sec.~\ref{VIJ2}.
In addition, meters with configuration 1 in Table \ref{t2} are also optimal for weak standard measurements, since they satisfy Eq.~\rqn{2.6}.
Moreover, meters with configurations 4 and 7 in Table \ref{t2} are regular meters, i.e., they are effective for weak PPS measurements, since they satisfy the condition \rqn{334}, with the exception of the cases $|\bar{F}_1|\approx1$ for meters with configuration 4 and $|\cos\eta|\ll1$ for meters with configuration 7.

The above results show that for weak PPS measurements, the qubit
meter is at least as versatile as the continuous-variable meter.
One or more of the parameters entering Table \ref{t2} can be varied
in experiments in order to perform tomography of weak values or
optimize the measurements.

\subsection{Experiments where the average input variable is nonzero, $\bar{F}\ne0$}

In previous theoretical and experimental studies of weak PPS measurements, $\bar{F}$ has been always set to zero, exactly or effectively.
In the present theory we do not make this assumption, i.e., now generally $\bar{F}\ne0$.
Still, as shown above, the linear response does not depend on $\bar{F}$ [see Eq.~\rqn{14''}]; however, its validity condition generally does depend on $\bar{F}$  [see Eq.~\rqn{3.33}].

What is more important is that, as shown above, weak PPS measurements depend significantly on the value of $\bar{F}$ in the nonlinear regime.
In particular, a nonzero $\bar{F}$ can facilitate measurements of
$A_w$ and $\gamma$ (Secs.~\ref{IIIA8d} and \ref{VIB3}), whereas the
optimal regime in the peculiar case $|\bar{F}|\gg\Delta F$ has some advantages, as discussed in Secs.~\ref{IIIA6} and \ref{IVC3}.
Here we mention some systems for which the above effects can be
checked experimentally.

Qubit meter is a simple example of a meter for which generally $\bar{F}\ne0$ (Sec.~\ref{VIC}).
As follows from Eq.~\rqn{222}, for qubit meters the ratio
$\bar{F}/\Delta F$ can be easily tuned by changing $\bar{F}_1$, i.e.,
by changing the initial meter state $\rho_{\rm M}$ (for specific
examples, see the values of $\bar{F}_1$ in Table \ref{t2}).
Moreover, $\bar{F}$ is always nonzero when $\hat{F}$ is a projector; it can also be shown that this is the case for the experiments \cite{sut95,wan06}.

The quantity $|\bar{F}|$ can be very large, as in the proposed Stern-Gerlach experiment \cite{aha88} and in the actual optical experiments using birefringent elements \cite{rit91,sut95,par98,bru04,wan06,cho10}.
In this case, under certain conditions the effects of $\bar{F}$ can be often eliminated \cite{aha88}, using the invariance of PPS measurements under gauge transformations of the system, see Sec.~\ref{As1}, especially Eqs.~\rqn{A6} and \rqn{A7}.
Indeed, in a typical case of a two-level system with $\hat{A}=\sigma_z$, Eqs.~\rqn{A6} and \rqn{A7} imply that $\bar{F}$ is effectively zero in PPS measurements when \cite{aha88} 
 \be
\gamma\bar{F}=n\pi\quad\quad{\rm for}\ \ n=0,\pm1,\pm2,\dots.
 \e{6.2}
In the above optical experiments $\gamma$ was not varied, since it was fixed by the condition \rqn{6.2} with some value of $n$.

As discussed above, a nonzero $\bar{F}$ can be useful in the nonlinear regime.
In the case of very large $\bar{F}$, one can obtain an effective $\bar{F}$ of an arbitrary magnitude by making the value of $\gamma$ or $\bar{F}$ slightly differing from that fixed by Eq.~\rqn{6.2}.
Then, in view of Eqs.~\rqn{6.2} and \rqn{A7}, in the results for PPS measurements obtained in the present paper, the average of $F$ should be substituted by its effective value,
 \be
\bar{F}\ \rightarrow\ \bar{F}-\frac{n\pi}{\gamma},
 \e{6.3}
where $n$ is the integer minimizing $|\bar{F}-n\pi/\gamma|$.
In particular, inserting Eq.~\rqn{6.3} into the validity conditions of the present theory and of different regimes, such as, e.g., Eqs.~\rqn{12} and \rqn{63}, provides the limits for the allowed values of the quantity $\bar{F}-n\pi/\gamma$.

Finally, we note that $\bar{F}$ can be tuned also by performing in any part of the interval $(0,t_{\rm S})$ an additional unitary transformation $U'=\exp(-i\alpha\hat{A})$ on the system, where $\alpha$ is a real number.
This will replace the transformation $U$ \rqn{2} by $UU'$, which is equivalent to the replacement 
 \be
\bar{F}\ \rightarrow\ \bar{F}+\frac{\alpha}{\gamma}.
 \e{6.11}
\section{Distribution of the pointer values}
\label{V'}

Higher-order moments $\overline{R^n_s}$ of $R$ can be obtained by
substituting $\hat{R}\rightarrow\hat{R}^n$ in Eq.~\rqn{13} or
\rqn{61}.
 These moments can be written in the form
 \be
\overline{R^n_s}\,=\,\sum_RR^n\,\Phi_s(R),
 \e{279}
where $\Phi_s(R)$ is the distribution of the eigenvalues $R$ of
$\hat{R}$ for $t\ge t_{\rm f}$.
 Hence, the maximum information is provided by the distribution
$\Phi_s(R)$, discussed in this section.

\subsection{General meter}
\label{VA}

Here we discuss the case of a general meter which can be a system with
a finite number of states or a continuous-variable system.
For simplicity, we will consider $\Phi_s(R)$ for a nondegenerate
$\hat{R}$; then
 \be
\Phi_s(R)\:=\:\overline{|R\rangle\,\langle R|}_s,
 \e{184}
where $|R\rangle$ is the eigenvector of $\hat{R}$ with the
eigenvalue $R$.
 Substituting $\hat{R}\rightarrow|R\rangle\langle R|$ into
Eq.~\rqn{182} yields
 \be
\Phi_s(R)\:=\:\{\Phi(R)+2\gamma\,{\rm Im}\,[A_w\Phi_1(R)]
+\gamma^2\,|A_w|^2\,\Phi_2(R)\}/Q_0.
 \e{53}
Here $\Phi(R)=\langle R|\rho_{\rm M}|R\rangle$ is the initial
distribution of $R$, $\Phi_1(R)=\langle R|\hat{F}\rho_{\rm
M}|R\rangle$ is generally complex, whereas $\Phi_2(R)=\langle
R|\hat{F}\rho_{\rm M}\hat{F}|R\rangle$ is real; finally, $Q_0$ equals 
 \be
Q_0\,=\,1+2\gamma\,\bar{F}\:{\rm Im}\,A_w
+\gamma^2\,\overline{F^2}\,|A_w|^2.
 \e{161}
Here and below in Sec.~\ref{V'} we assume that the initial state of
the system is pure; if this is not the case, the results obtained
still hold under the replacement \rqn{57a}.
 Consider several important cases.

When the meter is initially in a pure state $|\psi_{\rm M}\rangle$,
then in Eq.~\rqn{53}
 \be
\Phi(R)=|\psi_{\rm M}(R)|^2,\quad\quad 
\Phi_1(R)=\psi_{\rm M}^*(R)\,d_R, \quad\quad \Phi_2(R)=|d_R|^2,
 \e{54}
where
 \be
\psi_{\rm M}(R)=\langle R|\psi_{\rm M}\rangle,\quad\quad d_R=\langle
R|\hat{F}|\psi_{\rm M}\rangle.
 \e{127}

The function $\Phi_s(R)$ in Eq.~\rqn{53} simplifies when $F$ is a
function of $R$, $F=h(R)$.
The latter implies that $[\hat{F},\hat{R}]=0$.
 [For nondegenerate $\hat{F}$ and $\hat{R}$, also the converse is true, i.e., the equality $[\hat{F},\hat{R}]=0$ implies that $F=h(R)$.]
 When $F=h(R)$, in Eq.~\rqn{53}
 \be
\Phi_n(R)\:=\:h^n(R)\,\Phi(R)\quad (n=1,2),
 \e{259}
$\Phi_1(R)$ now being real.

It is interesting that in the case $F=h(R)$ the final pointer distribution
\rqn{53} depends on the initial probability distribution $\Phi(R)$
but not on the coherent properties of the initial state $\rho_{\rm
M}$ [cf.\ Eq.~\rqn{259}], whereas for noncommuting $\hat{F}$ and
$\hat{R}$, Eq.~\rqn{53} generally depends on the phase of the
initial state [cf.\ Eq.~\rqn{54}].

\subsection{Continuous-variable meter}
\label{VIB}

Consider now in more detail the distribution of the values of a
continuous pointer variable (e.g., $p$ or $q$).
In the previous studies, it was assumed that the initial meter state
is a real Gaussian in the $F$ or $R$ representation \cite{aha88}.
Here we only assume that the initial pointer distribution $\Phi(R)$
has a bell-like shape (e.g., Lorentzian or Gaussian).

Note that the validity condition for the result \rqn{53} for $\Phi_s(R)$ generally depends on $R$.
In the main part of the peak $\Phi_s(R)$, i.e., in the interval within the peak width, the validity condition is the same as for $\bar{R}_s$ (see Secs.~\ref{IIIA3'} and \ref{IVB}).
However, for far tails of $\Phi_s(R)$ the present theory can fail, as illustrated by examples shown below.
This is explained by the fact that the validity conditions of the present theory can become much stricter for the tails than for the central part of $\Phi_s(R)$.
Since far tails of $\Phi_s(R)$ are of little interest, we do not go further into this point.

A weak PPS measurement can change the distribution of $R$ significantly or slightly, depending on the values of the parameters, as discussed below.
When the effect of the measurement is not too strong, an initially bell-shaped distribution can remain bell-shaped with the maximum generally shifted from the initial position.
This shift is important in some applications, such as superluminal propagation and slow light \cite{aha90,ste95,aha02,sol04,bru04,wan06}. 
If this shift is sufficiently small, simple formulas for the shift can be derived, as shown below.
These formulas hold regardless of whether the shape of the distribution changes or remains the same.
Recall that some cases where the distribution is shifted practically without a change of the shape are listed in Sec.~\ref{IC6}.

\subsubsection{Coinciding meter variables, $R=F$}
\label{VIIB1}

We begin with the simple case $R=F$.
 Then Eqs.~\rqn{53} and \rqn{259} yield
 \be
\Phi_s(F)\:=\:\Phi(F)\,[1+2\gamma\,({\rm Im}\,A_w)F +\gamma^2\,|A_w|^2F^2]/Q_0,
 \e{25}
where $\Phi(F)=\langle F|\rho_{\rm M}|F\rangle$ is the distribution
of $F$ before the measurement (at $t=0$).
 Thus, $\Phi_s(F)/\Phi(F)$ is a quadratic polynomial in $F$.

Note that the linear-response approximation provides a wrong result for the tails of the distribution \rqn{25} even in the linear-response regime \rqn{113}, since for large $|F|$ the nonlinear term dominates in Eq.~\rqn{25}.
This is an indication that for the case $R=F$ the present theory does
not describe the far tails of $\Phi_s(F)$, as discussed above.

The quantity $\Phi_s(F)/\Phi(F)$ is minimal at
 \be
F_{\rm min}\:=\:-\frac{{\rm Im}\,A_w}{\gamma\,|A_w|^2},
 \e{162}
where
 \be
\frac{\Phi_s(F_{\rm min})}{\Phi(F_{\rm min})}\:=\:\frac{({\rm
Re}\,A_w)^2} {|A_w|^2Q_0}.
 \e{163}
Thus, $\Phi_s(F)$ is always positive, except for the case of a
purely imaginary weak value, ${\rm Re}A_w=0$, when
 \be
\Phi_s(F_{\rm min})=0,\quad\quad F_{\rm min}\:=\:-(\gamma\,{\rm Im}\,A_w)^{-1}.
 \e{164}

Consider now the typical case $|\bar{F}|\lesssim\Delta F$ (the other
case $|\bar{F}|\gg\Delta F$ is discussed in the last but one paragraph of this subsection).
 In the linear regime [Eq.~\rqn{22} or \rqn{100}] the main part of
$\Phi_s(F)$, except for the far tails, is given by
 \be
\Phi_s(F)\:\approx\:\Phi(F)\,[1 +2\gamma\,({\rm Im}\,A_w)F]\quad\quad{\rm for}\ \ |F-\bar{F}|\lesssim\Delta F.
 \e{26}
This equation implies that, like $\Phi(F)$, the function $\Phi_s(F)$
\rqn{26} has a bell-like shape with the maximum of $\Phi_s(F)$
shifted from the maximum $F_{\rm max}$ of $\Phi(F)$ by
 \be
\Delta F_{\rm max}\:=\:\beta\,(\bar{F}_s-\bar{F})\:=\:2\beta\gamma\,(\Delta F)^2\,{\rm Im}\,A_w.
 \e{28}
Here $\bar{F}_s$ is given by Eq.~\rqn{27} and
 \be
\beta\:=\:\frac{\Phi(R_{\rm max})}{|\Phi''(R_{\rm max})|\,(\Delta R)^2},
 \e{45}
where now $R=F$, the primes denoting the second derivative.
 In the derivation of Eq.~\rqn{28} we assumed that the peak top
has a parabolic shape,
 \be
\Phi(F)\:\approx\: \Phi(F_{\rm max})-|\Phi''(F_{\rm max})|(F-F_{\rm
max})^2/2\quad\quad{\rm for}\ \ |F-F_{\rm max}|\ll\Delta F.
 \e{46}

When $\Phi(F)$ is Gaussian, 
 \be
\Phi(F)\:=\:\frac{1}{\sqrt{2\pi}\,\Delta F} \exp\left[-\frac{(F-\bar{F})^2}{2(\Delta F)^2}\right],
 \e{9.24}
then $\beta=1$ in Eq.~\rqn{28}, yielding again Eq.~\rqn{160}.
Thus, we extended Eq.~\rqn{160}, derived in Sec.~\ref{IC6} for the case of a pure meter state $\psi_{\rm M}$, to the case of an arbitrary meter state $\rho_{\rm M}$ with a Gaussian $\Phi(F)$.
Moreover, Eq.~\rqn{28} shows that for a general non-Gaussian $\Phi(F)$, the shift of the maximum $\Delta F_{\rm max}$ differs from the average pointer deflection $(\bar{F}_s-\bar{F})$ by a dimensionless factor $\beta$, which depends on the shape of $\Phi(F)$.

In the opposite limit $A_w\rightarrow\infty$, i.e., for mutually
orthogonal
$|\psi\rangle$ and $|\phi\rangle$, Eq.~\rqn{25} yields
 \be
\Phi_s(F)\:=\:\frac{F^2\Phi(F)}{\overline{F^2}}.
 \e{29}
This equality holds approximately also for
$\gamma^2|A_w|^2\gg1/\,\overline{F^2}$.
 Now the function $\Phi_s(F)$ has two peaks of comparable heights,
at least, for $|\bar{F}|\lesssim\Delta F$.

In contrast, when $|\bar{F}|\gg\Delta F$, $\Phi_s(F)$ is a
bell-shaped function, except for the case \rqn{173}-\rqn{168},
where the narrow resonance \rqn{169} occurs.
 The shift of the maximum of this bell-shaped function from the
maximum of $\Phi(F)$ can be shown to be given approximately by
the first equality in Eq.~\rqn{28}.
However, now the average pointer deflection $(\bar{F}_s-\bar{F})$ is described by a {\em nonlinear formula}, so that
 \be
\Delta F_{\rm max}\:=\:\beta(\bar{F}_s-\bar{F})\:=\:\beta\frac{2\gamma\,(\Delta F)^2\,(\,{\rm Im}\,A_w
+\gamma\bar{F}|A_w|^2)} {1+2\gamma\,\bar{F}\:{\rm Im}\,A_w
+\gamma^2\,\bar{F}^2\,|A_w|^2},
 \e{166}
which follows from Eq.~\rqn{21} for $|\bar{F}|\gg\Delta F$.
 In Eq.~\rqn{166} we took into account that for $|\bar{F}|\gg\Delta
F$ one has
 \be
\overline{F^2}\:=\:(\Delta F)^2+\bar{F}^2\:\approx\:\bar{F}^2
 \e{330}
and [cf.\ Eq.~\rqn{167}]
 \be
\overline{F^3}- \overline{F^2}\bar{F}\:=\:2(\Delta F)^2\bar{F}+
\overline{F_c^3}\:\approx\:2(\Delta F)^2\bar{F}.
 \e{172}
 This result is obtained if the condition \rqn{46} holds and if $F_{\rm max}\sim\bar{F}$, which holds when $\Phi(F)$ is not too asymmetric.
Note that for a Gaussian $\Phi(F)$, $\beta=1$ in Eq.~\rqn{166}.

Finally, we note that the general Eq.~\rqn{25} simplifies for an
imaginary weak value,
 \be
\Phi_s(F)\:=\:\Phi(F)[1+\gamma\,({\rm Im}\,A_w)F]^2/Q_0.
 \e{362}
In particular, the intensity distribution of the ``split-Gaussian mode'' obtained in Ref.~\cite{sta10x} can be interpreted quantum-mechanically as a quantity proportional to the special case of Eq.~\rqn{362} where $\Phi(F)$ is Gaussian and $\bar{F}=0$ (see Sec.~\ref{XI}).

\subsubsection{Canonically conjugate $R$ and $F$}

Here we study a meter with canonically conjugate variables $R$ and $F$.
For such a meter, the shift of the pointer distribution in the linear regime is known to be proportional to ${\rm Re}\,A_w$, at least, when the initial meter state in the pointer representation is a real Gaussian \cite{aha88,ste10,aha10,hos08}.
However, for the general case the shift has not been discussed yet.

Here we assume that the meter is initially in a pure state
$|\psi_{\rm M}\rangle$, the pointer distribution possessing an {\em
arbitrary} bell-like shape.
In particular, we will show that for a complex Gaussian and for
non-Gaussian states, the shift of the maximum of the pointer
distribution generally depends on both the real and imaginary parts of
the weak value.

For the canonically conjugate variables \rqn{97}, the second Eq.~\rqn{127} yields
$d_q=-i\psi_{\rm M}'(q)$, where the prime denotes differentiation.
 Then Eqs.~\rqn{53}-\rqn{54} yield that
 \be
\Phi_s(q)\:=\:\frac{\Phi(q)-2\gamma\,{\rm Re}[A_w\psi_{\rm M}^*(q)\psi_{\rm M}'(q)] +\gamma^2|A_w|^2\,|\psi_{\rm M}'(q)|^2}
{1+2\gamma\bar{p}\,{\rm Im}\,A_w +\gamma^2\,\overline{p^2}\,|A_w|^2}.
 \e{33}
This expression depends on the phase of the initial state, unlike
the results in Sec.~\ref{VIIB1}.

For $|\gamma A_w|\ll(\overline{p^2})^{-1/2}$ [cf.\ Eq.~\rqn{22}]
and $|q-\bar{q}|\lesssim\Delta q$ one can use in Eq.~\rqn{33} the
approximation linear in $\gamma$, yielding
 \bea
&\Phi_s(q)\:&\approx\:\Phi(q)-2\gamma\,{\rm Re}[A_w\psi_{\rm M}^*(q)\psi_{\rm M}'(q)]\nonumber\\
&&=\:\Phi(q)-\gamma\,({\rm Re}\,A_w)\,\Phi'(q)+2\gamma\,({\rm Im}\,A_w)\,\xi'(q)\Phi(q),\quad\quad
 \ea{34}
where in the last equality Eq.~\rqn{6.12} is taken into account.
We assume that $\Phi(q)$ is a bell-shaped function with the maximum at $q_{\rm max}$, so that for $|q-q_{\rm max}| \ll\Delta q$ it is expressed as
 \be
\Phi(q)\:\approx\:\Phi(q_{\rm max})-|\Phi''(q_{\rm max})|(q-q_{\rm
max})^2/2.
 \e{47}
Moreover, in some interval $|q-q_{\rm max}| \ll\Delta_\xi$, we have
 \be
\xi'(q)\approx\xi'(q_{\rm max})+\xi''(q_{\rm max})(q-q_{\rm max}),
 \e{48}
the double primes in Eqs.~\rqn{47} and \rqn{48} denoting the second derivative.
Then $\Phi_s(q)$ is also a bell-shaped function with the maximum at $q_{\rm max}+\Delta q_{\rm max}$, where
 \be
\Delta q_{\rm max}\:=\:\gamma\,[{\rm Re}\,A_w+2\beta\,\xi''(q_{\rm
max})\,(\Delta q)^2\,{\rm Im}\,A_w].
 \e{35}
Here $\beta$ is given by Eq.~\rqn{45} with $R=q$.
Equation \rqn{35} holds when
 \be
|\Delta q_{\rm max}|\ll\min\{\Delta q,\Delta_\xi\}.
 \e{175}
Equation \rqn{35} shows that generally {\em the shift of the maximum of the distribution of $q$ depends on both the real and imaginary parts of the weak value}.
Generally, the shift does not coincide with the average pointer deflection, $\Delta q_{\rm max}\ne\bar{q}_s-\bar{q}$.

However, there are cases when Eq.~\rqn{35} possesses the convenient property \rqn{1.25}, which now has the form
 \be
\Delta q_{\rm max}=\bar{q}_s-\bar{q}.
 \e{7.1}
In particular, Eq.~\rqn{7.1} holds in the following cases.\\
(a) The weak value is real.
In this case Eq.~\rqn{1.26} holds, as shown above.\\
(b) The phase of $\psi_{\rm M}(q)$ is vanishing or linear in $q$.
In this case, Eq.~\rqn{35} simplifies and becomes equal to the average pointer deflection \rqn{4}, 
 \be
\Delta q_{\rm max}=\bar{q}_s-\bar{q}=\gamma\,{\rm Re}\,A_w.
 \e{37}
 Previously in Ref.~\cite{aha88}, Eq.~\rqn{37} was obtained for the special case when $\psi_{\rm M}(q)$ is a real Gaussian.\\
(c) $\psi_{\rm M}(q)$ is a general complex Gaussian state.
For this state we obtained above Eq.~\rqn{1.27}.
Here we can derive Eq.~\rqn{1.27} in a different way.
Namely, taking into account that for a general Gaussian state, $\beta=1$ and the phase $\xi(q)$ is [see Eq.~\rqn{38}]
 \be
\xi(q)\:=\:\frac{b(q-\bar{q})^2}{4(\Delta q)^2} +\bar{p}q,
 \e{260}
we obtain that Eq.~\rqn{35} coincides with Eq.~\rqn{159}.

For orthogonal states $|\psi\rangle$ and $|\phi\rangle$ (when
$A_w=\infty$), Eq.~\rqn{33} yields
 \be
\Phi_s(q)\:=\:\frac{|\psi_{\rm M}'(q)|^2}{\overline{p^2}}\:=\:
\frac{[f_q'(q)]^2+[\xi'(q)]^2\Phi(q)}{\overline{p^2}},
 \e{40}
where Eq.~\rqn{6.12} was used.
 In particular, for a real $\psi_{\rm M}(q)$, the function \rqn{40}
becomes
 \be
\Phi_s(q)\:=\: [f_q'(q)]^2\,/\:\overline{p^2};
 \e{177}
it has two peaks of comparable heights with the minimum
$\Phi_s(q_{\rm max})=0$ at the maximum of $\Phi(q)$.
 However, for a complex $\psi_{\rm M}(q)$, Eq.~\rqn{40} generally
does not vanish at any point.

As an example, for a real Gaussian state Eq.~\rqn{177} becomes
 \be
\Phi_s(q)\:=\:\frac{Z_q^2(q-\bar{q})^2}{(\Delta q)^2}
\exp\left[-\frac{(q-\bar{q})^2}{2(\Delta q)^2}\right].
 \e{41}
This is a two-peak function, symmetric with respect to $\bar{q}$ and
vanishing at $q=\bar{q}$. 
Previously, an unnormalized distribution proportional to Eq.~\rqn{41} was obtained numerically (Fig.~4(b) in Ref.~\cite{duc89}) and experimentally (Fig.~2(c) in Ref.~\cite{rit91}).
 In contrast to Eq.~\rqn{41}, for the complex Gaussian meter state \rqn{38} the distribution \rqn{40} vanishes nowhere and is generally not symmetric.

Note that the general Eq.~\rqn{33} can be written in an explicit
form for the complex Gaussian state \rqn{38}.
 In this case $\Phi_s(q)/\Phi(q)$ is a quadratic function of $q$,
where 
 \be
\Phi(q)\:=\:Z_q^2\exp\left[-\frac{(q-\bar{q})^2 }{2(\Delta q)^2}\right].
 \e{7.2}
 In particular, for a real Gaussian state, Eq.~\rqn{38} with
$\bar{p}=\bar{q}=b=0$, and a real $A_w$ we obtain  a simple result,
 \be
\Phi_s(q)\:=\:\frac{\Phi(q)[1+(\gamma A_w\Delta p)(q/\Delta q)]^2}{1
+(\gamma A_w\Delta p)^2},
 \e{43}
where $\Delta p=(2\Delta q)^{-1}$ [cf. Eq.~\rqn{49} with $b=0$].

In the case of a complex Gaussian state, as in Sec.~\ref{VIIB1},
the present theory is not applicable for the far tails of $\Phi_s(q)$.
An indication to this is the fact that the far tails in Eq.~\rqn{43}
cannot be described by the linear-response approximation, irrespective of how weak the coupling is, since the term $\propto\gamma^2q^2$ in the numerator of Eq.~\rqn{43} always dominates for sufficiently large $|q|$.

\section{Weak values for a qubit}
\label{VI}

Weak values $A_w$ for a qubit were calculated previously for a number
of special cases, usually with a pure preselected state
\cite{aha88,lor08,wum09,hos08,dix09}.
Here we provide a general study of the standard and associated weak
values $A_w$ and $A_w^{(1,1)}$ for a qubit, with an arbitrary
preselected state.

\subsection{General formulas}

 We assume that the measured system operator is
 \be
\hat{A}=\vec{\sigma}\cdot\vec{n}_A,
 \e{178}
where $\vec{\sigma}=(\sigma_x,\sigma_y,\sigma_z)$ is the vector of
the Pauli matrices and $\vec{n}$ is a unit vector in the Bloch sphere.
 The operator \rqn{178} has the eigenvalues $\pm1$.
Generally, the pre- and post-selected states of the qubit are,
respectively,
 \bea
&&\rho\:=\:(I+P_{\rm in}\,\vec{\sigma}\cdot\vec{n}_{\rm
in})/2\quad\quad (0\le P_{\rm in}\le1),\nonumber\\
&&\Pi_\phi\:\equiv\:|\phi\rangle\langle\phi|
\:=\:(I+\vec{\sigma}\cdot\vec{n}_f)/2.
 \ea{208}
Here $\vec{n}_{\rm in}$ and $\vec{n}_f$ are unit vectors, $P_{\rm in}\,\vec{n}_{\rm in}$ and $\vec{n}_f$ being the pseudospins of $\rho$ and $|\phi\rangle$, respectively, whereas $P_{\rm in}$ is the
length of the initial-state pseudospin. 

The initial state $\rho$ can also be written in the form of the spectral expansion \rqn{59},
 \be
\rho\:=\:\frac{1+P_{\rm in}}{2}\:|\vec{n}_{\rm in}\rangle\langle\vec{n}_{\rm in}| +\frac{1-P_{\rm in}}{2}\:|-\vec{n}_{\rm in}\rangle \langle-\vec{n}_{\rm in}|,
 \e{8.1}
where $|\vec{n}_{\rm in}\rangle$ is a state with the pseudospin given by the unit vector $\vec{n}_{\rm in}$.
Equation \rqn{8.1} implies that $P_{\rm in}$ is expressed through the eigenvalues of $\rho$ by the relation
 \be
P_{\rm in}=|\lambda_1-\lambda_2|.
 \e{9.11}
$P_{\rm in}$ characterizes the purity of the initial state; $P_{\rm in}$ varies from 1, corresponding to a pure state, to 0 for a maximally mixed state.
Correspondingly $1-P_{\rm in}$ is a measure of the impurity of the initial state.

 Taking into account that $O_{\phi\phi} ={\rm Tr}\,(O\Pi_\phi)$ for any operator $O$ of a qubit, we obtain from Eqs.~\rqn{56}, \rqn{57a}, \rqn{178}, \rqn{208}, and the formula \cite{lan77}
 \be
(\vec{\sigma}\cdot\vec{n})(\vec{\sigma}\cdot\vec{n}') =(\vec{n}\cdot\vec{n}')I_{\rm S} +i\vec{\sigma}\cdot(\vec{n}\times\vec{n}')
 \e{9.12}
that
 \bea
&&A_w\:=\:\frac{\vec{n}_A\cdot\vec{n}_f+P_{\rm in}
(\vec{n}_A\cdot\vec{n}_{\rm in}+i\vec{n}_A\cdot\vec{n}_{\rm in}
\times \vec{n}_f)}{1+P_{\rm in}\vec{n}_{\rm in}\cdot\vec{n}_f},
\label{209'}\\
&&A_w^{(1,1)}\:=\:\frac{1-\vec{n}_{\rm in}\cdot\vec{n}_f+P_{\rm in}
(\vec{n}_A\cdot\vec{n}_{\rm in})(\vec{n}_A\cdot\vec{n}_f)}
{1+P_{\rm in}\vec{n}_{\rm in}\cdot\vec{n}_f}.
 \ea{209}
According to Eq.~\rqn{209'}, $A_w$ is generally complex.
 $A_w$ is real if the vectors $\vec{n}_A,\ \vec{n}_{\rm
in},$ and $\vec{n}_f$ lie in the same plane or for $P_{\rm in}=0$
(the completely mixed initial state), whereas $A_w$ is purely
imaginary when $\vec{n}_A$ is perpendicular to the sum of the pre-
and post-selected pseudospins $P_{\rm in}\,\vec{n}_{\rm in}+\vec{n}_f$
and, in addition, the vectors $\vec{n}_{\rm in}$ and $\vec{n}_f$ are
not collinear.

\subsection{Conditions for maximizing weak values}
\label{VB}

Equation \rqn{209'} shows that a necessary condition for $|A_w|$ to
be large is $P_{\rm in}\vec{n}_{\rm in}\cdot\vec{n}_f \simeq-1$
or, equivalently, the simultaneous relations
 \be
P_{\rm in}\simeq1,\quad \vec{n}_{\rm in}\simeq-\vec{n}_f.
 \e{210}
The conditions \rqn{210} ensure that the overlap between the initial
and final states is small,
 \be
\rho_{\phi\phi}\ll1.
 \e{266}
In the case of a pure preselected state, $P_{\rm in}=1$, the condition \rqn{210} or \rqn{266} requires that the pre- and post-selected states be almost orthogonal.
In the case of a mixed preselected state, the condition \rqn{210} requires that the preselected state $\rho$ be almost pure and that its eigenstate corresponding to the greater eigenvalue, i.e., $|\vec{n}_{\rm in}\rangle$ [cf.\ Eq.~\rqn{8.1}], be almost (or completely) orthogonal to the post-selected state.

The condition \rqn{210} is necessary but not sufficient to maximize the weak value.
In the further study of conditions under which the weak value is
maximal, we consider separately the cases of pure and mixed preselected states.
As mentioned above, in the case of a pure preselected state, Eq.~\rqn{266} means that the initial and final
states are almost orthogonal, $|\langle\phi|\psi\rangle|\ll1$.
 This condition is equivalent to
 \be
|\psi\rangle\simeq|\phi_1\rangle,\quad
|\phi\rangle\simeq|\phi_2\rangle,
 \e{211}
where $\{|\phi_1\rangle,|\phi_2\rangle\}$ is an orthonormal basis,
 \be
|\phi_1\rangle=|\vec{n}_0\rangle,\quad|\phi_2\rangle=|-\vec{n}_0\rangle.
 \e{9.6}
Here $\vec{n}_0$ is a unit vector in the Bloch sphere.
Note that the pseudospin $\vec{n}_0$ is not uniquely determined by the condition \rqn{211}, and actually there is a narrow cone of possible values of $\vec{n}_0$.

Anyhow, in the case \rqn{211} for any allowed value of $\vec{n}_0$, Eq.~\rqn{6} yields
 \be
A_w\approx \frac{A_{\phi_2\phi_1}}{\langle\phi|\psi\rangle}.
 \e{264}
For a given magnitude of the overlap $|\langle\phi|\psi\rangle|$,
the magnitude of the weak value $|A_w|$ is maximal when
$|A_{\phi_2\phi_1}|$ is maximal.
Let us obtain $|A_{\phi_2\phi_1}|$ for the operator $\hat{A}$ in Eq.~\rqn{178}.
Since $\hat{A}^2=I_{\rm S}$ [cf.\ Eq.~\rqn{9.12}], we obtain that 
 \be
A_{\phi_1\phi_1}A_{\phi_1\phi_1}+A_{\phi_1\phi_2}A_{\phi_2\phi_1} =(A^2)_{\phi_1\phi_1}=1,
 \e{9.13}
and hence
 \be
|A_{\phi_2\phi_1}|^2=1-(A_{\phi_1\phi_1})^2 =1-(\vec{n}_A\cdot\vec{n}_0)^2=1-\cos^2\eta_1=\sin^2\eta_1.
 \e{9.14}
Here $\eta_1$ is the angle between $\vec{n}_A$ and $\vec{n}_0$; we used Eq.~\rqn{9.12} in the second equality in Eq.~\rqn{9.14}.
Finally, Eq.~\rqn{9.14} yields
 \be
|A_{\phi_2\phi_1}|=\sin\eta_1.
 \e{219}

 Hence, the maximal $|A_{\phi_2\phi_1}|=1$ is obtained for any basis
$\{|\phi_1\rangle,|\phi_2\rangle\}$ corresponding to a
vector $\vec{n}_0$ in Eq.~\rqn{9.6} perpendicular to $\vec{n}_A$,
 \be
\vec{n}_0\perp\vec{n}_A.
 \e{9.7}
Accordingly, the magnitude of $A_w$ in Eq.~\rqn{264} is maximum for a given $|\langle\phi|\psi\rangle|\ll1$, 
 \be
|A_w|_{\rm max}=|\langle\phi|\psi\rangle|^{-1},
 \e{9.1}
under the condition Eq.~\rqn{9.7}.
Note that the weak value in Eq.~\rqn{9.1} is unusual.

In the case of a mixed preselected state, Eq.~\rqn{9.1} does not hold, however the conditions for maximizing $|A_w|$ are the same as above, with the only difference that now $|\psi\rangle$ in Eq.~\rqn{211} should be replaced by $|\vec{n}_{\rm in}\rangle$.
In the next subsection, we provide explicit formulas for weak values of a qubit.

\subsection{Explicit formulas for a typical case}
\label{VC}

Let us consider in detail a typical case.
First, we recall that a unit vector in the Bloch sphere has the form
 \be
\vec{n}=(\sin\kappa\cos\nu,\sin\kappa\sin\nu ,\cos\kappa),\quad\quad{\rm where}\ \ 0\le\kappa<\pi;\ -\pi<\nu\le\pi.
 \e{212}
Here $\kappa$ and $\nu$ are the usual spherical coordinates of the
pseudospin, i.e., $\kappa$ is the angle between the pseudospin and
the $z$ axis, whereas the projection of $\vec{n}$ on the $xy$
plane forms the polar angle $\nu$ with $\vec{x}$.
A pure state with the pseudospin $\vec{n}$ is given by
 \be
|\vec{n}\rangle=\cos(\kappa/2)|\vec{z}\rangle +\exp(i\nu)\sin(\kappa/2)|-\vec{z}\rangle.
 \e{213}

 We assume here that $\hat{A}=\sigma_x$, i.e., $\vec{n}_A=\vec{x}$.
Moreover, we set $\vec{n}_{\rm in}=\vec{n}$ and $\vec{n}_f=-\vec{z}$, i.e., $|\phi\rangle=|-\vec{z}\rangle$.
Hence, for a pure preselected state, we have $|\psi\rangle=|\vec{n}\rangle$; in this case the angle $\kappa$ in Eq.~\rqn{213} is a measure of the overlap of the pre- and post-selected states, since
 \be
|\langle\phi|\psi\rangle|=\sin\frac{\kappa}{2}.
 \e{9.8}
In particular, for small $\kappa$, we have $\sin(\kappa/2)\approx\kappa/2$, implying
 \be
\kappa=2|\langle\phi|\psi\rangle|\quad\quad{\rm for}\ \ \kappa\ll1.
 \e{9.9}

Now Eqs.~\rqn{209'} and \rqn{209} yield
 \be
A_w\:=\:\frac{P_{\rm in}\sin\kappa}{1-P_{\rm in}\cos\kappa}
\exp(-i\nu),\quad\quad A_w^{(1,1)}\:=\:\frac{1+\cos\kappa} {1-P_{\rm
in}\cos\kappa}.
 \e{226}
In particular, for a pure initial state ($P_{\rm in}=1$) Eq.~\rqn{226} yields
 \be
A_w\:=\:\cot\left(\frac{\kappa}{2}\right)\exp(-i\nu).
 \e{227}
As shown by Eqs.~\rqn{226} and \rqn{227}, the phase of $A_w$ can be chosen at will by an appropriate choice of the initial and final states.
In particular, the weak value in Eqs.~\rqn{226} and \rqn{227} is real (imaginary) when $\vec{n}$, Eq.~\rqn{212}, lies in the $xz$ ($yz$) plane.
For a mixed initial state, the magnitude of the weak value in Eq.~\rqn{226} decreases with the decrease of the purity $P_{\rm in}$. 

In the important case $\kappa\ll 1$, the expressions for $A_w$ and $A_w^{(1,1)}$ simplify.
In particular, for a pure initial state, the weak value in Eq.~\rqn{227} is large, tending to infinity for $\kappa\rightarrow0$,
 \be
A_w\:\approx\:\frac{2}{\kappa}\exp(-i\nu)\quad\quad(\kappa\ll 1).
 \e{9.4}
Now for $\kappa\ll 1$ the condition \rqn{9.7} for maximizing $|A_w|$ is satisfied, since in the present case $\vec{n}_0=\vec{z}$ and $\vec{n}_A=\vec{x}$.
Correspondingly, Eq.~\rqn{9.4} yields the same $|A_w|$ as in Eq.~\rqn{9.1}, in view of Eq.~\rqn{9.9}.

For a mixed initial state, the necessary conditions for a large weak
value \rqn{210} become now
 \be
P_{\rm in}\approx1,\quad\quad\kappa\ll 1.
 \e{10.10}
Therefore, below we assume the initial state-impurity is small, $1-P_{\rm in}\ll1$, i.e., $P_{\rm in}\approx1$.
In the case \rqn{10.10}, we obtain that Eq.~\rqn{226} becomes approximately
 \be
A_w\:=\:\frac{2\kappa\exp(-i\nu)}{\kappa^2+2(1-P_{\rm in})},\quad\quad A_w^{(1,1)}\:=\:\frac{4}{\kappa^2+2(1-P_{\rm in})}.
 \e{263}
Now $|A_w|$ and $A_w^{(1,1)}$ as functions of $\kappa$ have
dispersive and Lorentzian shapes, respectively.
Equation \rqn{263} shows explicitly that the weak values for a qubit with a mixed preselected state are always finite.

\begin{figure}[htb]
\centerline{\includegraphics[width=11cm]{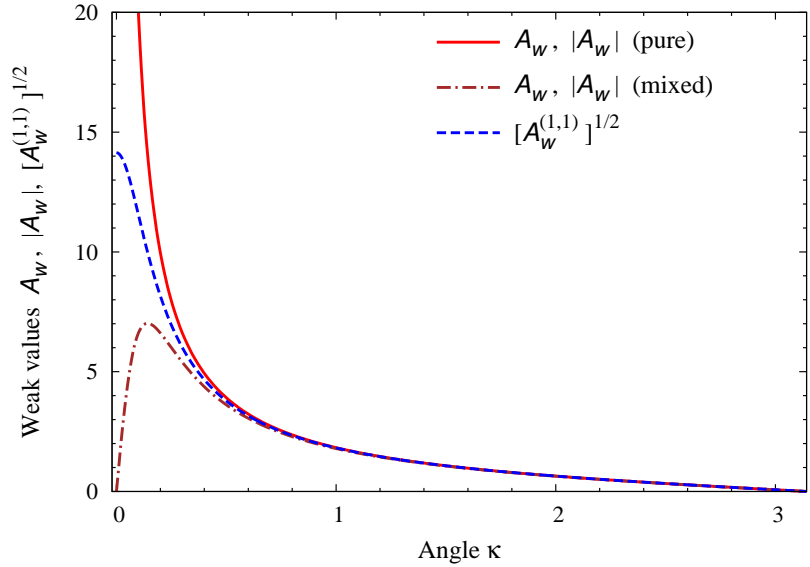}}
\caption{(color online). 
Weak values for a qubit versus the angle $\kappa$.
We show $A_w$ and $|A_w|$ for a pure preselected state, Eq.~\rqn{227}, and $A_w$, $|A_w|$, and $[A_w^{(1,1)}]^{1/2}$ for a mixed preselected state with $P_{\rm in}=0.99$, Eq.~\rqn{226}.
The quantity $A_w$ is shown for $\nu=0$.
Since $|A_w|$ corresponding to an arbitrary $\nu$ equals $A_w$ corresponding to $\nu=0$, the solid and and dot-dashed curves show both $A_w$ and $|A_w|$.
}
 \label{f14}\end{figure}

In Fig.~\ref{f14}, the weak values $A_w$ for pure and mixed preselected states with $\nu=0$ as well as the quantity $[A_w^{(1,1)}]^{1/2}$ for a mixed state with an arbitrary $\nu$ [see Eq.~\rqn{227} and  \rqn{226}] are plotted as a function of the angle $\kappa$, i.e., essentially as a function of the overlap $|\langle\phi|\psi\rangle|$ [cf.\ Eqs.~\rqn{9.8} and \rqn{9.9}].
Note that the quantity $A_w$ corresponding to $\nu=0$ coincides with $|A_w|$ corresponding to an arbitrary $\nu$, as seen from the first equality in Eq.~\rqn{226}.
Hence, the solid and and dot-dashed curves in Fig.~\ref{f14} show both $A_w$ and $|A_w|$.

For
 \be
\kappa^2\gg2(1-P_{\rm in}),
 \e{9.2}
the weak values for the pure and mixed preselected states [Eqs.~\rqn{227} and \rqn{226}, respectively] are approximately equal, and $A_w^{(1,1)}\approx|A_w|^2$ (cf.\ Fig.~\ref{f14}).
However, for 
 \be
\kappa^2\alt2(1-P_{\rm in}) 
 \e{9.3}
the magnitude of the weak value for a mixed preselected state is significantly less than that for a pure preselected state and also than $[A_w^{(1,1)}]^{1/2}$ (cf.\ Fig.~\ref{f14}).
 Both $|A_w|$ and $A_w^{(1,1)}$ have large but {\em finite} maxima:
 \be
|A_w|\:=\:[2(1-P_{\rm in})]^{-1/2}\quad\text{at}\ \ \kappa=[2(1-P_{\rm in})]^{1/2}
 \e{331}
and
 \be
A_w^{(1,1)}\:=\:\frac{2}{1-P_{\rm in}}\quad\text{at}\ \ \kappa=0.
 \e{332}
It is interesting that now $\lim_{\kappa\rightarrow0}A_w=0$ for a mixed preselected state.
In contrast, for a pure preselected state, $\lim_{\kappa\rightarrow0}A_w=\infty$, as one would expect.

Equation \rqn{331} shows that the maximum weak-value magnitude increases with the purity of the preselected state.
This dependence is rather slow; e.g., for a state with a high purity, $P_{\rm in}=0.99$, the maximum $|A_w|$ is only $\sqrt{50}\approx7.1$ (see the dot-dashed curve in Fig.~\ref{f14}).
Generally, the greater $|A_w|$, the more stringent is the condition on the purity of $\rho$.
Namely, to achieve a given large weak-value magnitude $|A_w|$, the parameter $P_{\rm in}$ should satisfy the condition
 \be
1-P_{\rm in}\le(2|A_w|^2)^{-1}\quad\quad(|A_w|\gg1).
 \e{9.10}
In other words, the allowed impurity of the preselected state decreases quadratically with the target weak-value magnitude.
For example, to obtain $|A_w|=100$, it is necessary that $1-P_{\rm in}\le5\times10^{-5}$.

Note that Eq.~\rqn{9.10} holds for the operator $\hat{A}$ in Eq.~\rqn{178}.
A general observable $O$ for a qubit is a linear function of the physical variable with the operator given by Eq.~\rqn{178}.
Therefore, $A_w$ is a linear function of $O_w$.
Replacing $A_w$ by this function in Eq.~\rqn{9.10} yields the upper value of the allowed impurity of the preselected state for a given weak value $O_w$ (assuming that $|A_w|\gg1$).

Here, as almost everywhere in the present paper, we assume that the post-selection measurement is ideal.
However, a non-ideal post-selection measurement can affect PPS measurements similarly to a mixed preselected state, as shown below in Sec.~\ref{XIIIE}.

\subsection{Effects of preselected-state impurity on weak PPS measurements}
\label{IXD}

The results of the previous subsection allow us to analyze the effects of the impurity of the preselected state on weak PPS measurements.
In the linear-response regime, the pointer deflection is proportional to the weak value, Eq.~\rqn{14''}, and hence the dependence of the pointer deflection on the impurity of the preselected state is the same as that of $A_w$, see the discussion in the previous subsection.

Consider now the nonlinear behavior, which becomes important under the conditions in Eq.~\rqn{10.10}.
In this case, inserting Eq.~\rqn{263} into Eq.~\rqn{61} yields
 \be
\bar{R}_s-\bar{R}\:=\:\frac{\gamma\kappa\,{\rm Im}\,(\,\overline{R_cF}e^{-i\nu})+\gamma^2\,\overline{FR_cF}} 
{\kappa^2/4+(1-P_{\rm in})/2-\gamma\kappa\bar{F}\sin\nu +\gamma^2\,\overline{F^2}}.
 \e{9.15}
The impurity of the preselected state contributes in Eq.~\rqn{9.15} only the term $2(1-P_{\rm in})$, which disappears for a pure preselected state.
Thus, generally the main effect of the impurity of the preselected state on weak PPS measurements of a qubit is to decrease the average pointer deflection.

Let us discuss the conditions under which the initial-state impurity effects can be neglected.
For the linear-response and strongly-nonlinear regimes, i.e., in the interval [cf.\ Eqs.~\rqn{100} and \rqn{282}]
 \be
\gamma^2\,[(\Delta F)^2+\bar{F}^2]\:\alt\:\kappa^2/4,
 \e{9.16}
the impurity is negligible under the condition given in Eq.~\rqn{9.2}, i.e., in the case when the weak values for the pure and mixed preselected states are approximately equal.

A more interesting situation arises in the inverted region,
 \be
\kappa^2/4\:\ll\:\gamma^2\,[(\Delta F)^2+\bar{F}^2]\:\ll\:1
 \e{9.17}
(here the right inequality is the weak-measurement condition).
Now for the case
 \be
(1-P_{\rm in})/2\:\ll\:\gamma^2\,[(\Delta F)^2+\bar{F}^2],
 \e{9.18}
Eq.~\rqn{9.15} becomes in the first approximation
 \bea
&\bar{R}_s-\bar{R}&=\:\frac{\overline{FR_cF}}{\overline{F^2}} \left[1- \frac{1-P_{\rm in}}{2\gamma^2\overline{F^2}}\right] +\frac{\kappa\,{\rm Im}\,(\,\overline{R_cF}e^{-i\nu})}{2\gamma\overline{F^2}} +\frac{\kappa\,\overline{FR_cF}\,\bar{F}\sin\nu}{\gamma(\overline{F^2})^2}\nonumber\\
&&\approx\frac{\overline{FR_cF}}{\overline{F^2}} +\frac{\kappa\,{\rm Im}\,(\,\overline{R_cF}e^{-i\nu})}{2\gamma\overline{F^2}} +\frac{\kappa\,\overline{FR_cF}\,\bar{F}\sin\nu}{\gamma(\overline{F^2})^2},
 \ea{9.19}
where the final expression is the same as for a pure preselected state, Eq.~\rqn{15}.
This expression is obtained by neglecting in the first line the small term in the brackets, which slightly changes the value of the pointer deflection in the limit $\kappa\rightarrow0$.
Thus, in the cases when Eqs.~\rqn{9.17} and \rqn{9.18} are satisfied the initial-state impurity has a negligible effect on weak PPS measurements, though in this case it is possible that $\kappa^2\alt2(1-P_{\rm in})$ and hence {\em the weak value can be much less than that for a pure preselected state}.
The above discussion shows that in the inverted region weak PPS measurements are less sensitive to the impurity effects than in other regimes; this can be advantageous for precision measurements.

Finally, we consider the resonance given in Eq.~\rqn{221} for the case $|\bar{F}|\gg\Delta F$.
It is easy to see that now the resonance conditions given in Eqs.~\rqn{173} and \rqn{168} have the form
 \be
\nu\approx\pm\frac{\pi}{2},\quad\quad2\gamma\bar{F}\approx\pm\kappa,
 \e{9.20}
where the upper or the lower signs should be considered simultaneously.
Moreover, the parameter $v$ in Eq.~\rqn{228} becomes
 \be
v\:=\:\frac{\sqrt{2(1-P_{\rm in})}}{\kappa}\ll1.
 \e{9.21}
As mentioned in Sec.~\ref{IVC3}, this parameter should be very small for the resonance not to be smeared out.
In particular, as follows from Eq.~\rqn{221}, the effects of the impurity can be neglected when
 \be
\frac{2(1-P_{\rm in})}{\kappa^2}\ll\left(\frac{\Delta F}{\bar{F}}\right)^2\ll1,
 \e{9.22}
which is much stronger than the condition in Eq.~\rqn{9.2}.
This fact can be used to measure very small impurities of quantum states.

It is remarkable that, although the three conditions given in Eqs.~\rqn{9.2}, \rqn{9.18}, and \rqn{9.22} are very different, they are special cases of the simple condition
 \be
(1-P_{\rm in})/2\ll\langle\Pi_\phi\rangle_f.
 \e{9.23}
Thus, the initial-state impurity has negligible effects on weak PPS measurements of a qubit, when the impurity is much less than the post-selection probability.
To understand why the condition \rqn{9.23} arises, we note that it is easy to show that the denominator in Eq.~\rqn{9.15} equals $\langle\Pi_\phi\rangle_f$ in Eq.~\rqn{7.9b}.
Therefore, the effects of the impurity of the preselected state on weak PPS measurements of a qubit, at least, in the most interesting case $|\kappa|\ll1$, can be taken into account in Eq.~\rqn{44}, written for a pure initial state ($P_{\rm in}=1$), by the substitution
 \be
\langle\Pi_\phi\rangle_f=\langle\Pi_\phi\rangle_{f0}\ \rightarrow\ \langle\Pi_\phi\rangle_f=\langle\Pi_\phi\rangle_{f0}+(1-P_{\rm in})/2,
 \e{10.9}
where $\langle\Pi_\phi\rangle_{f0}$ is the post-selection probability for a pure state.
This explains why the condition allowing to neglect the impurity effects has the form of Eq.~\rqn{9.23}.

\section{Exact solutions for arbitrary-strength PPS and standard measurements}
 \label{VIII}

Exact solutions for PPS measurements of arbitrary strength were obtained in a number of papers, \cite{duc89,kni90,rit91,sut93,sut95,%
bru03,bru04,sol04,pry05,mir07,wil08,koi11,nak12}.
Here we obtain exact results for PPS measurements of arbitrary strength in the case where
 \be
\hat{A}^2=C_0I_{\rm S},
 \e{11.29}
$C_0$ being an arbitrary positive number.
The Hermitian operator $\hat{A}$ in Eq.~\rqn{11.29} can have an arbitrary dimension, but the eigenvalues of $\hat{A}$, possibly degenerate, can assume only two values that have equal magnitudes and opposite signs, namely, $\pm\sqrt{C_0}$.

Without loss of generality, we will assume that
 \be
\hat{A}^2=I_{\rm S}.
 \e{11.30}
Indeed, as implied by Eq.~\rqn{2}, the results obtained for the case \rqn{11.30} become valid for the case \rqn{11.29} under the substitutions
 \be
\hat{A}\rightarrow\hat{A}/\sqrt{C_0},\quad\quad
\gamma\rightarrow\sqrt{C_0}\,\gamma.
 \e{11.31}
The eigenvalues of the operator $\hat{A}$ in Eq.~\rqn{11.30} can equal 1 or $-1$.

The case \rqn{11.29} includes many useful situations.
In particular, the set of operators with the property \rqn{11.30} includes the Pauli matrices, qubit operators of the form \rqn{178}, and direct products of them.

Our results are very general.
In particular, we consider arbitrary initial states of the system and the meter.
 First we consider the case of a general meter and then discuss
specific examples of the meter.

\subsection{Formulas for a general meter}
\label{VIIA}


{\em a. Average pointer value.}
It is easy to see that in the case \rqn{11.30}, Eq.~\rqn{2} yields
 \be
U=\cos(\gamma\hat{F})-i\hat{A}\sin(\gamma\hat{F}).
 \e{179}
On inserting Eq.~\rqn{179} into Eqs.~\rqn{112} and \rqn{3.4}, we obtain from Eq.~\rqn{241} that
 \bes{180}
 \be
\bar{R}_s\:=\:\frac{G_{\rm cc} +2{\rm Im}(A_wG_{\rm cs}) +A_w^{(1,1)}G_{\rm ss}}{Q_1},
 \e{180a}
where
 \be
Q_1\:=\:[1+M_c+2M_s{\rm Im}\,A_w+(1-M_c)A_w^{(1,1)}]/2.
 \e{180c}
 \ese
In Eqs.~\rqn{180} we denoted
 \bea
&&G_{\rm cc}=\overline{\cos(\gamma F)R\cos(\gamma F)},\quad\quad G_{\rm cs}=\overline{\cos(\gamma F)R\sin(\gamma F)},\quad\quad 
G_{\rm ss}=\overline{\sin(\gamma F)R\sin(\gamma F)},
\label{180'}\\
&&M_c=\overline{\cos(2\gamma F)},\quad\quad 
M_s=\overline{\sin(2\gamma F)}
 \ea{180b}
It is interesting that now the system parameters enter the exact
Eqs.~\rqn{180} through the same weak values $A_w$ and $A_w^{(1,1)}$
as in the weak-coupling case, Eq.~\rqn{183}.
 Equation \rqn{183} results on expanding Eqs.~\rqn{180} in powers of
$\gamma$ up to $\gamma^2$ and neglecting small terms as discussed in
Sec.~\ref{IIIA3}.
Thus, in the case \rqn{11.30} [and hence in the case \rqn{11.29}], weak values can be obtained from PPS measurements of an arbitrary strength and not only from weak PPS measurements.


{\em b. Exact pointer distribution for PPS measurements.}
An exact expression for the pointer-value distribution is obtained
from Eq.~\rqn{180a} under the substitution
$\hat{R}\rightarrow|R\rangle\langle R|$, using Eq.~\rqn{184}.
 This yields
 \be
\Phi_s(R)\:=\:\frac{\Phi_{\rm cc}(R)+2{\rm Im}[A_w\Phi_{\rm sc}(R)]
+A_w^{(1,1)}\Phi_{\rm ss}(R)}{Q_1},
 \e{186}
where
 \be
\Phi_{\rm cc}(R)=\langle R|\cos(\gamma F)\rho_{\rm M}\cos(\gamma
F)|R\rangle,\quad 
\Phi_{\rm sc}(R)=\langle R|\sin(\gamma F)\rho_{\rm M}\cos(\gamma F)|R\rangle,\quad 
\Phi_{\rm ss}(R)=\langle R|\sin(\gamma F)\rho_{\rm M}\sin(\gamma F)|R\rangle.
 \e{187}
When the initial state of the meter is pure, $|\psi_{\rm M}\rangle$,
then Eq.~\rqn{187} becomes,
 \be
\Phi_{\rm cc}(R)=|\psi_c(R)|^2,\quad\quad
\Phi_{\rm sc}(R)=\psi_s(R)\psi_c^*(R),\quad\quad
\Phi_{\rm ss}(R)=|\psi_s(R)|^2,
 \e{188}
where
 \be
\psi_c(R)=\langle R|\cos(\gamma F)|\psi_{\rm M}\rangle,\quad\quad
\psi_s(R)=\langle R|\sin(\gamma F)|\psi_{\rm M}\rangle.
 \e{189}

\subsection{Specific types of meters}

The results in Sec.~\ref{VIIA} are exact and can be applied for arbitrary coupling strength and an arbitrary meter.
 Here we specify these results for several important types of the meter.

\subsubsection{Coinciding meter variables, $F=R$}
\label{VIIA3}

Equations \rqn{180a} and \rqn{186} simplify when $F$ is a function
of $R$.
 For example, when $F=R$, Eq.~\rqn{180'} implies that
 \be
G_{\rm cc}\:=\: \overline{F\cos^2(\gamma F)}\:=\:[\bar{F}+\overline{F\cos(2\gamma F)}]/2\:=\:\bar{F}/2+M_s'/2
 \e{201}
and, similarly,
 \be
G_{\rm cs}\:=\:M_c'/2,\quad\quad G_{\rm ss}\:=\: \bar{F}/2-M_s'/2,
 \e{225}
where 
 \be
M_s'=\overline{F\cos(2\gamma F)},\quad\quad M_c'=\overline{F\sin(2\gamma F)}
 \e{10.2}
 As a result, Eq.~\rqn{180} becomes
 \be
\bar{F}_s\:=\:\frac{\bar{F}+M_s' +2M_c'\,{\rm Im}\,A_w
+(\bar{F}-M_s')A_w^{(1,1)}}{1+M_c+2M_s{\rm Im}\,A_w+(1-M_c)A_w^{(1,1)}}.
 \e{199}

Furthermore, in the present case $F=R$, the exact pointer distribution \rqn{186} with the account of Eq.~\rqn{187} becomes
 \be
\Phi_s(F)\:=\:2\Phi(F)\,\frac{\cos^2(\gamma F)+({\rm Im}\,A_w)\sin(2\gamma F)+A_w^{(1,1)}\sin^2(\gamma F)}{1+M_c+2M_s{\rm Im}\,A_w+(1-M_c)A_w^{(1,1)}}.
 \e{200}

{\em a. Symmetric} $\Phi(F)$.
As an example, consider the case of a symmetric distribution $\Phi(F)$ centered at $F=0$, i.e., $\Phi(F)=\Phi(-F)$.
Now one has $\bar{F}=M_s=M_s'=0$, and Eq.~\rqn{199} acquires a simple form,
 \be
\bar{F}_s\:=\:\frac{2M_c'\,{\rm Im}\,A_w}{1+M_c+(1-M_c)A_w^{(1,1)}}.
 \e{10.3}
If, in addition, the weak value is purely imaginary and the preselected state is pure, so that $A_w^{(1,1)}=|A_w|^2=({\rm Im}\,A_w)^2$, the exact pointer distribution in Eq.~\rqn{200} becomes especially simple,
 \be
\Phi_s(F)\:=\:2\Phi(F)\,\frac{[\cos(\gamma F)+({\rm Im}\,A_w)\sin(\gamma F)]^2}{1+M_c+(1-M_c)|A_w|^2}.
 \e{11.34}

{\em b. Gaussian} $\Phi(F)$.
Consider the case of a continuous-variable meter, so that the variable $F=R$ can be, e.g., the coordinate or the momentum.
Assume that the distribution $\Phi(F)$ is Gaussian, Eq.~\rqn{9.24}.
Then Eqs.~\rqn{180b} and \rqn{10.2} yield
\be
M_c\:=\:\cos(2\gamma\bar{F})\,\exp[-2(\gamma\Delta F)^2],\quad\quad
M_s\:=\:\sin(2\gamma\bar{F})\,\exp[-2(\gamma\Delta F)^2],
\e{204'}
 \bea
&&M_c'\:=\:[\bar{F}\sin(2\gamma\bar{F}) +2\gamma(\Delta F)^2
\cos(2\gamma\bar{F})]\,\exp[-2(\gamma\Delta F)^2],\nonumber\\
&&M_s'\:=\:[\bar{F}\cos(2\gamma\bar{F}) -2\gamma(\Delta F)^2
\sin(2\gamma\bar{F})]\,\exp[-2(\gamma\Delta F)^2].
 \ea{204}
Now the average pointer value is given by Eq.~\rqn{199} or \rqn{10.3} with the account of Eqs.~\rqn{204'} and \rqn{204}.
Moreover, the pointer distribution after the measurement is given by Eq.~\rqn{200} or \rqn{11.34} with the account of Eq.~\rqn{204'}.

In particular, for a zero-mean Gaussian $\Phi(F)$, $\bar{F}=0$, Eqs.~\rqn{204'} and \rqn{204} become
\bea
&&M_c\:=\:\exp[-2(\gamma\Delta F)^2],\quad\quad M_s\:=\:0, \label{11.42}\\
&&M_c'\:=\:2\gamma(\Delta F)^2\exp[-2(\gamma\Delta F)^2],\quad\quad 
M_s'\:=\:0.
 \ea{11.43}
Then Eq.~\rqn{10.3} with the account of Eqs.~\rqn{11.42} and \rqn{11.43} yields the following explicit expression,
 \be
\bar{F}_s\:=\:\frac{4\gamma\,(\Delta F)^2\,{\rm Im}\,A_w}{1-A_w^{(1,1)}+(1+A_w^{(1,1)})\exp[2(\gamma\Delta F)^2]}.
 \e{11.37}
For the special case, where $F=p$ is the meter momentum and the preselected state is pure, so that $A_w^{(1,1)}=|A_w|^2$, Eq.~\rqn{11.37} and Eq.~\rqn{200} with the account of Eq.~\rqn{11.42} were obtained in Ref.~\cite{nak12}.

\subsubsection{Conjugate meter variables}
\label{XB3}

Consider a meter with continuous variables.
 The case $F=R$ was discussed in Sec.~\ref{VIIA3}.
Here we consider the case of the canonically conjugate variables \rqn{97}, $F=p$ and $R=q$.

{\em a. Arbitrary pure meter state.}
For a pure meter state, as shown in Appendix \ref{C}, we have
 \bea
&&G_{\rm cc}\:=\:\overline{\cos(\gamma p)\,q\,\cos(\gamma p)}\:=\:
(\bar{q}+G_1)/2,
\nonumber\\
&&G_{\rm ss}\:=\:\overline{\sin(\gamma p)\,q\,\sin(\gamma p)}\:=\:
(\bar{q}-G_1)/2,
\nonumber\\
&&G_{\rm cs}\:=\:\overline{\cos(\gamma p)\,q\,\sin(\gamma p)}\:=\:
(i\gamma+G_2)/2,
 \ea{198}
where
 \be
G_1=\overline{\zeta'(p)\cos(2\gamma p)},\quad\quad
G_2=\overline{\zeta'(p)\sin(2\gamma p)}.
 \e{10.4}

Furthermore, consider the exact pointer distribution. 
In the present case of the meter variables \rqn{97}, it is given in paragraph {\em b} in Sec.~\ref{VIIA}, where now, using the expression $\hat{F}=p=-i\partial/\partial q$, we obtain that Eq.~\rqn{189} becomes
 \be
\psi_c(q)=[\psi_{\rm M}(q+\gamma)+\psi_{\rm M}(q-\gamma)]/2,\quad\quad
\psi_s(q)=[\psi_{\rm M}(q+\gamma)-\psi_{\rm M}(q-\gamma)]/(2i).
 \e{203}

{\em b. Real even meter wavefunction.}
Consider a simple example.
Let the initial meter state be pure and such that $\psi_{\rm M}(p)$ and $\psi_{\rm M}(q)$ are real, even functions, i.e.,
\bes{10.6}
 \bea
&&\psi_{\rm M}(p)=\psi_{\rm M}^*(p),\quad\quad
\psi_{\rm M}(p)=\psi_{\rm M}(-p);\label{10.6a}\\
&&\psi_{\rm M}(q)=\psi_{\rm M}^*(q),\quad\quad
\psi_{\rm M}(q)=\psi_{\rm M}(-q).
 \ea{10.6b}
\ese
Note that the condition \rqn{10.6a} implies Eq.~\rqn{10.6b} and vice versa.
In the case \rqn{10.6}, we have $\bar{q}=\bar{p}=M_s=0$.
As a result, in particular, Eq.~\rqn{180c} becomes
 \be
Q_1\:=\:[1+M_c+(1-M_c)A_w^{(1,1)}]/2.
 \e{11.32}
Moreover, now $\zeta(p)=0$, yielding $G_{\rm cc}=G_{\rm ss}=0$ and $G_{\rm cs}=i\gamma/2$.
Then Eqs.~\rqn{180a} and \rqn{11.32} yield the following simple formula,
 \be
\bar{q}_s\:=\:\frac{2\gamma\,{\rm Re}\,A_w}{1+M_c+(1-M_c)A_w^{(1,1)}}.
 \e{10.5}

The pointer distribution after the measurement is obtained from Eq.~\rqn{186} with the account of Eqs.~\rqn{188}, \rqn{203}, and \rqn{10.6b},
 \be
\Phi_s(q)\:=\:[(1-2{\rm Re}\,A_w+A_w^{(1,1)})\,\Phi(q+\gamma)+(1+2{\rm Re}\,A_w+A_w^{(1,1)})\,\Phi(q-\gamma)+2(1-A_w^{(1,1)})\,\psi_{\rm M}(q+\gamma)\psi_{\rm M}(q-\gamma)]/(4Q_1),
 \e{11.35}
where $Q_1$ is given in Eq.~\rqn{11.32}.
If, moreover, the preselected state is pure and $A_w$ is real, so that $A_w^{(1,1)}=A_w^2$, Eq.~\rqn{11.35} simplifies,
 \be
\Phi_s(q)\:=\:[(1-A_w)\,\psi_{\rm M}(q+\gamma)+(1+A_w)\,\psi_{\rm M}(q-\gamma)]^2/(4Q_1).
 \e{11.36}

{\em c. General Gaussian meter state.}
When the meter initial state is a general complex Gaussian state
\rqn{103}, all the relevant averages can be expressed in a simple
form.
 Taking into account Eqs.~\rqn{216} and \rqn{10.4}, we obtain that in Eq.~\rqn{198}
 \bea
&&G_1\:=\:[\bar{q}\cos(2\gamma\bar{p})
-\gamma b\sin(2\gamma\bar{p})]\,\exp[-2(\gamma\Delta p)^2],\nonumber\\
&&G_2=[\bar{q}\sin(2\gamma\bar{p})
+\gamma b\cos(2\gamma\bar{p})]\,\exp[-2(\gamma\Delta p)^2].
 \ea{202}
Now $M_c$ and $M_s$ are given by Eq.~\rqn{204'} with $F=p$.

{\em d. Real Gaussian meter state.}
In particular, for a real and zero-mean Gaussian meter state, Eq.~\rqn{10.5} with the account of Eq.~\rqn{204'} yields the following explicit expressions,
 \be
\bar{q}_s\:=\:\frac{2\gamma\,{\rm Re}\,A_w}{1+A_w^{(1,1)} +(1-A_w^{(1,1)})\exp[-2(\gamma\Delta p)^2]}.
 \e{10.7}

Now the pointer distribution is given by Eq.~\rqn{11.35} or \rqn{11.36}, where
 \bea
&&\psi_{\rm M}(q)=(\sqrt{2/\pi}\,\Delta p)^{1/2} \exp[-(\Delta p)^2q^2],\label{11.40}\\
&&\Phi(q)=\psi_{\rm M}^2(q)=\sqrt{2/\pi}\,\Delta p\, \exp[-2(\Delta p)^2q^2],\label{11.41}\\
&&Q_1=\{1+A_w^{(1,1)} +(1-A_w^{(1,1)})\exp[-2(\gamma\Delta p)^2]\}/2.
 \ea{11.38}
In particular, Eq.~\rqn{11.35} becomes
 \be
\Phi_s(q)\:=\:\frac{(1-2{\rm Re}\,A_w+A_w^{(1,1)})\,\Phi(q+\gamma)+(1+2{\rm Re}\,A_w+A_w^{(1,1)})\,\Phi(q-\gamma)+2(1-A_w^{(1,1)})\,\exp[-2(\gamma\Delta p)^2]\,\Phi(q)}{2\{1+A_w^{(1,1)} +(1-A_w^{(1,1)})\exp[-2(\gamma\Delta p)^2]\}}.
 \e{11.39}

A result similar to that in Eq.~\rqn{11.36} with the account of Eq.~\rqn{11.40} was obtained (without a normalization factor) in Ref.~\cite{duc89} for the case $F=q,\ R=p$.
Moreover, for the special case, where the preselected state is pure, so that $A_w^{(1,1)}=|A_w|^2$, Eqs.~\rqn{10.7} and \rqn{11.39} were obtained in Ref.~\cite{nak12}.

\subsubsection{Two-level meter}
\label{IXB2}

 Consider now a two-level meter (qubit).
 Using Eq.~\rqn{6.16} and taking into account that $\hat{F}_1^2=I$,
we obtain that
 \be
\cos\gamma\hat{F}=\cos\gamma\cos(\gamma f_0) -\hat{F}_1\sin\gamma\sin(\gamma f_0),\quad\quad
\sin\gamma\hat{F}=\cos\gamma\sin(\gamma f_0) +\hat{F}_1\sin\gamma\cos(\gamma f_0).
 \e{191}
Inserting Eq.~\rqn{191} into Eqs.~\rqn{180'} and \rqn{180b} yields
 \bea
&&G_{\rm cc}\:=\:C_{10}^2C_{11}^2\bar{R}-S_{20}S_{21}M_R/2
+S_{10}^2S_{11}^2M,\nonumber\\
&&G_{\rm cs}\:=\:[C_{10}^2S_{21}\bar{R}+S_{20}(C_{21}M_R+iM_I)
-S_{10}^2S_{21}M]/2,\nonumber\\
&&G_{\rm ss}\:=\:C_{10}^2S_{11}^2\bar{R}+S_{20}S_{21}M_R/2 +
S_{10}^2C_{11}^2M,\nonumber\\
&&M_c\:=C_{20}C_{21}-S_{20}S_{21}\bar{F}_1,\quad
M_s= C_{20}S_{21}+S_{20}C_{21}\bar{F}_1.
 \ea{192}
Here $M_R,\ M_I$, and $M$ are given by Eqs.~\rqn{195}, \rqn{206},
and \rqn{197}, whereas ($m=1,2$)
 \be
C_{m0}=\cos(m\gamma),\quad S_{m0}=\sin(m\gamma),\quad
C_{m1}=\cos(m\gamma f_0),\quad S_{m1}=\sin(m\gamma f_0).
 \e{193}

\subsection{Exact solution for standard measurements}

Though in this section we focus mainly on PPS measurements, here we also briefly mention the exact result for standard (i.e., not post-selected) measurements with an arbitrary measurement strength for the case \rqn{11.30}.
On inserting Eq.~\rqn{179} into Eq.~\rqn{74}, it is easy to obtain that
 \be
\bar{R}_f\:=\:G_{\rm cc}+G_{\rm ss} +2\bar{A}\,{\rm Im}\,G_{\rm cs}.
 \e{10.8}
Equation \rqn{10.8} provides an extension of Eq.~\rqn{75}; expanding Eq.~\rqn{10.8} up to first order in $\gamma$ yields Eq.~\rqn{75}.
Thus, the average values of operators satisfying Eq.~\rqn{11.30} or, more generally, Eq.~\rqn{11.29} can be provided directly by standard measurements of an arbitrary strength and not only by weak measurements.

\begin{figure}[htb]
\centerline{\includegraphics[width=15cm]{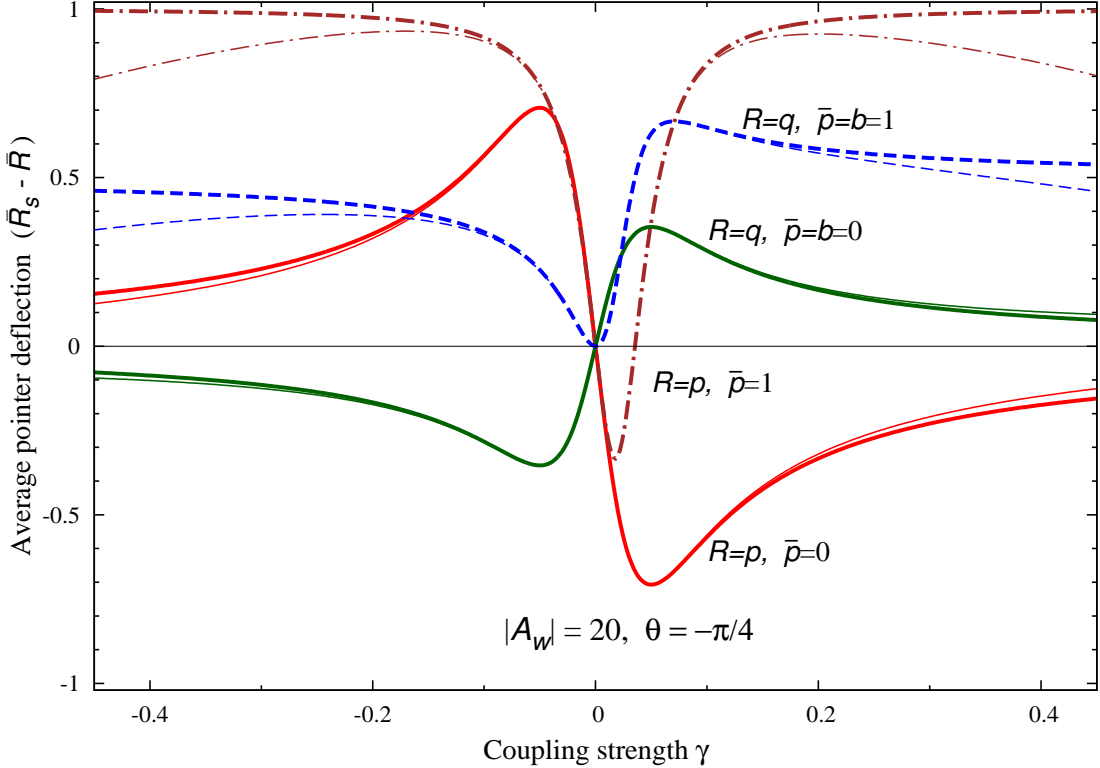}}
\caption{(color online). The average pointer deflection
$(\bar{R}_s-\bar{R})$ versus the coupling strength $\gamma$.
Here $\Delta p=1$, $P_{\rm in}=1$ (a pure initial state); thick
lines: Eqs.~\rqn{359}; thin lines: the exact solution \rqn{180}.
Solid red lines: $R=p,\ \bar{p}=0$; dot-dashed brown lines: $R=p,\ \bar{p}=1$; solid dark-green lines: $R=q,\ \bar{p}=b=0$; dashed blue lines: $R=q,\ \bar{p}=b=1$.
The figure shows several lineshapes which are possible in weak PPS measurements.
Notice that our general formulas \rqn{359} approximate very well the exact solutions in the region where the measurements are weak, see the inequality \rqn{229}, which now is $|\gamma|\ll1$.}
 \label{f1}\end{figure}

\begin{figure}[htb]
\centerline{\includegraphics[width=15cm]{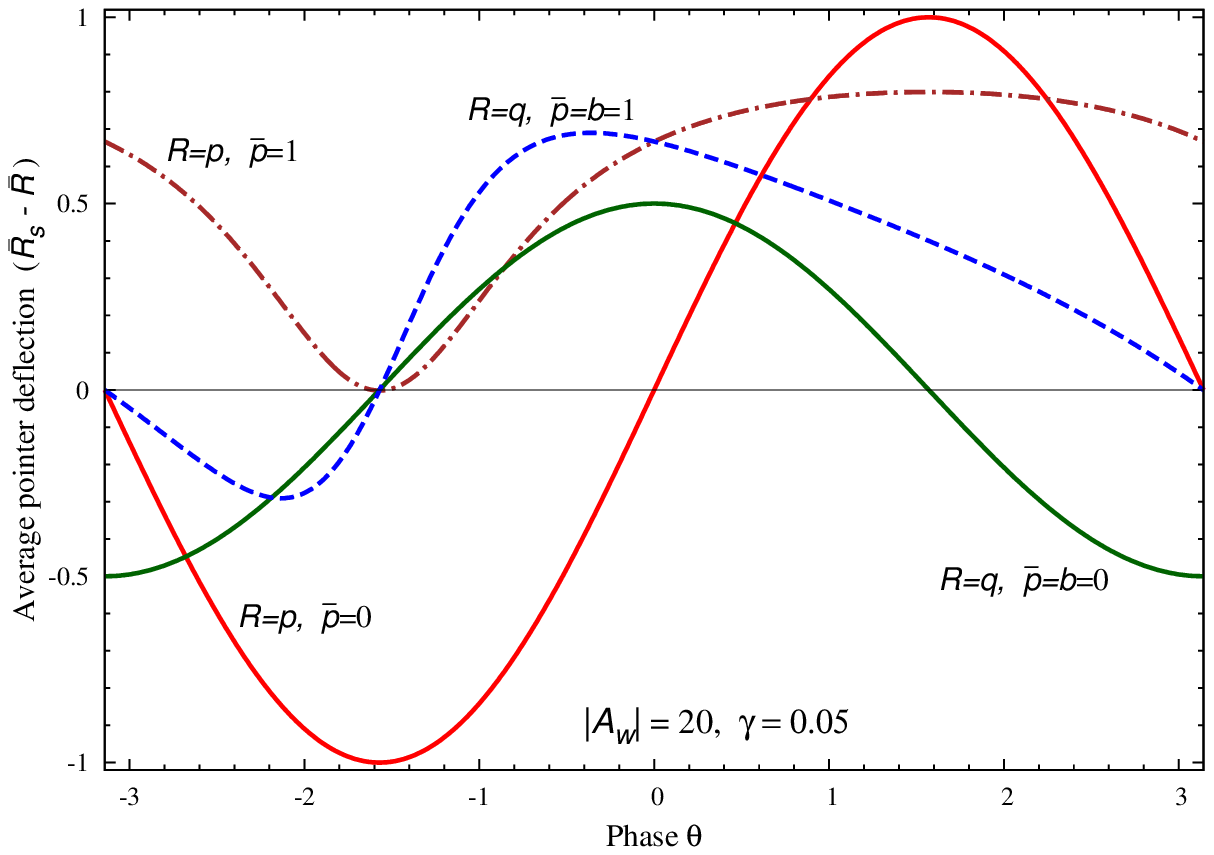}}
\caption{(color online). The average pointer deflection
$(\bar{R}_s-\bar{R})$ versus the phase $\theta$ of $A_w$.
Here $\Delta p=1$, $P_{\rm in}=1$; thick lines: Eqs.~\rqn{359}; thin
lines: the exact solution \rqn{180}.
The thin lines are not seen since they practically coincide with the thick lines.
The four cases are plotted with the same line styles as in Fig.~\ref{f1}.
}\label{f2}\end{figure}

\section{Numerical results and discussion}
 \label{IX}

Here we present results of calculations for weak PPS measurements of
a qubit with two types of a continuous-variable meter.
 We assume that for both meters, $F=p$ and $\overline{p_c^3}=0$.
For meter 1, $R=p$, whereas for meter 2, $R=q$ and the phase
$\zeta(p)$ is quadratic.
 In this section we set $\Delta p=1$.

\begin{figure}[htb]
\centerline{\includegraphics[width=15cm]{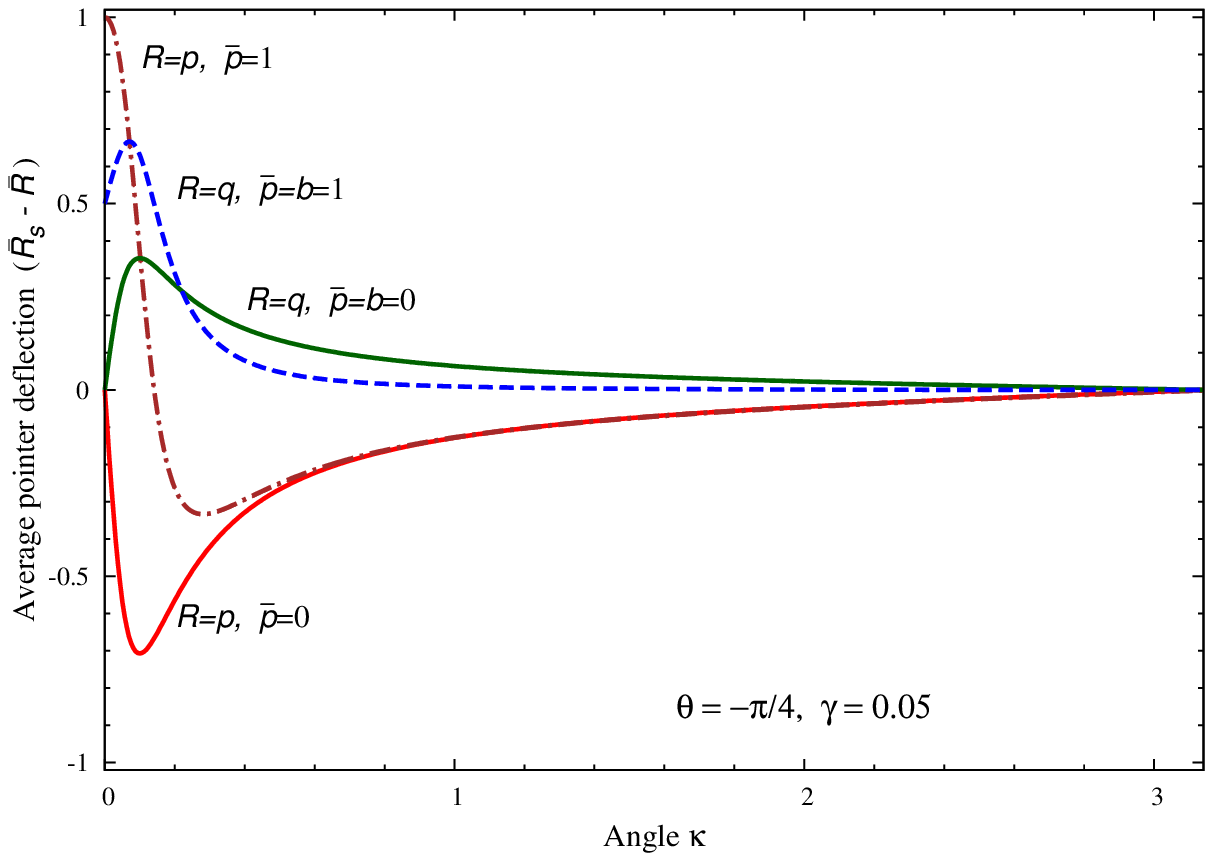}}
\caption{(color online). The average pointer deflection
$(\bar{R}_s-\bar{R})$ versus the angle $\kappa$ in Eq.~\rqn{227}.
Here $\Delta p=1$, $P_{\rm in}=1$; thick lines: Eqs.~\rqn{359}; thin
lines: the exact solution \rqn{180}.
The thin lines are not seen since they practically coincide with the thick lines.
The four cases are plotted with the same line styles as in Fig.~\ref{f1}.
} \label{f3}\end{figure}

\begin{figure}[htb]
\centerline{\includegraphics[width=15cm]{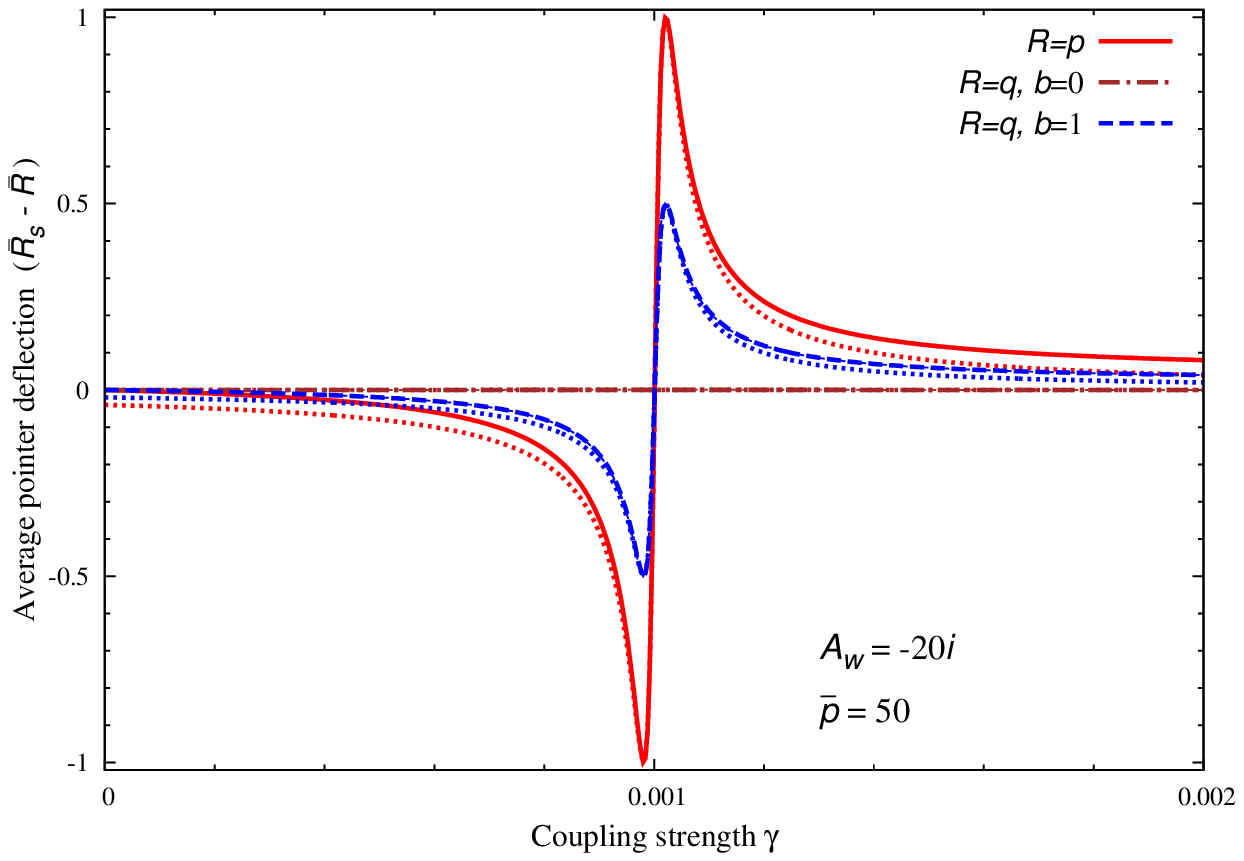}}
\caption{(color online). The average pointer deflection
$(\bar{R}_s-\bar{R})$ versus the coupling strength $\gamma$.
Here $\Delta p=1$, $P_{\rm in}=1$; thick lines: Eqs.~\rqn{359}; thin
lines: the exact solution \rqn{180}; dotted lines: the simplified
Eqs.~\rqn{360}.
The thin lines are not seen since they practically coincide with the thick lines.
The figure shows the regime of a narrow resonance obtainable for $\bar{F}\gg\Delta F$, i.e., now for $\bar{p}\gg1$.
}\label{f4}\end{figure}

\begin{figure}[htb]
\centerline{\includegraphics[width=15cm]{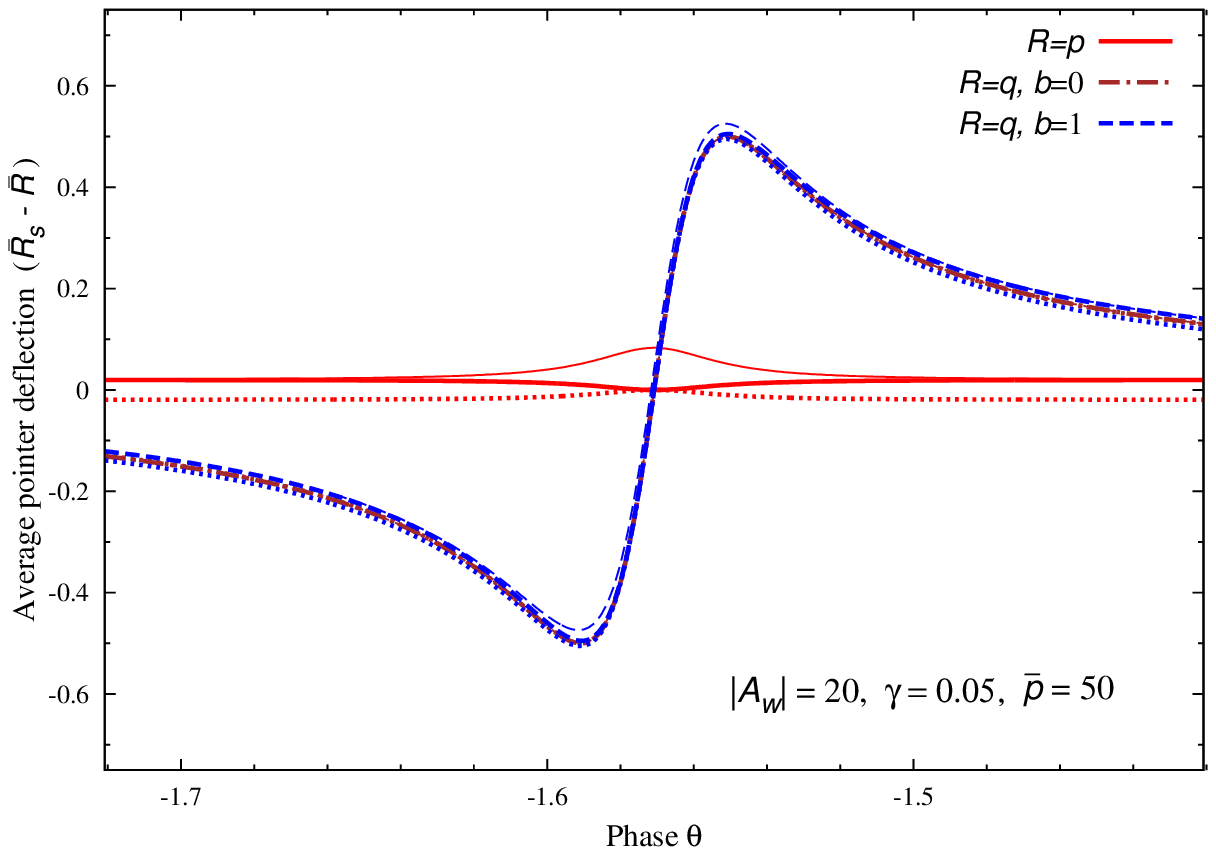}}
\caption{(color online). The average pointer deflection
$(\bar{R}_s-\bar{R})$ versus the phase $\theta$ of $A_w$.
Here $\Delta p=1$, $P_{\rm in}=1$; thick lines: Eqs.~\rqn{359}; thin
lines: the exact solution \rqn{180}; dotted lines: the simplified
Eqs.~\rqn{360}.
The figure shows the regime of narrow resonance obtainable for $|\bar{p}|\gg1$.
}\label{f5}\end{figure}

\begin{figure}[htb]
\centerline{\includegraphics[width=15cm]{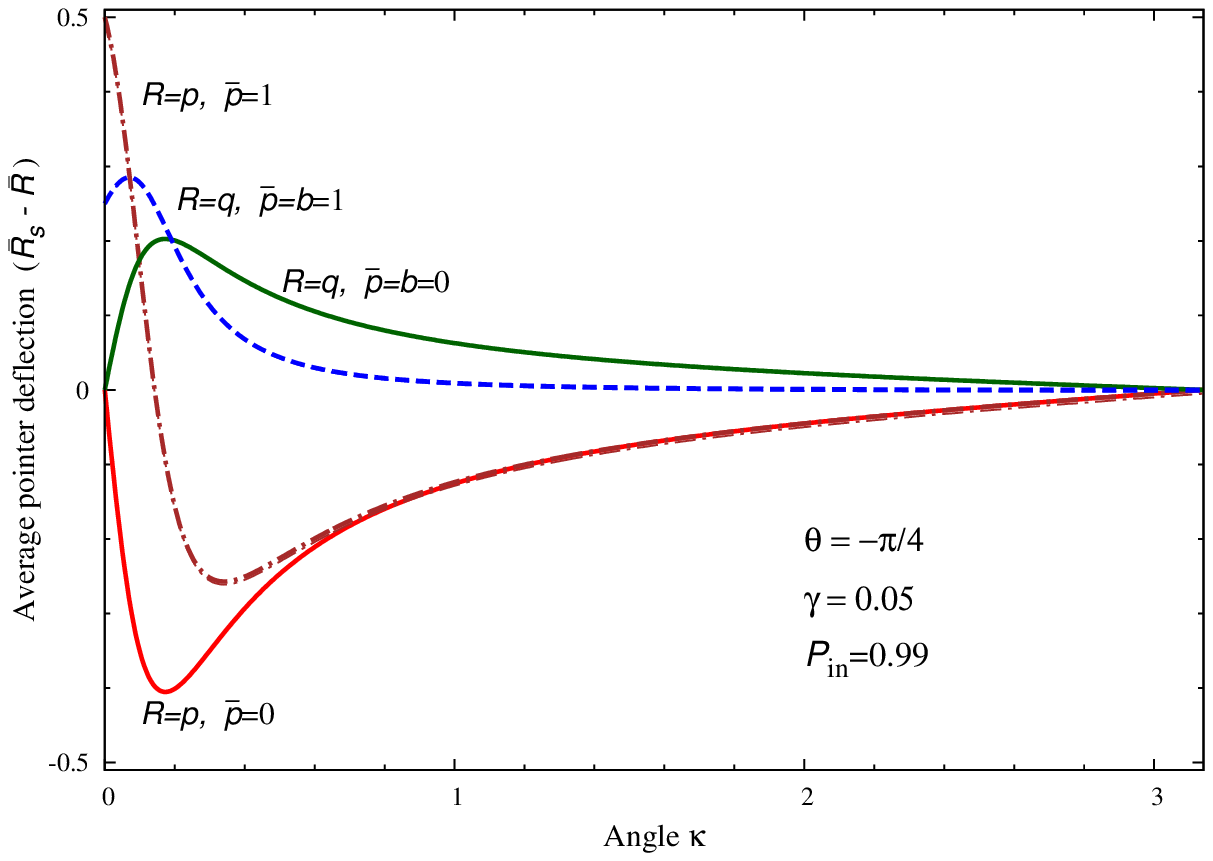}}\caption{(color online). The average pointer deflection
$(\bar{R}_s-\bar{R})$ versus the angle $\kappa$ in Eq.~\rqn{227}, for
a mixed initial state.
Here $\Delta p=1$; thick curves: Eqs.~\rqn{359} with the substitution
\rqn{57a}; thin lines: the exact solution \rqn{180}.
The thin curves are not seen since they practically coincide with the thick lines.
The four cases are plotted with the same line styles as in Fig.~\ref{f1}.
The figure shows that adding a small impurity in the preselected state  can result in a significant decrease of the maximum magnitude of the average pointer deflection.
}\label{f7}\end{figure}

For the measured qubit we take $\hat{A}=\sigma_x$, $|\psi\rangle=|\vec{n}\rangle$, Eq.~\rqn{213}, and
$|\phi\rangle=|-\vec{z}\rangle$.
Correspondingly, in Figs.~\ref{f1}-\ref{f5}, $A_w$ is given by
Eq.~\rqn{227}, whereas in Fig.~\ref{f7}, $A_w$ and $A_w^{(1,1)}$ are
given by Eq.~\rqn{226}.
Equations \rqn{226} and \rqn{227} imply that now the weak-value argument is given by $\theta=-\nu$.
Note that $|A_{\phi\psi}|\le1$, and hence the validity condition
\rqn{12'} of Eq.~\rqn{13} certainly holds for
 \be
|\gamma|\ll(1+|\bar{p}|)^{-1}.
 \e{229}

To get a better understanding of weak PPS measurements, below we show
plots of our general formulas \rqn{13}, \rqn{169}, and \rqn{61}
versus various parameters.
The values of the meter parameters in Eqs.~\rqn{13}, \rqn{169}, and
\rqn{61} are given by configuration 1 in Table \ref{t3} and configuration 2 in Table \ref{t1} for meters 1 and 2, respectively.
In particular, for meters 1 and 2, respectively, Eq.~\rqn{13} can be rewritten as
\bes{359}
 \bea
&&\bar{p}_s-\bar{p}\:=\:\frac{2\gamma\,(\Delta p)^2\,({\rm Im}\,A_w
+\gamma\bar{p}\,|A_w|^2)}{1+2\gamma\bar{p}\,{\rm Im}\,A_w
+\gamma^2(1+\bar{p}^2)\,|A_w|^2},\label{359a}\\
&&\bar{q}_s-\bar{q}\:=\:\frac{\gamma\,{\rm Re}\,A_w+\gamma b\,{\rm Im}\,A_w+\gamma^2b\bar{p}\,|A_w|^2}{1+2\gamma\bar{p}\,{\rm Im}\,A_w
+\gamma^2(1+\bar{p}^2)\,|A_w|^2}.
 \ea{359b}
\ese
Equations \rqn{359} are valid for a pure preselected state.
They hold also for a mixed preselected state when the substitution \rqn{57a} is performed.
In Figs.~\ref{f1}-\ref{f7} the plots of Eqs.~\rqn{359} are shown by
thick lines.
These results are verified against the exact solution specified for a Gaussian meter state; this solution is plotted by thin lines.
The exact solution is given by Eqs.~\rqn{180}, with the account of Eqs.~\rqn{198}, \rqn{202}, and \rqn{204'} for $R=q$ or Eqs.~\rqn{201}, \rqn{225}, \rqn{204'}, and \rqn{204} for $R=p$.

Figure \ref{f1} presents the average pointer deflection
$(\bar{R}_s-\bar{R})$ versus the coupling strength for different
values of $\bar{p}$ and $b$.
One can see that Eqs.~\rqn{359} approximate the exact solutions very
well when the condition \rqn{229} holds (i.e., now for
$|\gamma|\ll1$).
 In this interval, Fig.~\ref{f1} shows a variety of lineshapes,
which, as discussed above, include Lorentzian-like,
dispersive-like, as well as similar, though more complicated,
lineshapes.
The main features of the curves agree with the analysis in
Sec.~\ref{III'}.
Namely, when $\gamma$ is very close to zero, the dependence is
described by the linear Eq.~\rqn{27} (for $R=p$) or \rqn{159}
(for $R=q$).
The exception is the case $R=q,\ \bar{p}=b=1$ (the blue dashed lines),when the linear response vanishes; this plot shows clearly that weak PPS measurements are possible even in the absence of a linear
response.
With a further increase of $|\gamma|$, the quantity
$|\bar{R}_s-\bar{R}|$ increases to a value of the order of the
maximum, i.e., of order $\Delta R$ [see Eq.~\rqn{238}], for
 \be
|\gamma A_w|\sim(1+|\bar{p}|)^{-1}
 \e{357}
[cf.\ Eq.~\rqn{132}], i.e., now for $|\gamma|\sim0.05/(1+|\bar{p}|)$.
 Note that in the present case 
\bes{10.1}
 \bea
&&\Delta R=\Delta p=1\quad\quad\text{for}\ \ \ R=p,\label{10.1a}\\
&&\Delta R=\Delta q=\frac{\sqrt{1+b^2}}{2}\quad\quad\text{for}\ \ \ R=q,
 \ea{10.1b}
\ese
where Eq.~\rqn{10.1b} follows from Eq.~\rqn{49}.

Figure \ref{f2} shows the dependence of the average pointer deflection $(\bar{R}_s-\bar{R})$ on the phase of the weak value for
$\gamma=0.05$, the other parameters being as in Fig.~\ref{f1}.
Now the coupling strength $\gamma$ satisfies the condition \rqn{229};
correspondingly, there is no discernable difference between the exact
and approximate formulas.
Moreover, now we set $|\gamma A_w|=1$ to satisfy the condition
\rqn{357}; hence Fig.~\ref{f2} corresponds to a significantly
nonlinear regime.
Therefore, the maximum values of $|\bar{R}_s-\bar{R}|$ in
Fig.~\ref{f2} are of order $\Delta R$.
For curves with $\bar{p}=0$, the $\theta$ dependence is sinusoidal,
and the maxima and zeros of $|\bar{R}_s-\bar{R}|$ occur at the same
values of $\theta$ as for the linear response, Eqs.~\rqn{356} and
\rqn{358}, since the last term in the numerator and the second term in
the denominator of Eqs.~\rqn{359} disappear now.
Note that for meter 1, $\theta_0=0$, whereas for meter 2,
$\theta_0=\pi/2$ for $\bar{p}=0$ and $\theta_0=\pi/4$ for $\bar{p}=1$.
For curves with $\bar{p}\ne0$, the $\theta$ dependence is not
sinusoidal, and the positions of the maxima and zeros of
$|\bar{R}_s-\bar{R}|$ differ from those for the linear response,
since then, in contrast to the case $\bar{p}=0$, generally all the
terms in Eqs.~\rqn{359} are nonzero.

Figure \ref{f3} demonstrates the dependence of $(\bar{R}_s-\bar{R})$
on the angle $\kappa$ which determines $|A_w|=\cot(\kappa/2)$, see
Eq.~\rqn{227}.
 Actually for $\kappa\ll2$, Fig.~\ref{f3} shows the dependence on the quantity $2/|A_w|\ (\approx\kappa$).
In Fig.~\ref{f3}, the exact and approximate formulas practically
coincide.
 Figure \ref{f3} illustrates the fact that $|\bar{R}_s-\bar{R}|$
becomes significant, i.e., of order $\Delta R$, when $A_w$ is
sufficiently large to satisfy Eq.~\rqn{357}.
The latter condition is now equivalent to $\kappa\sim0.1(1+|\bar{p}|)$.

In Figs.~\ref{f1}-\ref{f3} above we illustrated the behavior of the average pointer deflection $(\bar{R}_s-\bar{R})$ for the case
$\bar{F}\lesssim\Delta F$ ($\bar{p}\lesssim1$ now).
 Consider now the case $\bar{F}\gg\Delta F$ (i.e., $\bar{p}\gg1$
in this section), which, as mentioned in Sec.~\ref{III'}, is quite
different from the case $\bar{F}\lesssim\Delta F$.
 In this case $(\bar{R}_s-\bar{R})$ is significant only when
the parameters $x$ and $\epsilon$ in Eqs.~\rqn{174}-\rqn{218}
are small.
Thus, as shown by Figs.~\ref{f4} and \ref{f5}, $(\bar{R}_s-\bar{R})$
is a narrow resonance as a function of different parameters in the
problem.
This case is approximated by Eq.~\rqn{169}, which for meters 1 and 2
becomes, respectively,
\bes{360}
 \bea
&&\bar{p}_s-\bar{p}\:=\:\frac{-2(\Delta p)^2x}
{\bar{p}\,(x^2+\epsilon^2 +\bar{p}^{-2})},\label{360a}\\
&&\bar{q}_s-\bar{q}\:=\:\frac{-\epsilon-bx}
{\bar{p}\,(x^2+\epsilon^2 +\bar{p}^{-2})}.
 \ea{360b}
\ese
Figures \ref{f4} and \ref{f5} show that the simple Eqs.~\rqn{360}
describe well the resonance in the pointer deflection.
In Fig.~\ref{f4}, $\epsilon=0$, whereas the varied parameter $\gamma$
is linearly related to $x$.
In contrast, in Fig.~\ref{f5}, $x$ is negligibly small, whereas
$\theta$ is linearly related to $\epsilon$.
In this case, as shown by Eq.~\rqn{360b}, $(\bar{q}_s-\bar{q})$ is
practically independent of $b$; correspondingly, the plots for $b=0$
and $b=1$ practically coincide in Fig.~\ref{f5}.
One can see that $(\bar{R}_s-\bar{R})$ is almost zero for the case
$R=q,\ b=0$, when Re$\,A_w=\epsilon=0$ (Fig.~\ref{f4}), and for $R=p$
when $x$ practically vanishes (Fig.~\ref{f5}).
The reason for this is clear from the simplified Eqs.~\rqn{360b} and
\rqn{360a}, respectively.

 Note that there is a slight discrepancy between Eqs.~\rqn{359a} and
\rqn{180} for $R=p$ when $\theta\approx-\pi/2$ (see the thick and
thin red solid lines in Fig.~\ref{f5}).
 This is explained by the fact that in this case the two terms in
the numerator of Eq.~\rqn{359a} practically cancel, so the
higher-order terms neglected in Eq.~\rqn{13} become noticeable.
 However, this discrepancy is rather insignificant, since
it occurs for the not very interesting case of a small
$(\bar{R}_s-\bar{R})$ [see the remark after Eq.~\rqn{13}].

The effect of a mixed initial state is illustrated in Fig.~\ref{f7},
where $P_{\rm in}=0.99$.
 A comparison of Figs.~\ref{f3} and \ref{f7} shows that even a small
impurity of the initial state can significantly decrease the maximum
magnitude of the average pointer deflection; see also the discussion in Sec.~\ref{IXD}.

\section{Discussion of two recent interferometric experiments}
\label{XI}

Starling et al. \cite{sta10x} experimentally demonstrated an optical-phase measurement technique based on phase amplification.
Similar sensitivity to balanced homodyne detection was obtained.
In Ref.~\cite{sta10x}, the experiment was explained on the basis of classical wave optics.
A similar experiment was performed also in Ref.~\cite{dix09}, but there the explanation was given in terms of the weak value.
Modified versions of the experiment in Ref.~\cite{dix09} were performed in Refs.~\cite{tur11,hog11} in connection with potential metrology applications.
The experiment in Ref.~\cite{dix09} was analyzed in Refs.~\cite{sta09,how10}, and a nonlinear version of this experiment was discussed in Ref.~\cite{koi11}.

Below we provide a unified description of both experiments \cite{dix09,sta10x} on the basis of the present nonperturbative theory of weak PPS measurements.
In particular, we show that the results of Refs.~\cite{dix09,sta10x} are described by two opposite limits of the same nonlinear formula.
The correspondence between the notation here and in Refs.~\cite{dix09,sta10x} is presented in Table \ref{t5}.

\begin{table}[tb]
\begin{center}
\begin{tabular}{c|cccc}
\hline
 &$\gamma$&$q$&$\Delta q$&$\varphi$\bigstrut\\
\hline\\[-1.8ex]
\protect\cite{dix09}&$k$&$x$&$a$&$-\phi$\\
\protect\cite{sta10x}&$k$&$x$&$\sigma$&$\phi$\bigstrut[t]\\[.5ex]
\hline
\end{tabular}
\end{center}
\caption{The correspondence between the notation used here and the one used in Refs.~\cite{dix09,sta10x}.}
 \label{t5}\end{table}

\begin{figure}[htb]
\centerline{\includegraphics[width=8cm]{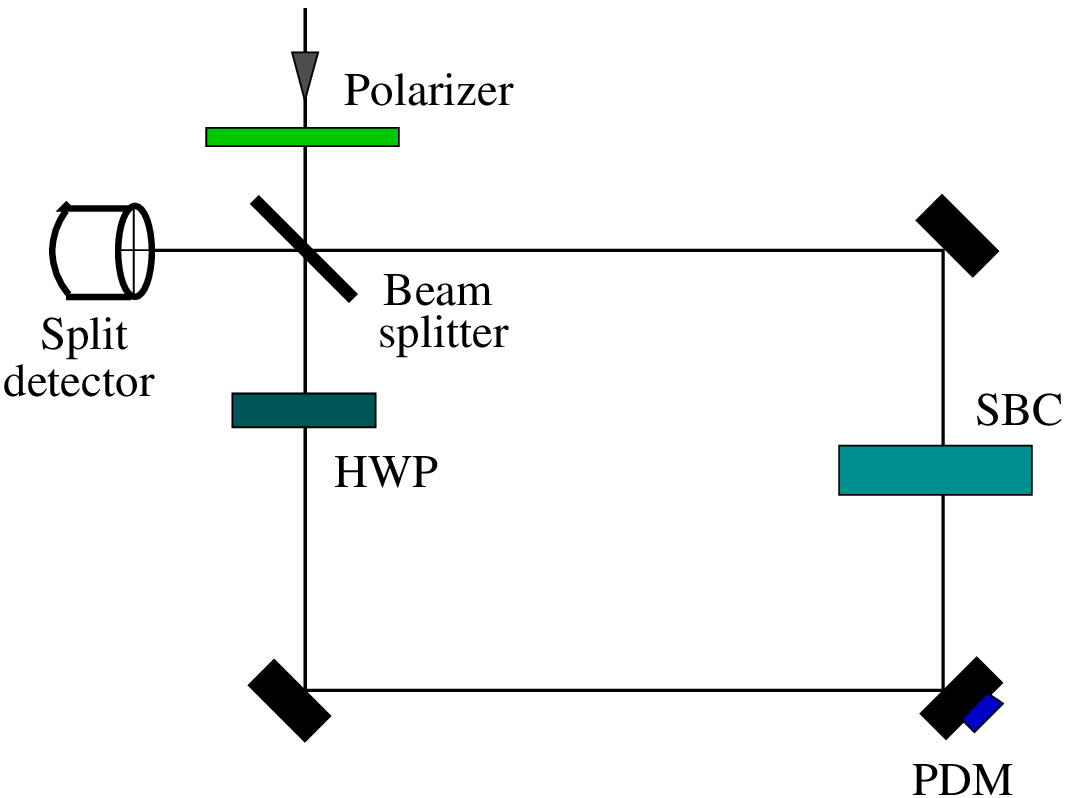}}
\caption{(color online). Schematic diagram of the Sagnac interferometer for weak PPS measurements used in Refs.~\protect\cite{dix09,sta10x}.
Here HWP is a half-wave plate, SBC is a Soleil-Babinet compensator, and PDM is a piezo-driven mirror.}
 \label{f10}\end{figure}

\subsection{Unified theory of two interferometric measurement schemes}

We begin with a brief description of the schematic of the weak PPS measurement in Refs.~\cite{dix09,sta10x} shown in Fig.~\ref{f10} (the details can be found in Refs.~\cite{dix09,sta10x}).
A photon enters a Sagnac interferometer, composed of a 50/50 beam splitter and mirrors, and eventually exits the same beam splitter.
The measured quantum system consists of the clockwise and counterclockwise paths of a photon in the interferometer, denoted by $|1\rangle$ and $|2\rangle$, respectively.
The system is coupled to the meter (a transverse degree of freedom of the photon) by a controlled tilt given to the piezo-driven mirror (PDM), resulting in the transverse-momentum shifts $\gamma$ and $-\gamma$ of the photon in the paths $|1\rangle$ and $|2\rangle$, respectively.
The coupling unitary operator is given \cite{dix09} by Eq.~\rqn{2}, where $\hat{F}=q$ is the transverse coordinate and
\be
\hat{A}=|1\rangle\langle1|-|2\rangle\langle2|.
\e{11.1}

Moreover, due to the polarizer and the half-wave plate (HWP), the photon passes the Soleil-Babinet compensator (SBC) in the polarization state depending on the path.
As a result, after passing the SBC, the photon acquires different phases $\varphi_1$ and $\varphi_2$ in the paths $|1\rangle$ and $|2\rangle$, respectively, so that a relative phase
\be
\varphi=\varphi_1-\varphi_2
\e{11.7}
is introduced between the paths.
In the clockwise (counterclockwise) path the photon passes the SBC before (after) the PDM, therefore the photon state before the exit of the photon from the interferometer is
\be
U_2UU_1|\psi_0\rangle\psi_{\rm M}(q)
\:=\:UU_2U_1|\psi_0\rangle\psi_{\rm M}(q).
\e{11.6}
Here $|\psi_0\rangle$ is the photon state immediately after the photon enters the interferometer, 
\be
|\psi_0\rangle=\frac{1}{\sqrt{2}}(|1\rangle+i|2\rangle),\quad\quad
U_j=\exp(i\varphi_j|j\rangle\langle j|).
\e{11.8}
In Eq.~\rqn{11.6} we changed the order of the operators $U_2$ and $U$, since $U_2$ commutes with $\hat{A}$ and hence with $U$.
Equation \rqn{11.6} implies that the effective preselected state is \cite{dix09}
\be
|\psi\rangle\:=\:U_2U_1|\psi_0\rangle\:=\:\frac{1}{\sqrt{2}}(e^{i\varphi}|1\rangle+i|2\rangle).
\e{11.2}
Here the last equality holds with the accuracy to an irrelevant total phase.

The post-selection is performed by detecting (with the split detector) the photons exiting only the ``dark port'' of the beam splitter, the post-selected state being \cite{dix09}
\be
|\phi\rangle=\frac{1}{\sqrt{2}}(|1\rangle-i|2\rangle).
\e{11.3}
Equations \rqn{11.1}, \rqn{11.2}, and \rqn{11.3} imply that
\be
\langle\phi|\psi\rangle\:=\:\frac{e^{i\varphi}-1}{2}\:
\approx\:\frac{i\varphi}{2},\quad\quad
A_{\phi\psi}\:=\:\frac{e^{i\varphi}+1}{2}\:\approx\:1.
\e{11.9}
Here and below the approximations hold for $|\varphi|\ll1$.
Using Eq.~\rqn{11.9} in Eq.~\rqn{6} yields the weak value \cite{dix09}
\be
A_w\:=\:-i\cot\left(\frac{\varphi}{2}\right)\:\approx\:-\frac{2i}{\varphi}.
\e{11.4}

The split detector measures the average transverse coordinate of the photon, which means that $R=F=q$.
Moreover, in Refs.~\cite{dix09,sta10x} the initial meter state is a Gaussian with $\bar{q}=0$, whereas the beam divergence is negligible, i.e., the meter Hamiltonian is zero.
Hence, from Eq.~\rqn{13} and line 1 in Table \ref{t3} we obtain that
 \be
\bar{q}_s\:=\:\frac{2\gamma\,(\Delta q)^2\,{\rm Im}\,A_w} {1+\gamma^2\,(\Delta q)^2\,|A_w|^2}
\:=-\:\frac{2\gamma\,(\Delta q)^2\tan(\varphi/2)} {\tan^2(\varphi/2)+\gamma^2\,(\Delta q)^2}
\:\approx\:-\frac{4\gamma\,(\Delta q)^2\varphi} {\varphi^2+4\gamma^2\,(\Delta q)^2}.
 \e{11.5}
The last expression in Eq.~\rqn{11.5} as a function of $\gamma$ or $\varphi$ has the dispersion shape, with the maximum and minimum given by
 \be
\bar{q}_s\:=\:\pm\Delta q\quad\quad{\rm at}\ \ \varphi=\mp2\gamma\,(\Delta q).
 \e{11.28}
The formula \rqn{11.5} holds under the condition \rqn{12a}, which now becomes
 \be
|\gamma|\,\Delta q\ll1.
 \e{11.26}

The formula \rqn{11.5} simplifies in two limits. 
In the linear-response regime,
 \be
|\gamma A_w|\,\Delta q\ll1,\quad{\rm i.e.,}\quad
2|\gamma|\,(\Delta q)\ll|\varphi|,
 \e{11.12}
Eq.~\rqn{11.5} yields (cf.\ \cite{dix09}, Eq.~(5))
 \be
\bar{q}_s\:=\:2\gamma\,(\Delta q)^2\,{\rm Im}\,A_w
\:=\:-2\gamma\,(\Delta q)^2\cot\left(\frac{\varphi}{2}\right)\:\approx\:-\frac{4\gamma\,(\Delta q)^2}{\varphi},
\e{11.10}
whereas in the inverted region,
 \be
|\gamma A_w|\,\Delta q\gg1,\quad{\rm i.e.,}\quad
2|\gamma|\,\Delta q\gg|\varphi|,
 \e{11.13}
Eq.~\rqn{11.5} yields
 \be
\bar{q}_s\:=\:-\frac{2}{\gamma}\,{\rm Im}\,A_w^{-1},
 \e{11.11'}
i.e. (cf.\ \cite{sta10x}, Eq.~(9)),
 \be
\bar{q}_s\:=\:-\frac{2}{\gamma}\tan\left(\frac{\varphi}{2}\right)\:
\approx\:-\frac{\varphi}{\gamma}.
 \e{11.11}

Equations \rqn{11.10} and \rqn{11.11} constitute the central results\footnote{
Dixon et al. \cite{dix09} demonstrated also an enhancement of the pointer deflection due to propagation effects, see also the discussion in Refs.~\cite{how10,koi11}.
These propagation effects are completely analogous to the time evolution due to a meter Hamiltonian \cite{hos08}.
Hence, they can be explained also with the help of the present theory of the effects of the meter Hamiltonian (Secs.~\ref{IIID'} and \ref{VIB4}).} of Refs.~\cite{dix09} and \cite{sta10x}, respectively.\footnote{
Equation \rqn{11.11'} is absent in Ref.~\cite{sta10x}, since this paper describes a classical theory.
However, a formula, which differs from Eq.~\rqn{11.11'} by the absence of the factor $\gamma^{-1}$, is mentioned in an earlier version of the above paper: Ref.~\cite{sta09ar}, Eq.~(9).}
As shown above, these two experiments were performed in different regimes of weak PPS measurements, described by two limits of our general nonlinear formula, which now has the form \rqn{11.5}.

It is of interest also to consider the distribution of the pointer values.
It is given by Eq.~\rqn{362} taking into account Eqs.~\rqn{161} and \rqn{11.4},
 \be
\Phi_s(q)\:=\:\frac{[\gamma q-\tan(\varphi/2)]^2\,\Phi(q)}
{\tan^2(\varphi/2)+\gamma^2\,(\Delta q)^2}.
 \e{11.14}
In particular, in the limit \rqn{11.13}
 \be
\Phi_s(q)\:\approx\:\left[\frac{q}{\Delta q}
-\frac{\tan(\varphi/2)}{\gamma\Delta q}\right]^2\,\Phi(q).
 \e{11.15}
For a Gaussian $\Phi(q)$, Eq.~\rqn{11.15} coincides, up to a constant factor, with the intensity at the dark port (i.e., the unnormalized pointer distribution) given by Eq. (7) in Ref.~\cite{sta10x} (cf.\ Table \ref{t5}).
In Ref.~\cite{sta10x} the light with the intensity distribution \rqn{11.15} is called the ``split-Gaussian mode'', since this distribution has a slightly asymmetric two-peak shape, shown in Fig.~2 in Ref.~\cite{sta10x}.

Thus, we have derived the results of Ref.~\cite{sta10x} quantum-mechanically.
We have shown that the experiment in Ref.~\cite{sta10x} is a weak PPS measurement in the inverted region.

\subsection{Amplification coefficient for phase measurements}

Starling et al. \cite{sta10x} showed that their experiment is well suited for precision measurements of the phase $\varphi$.
They claimed that their technique involves phase amplification with the coefficient proportional to $\gamma^{-1}$, but they did not provide the exact value of the amplification coefficient.
Let us obtain the latter.
Note that, in view of the first of Eqs.~\rqn{11.9}, 
 \be
|\varphi|=2|\langle\phi|\psi\rangle|.
 \e{11.20}
As follows from the general discussion in Sec.~\ref{IVF4}, in measurements of the overlap the proper amplification coefficient is [cf.\ Eq.~\rqn{3.16}]
 \be
{\cal A}'\,=\,(|\gamma|\,\Delta q)^{-1},
 \e{11.16}
where we took into account the second of Eqs.~\rqn{11.9}.
The amplification coefficient ${\cal A}_\varphi$ for the phase $\varphi$ is related to ${\cal A}'$ by the relation ${\cal A}'|\langle\phi|\psi\rangle|={\cal A}_\varphi|\varphi|$, i.e., in view of Eq.~\rqn{11.20},
 \be
{\cal A}_\varphi=\frac{{\cal A}'}{2}\,=\,(2|\gamma|\,\Delta q)^{-1}.
 \e{11.17}

This can be compared with the linear-response result in Eq.~\rqn{11.10}, which was used in Ref.~\cite{dix09} for an amplification of the pointer deflection (i.e., the beam deflection) or for measuring $\gamma$, the proper amplification coefficient being [cf.\ Eqs.~\rqn{301} and \rqn{11.20}]
 \be
{\cal A}\,\sim\,|\varphi|^{-1}.
 \e{11.21}
In the case \rqn{11.12} [\rqn{11.13}] the magnitude of the pointer deflection \rqn{11.10} [\rqn{11.11}] increases, when the ratio $\gamma/\varphi$ increases (decreases).
For both cases, the amplification is maximal in the strongly-nonlinear regime,
 \be
|\gamma A_w|\,\Delta q\sim1,\quad{\rm i.e.,}\quad
2|\gamma|\,\Delta q\sim|\varphi|,
 \e{11.18}
where, in view of Eqs.~\rqn{11.17} and \rqn{11.21}, we have
 \be
{\cal A}_\varphi\,\sim\,{\cal A}\,\sim\,|\varphi|^{-1}.
 \e{11.19}
In this regime, the nonlinear Eq.~\rqn{11.5} should be used.

\subsection{A comparison of the phase-amplification technique with projective measurements}
\label{XIC}

In Sec.~\ref{IIIK3}, we obtained an estimation of the SNR with respect to the quantum noise for weak PPS measurements in the regime of very large weak values [see Eq.~\rqn{3.21}].
The above estimation holds for the general case.
However, for the special case of the phase-amplification technique \cite{sta10x} (i.e., the weak PPS measurement described above), the quantum SNR can be obtained exactly.
Indeed, in view of Eq.~\rqn{4.2} and the second equation in Eq.~\rqn{11.9}, we obtain that now the post-selection probability is given by the value \cite{sta10x}
 \be
\langle\Pi_\phi\rangle_f\,=\,\gamma^2(\Delta q)^2\ll1,
 \e{11.25}
which is small due to Eq.~\rqn{11.26}.
Taking into account also that the pointer deflection is given by Eq.~\rqn{11.11}, we obtain from Eq.~\rqn{6.28} that the quantum SNR is given by
 \be
{\cal R}\:=\:3^{-1/4}|\varphi|\sqrt{N}\:\approx\:0.76\,|\varphi|\sqrt{N}.
 \e{11.23}
Here we took into account that now $R=q$ and that Eq.~\rqn{11.15} with a Gaussian $\Phi(q)$ implies the pointer uncertainty $\Delta q_s\approx3^{1/4}\Delta q$.
[The latter result follows from the fact, which can be easily shown, that one can approximately neglect the small second term in the brackets in Eq.~\rqn{11.15}.]

Equation \rqn{11.23} is derived assuming that the average position is obtained by a statistical analysis of the measurement results.
For comparison, in the split-detection method the average position is deduced from the difference between the integrated intensities on the left and right sides of the detector, resulting in a somewhat higher SNR than Eq.~\rqn{11.23} \cite{sta10x},
 \be
{\cal R}=\sqrt{\frac{2}{\pi}}\,|\varphi|\sqrt{N}\approx0.80\,|\varphi|\sqrt{N}.
 \e{11.27}

The phase $\varphi$ can be measured also with the help of strong (projective) quantum measurements.
The projective measurement described in Sec.~\ref{IIIK3} can be implemented in the present case by setting $\gamma=0$ and measuring the statistics of photons exiting the dark and bright output ports of the beam splitter in Fig.~\ref{f10}, since the exit probabilities equal $P_1=|\langle\phi|\psi\rangle|^2\approx\varphi^2/4$ and $P_0=1-P_1$, respectively.
As follows from Eqs.~\rqn{3.22}, for this method, the SNR with respect to the quantum noise is 
 \be
{\cal R}_1=\frac{|\varphi|\sqrt{N}}{2}.
 \e{11.22}

In fact, in Ref.~\cite{sta10x} a more sophisticated version of strong measurements was implemented, the so called balanced homodyne detection.
In this scheme a unitary transformation of $|\psi\rangle$ is performed, so that $\varphi\rightarrow\pi/2+\varphi$ in Eq.~\rqn{11.2}, and then the integrated intensities of both output ports are subtracted, resulting in the homodyne signal per one photon $\sin\varphi\approx\varphi$.
As shown in Ref.~\cite{sta10x}, the quantum SNR for the balanced homodyne detection is 
 \be
{\cal R}_2=|\varphi|\sqrt{N}, 
 \e{11.24}
two times greater than for the above simple scheme of projective measurements [cf.\ Eq.~\rqn{11.22}].

A comparison of Eqs.~\rqn{11.23} and \rqn{11.27}, on one hand, with Eq.~\rqn{11.24}, on the other hand, shows that the phase-amplification technique has similar sensitivity to balanced homodyne detection with respect to quantum noise \cite{sta10x}. 
This is in agreement with the discussion in Sec.~\ref{IIIK3}, where the quantum SNR was shown to be generally of the same order for projective and weak PPS measurements. 

As noted in Ref.~\cite{sta10x}, the phase-amplification technique is a robust, low-cost alternative to balanced homodyne phase detection, since one can use a low-cost split detector with a low saturation intensity, owing to the large attenuation [cf.\ Eq.~\rqn{11.25}].
In this case, an increase of the attenuation does not decrease the quantum SNR \rqn{11.27} due to the simultaneous increase of the amplification coefficient \rqn{11.17}.

\section{PPS measurements with a general post-selection measurement}
 \label{XII}

\subsection{General formulas}
\label{XIIIA}

Until now (except for some remarks in Sec.~\ref{IC1}), we considered PPS measurements in which the post-selection is performed by a measurement projecting the system state on a discrete, nondegenerate eigenstate of some variable $B$.
In this section, we discuss the general case, where the post-selection is performed by means of a general measurement described by an arbitrary POVM.
This case includes different possible situations, such as, e.g., a projection on a degenerate eigenvalue of $B$.
Another situation, where this case may be relevant, arises when one takes into account errors in the post-selection measurement.
Indeed, in the presence of measurement errors, a projective measurement can be described as a general measurement characterized by a POVM \cite{bra92,hel76,kof08} rather than by a projection-valued measure.

In the general case, a PPS ensemble consists of systems for which the post-selection measurement yields a specific outcome.
The POVM operator corresponding to this measurement outcome is denoted here by $E$.
By repeating the derivation in Sec.~\ref{IIIA1} with the change 
 \be
\Pi_\phi\ \rightarrow\ E,
 \e{12.1}
it is easy to obtain that in the general case [cf.\ Eqs.~\rqn{241} and \rqn{44}]
 \be
\bar{R}_s\:=\:\frac{\langle ER\rangle_f}{\langle E\rangle_f}
 \e{12.2}
and
 \be
\bar{R}_s-\bar{R}\:=\:\frac{\langle ER_c\rangle_f}{\langle E\rangle_f},
 \e{12.3}
where $\langle E\rangle_f$ is the post-selection probability and the averages are given by [cf.\ Eqs.~\rqn{112} and \rqn{3.4}]
 \be
\langle ER\rangle_f={\rm Tr}\,[(E\otimes\hat{R})\,\rho_f],\quad\quad
\langle E\rangle_f={\rm Tr}\,[(E\otimes I_{\rm M})\,\rho_f].
 \e{12.4}

\subsection{Relation between PPS and standard measurements of any strength}
\label{XIIIB}

Equation \rqn{12.2} allows us to connect PPS and standard quantum measurements of arbitrary strength.
In the limiting case when $E$ is just the unity operator $I_{\rm S}$, we obtain that $\langle E\rangle_f=\langle I_{\rm S}\rangle_f=1$ and hence Eq.~\rqn{12.2} reduces to Eq.~\rqn{74}.
Thus, in this case PPS measurements coincide with standard measurements for any measurement strength and any preselected state.
This statement extends the similar results obtained for the special cases of strong and weak measurements with a pure preselected state in Ref.~\cite{aha02} (cf.\ Sec.~\ref{IC1}).

The above statement is a limiting case of a more general relation between PPS and standard quantum measurements of any strength, as follows.
If $\hat{A}$ commutes with {\em either $\rho$ or} $E$, then an arbitrary-strength PPS measurement of $A$ provides formally the same results as a standard measurement with the preselected state described by the ``density matrix''
 \be
\rho'\:=\:\frac{E\rho+\rho E}{2\,{\rm Tr}\,(E\rho)}.
 \e{13.1}
To prove this statement, we use Eqs.~\rqn{12.4} and \rqn{3.2} to write
 \be
\langle ER\rangle_f\:=\:{\rm Tr}\,[(E\otimes\hat{R})\,U\,(\rho\otimes \rho_{\rm M})\,U^\dagger]
\:=\:\frac{1}{2}{\rm Tr}\,[(I_s\otimes\hat{R})\,U\,((E\rho+\rho E)\otimes \rho_{\rm M})\,U^\dagger],
 \e{13.2}
 \be
\langle E\rangle_f\:=\:{\rm Tr}\,[(E\otimes I_{\rm M})\,U\,(\rho\otimes \rho_{\rm M})\,U^\dagger]
\:=\:{\rm Tr}\,[U\,(E\rho\otimes \rho_{\rm M})\,U^\dagger]
\:=\:{\rm Tr}\,(E\rho\otimes \rho_{\rm M})
\:=\:{\rm Tr}\,(E\rho)\,{\rm Tr}\,\rho_{\rm M}\:=\:{\rm Tr}\,(E\rho).
 \e{13.3}
In the second equalities in Eqs.~\rqn{13.2} and \rqn{13.3} we used Eqs.~\rqn{B7'''} and \rqn{B7'}, respectively.
Inserting Eqs.~\rqn{13.2} and \rqn{13.3} into Eq.~\rqn{12.2} yields Eq.~\rqn{74} with $\rho$ given by Eq.~\rqn{13.1}.
This proves the statement in question.

When ${\rm Tr}\,(E\rho)=0$, $\rho'$ in Eq.~\rqn{13.1} does not exist.
However, in this case PPS measurements are not possible, since the post-selection probability \rqn{13.3} is zero.

Consider some consequences of the above equivalence of PPS and standard measurements of any strength.
For strong PPS measurements, the probabilities in Eq.~\rqn{81'} are given now by
 \be
P_{i|E}\,=\,{\rm Tr}\,(\Pi_i\,\rho').
 \e{14.3}
Indeed, it is easy to check that Eqs.~\rqn{81'} and \rqn{14.3} are equivalent now.
Correspondingly, the average of $A$ in strong PPS measurements given by Eq.~\rqn{142} or, more generally, by the formula
 \be
A_s=\sum_ia_i\,P_{i|E}
 \e{14.4}
coincides now with the average of $A$ resulting from the equivalent strong standard measurements: $A_s=\bar{A}'$, where
 \be
\bar{A}'={\rm Tr}\,(\hat{A}\rho').
 \e{14.1}
This equality results directly on inserting Eq.~\rqn{14.3} into Eq.~\rqn{14.4}.
Moreover, the equivalence of {\em weak} PPS and standard measurements 
means that $A_w=\bar{A}'$.
This equality is derived independently also in Sec.~\ref{XIIIC}.
The above results imply that now
 \be
A_w=A_s.
 \e{14.5}

It should be noted that $\rho'$ in Eq.~\rqn{13.1} is Hermitian and ${\rm Tr}\,\rho'=1$, however generally $\rho'$ is not a positive operator.
When $\rho'$ is a positive operator, both PPS and standard measurements can be performed.
However, when $\rho'$ has, at least, one negative eigenvalue, then $\rho'$ does not correspond to a physical state; in this case a standard measurements cannot be performed, and the above equivalence is formal.
For example, when $E=|\phi\rangle\langle\phi|$ and $\rho=|\psi\rangle\langle\psi|$, whereas $|\phi\rangle\ne|\psi\rangle$ and $\langle\phi|\psi\rangle\ne0$, then $\rho'$ can be shown to have, at least, one negative eigenvalue.
Still, now the results of measurements are usual values, irrespective of whether the operator $\rho'$ is positive or not.
In particular, $\bar{A}'$ in Eq.~\rqn{14.1} is {\em a usual value of} $A$.
This follows from the above result $\bar{A}'=A_s$ and the fact that $A_s$ is a usual value of $A$, since $A_s$ in Eq.~\rqn{14.4} is an average of $A$ over a classical (i.e., positive) probability distribution $P_{i|E}$.

Note that the quantity $\rho'$ in Eq.~\rqn{13.1} was considered previously in a different context in Ref.~\cite{hof10}, where it was called the ``transient density matrix''.
There $\rho'$ was discussed in connection with weak PPS measurements in the linear-response regime.

Now let us apply the above relation between PPS and standard measurements to prove the time-symmetry property, mentioned in Sec.~\ref{IE2}, that measurements in a preselected (only) ensemble and a post-selected (only) ensemble with the same pre- or post-selected state, respectively, produce the same results, irrespective of the measurement strength.
Consider measurements in a post-selected ensemble, i.e., an ensemble with the completely mixed preselected state $\rho_{\rm c.m.}$, Eq.~\rqn{1.10}, and a post-selected state $|\phi\rangle$.
Since $\rho_{\rm c.m.}$ commutes with any $\hat{A}$, a measurement in a post-selected ensemble is equivalent to a measurement in a preselected ensemble with the preselected state $\rho'$, Eq.~\rqn{13.1}.
Now $E=\Pi_\phi\equiv|\phi\rangle\langle\phi|$, and, in view of Eq.~\rqn{1.10}, Eq.~\rqn{13.1} yields
 \be
\rho'\:=\:\frac{\Pi_\phi\,\rho_{\rm c.m.}+\rho_{\rm c.m.}\Pi_\phi}{2\,{\rm Tr}\,(\Pi_\phi\,\rho_{\rm c.m.})}
\:=\:\frac{2\Pi_\phi/d}{2d^{-1}}\:=\:\Pi_\phi.
 \e{13.5}
This proves the statement in question.

\subsection{Weak PPS measurements}
\label{XIIIC}

Now the expansions in the coupling parameter can be obtained in the form [cf.\ Eqs.~\rqn{3.23}-\rqn{3.34}],
 \bea
&&\langle ER_c\rangle_f\:=\:\sum_{n=1}^\infty\frac{i^n\gamma^n}{n!} \sum_{k=0}^n(-1)^k{n\choose k}\,{\rm Tr}\,(\hat{A}^{n-k}E\hat{A}^k\rho)\, \overline{F^{n-k}R_cF^k},\label{13.15}\\
&&\langle E\rangle_f\:=\:\sum_{n=0}^\infty\frac{i^n\gamma^n\, \overline{F^n}}{n!}\,\sum_{k=0}^n(-1)^k{n\choose k}\,{\rm Tr}\,(\hat{A}^{n-k}E\hat{A}^k\rho),\label{13.16}\\
&&\langle ER\rangle_f\:=\:\sum_{n=0}^\infty\frac{i^n\gamma^n}{n!} \sum_{k=0}^n(-1)^k{n\choose k}\,{\rm Tr}\,(\hat{A}^{n-k}E\hat{A}^k\rho)\, \overline{F^{n-k}RF^k}.
 \ea{13.18}
Correspondingly, the expansions \rqn{18'} hold now with the changes [cf.\ Eq.~\rqn{325t}]
\bes{12.5}
 \bea
&&|\langle\phi|\psi\rangle|^2\ \rightarrow\ {\rm Tr}\,(E\rho),\label{12.5a}\\
&(A^{k})_w (A^{l})_w^*\ \rightarrow& A_w^{(k,l)}\:\equiv\:\frac{{\rm Tr}\,(\hat{A}^lE\hat{A}^k\rho)}{{\rm Tr}\,(E\rho)}\quad\quad (k,l\ge0).
 \ea{12.5b}
\ese

In the linear approximation in $\gamma$, Eqs.~\rqn{12.3}, \rqn{13.15}, and \rqn{13.16} yield the linear-response result \rqn{14''}, where in the present general case the weak value is given by
 \be
A_w\:=\:\frac{{\rm Tr}\,(E\hat{A}\rho)}{{\rm Tr}\,(E\rho)}.
 \e{12.6}
Equation \rqn{12.6} follows also from Eq.~\rqn{12.5b}, on taking into account that $A_w=A_w^{(1,0)}$.
The real part of the weak value \rqn{12.6} was obtained in Ref.~\cite{wis02}.
Note, however, that the real part of $A_w$ is generally not sufficient to describe the linear response \rqn{14''}.
Equation \rqn{12.6} was obtained in Ref.~\cite{joh04} and discussed in Ref.~\cite{wum09}.

It is easy to show that $A_w$ is a usual value, when any two of the operators $\hat{A},\ \rho$, and $E$ commute.
Indeed, then $A_w$ in Eq.~\rqn{12.6} can be written as the average of $A$ in the state \rqn{13.1},
 \be
A_w\:=\:{\rm Tr}\,(\hat{A}\rho').
 \e{13.4}
This situation includes two different cases:\\
(a) When $E$ or $\rho$ commutes with $\hat{A}$, then, as shown in Sec.~\ref{XIIIB}, the results of PPS and standard measurements of any strength coincide, Eq.~\rqn{13.4} being a consequence of this fact for weak measurements.
Moreover, as shown in Sec.~\ref{XIIIB}, $A_w$ in Eq.~\rqn{13.4} is a usual value, though $\rho'$ may have negative eigenvalues.
This case involves paragraphs {\em b, c, f}, and {\em g} in Sec.~\ref{IE2} as its special cases.\\
(b) When $E$ and $\rho$ commute, $\rho'$ in Eq.~\rqn{13.1} becomes $\rho'=E\rho/{\rm Tr}\,(E\rho)$, which is a positive operator, implying that $A_w$ in Eq.~\rqn{13.4} is a usual value.
This case is an extension of paragraphs {\em a} and {\em e} in Sec.~\ref{IE2}.

As shown above, in the nonlinear theory of weak PPS measurements, in addition to the weak value, also the associated weak value $A_w^{(1,1)}$ is generally required.
As follows from Eq.~\rqn{12.5b}, in the general case this is given by
 \be
A_w^{(1,1)}\:=\:\frac{{\rm Tr}\,(\hat{A}E\hat{A}\rho)}{{\rm Tr}\,(E\rho)}.
 \e{12.10}

\subsection{Time-symmetry properties for PPS measurements of any strength}
\label{XIIID}

A time-symmetry relation for measurements in ensembles of some special types was discussed in the last paragraph of Sec.~\ref{XIIIB}.
Here we consider a more general time-symmetry property.

The expansions \rqn{18'} with the changes \rqn{12.5} imply a time-symmetry property for PPS measurements of any strength.
Namely, it is easy to see that the above expansions remain invariant, except for the changes
 \be
A_w^{(k,l)}\ \rightarrow\ (A_w^{(k,l)})^*=A_w^{(l,k)},
 \e{12.12}
under the simultaneous substitutions
 \be
\rho\ \rightarrow\ \frac{E}{{\rm Tr}\,E},\quad\quad E\ \rightarrow\ e_1\rho.
 \e{12.7}
Here $e_1$ is any positive number such that $e_1\rho$ is an allowed POVM operator.
As implied by Eq.~\rqn{1.48}, this means that the maximal eigenvalue of $e_1\rho$ should not exceed one; hence,
 \be
0<e_1\le\lambda_{\rm max}^{-1},
 \e{12.8}
where $\lambda_{\rm max}$ is the maximal eigenvalue of $\rho$.

The quantities which interchange in the time-symmetry relation \rqn{12.7} are, with an accuracy up to numerical factors, the preselected state $\rho$ and the post-selection POVM operator $E$, rather than pre- and post-selected states, as one might expect naively.
The reason for this is that the density matrix $\rho$, on one hand, and the post-selection measurement outcome together with the corresponding POVM operator $E$, on the other hand, {\em provide the complete information about the pre-selection and the post-selection, accessible to an experimenter, and this information completely determines a given PPS ensemble}.

In this connection, we note that the terms ``pre-selection'' and ``post-selection'' are perhaps somewhat confusing.
Using the same word ``selection'' masks the fact that ``pre-selection'' and ``post-selection'' describe in principle different physical processes.
``Pre-selection'' means the process of preparation of the initial state of the quantum system.
Correspondingly, the ``preselected state'' is a well defined notion---it is simply the initial state of the system.

In contrast, ``post-selection'' is conditioning of the measured statistical data on acquired information from the final measurement of the system.
The only information required for the post-selection is the result of the final measurement and the corresponding POVM operator.
Hence, the notion of the ``post-selected state'' is generally meaningless, since the final state of the system is irrelevant.
Indeed, it is irrelevant whether the system is destroyed by the measurement or, if not, in what state it is.
The notion of the ``post-selected state'' has a physical meaning only when the post-selection is performed by a strong measurement corresponding to a nondegenerate eigenvalue of an observable. 
Then, according to the projection postulate, if the measurement is minimally disturbing, i.e., projective, the state of the system after the measurement coincides with the corresponding eigenstate.
A pre- and post-selected ensemble can be characterized by such a pure ``post-selected state'', even when the strong measurement is not minimally disturbing, i.e., when the system is destroyed or its final state differs from the ``post-selected state''.
In all other cases the final state generally depends on the state of the system before the final measurement even for minimally disturbing (e.g., projective) measurements [cf.\ Eq.~\rqn{1.29}].
As a result, the final state generally depends on both the preselected state and the measurement(s) performed in the PPS ensemble in between the pre-selection and the post-selection.
Therefore, identically pre- and post-selected ensembles, which have underwent different measurements, generally have different final states and hence cannot be characterized by a unique ``post-selected state''.
The above discussion implies that {\em the post-selection POVM operator is a more fundamental characteristic of a PPS ensemble than the final state of this ensemble}.

Consider the important case when the pre- and post-selected states are pure. 
(Here, as in many other places in this paper, we use the term ``pure post-selected state'', since it has a physical meaning in the sense discussed in the previous paragraph.)
In this case, we have $\rho=\Pi_\psi\equiv|\psi\rangle\langle\psi|$ and $E=e_0\Pi_\phi\ (0<e_0\le1)$.
When $e_0\ne1$, the POVM operator $E=e_0\Pi_\phi$ is not a projector, but still the post-selected state is $|\phi\rangle$.
Now the time-symmetry relation \rqn{12.7} becomes 
 \be
\rho=\Pi_\psi\ \rightarrow\ \Pi_\phi,\quad\quad E=e_0\Pi_\phi\ \rightarrow\ e_1\Pi_\psi,
 \e{12.15}
where $e_1$ is an arbitrary number satisfying $0<e_1\le1$.
Bearing in mind that the POVM operator corresponding to a pure post-selected state $|\phi\rangle$ is generally proportional to the projector $\Pi_\phi$ with the coefficient which may differ from 1, we can replace Eq.~\rqn{12.15} by a simple relation,
 \be
|\psi\rangle\ \leftrightarrow\ |\phi\rangle.
 \e{12.16}
Thus, in the case of pure pre- and post-selected states, the time-symmetry relation is conceptually simple: Eq.~\rqn{12.12} holds under the exchange of the pre- and post-selected states.
This is an extension of the time-symmetry relation for strong PPS measurements \cite{aha64,aha02,aha05} (see Sec.~\ref{IC2}) to PPS measurements of an arbitrary strength.

Consider the important special cases of strong and weak PPS measurements.
Strong PPS measurements are not affected by the change \rqn{12.7}, as implied by Eq.~\rqn{81'}.
This is an extension of the time-symmetry relation for strong PPS measurements \cite{aha64,aha02,aha05} (see Sec.~\ref{IC2}) from the case of pure pre- and post-selected states to the general case of strong PPS measurements.

For weak PPS measurements the above time-symmetry property \rqn{12.12}-\rqn{12.7} means that the weak value and the associated weak value satisfy the relations [cf.\ Eq.~\rqn{12.12}]
 \be
A_w\ \rightarrow\ A_w^*,\quad\quad A_w^{(1,1)}\:=\:\mbox{invariant}
 \e{12.11}
under the simultaneous substitutions \rqn{12.7}.
Thus, the results of weak PPS measurements generally are changed by the substitutions \rqn{12.7}, unless $A_w$ is real.

Consider a simple example.
When the pre- and post-selected state are pure, Eqs.~\rqn{12.16} and \rqn{12.11} yield the following symmetry relation,
 \be
A_w\ \rightarrow\ A_w^*\quad\quad\mbox{for}\quad|\psi\rangle\ \leftrightarrow\ |\phi\rangle.
 \e{12.9}
This relation also follows from Eq.~\rqn{6}.

\subsection{A pure preselected state}
\label{XIIIE}

Consider an important situation when in a weak PPS measurement the preselected state is pure, $\rho=|\psi\rangle\langle\psi|$, but the post-selection measurement is general.
As mentioned above, such a situation may arise, e.g., when measurement errors are to be taken into account.

Now the weak values \rqn{12.6} and \rqn{12.10} become
 \be
A_w=\frac{(EA)_{\psi\psi}}{E_{\psi\psi}},\quad\quad
A_w^{(1,1)}=\frac{(AEA)_{\psi\psi}}{E_{\psi\psi}}.
 \e{12.13}
The present situation is closely related to the case of a mixed preselected state and a pure post-selected state.
Indeed, the weak values \rqn{56} and \rqn{57a} are connected to the formulas \rqn{12.13} by the relation \rqn{12.11} under the substitutions 
 \be
\rho\ \rightarrow\ \frac{E}{{\rm Tr}\,E},\quad\quad
|\phi\rangle\ \rightarrow\ |\psi\rangle.
 \e{12.14}
Owing to this relation, one can use results of preceding sections in the present case.
Namely, in the present situation the nonlinear equations \rqn{61} and \rqn{183} hold provided the definitions \rqn{12.13} are used.
Moreover, the other results obtained above for the case of a mixed preselected state (see especially Secs.~\ref{V} and \ref{VI} and Fig.~\ref{f7}) are also valid now, provided the substitutions \rqn{12.14} and the definitions \rqn{12.13} are used.

\section{Conclusion}
 \label{X}

Weak pre- and post-selected measurements are important for studies of
the fundamentals of quantum mechanics.
They also hold promise for precision metrology, since they provide
significant amplification of the pointer deflection in comparison to
standard weak measurements.
This paper starts with a brief review of strong and weak PPS measurements (Sec.~\ref{I}).
Afterwards, we present original contributions, which generalize previous theoretical work. 

In particular, a nonperturbative theory of weak PPS measurements is developed.
The theory is applicable to an arbitrary quantum system and an
arbitrary meter, with arbitrary initial states for both of them.
 The results are expressed in simple analytical forms.
We have shown that weak values and the coupling strength can be
measured not only in the linear regime, as was done previously, but
also in two other regimes: the strongly-nonlinear regime and the inverted region (i.e., the limit of very large weak values, where the overlap of the pre- and post-selected states is very small).
We have verified our theory by showing that the optical experiment in
Ref.~\cite{sta10x} can be described quantum-mechanically as a weak PPS
measurement in the regime of large weak values.

Optimal conditions for measurements are obtained in the strongly
nonlinear regime, since there the pointer deflection is generally of
the order of the maximum value.
Correspondingly, under optimal conditions, the amplification is
stronger than in the linear regime by at least an order of
magnitude.
 The nonlinear regime can be achieved only for anomalously large
weak values, which implies the requirement that the overlap of the
pre- and post-selected states is small.
The optimal conditions are obtained when the above overlap is of the
order of the small parameter of the theory.

We have revealed that, in the nonlinear regime, weak PPS measurements
significantly depend on the value of $\bar{F}$.
In particular, a nonzero $\bar{F}$ may facilitate measurements of
weak values (Sec.~\ref{IIIA8d}) and the coupling strength $\gamma$
(Sec.~\ref{VIB3}).
Moreover, the optimal regime of measurements is qualitatively
different for $|\bar{F}|\lesssim\Delta F$ and for $|\bar{F}|\gg\Delta F$.
In the latter case, the optimal conditions are much stricter, but the
amplification is much stronger, than for $|\bar{F}|\lesssim\Delta F$.
This increase of the amplification may result in an increased
measurement precision.
The optimal regime for $|\bar{F}|\gg\Delta F$ is very sensitive to
small perturbations of several parameters; this property can be used
for various precision measurements.
We have indicated experimental schemes where $\bar{F}$ is nonzero and
tunable.

 We have derived exact solutions for PPS measurements of a qubit
with several types of meters and, using these solutions, verified
the present theory by numerical calculations.
The present theory can be verified experimentally in many physical systems, including optical experiments and experiments with various types of qubits (such as qubits in solid state, atoms, NMR, etc.).
Moreover, the present results can be applied to improve the accuracy
of precision measurements.
In particular, the present theory can be applied to existing
experimental setups, such as those in Refs.~\cite{hos08,dix09,how10,
sta09,sta10x,sta10a}, where using the optimal regime can increase the
amplification by, at least, an order of magnitude.

In recent years, research on weak PPS measurements and weak values has been expanding with an increasing rate.
In spite of the initial controversy, weak values have demonstrated to be a fruitful concept both for fundamental studies and for designing novel experimental techniques.
Potential applications of weak values include such diverse topics as optical communications, metrology, and quantum information processing. 
The general theory of weak PPS measurements developed here will provide insights and a useful guide for further applications of weak values.

\addcontentsline{toc}{section}{Acknoledgements}
\section*{Acknowledgements}

 We thank Prof.\ S.\ A.\ Gurvitz for fruitful discussions and especially Dr.\ S.\ Ozdemir for his very careful reading of the manuscript and numerous very useful suggestion​s.
FN is partially supported by the Army Research Office, National Science Foundation under grant No. 0726909, JSPS-RFBR contract No. 12-02-92100, Grant-in-Aid for Scientific Research (S), MEXT Kakenhi on Quantum Cybernetics, and the JSPS via its FIRST program. 

\begin{appendices}

\section{Formula for embedded commutators}
\label{B}

Here we derive a formula for $n$ consecutively embedded commutators,
 \be
\underbrace{[D,\dots[D,}_nC\underbrace{]\dots]}_n= \sum_{k=0}^na_{nk}\,D^{n-k}CD^k,
 \e{B1}
where the coefficients $a_{nk}$ are to be determined.
 It is easy to see that the latter satisfy the recursive formula
$a_{n+1,k}=a_{nk}-a_{n,k-1}$, which by the change
 \be
a_{nk}=(-1)^k\,a_{nk}'
 \e{B5}
becomes
 \be
a_{n+1,k}'=a_{nk}'+a_{n,k-1}'\quad\quad(0\le k\le n+1;\ n\ge1)
 \e{B2}
with $a_{n,-1}'=a_{n,n+1}'=0$.
 As follows from Eq.~\rqn{B1} with $n=1$ and Eq.~\rqn{B5},
 \be
a_{10}'=a_{11}'=1.
 \e{B3}
Equation \rqn{B2} with the initial conditions \rqn{B3} has a unique
solution given by the binomial coefficients \cite{kor68},
$a_{nk}'={n\choose k}$.
 Combining the latter formula with Eqs.~\rqn{B5} and \rqn{B1} yields
finally Eq.~\rqn{B4}.

\section{Generalized uncertainty relation and estimation of moments of meter variables}
\label{D}

In Sec.~\ref{AC1}, we derive the generalized uncertainty relation for a quantum system in a mixed state and prove several inequalities required in Sec.~\ref{AC2}.
In Sec.~\ref{AC2}, we estimate the magnitude of the moments $\overline{F^kR_cF^{n-k}}\ (0\le k\le n)$ for a system in a mixed state [cf.\ Eq.~\rqn{59}]
 \be
\rho_{\rm M}=\sum_i\tilde{\lambda}_i\,|\tilde{\psi}_i\rangle\,
\langle\tilde{\psi}_i|,
 \e{D7}
where $\langle\tilde{\psi}_i|\tilde{\psi}_j\rangle=\delta_{ij}$,
$\tilde{\lambda}_i\ge0$, and $\sum_i\tilde{\lambda}_i=1$.

\subsection{Generalized uncertainty relation}
\label{AC1}

First, we prove the following useful inequality for arbitrary operators $O_1$ and $O_2$,
 \be
|\,\overline{O_1O_2}\,|^2\,\le\,\overline{O_1O_1^\dagger}\ \
\overline{O_2^\dagger O_2}.
 \e{D12}
When the averages here are taken over a pure state, Eq.~\rqn{D12} was
shown to be a direct consequence of the Cauchy-Schwarz inequality for Hermitian $O_1$ and $O_2$ in Ref.~\cite{sch30} and for general non-Hermitian $O_1$ and $O_2$ in Ref.~\cite{kof08}.
Here we extend Eq.~\rqn{D12} to the case of a general mixed state \rqn{D7}, by writing
 \bea
&|\,\overline{O_1O_2}\,|^2&=\ |{\rm Tr}\,(O_1O_2\,\rho_{\rm M})|^2
\:=\:\left|\sum_i\tilde{\lambda}_i\langle\tilde{\psi}_i|O_1O_2
|\tilde{\psi}_i\rangle\right|^2\
\le\ \left(\sum_i\tilde{\lambda}_i|\langle\tilde{\psi}_i|O_1O_2
|\tilde{\psi}_i\rangle|\right)^2\nonumber\\
&&\le\ \left[\sum_i\tilde{\lambda}_i(\langle\tilde{\psi}_i|O_1
O_1^\dagger|\tilde{\psi}_i\rangle\langle\tilde{\psi}_i|O_2^\dagger
O_2|\tilde{\psi}_i\rangle)^{1/2}\right]^2\
\le\ \sum_i\tilde{\lambda}_i\langle\tilde{\psi}_i|O_1
O_1^\dagger|\tilde{\psi}_i\rangle
\sum_j\tilde{\lambda}_j\langle\tilde{\psi}_j|O_2^\dagger
O_2|\tilde{\psi}_j\rangle\nonumber\\
&&=\:\overline{O_1O_1^\dagger}\ \ \overline{O_2^\dagger O_2},
 \ea{D13}
which proves Eq.~\rqn{D12} for the general case.
In Eq.~\rqn{D13} the second inequality follows from Eq.~\rqn{D12} for
a pure state, and the third inequality results from the
Cauchy-Schwarz inequality.

The inequality \rqn{D12} implies that
 \be
|\,\overline{R_cF_c^l}\,|^2\ \le\ \overline{R_c^2}\ \overline{F_c^{2l}}\quad\quad(l\ge1).
 \e{D8}
Equation \rqn{D8} with $l=1$ yields the generalized uncertainty relation for the variables $F$ and $R$
 \be
\Delta R\,\Delta F\ \ge\ |\,\overline{R_c\,F_c}\,|.
 \e{D9}
Since
 \be
|\,\overline{R_c\,F_c}\,|\ =\ |\,\overline{R_c\,F}\,|\ =\ |\,\overline{R\,F_c}\,|\ =\ |\,\overline{RF}\,-\bar{R}\bar{F}|,
 \e{C13}
the generalized uncertainty relation \rqn{D9} can be rewritten in different forms, e.g., in the form \rqn{C14}.
The generalized uncertainty relation \rqn{D9} [or \rqn{C14} or \rqn{261}] was derived by Schr\"{o}dinger \cite{sch30} for quantum systems in a pure state (see also Refs.~\cite{gar00,chi01}).
Here it is proved for the general case of an arbitrary (pure or mixed) state.

Combining Eqs.~\rqn{167} and \rqn{D8} results in the relation
 \be
|\,\overline{F_c^l\,R_c}\,|\:=\:|\,\overline{R_c\,F_c^l}\,|\:\lesssim\:\Delta R\,(\Delta F)^l \quad\quad(l\ge1).
 \e{D3}
Using the relation
 \be
|\,\overline{F^nR_c}\,|\:=\:|\,\overline{(F_c+\bar{F})^nR_c}\,|\:\le\:\sum_{l=0}^n\,{n\choose l}\,|\bar{F}^{n-l}\,\overline{F_c^lR_c}\,|
 \e{D11}
and Eq. \rqn{D3}, we obtain 
 \be
|\,\overline{F^n\,R_c}\,|\:=\:|\,\overline{R_c\,F^n}\,|\:\lesssim\:\Delta R\,\Delta F\,(\Delta F+|\bar{F}|)^{n-1}\quad\quad(n\ge1).
 \e{D10}

\subsection{Estimating the moments}
\label{AC2}

Let us now estimate the magnitude of the moments
$\overline{F^kR_cF^{n-k}}\ (0\le k\le n;\ n\ge1)$.

We begin with two important cases, where the calculations are simple.
First, let $F$ and $R$ be canonically conjugate variables.
Then, using the commutation relation $\hat{R}_c\hat{F}= \hat{F}\hat{R}_c\pm i$, we can move $\hat{R}_c$ to the last place in the product $\hat{F}^k\hat{R}_c\hat{F}^{n-k}$, so that
 \be
|\,\overline{F^k\,R_c\,F^{n-k}}\,|\ \alt\ |\,\overline{F^{n-1}}\,|
+|\,\overline{F^nR_c}\,|\ \lesssim\ \Delta R\,\Delta
F\,(|\bar{F}|+\Delta F)^{n-1}.
 \e{D14}
Here in the last inequality we used Eqs.~\rqn{D10}, \rqn{232}, and the Heisenberg uncertainty relation $\Delta R\,\Delta F\,\ge1/2$.
Thus, Eq.~\rqn{D14} yields the estimate
 \be
|\,\overline{F^k\,R_c\,F^{n-k}}\,|\ \alt\ \Delta R\,\Delta
F\,(|\bar{F}|+\Delta F)^{n-1}.
 \e{B16}
Second, we note that the estimate \rqn{B16} holds also for commuting $F$ and $R$.
Indeed, then
 \be
|\,\overline{F^k\,R_c\,F^{n-k}}\,|\ =\ |\,\overline{R_c\,F^{n}}\,|\ \lesssim\ \Delta R\,\Delta F\,(|\bar{F}|+\Delta F)^{n-1},
 \e{D15}
where the inequality results from Eq.~\rqn{D10}.

The general case, where $F$ and $R$ are not necessarily canonically conjugate or commuting, is more complicated, since then the estimate \rqn{B16} does not generally hold.
Now we use the equality $F=\bar{F}+F_c$ to write that
 \be
\overline{F^k\,R_c\,F^{n-k}}\:=\:\sum_{l=0}^k\sum_{m=0}^{n-k} {k\choose l}{n-k\choose m}\bar{F}^{n-l-m}\:\overline{F_c^l\,R_c\,F_c^m}.
 \e{D1}
This reduces the problem to the estimation of the moments $\overline{F_c^k\,R_c\,F_c^{n-k}}$ where $0\le k\le n$.

When $\Delta R\rightarrow0$, whereas $\Delta F$ is bounded (which is possible, in particular, for finite-dimensional Hilbert spaces), the moments $\overline{F_c^lR_c}$ and $\overline{R_cF_c^l}$ vanish [cf.\ Eq.~\rqn{D3}].
In contrast, the moments $\overline{F_c^lR_cF_c^m}$ with $l,m\ge1$
generally do not vanish in the limit $\Delta R\rightarrow0$.
To proceed further, we make the simplifying assumption
 \be
\max_{1\le l\le m-1} |\,\overline{F_c^l\,R_c\,F_c^{m-l}}\,|
\:\sim\:\tilde{R}\,(\Delta F)^{m},
 \e{D4}
where $\tilde{R}$ does not depend very significantly on $m$ and generally does not vanish in the limit $\Delta R\rightarrow0$.
With the help of Eqs.~\rqn{D3} and \rqn{D4}, the quantity \rqn{D1}
can be estimated by the relation 
 \be
|\,\overline{F^k\,R_c\,F^{n-k}}\,|\ \lesssim\ \Delta R\,\Delta F\,\bar{F}^{k'}(|\bar{F}|+\Delta F)^{n-k'-1}+\tilde{R}\,(\Delta F)^2(|\bar{F}|+\Delta F)^{n-2}\quad\quad(1\le k\le n-1),
 \e{D5}
where $k'=\min\{k,n-k\}$.

Equation \rqn{D5} is equivalent to two simpler inequalities, which are obtained in two possible cases.
 First, in a typical situation, when $\Delta R$ is not too
small, Eqs.~\rqn{D5} and \rqn{D10} yield for $0\le k\le n$
 \be
|\,\overline{F^k\,R_c\,F^{n-k}}\,|\:\lesssim\:\Delta R\,\Delta
F\,(|\bar{F}|+\Delta F)^{n-1}\quad(\Delta R\agt\tilde{R}),
 \e{70}
which coincides with the above estimate \rqn{B16}.
Second, when $\Delta R$ vanishes or is very small, the first term on
the right-hand side of Eq.~\rqn{D5} can be neglected, yielding
 \be
|\,\overline{F^k\,R_c\,F^{n-k}}\,|\:\lesssim\:\tilde{R}\,(\Delta
F)^2\,(|\bar{F}|+\Delta F)^{n-2}\quad(\Delta R\ll\tilde{R})
 \e{D6}
for $1\le k\le n-1$, whereas $|\,\overline{F^n\,R_c}\,|=
|\,\overline{R_c\,F^n}\,|$ are zero or negligibly small [cf.\
Eq.~\rqn{D10}].

In either case \rqn{70} or \rqn{D6}, the terms of orders higher
than two in Eq.~\rqn{18} can be neglected under the condition
\rqn{12'}, when, at least, one of the two following cases takes
place:\\
(a) The quadratic term in Eq.~\rqn{18} is not anomalously small,
i.e., $\overline{F\,R_c\,F}$ is of the order of the right-hand side of Eq.~\rqn{70} or \rqn{D6} with $n=2$.\\
(b) The first-order term in Eq.~\rqn{18} is not anomalously small,
i.e., it is of order $|\gamma A_{\phi\psi}|\,\Delta R\,\Delta F$.
Then, even when the quadratic term is vanishing or small, it can be
shown that under the condition \rqn{12'} the contribution of the
third- and higher-order terms into the pointer deflection \rqn{44} is
negligibly small.\\

When the first- and second-order terms in Eq.~\rqn{18} vanish or are
anomalously small or cancel each other, the validity condition
\rqn{12'} may be inapplicable.
However, generally such cases are of little interest, since then the
pointer deflection is very small.

\section{Calculation of moments for canonically conjugate variables}
\label{C}

Here we derive formulas for moments of canonically conjugate variables used in the main text.
Let $G(p)$ be an arbitrary function, such that the integrals below in
Eq. \rqn{C1} converge and that
 \be
\lim_{p\rightarrow\pm\infty}G(p)\,\psi_{\rm M}(p)=0.
 \e{C11}
Using the expression $q=i\partial/\partial p$ and
Eq.~\rqn{96}, we obtain that
 \bea
&\overline{G(p)\,q\,G(p)}&=(2\pi)^{-1}\int_{-\infty}^\infty
dp\,\psi_{\rm M}^*(p)\,G(p)i[G(p)\,\psi_{\rm M}(p)]'\nonumber\\
&&=(2\pi)^{-1}\int_{-\infty}^\infty dp\,f_p(p)\,G(p)\,\{i[G(p)\,f_p(p)]'+\zeta'(p)\,G(p)\,f_p(p)\}\nonumber\\
&&=i(4\pi)^{-1}[G(p)\,f_p(p)]^2|_{-\infty}^\infty+ (2\pi)^{-1}\int_{-\infty}^\infty dp\,f_p^2(p)\,G^2(p)\,\zeta'(p),
 \ea{C1}
where the prime denotes differentiation over $p$.
 The first term in the last expression in Eq.~\rqn{C1} vanishes in
view of Eq.~\rqn{C11}, and we obtain that
 \be
\overline{G(p)\,q\,G(p)}\:=\:\overline{\zeta'(p)G^2(p)}.
 \e{C2}
In particular, Eq.~\rqn{C2} implies the formulas \rqn{99} and
\rqn{C3}, whereas Eqs.~\rqn{C2} and \rqn{C3} imply the first two
lines in Eq.~\rqn{198}.

In a similar fashion, it is not difficult to obtain that
\bes{C4}\bea
&K(\chi_1,\chi_2)&\equiv\:\overline{\exp(i\chi_1p)\,q\,\exp(i\chi_2p)}\label{C4a}\\
&&=\:\overline{[(\chi_1-\chi_2)/2+\zeta'(p)]\,\exp[i(\chi_1+\chi_2)p]}.
 \ea{C4b}\ese
The characteristic function $K(\chi_1,\chi_2)$ provides mixed moments of $p$ and $q$, linear in $q$, by the formula
 \be
\overline{p^n\,q\,p^m}\:=\:\left.(-i)^{n+m}\,\frac{\partial^{n+m}
K(\chi_1,\chi_2)}{\partial\chi_1^n\,\partial\chi_2^m}
\right|_{\chi_1=\chi_2=0}.
 \e{C5}
Equation \rqn{C4a} implies that
 \be
\overline{\cos(\gamma p)\,q\,\sin(\gamma p)}\:=\:\frac{K(\gamma,\gamma)-K(\gamma,-\gamma)+K(-\gamma,\gamma) -K(-\gamma,-\gamma)}{4i}.
 \e{C10}
Inserting Eq.~\rqn{C4b} into Eq.~\rqn{C10} yields the third line of
Eq.~\rqn{198}.

It is easy to check that the expressions for averages of functions
of $p$ and $q$ derived above hold also under the simultaneous
replacements
 \be
q\ \leftrightarrow\ p,\quad\quad\zeta(p)\ \rightarrow\ \xi(q).
 \e{C6}
In particular, Eqs.~\rqn{C2}-\rqn{C5} yield respectively
 \be
\overline{G(q)\,p\,G(q)}\:=\:\overline{\xi'(q)G^2(q)},
 \e{C7}
 \be
\tilde{K}(\chi_1,\chi_2)\:\equiv\:\overline{\exp(i\chi_1q)\,p\,\exp(i\chi_2q)}
\:=\:\overline{[(\chi_1-\chi_2)/2+\xi'(q)]\,\exp[i(\chi_1+\chi_2)q]},
 \e{C8}
 \be
\overline{q^n\,p\,q^m}\:=\:\left.(-i)^{n+m}\,\frac{\partial^{n+m}
\tilde{K}(\chi_1,\chi_2)}{\partial\chi_1^n\,\partial\chi_2^m}
\right|_{\chi_1=\chi_2=0}.
 \e{C9}
Equation \rqn{C7} implies Eq.~\rqn{C12}.
Taking into account that $\overline{\{q,p\}}= 2{\rm
Re}\,\overline{qp}$, Eq.~\rqn{C5} calculated for $n=0,m=1$ and
Eq.~\rqn{C9} with $n=1,m=0$ yield Eq.~\rqn{36}.

\section{Operator identities}
\label{E}

Here we prove several operator identities used in the present paper.
Let $O_{\rm S}$ and $O_{\rm S}'$ ($O_{\rm M}$ and $O_{\rm M}'$) be arbitrary operators in the Hilbert space ${\cal H}_{\rm S}$ (${\cal H}_{\rm M}$) of system $S$ ($M$), whereas $O$ and $O'$ are operators in the Hilbert space ${\cal H}_{\rm S}\otimes{\cal H}_{\rm M}$.
If $O$ and $O'$ can be written as the sums
 \be
O=\sum_i O_{{\rm S}i}\otimes O_{{\rm M}i},\quad 
O'=\sum_j O_{{\rm S}j}'\otimes O_{{\rm M}j}',
 \e{B11}
where $O_{{\rm S}i}$, $O_{{\rm S}j}'$, and {\em either $O_{\rm S}$ or} $O_{\rm S}'$ commute pairwise for all $i$ and $j$, then the following identities hold
 \bea
&{\rm Tr}\,[(O_{\rm S}\otimes O_{\rm M})O(O_{\rm S}'\otimes O_{\rm
M}')O']&=
{\rm Tr}\,[(I_{\rm S}\otimes O_{\rm M})O(O_{\rm S}O_{\rm S}'\otimes O_{\rm M}')O']\nonumber\\
&&={\rm Tr}\,[(I_{\rm S}\otimes O_{\rm M})O(O_{\rm S}'O_{\rm S}\otimes O_{\rm M}')O'],\nonumber\\
&&=\frac{1}{2}{\rm Tr}\,[(I_{\rm S}\otimes O_{\rm M})O((O_{\rm S}O_{\rm S}'+O_{\rm S}'O_{\rm S})\otimes O_{\rm M}')O'].
 \ea{B9}
Let us consider the above two cases separately.
First, when $O_{\rm S}$ commutes with $O_{{\rm S}i}$, it is easy to see that $O_{\rm S}\otimes I_{\rm M}$ commutes with $O$; in this case the first equality in Eq.~\rqn{B9} is obvious.
Second, let us prove the first equality in Eq.~\rqn{B9} for the case when $O_{\rm S}'$, $O_{{\rm S}i}$, and $O_{{\rm S}j}'$ commute pairwise.
The left-hand side of Eq.~\rqn{B9} can be recast as
 \bea
&&{\rm Tr}\,[(O_{\rm S}\otimes O_{\rm M})O(O_{\rm S}'\otimes O_{\rm
M}')O']=
\sum_{i,j}{\rm Tr}\,[(O_{\rm S}O_{{\rm S}i}O_{\rm S}'O_{{\rm S}j}')\otimes(O_{\rm M}O_{{\rm M}i}O_{\rm M}'O_{{\rm M}i'}')]\nonumber\\
&&=\sum_{i,i'}{\rm Tr}\,(O_{\rm S}O_{{\rm S}i}O_{\rm S}'O_{{\rm S}j}'){\rm Tr}\,(O_{\rm M}O_{{\rm M}i}O_{\rm M}'O_{{\rm M}i'}')
=\sum_{i,i'}{\rm Tr}\,(O_{{\rm S}i}O_{\rm S}O_{\rm S}'O_{{\rm S}j}'){\rm Tr}\,(O_{\rm M}O_{{\rm M}i}O_{\rm M}'O_{{\rm M}i'}').
 \ea{B12}
Here in the last equality we used the fact that $O_{{\rm S}i}$ commutes with $O_{\rm S}'$ and $O_{{\rm S}j}'$.
The substitutions $O_{\rm S}\rightarrow I_{\rm S},\ O_{\rm S}'\rightarrow O_{\rm S}O_{\rm S}'$ change the left-hand side of Eq.~\rqn{B12} into the right-hand side of the first equality in Eq.~\rqn{B9} but do not change the right-hand side of Eq.~\rqn{B12}, which proves the first equality in Eq.~\rqn{B9}.

The second equality in Eq.~\rqn{B9} follows from the fact that the above validity conditions for Eq.~\rqn{B9} imply that either $O_{\rm S}\otimes I_{\rm M}$ or $O_{\rm S}'\otimes I_{\rm M}$ commutes with $O$ and $O'$.
For example, when $O_{\rm S}'\otimes I_{\rm M}$ commutes with $O$ and $O'$, we obtain that
 \bea
&&{\rm Tr}\,[(I_{\rm S}\otimes O_{\rm M})O(O_{\rm S}O_{\rm S}'\otimes O_{\rm M}')O']
={\rm Tr}\,[(I_{\rm S}\otimes O_{\rm M})O(O_{\rm S}\otimes O_{\rm M}')O'(O_{\rm S}'\otimes I_{\rm M})]\nonumber\\
&&={\rm Tr}\,[(O_{\rm S}'\otimes I_{\rm M})(I_{\rm S}\otimes O_{\rm M})O(O_{\rm S}\otimes O_{\rm M}')O']
={\rm Tr}\,[(I_{\rm S}\otimes O_{\rm M})O(O_{\rm S}'O_{\rm S}\otimes O_{\rm M}')O'].
 \ea{B10}
A similar argument holds also when $O_{\rm S}\otimes I_{\rm M}$ commutes with $O$ and $O'$.
Finally, the third equality in Eq.~\rqn{B9} follows from the previous equality.

As an example, consider the case when $O=U$, $O'=U^\dagger$, and $\hat{A}$ commutes with either $O_{\rm S}$ or $O_{\rm S}'$. 
Here $U$ is given by Eq.~\rqn{2}.
In this case, the sums of the form \rqn{B11} are obtained by expanding $U$ and $U^\dagger$ in powers of $\gamma$, yielding
 \be
O_{{\rm M}j}'=O_{{\rm M}j}=\hat{A}^j\quad\quad(j\ge0).
 \e{B15}
Thus, the validity conditions for Eq.~\rqn{B9} hold now.
Consequently, when $\hat{A}$ commutes with either $O_{\rm S}$ or $O_{\rm S}'$, we obtain the identities
 \bea
&&{\rm Tr}\,[(O_{\rm S}\otimes O_{\rm M})U(O_{\rm S}'\otimes O_{\rm
M}')U^\dagger]=
{\rm Tr}\,[(I_{\rm S}\otimes O_{\rm M})U(O_{\rm S}O_{\rm S}'\otimes O_{\rm M}')U^\dagger],\label{B7'}\\
&&{\rm Tr}\,[(O_{\rm S}\otimes O_{\rm M})U(O_{\rm S}'\otimes O_{\rm
M}')U^\dagger]=
{\rm Tr}\,[(I_{\rm S}\otimes O_{\rm M})U(O_{\rm S}'O_{\rm S}\otimes O_{\rm M}')U^\dagger],\label{B7''}\\
&&{\rm Tr}\,[(O_{\rm S}\otimes O_{\rm M})U(O_{\rm S}'\otimes O_{\rm
M}')U^\dagger]=\frac{1}{2}{\rm Tr}\,[(I_{\rm S}\otimes O_{\rm M})U((O_{\rm S}O_{\rm S}'+O_{\rm S}'O_{\rm S})\otimes O_{\rm M}')U^\dagger].
 \ea{B7'''}

\end{appendices}

\phantomsection
\addcontentsline{toc}{section}{References}
\bibliographystyle{elsarticle-num}
\bibliography{weakval}

\end{document}